%% file: main.tex
\documentclass[12pt]{book}
\usepackage{epigraph}
\usepackage[english]{babel} 
\usepackage{lipsum} 
\usepackage{url} 
\usepackage[dvipsnames]{xcolor}
\usepackage{authblk}
\usepackage{tcolorbox}
\usepackage[load-configurations = abbreviations]{siunitx}
\usepackage{float}
\usepackage{mathtools}
\usepackage{stackengine}

\usepackage{booktabs} 
\usepackage{fancyhdr} 
\definecolor{Coolorange}{rgb}{0.9608, 0.5020, 0.1451}
\usepackage{titlesec}

\titleformat{\chapter}[display]
 {\bfseries}{}{0pt}{\Huge}

\usepackage{hyperref} 
\usepackage{amsmath} 
\usepackage[toc,page]{appendix}
\usepackage{bbold}
\usepackage{blindtext}
\usepackage{caption}
\usepackage{subcaption}
\usepackage{amsthm} 
\usepackage{mathtools} 
\usepackage{amsthm} 
\usepackage{bm} 
\usepackage{braket} 
\usepackage{tikz} 
\usetikzlibrary{positioning,arrows.meta,shapes.geometric,decorations.pathmorphing,decorations.markings,bending,calc,shadings} 
\usepackage{tikz-3dplot}
\usetikzlibrary{patterns}
\usepackage{tikz-feynman}
\usepackage{asypictureB}
\usepackage{amssymb} 
\usepackage{amsfonts} 
\usepackage{mathrsfs}
\newcommand{\R}{\mathbb{R}}

\newcommand{\Z}{\mathbb{Z}}
\newcommand{\T}{\mathbb{T}}
\newcommand{\de}{\partial}
\newcommand{\diff}{\mathrm{d}}

\newcommand{\ds}{\diff s}

\newcommand{\normord}[1]{{:}\!\mathrel{#1}\!{:}}

\DeclareMathOperator{\diag}{diag}

\newtheorem*{distance}{Swampland Distance Conjecture}
\newtheorem*{dS}{Swampland de Sitter Conjecture}
\newtheorem*{AdS}{AdS Distance Conjecture}
\newcommand{\be}{
\begin{equation}
}
\newcommand\quotient[2]{
        \mathchoice
            {
                \text{\raise1ex\hbox{$#1$}\Big/\lower1ex\hbox{$#2$}}%
            }
            {
                #1\,/\,#2
            }
            {
                #1\,/\,#2
            }
            {
                #1\,/\,#2
            }
    }
\newcommand{\beLabel}[1]{
\begin{equation}
\label{eq:#1}
}

\newcommand{\ee}{
\end{equation}
}

\newcommand{\eeLabel}{
\end{equation}
}

\newcommand{\bes}{
\begin{equation}
\begin{split}
}

\newcommand{\ees}{
\end{split}
\end{equation}
}

\usepackage{physics}
\newtheorem*{ricci}{Ricci Flow Conjecture}
\newtheorem*{dilmet}{Dilaton-Metric Flow Conjecture}

\let\oldref\ref
\renewcommand{\ref}[1]{(\oldref{#1})}

\title{\textbf{
\vskip-3cm
Though Tears No Longer Flow...}}
\date{}
\author[a]{Davide De Biasio}

\usepackage{cite}
\numberwithin{equation}{section}

\usepackage{a4wide}
\usepackage{fancyhdr}

\usepackage{afterpage}
\usepackage{asymptote}
\usepackage{graphicx}

\renewcommand{\chaptermark}[1]%
         {\markboth{\thechapter.\ #1}{}}
\renewcommand{\sectionmark}[1]%
         {\markright{\thesection\ #1}}
\newcommand{\LMUTitle}[9]{
  \thispagestyle{empty}
  \vspace*{\stretch{1}}
  {\parindent0cm
   \rule{\linewidth}{.7ex}}
  \begin{flushright}

    \vspace*{\stretch{1}}
    \sffamily\bfseries\Huge
    #1\\
    \vspace*{\stretch{1}}
    \sffamily\bfseries\large
    #2
    \vspace*{\stretch{1}}
  \end{flushright}
  \rule{\linewidth}{.7ex}
  \vspace*{\stretch{5}}
  \begin{center}
    \includegraphics[width=2in]{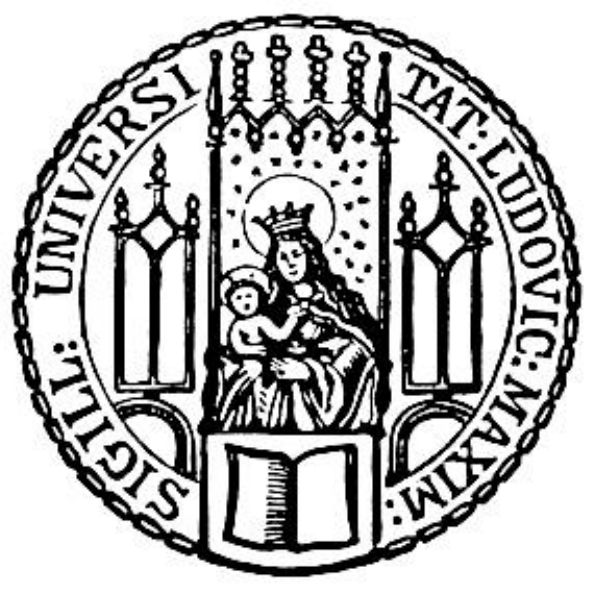}
  \end{center}
  \vspace*{\stretch{1}}
  \begin{center}\sffamily\LARGE{#5}\end{center}
  \newpage
  \thispagestyle{empty}

  \cleardoublepage
  \thispagestyle{empty}

  \vspace*{\stretch{1}}
  {\parindent0cm
  \rule{\linewidth}{.7ex}}
  \begin{flushright}
    \vspace*{\stretch{1}}
    \sffamily\bfseries\Huge
    #1\\
    \vspace*{\stretch{1}}
    \sffamily\bfseries\large
    #2
    \vspace*{\stretch{1}}
  \end{flushright}
  \rule{\linewidth}{.7ex}

  \vspace*{\stretch{3}}
  \begin{center}
    \Large Dissertation\\
    \Large an der #4\\
    \Large der Ludwig--Maximilians--Universit\"at\\
    \Large M\"unchen\\
    \vspace*{\stretch{1}}
    \Large vorgelegt von\\
    \Large #2\\
    \Large aus #3\\
    \vspace*{\stretch{2}}
    \Large M\"unchen, den #6
  \end{center}

  \newpage
  \thispagestyle{empty}

  \vspace*{\stretch{1}}

  \begin{flushleft}
    \large Erstgutachter:  #7 \\[1mm]
    \large Zweitgutachter: #8 \\[1mm]
    \large Tag der m\"undlichen Pr\"ufung: #9\\
  \end{flushleft}

  \cleardoublepage
}
\usepackage{pgfplots}
\pgfplotsset{compat=1.17}

\begin{document}

	\input{frontmatter/personalize}
	\maketitle
	\copyrightpage
	\abstractpage
	\tableofcontents
	
	\newpage \thispagestyle{fancy} \vspace*{\fill}
	\scshape \noindent \input{frontmatter/dedication}
	\vspace*{\fill} \newpage \rm

	\chapter*{Acknowledgments}
	\noindent
	\input{frontmatter/thanks}
	\vspace*{\fill} \newpage
	\setcounter{page}{1}
	\pagenumbering{arabic}

  \LMUTitle
      {Geometric flows and the Swampland}               
      {Davide De Biasio}                       
      {Palmanova, Italien}                             
      {Fakult\"at f\"ur Physik}                         
      {M\"unchen 2023}                          
      {30.05.2023}                            
      {Dieter L\"ust}                          
      {Michael Haack}                         
      {13.07.2023}                         

\chapter*{Abstract}
\input{Sections/AandA/abstract.tex}
\chapter*{Zusammenfassung}
\input{Sections/AandA/Zusammenfassung}
\chapter*{Literature}
\input{Sections/AandA/papers.tex}
\chapter*{Acknowledgements}
\input{Sections/AandA/thanks.tex}
\tableofcontents
\listoffigures
\mainmatter\setcounter{page}{1}
\chapter{Introduction}

\input{Sections/Introduction/introduction.tex}
\part{Preliminaries}
\chapter{Superstring Theory}\label{stringchap}
\input{Sections/Strings/stringsintro.tex}
\section{The classical superstring}
\input{Sections/Strings/superstrings.tex}

\section{The quantum superstring}
\input{Sections/Strings/quantum.tex}
\section{Low energy effective theories}\label{stringefts}
\input{Sections/Strings/EFT.tex}
\section{Compactification}\label{seccompact}
\input{Sections/Strings/compactification.tex}


\chapter{The Swampland Program}\label{swampchap}
\input{Sections/ASwamp/swampintro}\label{swamplandintro}
\section{Constraints on effective theories}
\input{Sections/ASwamp/eftsconstraints.tex}

\section{Swampland conjectures}
\input{Sections/ASwamp/swamplandconjectures.tex}


\chapter{Geometric Flows}\label{chap:geometricflows}
\input{Sections/GeometricFlows/intro.tex}
\section{AdS distance conjecture}\label{sec:adscon}
\input{Sections/GeometricFlows/ads.tex}
\section{The moduli space of metrics}
\input{Sections/GeometricFlows/moduli.tex}
\section{Ricci flow conjecture}\label{sec:ric}
\input{Sections/GeometricFlows/ricci.tex}
\section{Perelman's combined flow}\label{sec:perel}
\input{Sections/GeometricFlows/perelman.tex}

\part{Applications}\label{novel}

\chapter{Geometric flow of bubbles}\label{chap:bubblesflow}
\input{Sections/Bubbles/intro}
\section{Geometric flow of scalar bubbles}
\input{Sections/Bubbles/bubblesflow.tex}

\chapter{On-Shell flow}\label{chap:onshell}
\input{Sections/On-Shellflow/intro.tex}
\section{General flow equations}
\input{Sections/On-Shellflow/generalflow.tex}
\section{On-shell conditions}
\input{Sections/On-Shellflow/onshell}
\section{Scalar field example}\label{scaler}
\input{Sections/On-Shellflow/scalars.tex}


\part{Conclusions}
\chapter{Conclusive summary}
\input{Sections/Conclusions/conclusions.tex}
\section{Outlook}
\input{Sections/Conclusions/outlook.tex}
\backmatter
\bibliographystyle{utphys}
\bibliography{bibliogra.bib}

\end{document}

%% file: Theme/personalize.tex
\title{Title of the dissertation}
\author{Firstname M. Lastname}
\advisor{Bigname Scientist}

\degree{Doctor of Philosophy}
\field{Psychology}
\degreeyear{2024}
\degreemonth{May}
\department{Psychology}

\pdOneName{B.S.}
\pdOneSchool{Boston University}
\pdOneYear{2018}

\pdTwoName{M.A.}
\pdTwoSchool{Monster's Univeristy}
\pdTwoYear{2021}

%% file: Theme/dedication.tex

This is the dedication.

%% file: Sections/AandA/thanks.tex
If there is one thing harder than quantising the gravitational field and unifying it with the other fundamental forces of Nature, it is writing an \textit{acknowledgements} section. The process of selecting whom to thank after a journey, in particular when such journey is as long and tortuous as a doctorate can be, is somehow comparable to carving a statue out of marble. You find yourself in front a huge, tough block of events, memories and emotions, most of which must be removed in order to obtain a comprehensible shape. And every time the scalpel hits, a thick shower of fragments reminds you of all the precious material you are wasting. This strikes me as an unforgivable sin. Nevertheless, a choice ought to be made. Among all the people that encouraged and supported me, only a narrow minority will make it into the \textit{acknowledgements landscape}. The others, without whom this thesis could have never been completed, will be left in the muddy waters of the \textit{acknowledgements swampland}. I offer my most sincere apologies to all of them. That being said, I would like to thank Dieter L\"ust for having always guided me with patience, wisdom and kindness. Every single time I burst into his office with some bizarre scientific idea, he found the time for listening to me, making sense of my sketchy impressions and pointing me in the right direction. I am not sure what would have become of me, had I not been granted with the exceptional chance of joining his \textit{string theory} group at the \textit{Max Planck Institute for Physics}. There, I met some of the most impressive and fascinating people I will ever have the honour of discussing with. I thank Ralph Blumenhagen, Sven Krippendorf, Andriana, Andreas, Niccolò, Alessandra, Marco, Aleksandar, Christian, Thomas, Yixuan, Joaquin, Pouria, Saskia, Antonia, Carmine, Chuying and Ivano, for having spent some of their days with me discussing physics, standing in front of the coffee machine and complaining about the unpredictable Munich weather. I thank the International Max Planck Research School (IMPRS) of the Max Planck Society and Frank Steffen. Without him, none of this would have taken place. I thank Robert Helling and Markus Dierigl, for having given me a chance to explore how stimulating teaching during a global pandemic can be. I thank Julian, for having shared my fights, and Toby Wiseman, for having held my hand through uncharted and marvellous territories. I thank Michael Haack, for having been there for me from the start. Together with him, I deeply thank all the other members of my thesis committee, for having gifted me with some of their precious time. I thank David, for having been the best master student I could have ever asked for. I thank Matthias, Giacomo, Giulia and Sara, for always having an inestimable chunk of home in store for me. I thank Barbara, for having put her trust in my fragile words, Annette and Vera, without whom I would have drowned in the quicksand of bureaucracy. I thank Sébastien, for having fuelled my tendency to pursue philosophical matters my brain cannot fully grasp, and Daniele, who accepted -without apparent hesitation- all my outlandish proposals, while allowing me to embark on some of his projects. I thank my parents, with whom I share way more than my physical appearance, and Matte, the other (and more clever) half of my soul. I thank Alice, the sort-of-little-sister I feel lucky to have. I thank Alex, Andrea, Fede and Enri, my soon-to-be neighbour. I thank Kevin, with whom I shared most of my life, and my tribe: Tommy, Spi, Della, Flo, Ali, Mary, Freddy, Shani, Bea, G, Sofia, Boucha, Ele, Jo, Giamma, Pippo, Miky, Lara, Eli, Kev and France. One could never understand me, without first understanding you. I thank our eternal lighthouse: Rob. We'll have a lot to discuss, the day we meet again. For the time being, I guess we could say that \textit{this was a triumph}. In conclusion, I want to thank Carla. Nobody has done nearly as much as her, in keeping me together through the toughest days of my life. I wish you all, my fellow humans, to find your Carla. She truly is a porcelain supernova.
\textbf{ }\\
\textbf{ }\\
\textbf{ }\\
\textbf{ }\\
\textbf{ }\\
\textbf{ }\\
\textbf{ }\\
\textbf{ }\\
\textbf{ }\\
\textbf{ }\\
\begin{center}
\begin{figure}[h]
\centering
  \includegraphics[width=0.15\linewidth]{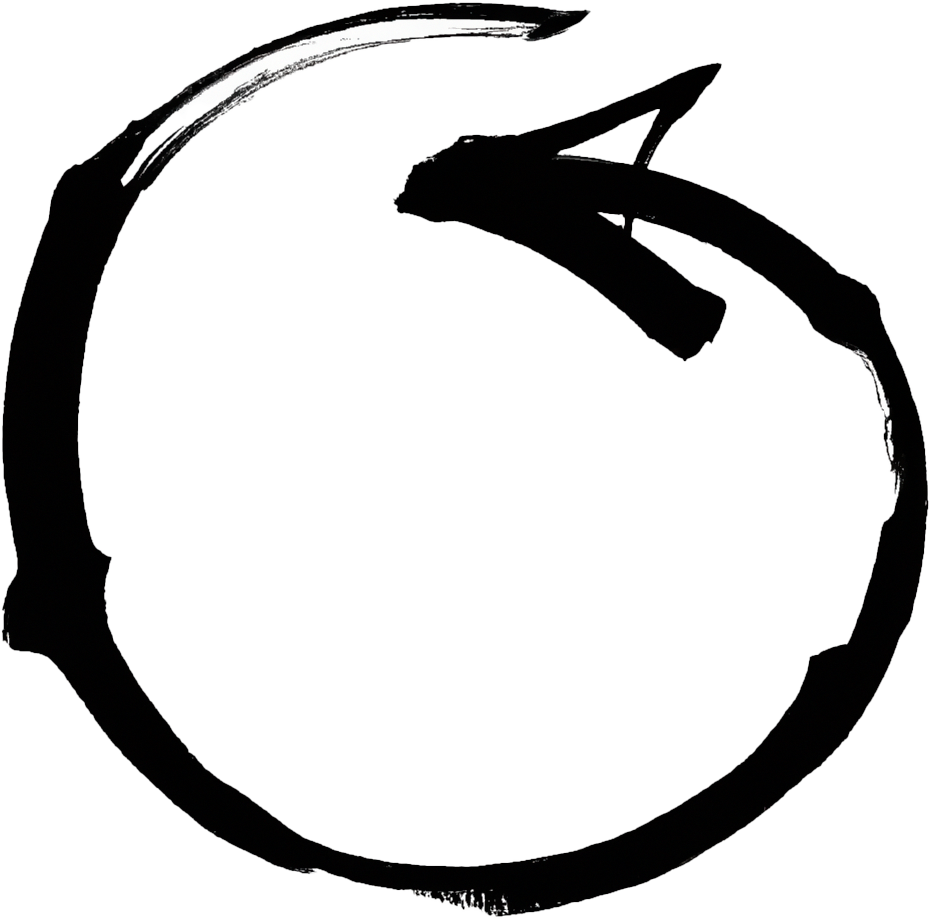}
\label{}
\end{figure}
\end{center}
\chapter*{}
\epigraph{\textit{If the current flow is taking you
where you want to go, don’t argue. }\\\textbf{ }}{--- Isaac Asimov, \textit{Fantastic Voyage II}}
\epigraph{\textit{A process cannot be understood by stopping it. Understanding must move with the flow of the process, must join it and flow with it.}\\\textit{  }}{--- Frank Herbert, \textit{Dune}}
\epigraph{\textit{The monkey replies only to past or present things, which is as far as the devil’s knowledge can go; future things cannot be known except through conjecture.}\\\textit{  }}{--- Miguel de Cervantes, \textit{Don Quijote}}

%% file: Sections/AandA/abstract.tex
After an introductory chapter on the quantum supersymmetric string, in which particular attention will be devoted to the techniques via which phenomenologically viable models can be obtained from the ultraviolet microscopic degrees of freedom, and a brief review of the swampland program, the technical tools required to deal with geometric flows will be outlined. The evolution of a broad family of scalar and metric bubble solutions under Perelman's combined flow will be then discussed, together with their asymptotic behaviour. Thereafter, the geometric flow equations associated to a generalised version of Perelman's entropy function will be derived and employed in defining the action-induced flow associated to a given theory for a scalar field and a dynamical metric. The problem of preserving Einstein field equations along the corresponding moduli space trajectories will be cured by allowing a supplementary energy-momentum tensor term to appear along the flow. In a particular example, such contribution will be shown to precisely reproduce the infinite tower of states with exponentially dropping masses postulated by the distance conjecture.

%% file: Sections/AandA/Zusammenfassung.tex
Nach einer Einführung in den Superstring, in der besonders auf die Methoden eingegangen wird, mit welchen man aus mikroskopischen Freiheitsgraden im ultravioletten Bereich phänomenologisch brauchbare Modelle erhalten kann und einem kurzen Überblick über das Swampland-Programm werden die mathematischen Methoden vorgestellt, die für die Beschreibung von geometrischem Fluss notwendig sind. Danach wird die Entwicklung einer breitgefächerten Familie von skalaren und metrischen Blasenlösungen unter Perelmans kombiniertem Fluss, zusammen mit deren asymptotischen Verhalten diskutiert. Anschließend werden die geometrischen Flussgleichungen, die im Zusammenhang mit einer verallgemeinerten Version der Perelman-Entropiefunktion stehen, hergeleitet und zur Definition des von der Wirkung induzierten Flusses verwendet. Dieser kann mit einer bestimmten Theorie für ein skalares Feld und eine dynamische Metrik in Verbindung gebracht werden. Es wird ein zusätzlicher Energie-Impuls-Tensor eingeführt, so dass während des geometrischen Flusses die Einstein’schen Feldgleichungen entlang der entsprechenden Trajektorie im Modulraum unverändert bleiben. In einem speziellen Beispiel wird gezeigt, dass ein solcher Beitrag einen Turm aus unendlich vielen Zuständen mit exponentiell abfallenden Massen, wie er von der Abstandsvermutung postuliert wird, exakt reproduziert.

%% file: Sections/AandA/papers.tex
This thesis is primarily based on the following works:
\begin{itemize}
    \item Davide De Biasio and Dieter L\"ust, “Geometric flow of bubbles,”\\ \textit{Nucl. Phys. B} \textbf{980} (2022)
115812, arXiv:2201.01679 [hep-th]
    \item Davide De Biasio, “On-Shell Flow,”\\ arXiv:2211.04231 [hep-th].
\end{itemize}
While working under the supervision of Professor Dieter L\"ust, I also had the opportunity to contribute to the publications:
\begin{itemize}
    \item Davide De Biasio and Dieter L\"ust, “Geometric Flow Equations for Schwarzschild-AdS
Space-Time and Hawking-Page Phase Transition,”\\ \textit{Fortsch. Phys.} \textbf{68} no. 8, (2020)
2000053, arXiv:2006.03076 [hep-th].
    \item Davide De Biasio, Julian Freigang, and Dieter L\"ust, “Geometric flow equations for the number
of space-time dimensions,”\\ \textit{Fortschritte der Physik} \textbf{70} no. 1, (2022) 2100171.
    \item Davide De Biasio, Julian Freigang, Dieter L\"ust, and Toby Wiseman, “Gradient flow of
Einstein-Maxwell theory and Reissner-Nordstr\"om black holes,”\\ \textit{JHEP} \textbf{03} (2023)
074, arXiv:2210.14705 [hep-th].
    \item David Martín Velázquez, D. De Biasio, and Dieter L\"ust, “Cobordism, singularities and the
Ricci flow conjecture,”\\ \textit{JHEP} \textbf{01} (2023) 126, arXiv:2209.10297 [hep-th].
\end{itemize}
The last paper in the above list, in particular, developed as a result of David Martín Velázquez's master thesis project, which I had the pleasure and honour to co-supervise. 

%% file: Sections/Introduction/introduction.tex
Had we a chance to explore our universe from scratch, without having formerly been exposed to the intricate apparatus of natural sciences, we would surely deem it to be arranged as a layered hierarchy of scales. If we happened to do so, moreover, at the specific spatio-temporal location at which this introduction is being written, we would rapidly conjecture said scales to be associated with levels of increasing complexity. The world would unveil itself as a nested architecture of structures within structures, themselves encysted inside wider, composite structures and so forth. It would be evident how the entities characterising a layer interact and combine, forming those pertaining to the subsequent one. At the same time, a careful enquiry would allow to disassemble them into their microscopic components, displaying a higher degree of simplicity. Regardless of how tortuous the endeavour might be, we would eventually figure out a collection of conceptual frameworks, each pertaining to a particular level and suitably describing its distinctive phenomena, and craft them appropriate names, such as \textit{particle physics}, \textit{chemistry}, \textit{biology}, \textit{anthropology}, \textit{economics} and \textit{cosmology}. Up to various extents, with an amount of rigour inversely related to the intrinsic complexity of a given layer, we could also manage to phrase them in terms of precisely defined, quantitative and unambiguous mathematical objects, suggested by observations and put together in a sequence of experimentally supported scientific theories. At that point, we could not avoid being struck by a sudden, unexpected and almost miraculous realisation: the levels in which the universe is organised are substantially -while not completely- independent from one another. This recognition would, in retrospect, shed light on the reason for which the previously mentioned disciplines could be studied separately, without evoking entities belonging to more fundamental descriptions of reality. Albeit rising from an enormous number of transactions between economic agents, themselves ultimately made of aggregated excitations of relativistic quantum fields, stock markets do not require to be understood with reference to path integrals, gluons and topologically protected superselection sectors. On the contrary, they are more naturally modelled in terms of stochastic price variables, drift rates and macroeconomic factors. Analogously, the spread of misinformation in enclosed communities of connected individuals, which can be successfully represented by epistemic networks, draws little benefit from a detailed account of the state of each and every neuron in their brains, let alone those of the elementary particles from which they are assembled. It appears to be a general feature of Nature that most of the subtleties of physical laws at a certain scale decrease in relevance when progressively longer distances are considered. Rather than a useful theoretical assumption, this is as much an empirical fact as the asymptotic value of the fine-structure constant, the almost complete inertness of noble gases or the equivalence between inertial and gravitational mass. Both when our gaze is pushed towards astrophysical events and all the way down to the micrometric resolutions of modern particle accelerators, the dynamical details of theories referring to separate scales seem to be, to a great extent, \textit{decoupled} from one another. Furthermore, populous clusters of interacting objects often display, when analysed from the point of view of a large scale observer, novel and emergent behaviours, which could have hardly been predicted from a naive extrapolation of the properties of their components. The \textit{reductionist hypothesis}, which assumes such a decomposition into more fundamental entities to fully exhaust the qualities of compound ones, has been a remarkably successful driving force for the scientific enterprise. Even so, it does not directly imply what Philip Anderson would have referred to as the \textit{constructionist hypothesis} \cite{Anderson1972MoreID}, claiming that all empirical data could be reconstructed from fundamental laws by following some sort of Leibnitzian principle of \textit{sufficient reason} \cite{sep-sufficient-reason}. In order to grasp the complexity of the universe, environmental, contingent and history-dependent factors, as the initial configuration from which a many-body system evolved, the outcome of a collection of quantum measurements or the choice of a specific symmetry-breaking pattern in particle physics, cannot be neglected. There is a plethora of distinct possible phenomenologies, equally compatible with our most profound physical theories. We might as well say, exploiting Anderson's renowned formula, that \textit{more is different}. The debate between emergentists \cite{EMMill1874-MILASO,EMbroad2002mechanism,EMMcLaughlin1992-MCLTRA} and reductionists \cite{REDNagel1961-NAGTSO-10,REDOppenheim1958-OPPUOS,REDWimsatt1972-WIMREA} around the nature of macroscopic properties, together with its contemporary developments \cite{NEWbrown2009physicalism,NEWButterfield2011-BUTLID,NEWkim1999making,NEWkim2002layered,NEWRueger2006-RUEFRA,NEWwimsatt1994ontology}, is a long-standing and elusive one, whose intricacy could not be exhausted by the current discussion. As far as the above-mentioned issues are concerned, it is nonetheless paramount to emphasise how profitably they can be captured and dealt with by exploiting \textit{effective field theories} \cite{EFTGeorgi:1993mps,EFTBurgess2007IntroductionTE,EFTPenco2020AnIT}. Such techniques grant us, first and foremost, with an appropriate mathematical framework for describing large scale limits of fundamental theories, along with the technical tools to assess how distinct scales decouple in a huge variety of settings. This striking aspect of natural phenomena is, therefore, perfectly reflected in the formulas. In essence, taking a given quantum field theory as describing short-range physics, its effective dynamics below an energy cut-off $\Lambda_{\text{EFT}}$ can be obtained by integrating out from the path integral all such degrees of freedom which require an energy $E\gg \Lambda_{\text{EFT}}$ to be excited. This procedure, which can be made systematic by employing \textit{renormalization group} techniques, allows to absorb the microscopic details of the Lagrangian describing the dynamics of a theory in a family of \textit{Wilson coefficients} $c_n$ and higher-order operators $\mathcal{O}_n$. Let's consider, for instance, the high-energy Lagrangian $\mathcal{L}\left(\Bar{\varphi},\Bar{\alpha}\right)$ for a family of fields $\left(\Bar{\varphi},\Bar{\alpha}\right)\equiv\left(\varphi_1,\dots,\varphi_N,\alpha_1,\dots,\alpha_M\right)$, where the $\Bar{\alpha}$ are significantly heavier and can be integrated out above an energy threshold $\Lambda_{\text{EFT}}$. Hence, the effective Lagrangian for the light fields $\Bar{\varphi}$ can be generally written as the sum of a renormalizable part and an infinite family of higher-order operators, constrained by an appropriate set of symmetries:
\[
    \mathcal{L}_{\text{EFT}}\left(\Bar{\varphi}\right)=\mathcal{L}_{\text{ren}}\left(\Bar{\varphi}\right)+\sum_{n}c_n\frac{\mathcal{O}_n\left(\Bar{\varphi}\right)}{\Lambda^n_{\text{EFT}}}\ .
\]
Decoupling is precisely achieved due to the fact that said operators are suppressed by powers of the cut-off scale $\Lambda_{\text{EFT}}$. In its contemporary understanding, the standard model of particle physics itself, which is one of the most accurate and predictive theories in the history of science, is often interpreted as the low energy effective limit of some more fundamental description \cite{SM:Brivio:2017vri,SM:Weinberg:2021exr,SM:Isidori:2023pyp}. This has proven itself to be an enormously successful approach in countless situations, from condensed matter systems \cite{CM:Brauner:2022rvf,CM:Kaneyoshi1990ComparisonOE,CM:shankar1999effective} to cosmological models of the early universe \cite{CO:Cabass:2022avo,CO:Cheung:2007st,CO:Davidson2020EffectiveFT}, from the analysis of quantum chromodynamics via chiral perturbation theory \cite{CHI:Bernard:1992mk,CHI:Machleidt:2011zz,CHI:Scherer:2002tk,CHI:Scherer2011APF} to that of the infrared dynamics of non-dissipative fluids \cite{HYDRO:Dubovsky:2011sj}. The idea of obtaining effective low energy descriptions of microscopic theories by integrating out all degrees of freedom lying above a suitably chosen cut-off, practically decoupling from one another most of the dynamical features of distinct length scales, has gained a central role in theoretical research, as it offers both the technical tools to construct functioning theories and the philosophical perspective within which they can be interpreted. It is part of the dominant \textit{paradigm} of contemporary physics \cite{kuhn2012structure}. Before moving on, it might be beneficial to appreciate how Albert Michelson, who would have later been awarded with the 1907 \textit{Nobel Prize in Physics} \cite{Moyer1987MichelsonI1}, described the state of fundamental research at his time \cite{MichelsonShort}:
\begin{center}
  \textit{The more important fundamental laws and facts of physical science have all been discovered, and these are now so firmly established that the possibility of their ever being supplanted in consequence of new discoveries is exceedingly remote.}  
\end{center}
A more comprehensive exposition of Michelson's argument was put forward in 1894, at the \textit{Ryerson Laboratory} dedication, and subsequently quoted in the University of Chicago 1896 Annual Register \cite{MichelsonLong}:
\begin{center}
  \textit{While it is never safe to affirm that the future of Physical Science has no marvels in store even more astonishing than those of the past, it seems probable that most of the grand underlying principles have been firmly established and that further advances are to be sought chiefly in the rigorous application of these principles to all the phenomena which come under our notice. It is here that the science of measurement shows its importance — where quantitative work is more to be desired than qualitative work. An eminent physicist remarked that the future truths of physical science are to be looked for in the sixth place of decimals.} 
\end{center}
Even though the identity of such \textit{eminent} scientist was never revealed, his or her alleged statement properly summarised the widespread perception of physics at the dawn of the twentieth century. The quest for the ultimate structure of Nature was deemed to have fulfilled its purpose. Hence, the only meaningful venture left was that of performing the most accurate possible measurements, allowing for equations to be refined. Remarkably, it was an experiment Michelson performed, in conjunction with Edward Morley \cite{MichelsonExperiment}, that provided the empirical backbone for the special theory of relativity, proposed by Albert Einstein in one of his four renowned \textit{annus mirabilis} papers \cite{EinsteinZurEB}. The advent of such a disruptively novel perspective, together with its subsequent extension to the general theory of relativity \cite{Einstein:1914bw,Einstein:1914bx,Einstein:1915ca,Einstein:1916vd} and the formulation of quantum mechanics \cite{QM1Planck:1901tja,QM2Bohr1913TheSO,QM3SchrdingerQuantisierungAE,QM4BornZurQ,QM5Heisenberg:1925zz,QM6Born1926ZurQI,QM7Bohr1928TheQP}, incontrovertibly established that the age of great theoretical discoveries was anything but over. The intention of this introductory section is by no means that of depicting modern day physicists as inherently less prone to absolute judgements with respect to their predecessors. On the contrary, our privileged hindsight point of view should serve as a sobering reminder that no theoretical framework, no matter how solid it may appear, is immune from being overthrown. Like any other paradigm that preceded it, even that of effective field theories might eventually be either partially or completely subverted. The \textit{swampland program}, within which this doctoral thesis finds its place, proposes one such subversion. The reason for doing so traces back to an almost obvious fact: \textit{our universe has gravity}. Gravity, moreover, can be best described as a deformation of the space-time geometry, whose metric gets promoted -up to mathematical redundancies- to a collection of \textit{physical}, field-theoretic degrees of freedom. Being more specific, low energy phenomenology can be pictured as taking place over a four-dimensional manifold $\mathcal{M}$ with a dynamical, general relativistic, Lorentzian metric tensor $g$, whose behaviour is controlled by the well-known Einstein equations  \cite{Misner:1973prb}:
\[
R_{\mu\nu}-\frac{1}{2}Rg_{\mu\nu}+\Lambda g_{\mu\nu}=\frac{8\pi G}{c^4}T_{\mu\nu}\ .
\]
In the above formula, $c$ is the speed of light in vacuum, $G$ is Newton's constant in four dimensions, $R_{\mu\nu}$ is the Ricci curvature tensor associated to the space-time metric, $R$ is its corresponding Ricci scalar, $T_{\mu\nu}$ is the overall energy-momentum contribution of all other matter fields and $\Lambda$ is a cosmological constant. Given the current experimental bounds, the value of $\Lambda$ characterising our universe appears to be roughly $\Lambda\sim 10^{-52}\ \text{m}^{-2}$, with a positive sign. Even at this stage, it is clear that those length scales we were arbitrarily tuning while working with standard effective models, as if they were perfectly controllable external parameters, are now deeply intertwined with the space-time degrees of freedom. Indeed, the very notion of distance is defined from the metric, which is now part of the field content of the theory. From the above equations, it is moreover clear that gravity couples to any sort of energy density. It is, in one word, \textit{universal}. Therefore, whatever the matter fields under scrutiny might be, non negligible contributions to the space-time curvature are bound to appear when probing extremely high energies. Distance measurements are not independent from the physical degrees of freedom, which are conversely expected to get highly excited when short length scales are being explored. Those are among the most groundbreaking teachings provided by general relativity. When pushed to its consequences, Einstein's theory forces us to question the foundations of our previous accounts of Nature, where space-time was taken to be a fixed Minkowski background structure. How can effective field theories, so greatly reliant on the traditionally innocuous idea of studying physics at certain length scales and below given energies, be affected by Einstein's revolution? It has been assessed that, fortunately, general relativistic gravity can be harmlessly merged with the effective field theory framework in a low energy regime \cite{Donoghue:2012zc}. Furthermore, in the last few decades, many insightful discoveries have been made by pursuing the investigation of quantum field theories in curved fixed backgrounds \cite{DeWitt:1975ys,Ford:1997hb,Lust:2018cvp,Wald:1975kc}. Nonetheless, the questions around what Nature's behaviour in the deep ultraviolet regime might be and how this may affect our low energy phenomenology, beyond the standard intuition outlined by effective field theories, are deeply unsettling, as well as yet to be exhausted. They represent the core focus of the swampland program. In order to address them, there is still a further, pivotal matter that requires to be brought up: the space-time metric should be quantised. The arguments in favour of such a view, spanning from consistency requirements to paradoxes that call for a solution, are overwhelming \cite{DW:DeWitt:1967yk,DW:DeWitt:1967ub,DW:DeWitt:1967uc,QGN:Butterfield:1999ah,QGN:Carlip:2008zf,QGN:Carlip:2015asa,QGN:Isham:1995wr,QGN:Kiefer:2005uk} and will not be discussed here. However, it must be pointed out that almost one century after Werner Heisenberg and Wolfgang Pauli\cite{HeisenbergZurQD,Heisenberg1930ZurQD,Rechenberg1993HeisenbergAP} proposed a first approach towards a quantum theory of gravity, the scientific debate around the fundamental, microscopic, quantum essence of space-time is still heated. Here, we will consider an approach to the quantisation of the gravitational field which can be arguably regarded as the most understood and well-developed one: \textit{superstring theory}. Taken at face value, superstring theory is the quantum theory of a supersymmetric, relativistic one-dimensional string, whose various excited states correspond to distinct space-time fields. Postponing a detailed account of the subject to chapter \ref{stringchap}, it should now be remarked that among them, whatever supplementary assumptions might be taken, there must always be a graviton, associated to local perturbations of the gravitational field. Moreover, imposing the quantised theory not to be anomalous, such field can be shown to satisfy Einstein's equations. Gravity, in the low energy limit of superstring theory, is unavoidable. We are therefore left with a framework able not only to construct phenomenologically interesting quantum field theories, but also to consistently merge them with general relativity. It goes without saying that the implications of such a discovery for effective field theories, both in refining their conceptual apparatus and in providing surprising results, is tremendous. More specifically, there is now a significant body of evidence, systematised in the context of the swampland program, suggesting that the quantum properties of the gravitational field should pose strict and previously unexpected constraints on the features of superstring low energy effective field theories. From a  practical perspective, this translates into the general expectation that just a small subset of the family of apparently consistent low energy effective field theories coupled to a dynamical space-time, which is typically referred to as the \textit{landscape}, can be completed towards superstring theory in the ultraviolet regime. The \textit{swampland}, on the other hand, is defined as the collection of those theories that, albeit being seemingly consistent below a certain cut-off, do not admit said completion. It is hence necessary to provide formal criteria of demarcation between the landscape and the swampland, usually stated in the form of the so-called \textit{swampland conjectures}. Among them, we will mostly focus our attention on the \textit{distance conjecture}. In its standard formulation, it corresponds to the claim that large displacements in the moduli space of an effective field theory should be accompanied by infinite towers of asymptotically massless fields, displaying an exponential mass drop with respect to a geodesic notion of moduli space distance. Some notable attempts at extending it to displacements of the space-time geometry itself will then be outlined, arguing that \textit{geometric flow equations} offer the most natural mathematical structures to achieve such a goal. Strikingly, it will be shown how Perelman's combined metric-scalar flow, introduced as a direct generalisation of the well-known Ricci flow, can be regarded as a  volume-preserving gradient flow for a particular entropy functional, which can in turn be employed in defining a proper distance along Perelman's combined flow trajectories. Having hence stated the \textit{dilaton-metric flow conjecture}, the evolution of a large class of scalar and metric bubble solutions under the previously discussed flow equations will be studied and proven to produce interesting paths in an extended moduli space. In the subsequent discussion, a new set of geometric flow equations will be derived from a more general entropy functional, which will reduce to Perelman's for a specific choice of some free parameters. By starting from the action for a scalar and a dynamical geometry and properly rescaling the fields, in order to match its expression to that of an entropy functional, a natural way of associating a set of geometric flow equations to a particular theory will be presented. In conclusion, the issue of preserving Einstein field equations along such action-induced flow trajectories will be dealt with by allowing an extra energy-momentum term to appear along the flow, so that any deformation of the metric will be reinterpreted in terms of the appearance of suitable additional matter contributions. This physical realisation of action-induced flow equations will be then applied to a particularly simple example, in which the supplementary energy-momentum tensor will be shown to be consistent with the gradual emergence of an infinite tower of fields with exponentially dropping masses, as those postulated by the distance conjecture. This will hence allow us to perform a non-trivial consistency check.

%% file: Sections/Strings/stringsintro.tex
In the following chapter, we will outline the main features of type II superstring theory. This will be achieved by introducing the classical supersymmetric string action, discussing its main properties and performing its quantisation. It is evident that the vastness of the subject prevents us from treating it in depth. In particular, we will not mention heterotic, type I or type 0 string theories, nor will we consider path integral quantisation. The interested reader is strongly encouraged to consult the standard references \cite{Blumenhagen:2013fgp,Polchinski:1998rq,Polchinski:1998rr,Becker:2006dvp,Ibanez:2012zz,Kiritsis:2019npv,Green:2012oqa,Green:2012pqa}, on which most of our discussion is grounded. It must be furthermore stressed that a wide variety of foundational topics in theoretical physics will be taken for granted. In that regard, we suggest to refer to \cite{Zee:2003mt,Schwartz:2014sze,Weinberg:1995mt,Weinberg:1996kr,Weinberg:2000cr,Sohnius:1985qm,Martin:1997ns} for quantum field theory and supersymmetry, to \cite{Misner:1973prb,Wald:1984rg,Hawking:1973uf,Freedman:2012zz,DallAgata:2021uvl} for general relativity and supergravity and to \cite{Zwiebach:2004tj,Tong:2009np} for graduate level introductions to bosonic string theory.

%% file: Sections/Strings/superstrings.tex
In its conventional conceptualisation, bosonic string theory is formulated by means of the \textit{Polyakov} action:
\begin{equation}\label{polyakovaction}
    S_{\text{Pol}}\equiv-\frac{1}{4\pi\alpha'}\int_{\Sigma_2}\diff\sigma\diff\tau\sqrt{-h}h^{\alpha\beta}\eta_{\mu\nu}\de_{\alpha}X^{\mu}\de_{\beta}X^{\nu}\ .  
\end{equation}
The integration domain $\Sigma_2$, referred to as the \textit{string world-sheet}, is the $2$-dimensional Lorentzian submanifold spanned by a string propagating in a $D$-dimensional space-time manifold. For a detailed analysis of the specificities of open strings, together with a description of the associated action boundary terms, we once more recommend to refer to \cite{Zwiebach:2004tj,Tong:2009np}. The world-sheet is charted by a time-like coordinate $\tau$ and a space-like coordinate $\sigma\in\left[0,l\right]$, parametrising the length of the string from $0$ to its total value $l$, and endowed with the metric tensor $h_{\alpha\beta}$. On top of that, it is the domain of definition of the scalar fields $X^{\mu}$. Taken at face value, \eqref{polyakovaction} describes a theory of $D$ free massless scalars in $2$ dimensions, with kinetic terms given by the diagonal matrix $\eta_{\mu\nu}=\diag\left(-1,+1,\dots,+1\right)$. From a complementary perspective, however, it defines a $\sigma$-model whose target space is the $D$-dimensional Minkowski space-time in which a string propagates, with coordinates $X^\mu$. The specific value of $D$ is remarkably fixed by consistency conditions. In the case of the covariant quantisation of bosonic strings, it must be set to $26$ in order for the resulting quantum theory not to break unitarity. Constructing the $26$-dimensional low energy effective quantum field theory in curved space-time coming from \eqref{polyakovaction} goes beyond the scope of the current chapter, let alone the innumerable $4$-dimensional theories one could obtain by dimensionally reducing it. Nevertheless, it is noteworthy that such a path encounters two major shortcomings. First of all, it does not allow for the existence of space-time fermions. This places it in a rough contradiction with one of the most elementary features of the real world, in which fermions are abundant and play a crucial phenomenological role. Furthermore, its spectrum unavoidably contains a tachyon, which is a transparent signal of vacuum instability. Both these flaws are tackled and solved by extending the Polyakov action \eqref{polyakovaction} to that of supersymmetric string theory. Before delving into the relevant mathematical details, it must be emphasised that we will express the superstring action in the \textit{Ramond-Neveu-Schwarz} formulation, in which supersymmetry is manifest on the world-sheet but not necessarily in space-time. The opposite is true for the \textit{Green-Schwarz} formulation, widely addressed in the above-mentioned references. Alternative approaches are the ones provided by \textit{pure-spinors} \cite{Berkovits:2021xwh,Berkovits:2022fth} and \textit{string fields} \cite{Erler:2019loq,Erbin:2021smf}. Picking up the threads of our discussion, the Ramond-Neveu-Schwarz formulation of type II superstring theory is defined as an extension of bosonic string theory, in which the action \eqref{polyakovaction} is supplemented with a fermionic sector. This allows to achieve supersymmetry on the world-sheet. For the sake of clarity, it is important to stress that this feature does not straightforwardly imply space-time supersymmetry, which will require us to introduce further structures. In order to explicitly write down the $N=1$ supersymmetric extension of \eqref{polyakovaction}, it is convenient to introduce a \textit{zwei-bein} 
$e^a_{\ \alpha}$ associated to the world-sheet metric $h_{\alpha\beta}$, transforming local Lorentz into Einstein indices and with determinant $e=\sqrt{-h}$. It is thus natural to construct the on-shell supergravity multiplet by defining a \textit{gravitino} $\chi_\alpha$ as a world-sheet vector of Majorana spinors. Furthermore, it is necessary to introduce a family of $D$ Majorana world-sheet fermions $\psi^\mu$. A discussion of the off-shell degrees of freedom goes beyond the scope of the current review chapter and can be found in \cite{Blumenhagen:2013fgp}. The overall expression for the action is
\begin{equation}\label{susypolyakovaction}
\begin{split}
        S_{\text{RNS}}&\equiv-\frac{1}{8\pi}\int_{\Sigma_2}\diff\sigma\diff\tau e\Biggl[\frac{2}{\alpha'}h^{\alpha\beta}\de_{\alpha}X^{\mu}\de_{\beta}X_{\mu}+2i\Bar{\psi}^\mu\rho^\alpha\de_\alpha\psi_\mu\\
        &\quad-i\Bar{\chi}_\alpha\rho^\beta\rho^\alpha\psi^\mu\left(\sqrt{\frac{2}{\alpha'}}\de_\beta X_\mu-\frac{i}{4}\Bar{\chi}_\beta\psi_\mu\right)\Biggr]\ , 
\end{split} 
\end{equation}
where the $\rho^\alpha$ are the world-sheet Dirac matrices associated to $h_{\alpha\beta}$, for which
\begin{equation}
    \left\{\rho^\alpha,\rho^\beta\right\}=2h^{\alpha\beta}\ .
\end{equation}
\subsection{Symmetries and gauge choice}
Before simplifying the expression \eqref{susypolyakovaction} by moving to \textit{superconformal gauge}, which is nothing more than the superstring analogue of the conformal gauge introduced in the bosonic case \cite{Tong:2009np}, we are required to briefly comment on the local world-sheet symmetries of the superstring action. In doing so, we will mostly follow the analysis performed in \cite{Blumenhagen:2013fgp} and list the various transformations that leave the theory unchanged separately. In all of them, it will be critical not to confuse the world-sheet perspective with the one associated to the $D$-dimensional space-time manifold.

\paragraph{Lorentz transformations}
First of all, we have \mbox{$2$-dimensional} world-sheet Lorentz transformations, which act on world-sheet indices and induce the infinitesimal field variations:
\begin{equation}
    \begin{split}
        \delta_l\psi^\mu&=-\frac{1}{2}l\Bar{\rho}\psi^\mu\ ,\quad\delta_l X^\mu=0\ ,\\
        \delta_l e_\alpha^{\ a}&=l\epsilon^a_{\ b}e_\alpha^{\ b}\ ,\quad \delta_l\chi_\alpha=-\frac{1}{2}l\Bar{\rho}\chi_\alpha\ .
    \end{split}
\end{equation}

\paragraph{Reparametrisations} 
Secondly, we must consider world-sheet reparametrisations induced by a vector $\xi^\alpha$, acting on the $\left(\tau,\sigma\right)$ coordinates and associated to the infinitesimal field variations:
\begin{equation}
    \begin{split}
\delta_\xi\psi^\mu&=-\xi^\alpha\de_\alpha\psi^\mu\ ,\\
\delta_\xi X^\mu&=-\xi^\alpha\de_\alpha X^\mu\ ,\\
\delta_\xi e_\alpha^{\ a}&=-\xi^\beta\de_\beta e_\alpha^{\ a}-e_\beta^{\ a}\de_\alpha\xi^\beta\ ,\\
\delta_\xi \chi_\alpha&=-\xi^\beta\de_\beta \chi_\alpha-\chi_\beta\de_\alpha\xi^\beta\ .
    \end{split}
\end{equation}

\paragraph{Weyl transformations}
As for the case of the bosonic string, the superstring action \eqref{susypolyakovaction} is invariant under Weyl rescalings. Infinitesimally, such transformations reduce to:
\begin{equation}
    \begin{split}
        \delta_\Lambda\psi^\mu&=-\frac{1}{2}\Lambda\psi^\mu\ ,\quad\delta_\Lambda X^\mu=0\ ,\\
        \delta_\Lambda e_\alpha^{\ a}&=\Lambda e_\alpha^{\ a}\ ,\quad \delta_\Lambda\chi_\alpha=\frac{1}{2}\Lambda\chi_\alpha\ .
    \end{split}
\end{equation}

\paragraph{Super-Weyl transformations}
On top of Weyl rescalings, we also have super-Weyl transformations. They act trivially on every field except for the gravitino, for which we have the infinitesimal variation
\begin{equation}
\delta_\eta\chi_\alpha=\rho_\alpha\eta\ ,
\end{equation}
where $\eta$ is a world-sheet Majorana spinor.

\paragraph{Supersymmetry}
Finally, we have that supersymmetry we constructed the action \eqref{susypolyakovaction} to accommodate for in the first place. Introducing 
\begin{equation}
    \omega_\alpha\equiv -\frac{1}{e}e_{\alpha a}\epsilon^{\beta\gamma}\de_\beta e_{\gamma}^{\ a}+\frac{i}{4}\Bar{\chi}_\alpha\Bar{\rho}\rho^\beta\chi_\beta\ ,
\end{equation}
we can define the covariant derivative of a Majorana spinor $\lambda$ in the presence of torsion
\begin{equation}
    D_\alpha\lambda\equiv \de_\alpha\lambda -\frac{1}{2}\omega_\alpha\Bar{\rho}\lambda
\end{equation}
and write the action of supersymmetry, for an infinitesimal Majorana spinor parameter $\epsilon$, as follows:
\begin{equation}
    \begin{split}
        \delta_\epsilon\psi^\mu&=\frac{1}{2}\rho^{\alpha}\left(\sqrt{\frac{2}{\alpha'}}\de_\alpha X^\mu-\frac{i}{2}\Bar{\chi}_\alpha\psi^\mu\right)\ ,\\
        \delta_\epsilon X^\mu&=\sqrt{\frac{\alpha'}{2}}i\Bar{\epsilon}\psi^\mu\ ,\quad\delta_\epsilon e_\alpha^{\ a}=\frac{i}{2}\Bar{\epsilon}\rho^a\chi_\alpha\ ,\\ 
       \delta_\epsilon\chi_\alpha&=2D_\alpha\epsilon\ .
    \end{split}
\end{equation}
\subsubsection{The superconformal gauge}
Exploiting world-sheet Lorentz transformations, reparametrisations and local supersymmetry, we can remove two degrees of freedom from the gravitino and as many from the zwei-bein. Without performing explicit computations, which can be found in the suggested superstring theory references, we state the superconformal gauge to correspond to:
\begin{equation}
    e_\alpha^{\ a}=e^\phi\delta_\alpha^{\ a}\ ,\quad\chi_\alpha=\rho_\alpha\lambda\ .
\end{equation}
At a classical level, we can set $\phi=\lambda=0$ with Weyl and super-Weyl rescalings. Such symmetries will require particular care at the quantum level, since they will be anomalous unless the number of space-time dimensions $D$ will be taken to have a specific value. Nonetheless, the action, containing now only degrees of freedom related to $X^\mu$ and $\psi^\mu$, takes the simple form:
\begin{equation}\label{superconformal}
    S_{SC}=-\frac{1}{8\pi}\int_{\Sigma_2}\diff\sigma\diff\tau\left(\frac{2}{\alpha'}\de_\alpha X^\mu\de^\alpha X_\mu+2i\Bar{\psi}^\mu\rho^\alpha\de_\alpha\psi^\mu\right)\ .
\end{equation}
The remaining local supersymmetry acts via the infinitesimal field displacements:
\begin{equation}
    \delta_\epsilon X^\mu=\sqrt{\frac{\alpha'}{2}}i\Bar{\epsilon}\psi^\mu\ ,\quad \delta_\epsilon\psi^\mu=\frac{1}{\sqrt{2\alpha'}}\rho^\alpha\de_\alpha X^\mu \epsilon\ .
\end{equation}
Furthermore, we have the gauge-preserving combinations of diffeomorphisms, Weyl rescalings and Lorentz transformations:
\begin{equation}
    \begin{split}
        \delta_\xi X^\mu&=\xi^\alpha\de_\alpha X^\mu\ ,\\
        \delta_\xi \psi^\mu&=\xi^\alpha\de_\alpha\psi^\mu+\frac{1}{4}\psi^\mu\de_\alpha\xi^\alpha-\frac{1}{4}\Bar{\rho}\psi^\mu\epsilon^{\alpha\beta}\de_\alpha\xi_\beta\ .
    \end{split}
\end{equation}
The equations of motion associated to \eqref{superconformal} are simply:
\begin{equation}\label{supconfeom}
    \de^\alpha\de_\alpha X^\mu=0\ ,\quad \rho^\alpha\de_\alpha\psi^\mu=0\ .
\end{equation}
The energy-momentum tensor and its associated supercurrent, which are defined as the variations
\begin{equation}
    T_{\alpha\beta}\equiv\frac{2\pi}{e}\frac{\delta S_{SC}}{\delta e_a^{\beta}}e_{\alpha a}\ ,\quad T_{F\alpha}\equiv\frac{2\pi}{ie}\frac{\delta S_{SC}}{\delta \Bar{\chi}^\alpha}\ ,
\end{equation}
take the explicit superconformal gauge forms:
\begin{equation}
    \begin{split}
        T_{\alpha\beta}&=\frac{1}{2\alpha'}\left(\eta_{\alpha\beta}\de_\gamma X^\mu\de^\gamma X_\mu-2\de_\alpha X^\mu\de_\beta X_\mu\right)\\
        &\quad-\frac{i}{4}\left(\Bar{\psi}^\mu\rho_\beta\de_\alpha\psi_\mu+\Bar{\psi}^\mu\rho_\alpha\de_\beta\psi_\mu\right)\ ,\\
        T_{F\alpha}&=-\frac{1}{\sqrt{8\alpha'}}\rho^\beta\rho_\alpha\psi^\mu\de_\beta X_\mu\ .
    \end{split}
\end{equation}
Such expressions vanish when $X^\mu$ and $\psi^\mu$ are imposed to satisfy the on-shell conditions \eqref{supconfeom}. Namely, as a direct generalisation of the bosonic case, we have constraints
\begin{equation}\label{nullenergymomentum}
    T_{\alpha\beta}=0\ ,\quad T_{F\alpha}=0
\end{equation}
that will require to be taken care of when quantising the theory. Moreover, by means of the conservation laws
\begin{equation}
    \de^\alpha T_{\alpha\beta}=0\ ,\quad \de^\alpha T_{F\alpha}=0\ ,
\end{equation}
an infinite number of conserved charges is generated. The tracelessness conditions
\begin{equation}
    T^{\alpha}_{\ \alpha}=0\ ,\quad \rho^\alpha T_{F\alpha}=0
\end{equation}
notably come from Weyl and super-Weyl invariance, respectively. Hence, they hold regardless of the equations of motion \eqref{supconfeom}. 
\subsubsection{Boundary conditions}
While performing variations of the superconformal gauge action \eqref{superconformal}, in order to derive the equations of motion \eqref{supconfeom}, one has to impose appropriate $\sigma$-boundary conditions to both the bosonic and the fermionic sector. Focusing on closed strings, we can straightforwardly observe that imposing $\sigma$-periodicity for the sake of consistency uniquely fixes the boundary behaviour of the $X^\mu$ fields. The fermions $\psi^\mu$, instead, simply have to satisfy the expression
\begin{equation}\label{fermionsboundarycondition}
    \int\diff\tau\Bigl[\psi_+\cdot\delta\psi_+-\psi_-\cdot\delta\psi_-\Bigr]_{\sigma=0}^{\sigma=l}=0\ ,
\end{equation}
where $\psi_+$ and $\psi_-$ are the Weyl components of the world-sheet Majorana spinors, with:
\begin{equation}
    \psi^\mu\equiv\begin{pmatrix}
\psi^\mu_+\\ \psi^\mu_-
    \end{pmatrix}\ .
\end{equation}
The condition \eqref{fermionsboundarycondition}, for the closed string, translate into $\psi_+\cdot\delta\psi_+-\psi_-\cdot\delta\psi_-$ being $\sigma$-periodic with period $l$. This can be achieved by the $\psi_+^\mu$ and $\psi^\mu_-$ being either periodic or anti-periodic in $\sigma$. In particular:
\begin{itemize}
    \item We refer to periodic boundary conditions as \textit{Ramond} (R) boundary conditions for closed strings.
    \item We refer to anti-periodic boundary conditions as \textit{Neveu-Schwarz} (NS) boundary conditions for closed strings.
\end{itemize}
Since such boundary conditions can be chosen independently for $\psi_+$ and $\psi_-$, we have that each of the $D$ world-sheet Majorana spinors $\psi^\mu$ can be taken to belong to four distinct sectors: (NS,NS), (NS,R), (R,NS) and (R,R). As for the open string fermionic sector, one obtains \textit{Dirichlet} (D) and \textit{Neumann} (N) boundary conditions similar to those which appear in the bosonic one \cite{Tong:2009np}. When imposing N boundary conditions to fermions at both ends of an open string, we can either impose periodicity or anti-periodicity. It turns out that the only relevant quantity is the relative sign between the two choices. Namely, when we have  
\begin{equation}
    \psi_+^\mu\left(0\right)=\alpha\psi_-^\mu\left(0\right)\ ,\quad \psi_+^\mu\left(l\right)=\beta\psi_-^\mu\left(l\right)
\end{equation}
with $\alpha,\beta\in\left\{-1,+1\right\}$, it only matters whether $\eta\equiv\alpha\cdot\beta$ is equal to $+1$ or $-1$. For what concerns the terminology: 
\begin{itemize}
    \item We refer to the relative sign choice $\eta=+1$ as \textit{Ramond} (R) boundary conditions for open strings.
    \item We refer to the relative sign choice $\eta=-1$ as \textit{Neveu-Schwarz} (NS) boundary conditions for open strings.
\end{itemize}
\subsection{Oscillator expansions}\label{oscillatorexpansions}
In order to solve the equations of motion \eqref{supconfeom}, both the bosonic and the fermionic degrees of freedom require to be expanded in modes, taken to satisfy specific algebras. In the following discussion, the main results of such procedure are outlined. A distinction is made between closed and open strings, following what was done in the main reference \cite{Blumenhagen:2013fgp}. It is strongly suggested to refer to such book for a more thorough discussion.
\subsubsection{Closed strings}
A parameter $\varphi$ is introduced to distinguish between R ($\varphi=0$) and NS ($\varphi=1/2$) boundary conditions for the fermionic degrees of freedom. Their modes expansion, together with the one related to the bosons, can be compactly expressed as follows
\begin{equation}
    \begin{split}
        \psi_+^\mu\left(\sigma,\tau\right)&=\sqrt{\frac{2\pi}{l}}\sum_{k\in\Z+\varphi}\Bar{b}_r^\mu e^{-2\pi ik\left(\tau+\sigma\right)/l}\ ,\\
        \psi_-^\mu\left(\sigma,\tau\right)&=\sqrt{\frac{2\pi}{l}}\sum_{k\in\Z+\varphi}b_r^\mu e^{-2\pi ik\left(\tau-\sigma\right)/l}\ ,\\
    X^\mu\left(\sigma,\tau\right)&=x^\mu+\frac{2\pi\alpha'}{l}p^\mu\tau+i\sqrt{\frac{\alpha'}{2}}\sum_{n\neq 0}\frac{1}{n}\alpha^\mu_n e^{-2\pi in\left(\tau-\sigma\right)/l}\\
    &\quad+i\sqrt{\frac{\alpha'}{2}}\sum_{n\neq 0}\frac{1}{n}\Bar{\alpha}^\mu_n e^{-2\pi in\left(\tau+\sigma\right)/l}\ ,
    \end{split}
\end{equation}
where the modes satisfy the reality and Majorana conditions:
\begin{equation}
    \begin{split}
        \alpha^\mu_{-n}&=\left(\alpha^\mu_n\right)^*\ ,\quad \Bar{\alpha}^\mu_{-n}=\left(\Bar{\alpha}^\mu_n\right)^*\ ,\\
        b^\mu_{-k}&=\left(b^\mu_k\right)^*\ ,\quad \Bar{b}^\mu_{-k}=\left(\Bar{b}^\mu_k\right)^*\ .
    \end{split}
\end{equation}
For what concerns the Poisson and Dirac brackets of the modes, which will then be promoted to commutators and anti-commutators, respectively, in the canonical quantum theory, we have:
\begin{equation}
    \begin{split}
        &\left\{\alpha^\mu_m,\alpha^\nu_m\right\}_{\text{PB}}=\left\{\Bar{\alpha}^\mu_m,\Bar{\alpha}^\nu_m\right\}_{\text{PB}}=in\delta_{m+n}\eta^{\mu\nu}\ ,\\
        &\left\{b^\mu_k,b^\nu_s\right\}_{\text{DB}}=\left\{\Bar{b}^\mu_k,\Bar{b}^\nu_s\right\}_{\text{DB}}=-i\delta_{k+s}\eta^{\mu\nu}\ ,\\
        &\left\{\alpha^\mu_m,\Bar{\alpha}^\nu_m\right\}_{\text{PB}}=\left\{b^\mu_k,\Bar{b}^\nu_s\right\}_{\text{DB}}=0\ ,\quad \left\{x^\mu,p^\nu\right\}_{\text{PB}}=\eta^{\mu\nu}\ .
    \end{split}
\end{equation}
By moving to light-cone coordinates $\sigma_\pm\equiv\tau\pm\sigma$ and decomposing the left-moving components of the energy-momentum tensor in modes $L_n$ and those of the supercurrent in modes $G_r$, we observe that the $L_n$ can be decomposed as
\begin{equation}
 L_n=L^{\left(\alpha\right)}_n+L^{\left(b\right)}_n
\end{equation}
in a contribution coming from the bosonic degrees of freedom and one coming from the fermionic ones. We obtain:
\begin{equation}
    \begin{split}
        L^{\left(\alpha\right)}_n&=\frac{1}{2}\sum_{p\in\Z}\eta_{\mu\nu}\alpha^{\mu}_{-p}\alpha^{\nu}_{p+n}\ ,\\
        L^{\left(b\right)}_n&=\frac{1}{2}\sum_{r\in\Z+\varphi}\left(r+\frac{n}{2}\right)\eta_{\mu\nu}b^{\mu}_{-r}b^{\nu}_{r+n}\ ,\\
        G_r&=\sum_{n\in\Z}\eta_{\mu\nu}\alpha^{\mu}_{-n}b^{\nu}_{r+n}\ .
    \end{split}
\end{equation}
The above expressions fulfil the reality conditions
\begin{equation}
    L^*_n=L_{-n}\ ,\quad G_r^*=G_{-r}
\end{equation}
and satisfy the centerless \textit{super-Virasoro} algebra:
\begin{equation}
    \begin{split}
        \left\{L_m,L_n\right\}_{\text{DB}}&=i\left(n-m\right)L_{m+n}\ ,\\
        \left\{L_m,G_r\right\}_{\text{DB}}&=i\left(r-\frac{m}{2}\right)G_{m+r}\ ,\\
        \left\{G_r,G_s\right\}_{\text{DB}}&=-2iL_{r+s}\ .
    \end{split}
\end{equation}
Since we are considering the case of a closed string, an analogous set of generators, respecting the same algebra, can be constructed for right-movers. 
\subsubsection{Open strings}
When it comes to open strings, things get slightly more complicated, as crucial distinctions must be made on the various choices of the boundary conditions. Once more, we introduce a parameter $\varphi$ to distinguish between R ($\varphi=0$) and NS ($\varphi=1/2$) boundary conditions for the fermionic degrees of freedom. Furthermore, the difference between (NN), (DN), (ND) and (DD) boundary conditions has to be taken into account for both bosons and fermions. Starting with the bosons, we have:
\begin{equation}
    \begin{split}
        \text{(NN)}\quad\quad X^{\mu}\left(\sigma,\tau\right)&=x^\mu+\frac{2\pi\alpha'}{l}p^\mu\tau\\
        &\quad+i\sqrt{2\alpha'}\sum_{n\neq 0}\frac{1}{n}\alpha_n^{\mu}e^{-i\pi n\tau/l}\cos{\left(\frac{n\pi\sigma}{l}\right)}\ ,
        \\
        \text{(DD)}\quad\quad X^{\mu}\left(\sigma,\tau\right)&=x_0^\mu+\frac{x_1^\mu-x_0^\mu}{l}\sigma\\
        &\quad+\sqrt{2\alpha'}\sum_{n\neq 0}\frac{1}{n}\alpha_n^{\mu}e^{-i\pi n\tau/l}\sin{\left(\frac{n\pi\sigma}{l}\right)}\ ,\\
        \text{(DN)}\quad\quad X^{\mu}\left(\sigma,\tau\right)&=x^\mu+\sqrt{2\alpha'}\sum_{k\in\Z+\frac{1}{2}}\frac{1}{k}\alpha^\mu_k e^{-i\pi k\tau/l}\sin{\left(\frac{k\pi\sigma}{l}\right)}\ ,\\
        \text{(ND)}\quad\quad X^{\mu}\left(\sigma,\tau\right)&=x^\mu+i\sqrt{2\alpha'}\sum_{k\in\Z+\frac{1}{2}}\frac{1}{k}\alpha^\mu_k e^{-i\pi k\tau/l}\cos{\left(\frac{k\pi\sigma}{l}\right)}\ .
    \end{split}
\end{equation}
The mode expansions of the Majorana spinors, instead, are:
\begin{equation}
    \begin{split}
        \text{(NN)}\quad\quad \psi^{\mu}_\pm\left(\sigma,\tau\right)&=\sqrt{\frac{\pi}{l}}\sum_{k\in\Z+\varphi}b^\mu_k e^{-i\pi k\left(\tau\pm\sigma\right)/l}\ ,\\
        \text{(DD)}\quad\quad \psi^{\mu}_\pm\left(\sigma,\tau\right)&=\pm\sqrt{\frac{\pi}{l}}\sum_{k\in\Z+\varphi}b^\mu_k e^{-i\pi k\left(\tau\pm\sigma\right)/l}\ ,\\
        \text{(DN)}\quad\quad \psi^{\mu}_\pm\left(\sigma,\tau\right)&=\sqrt{\frac{\pi}{l}}\sum_{k\in\Z+\varphi+\frac{1}{2}}b^\mu_k e^{-i\pi k\left(\tau\pm\sigma\right)/l}\ ,\\
    \text{(ND)}\quad\quad \psi^{\mu}_\pm\left(\sigma,\tau\right)&=\pm\sqrt{\frac{\pi}{l}}\sum_{k\in\Z+\varphi+\frac{1}{2}}b^\mu_k e^{-i\pi k\left(\tau\pm\sigma\right)/l}\ .
    \end{split}
\end{equation}
The expressions for the modes of the energy-momentum tensor and the supercurrent can be obtained in terms of the $\alpha^\mu_n$ and the $b^\mu_r$ modes. Distinguishing once more between the contributions $L^{\left(\alpha\right)}_n$ and $L^{\left(b\right)}_n$ to the modes of the energy-momentum tensor, we are left with:
\begin{equation}
    \begin{split}
        L^{\left(\alpha\right)}_n&=\frac{1}{2}\sum_{p\in\Z}\eta_{\mu\nu}\alpha^{\mu}_{-p}\alpha^{\nu}_{p+n}\ ,\\
        L^{\left(b\right)}_n&=\frac{1}{2}\sum_{r\in\Z+\varphi}\left(r+\frac{n}{2}\right)\eta_{\mu\nu}b^{\mu}_{-r}b^{\nu}_{r+n}\ ,\\
    G_r&=\sum_{n\in\Z}\eta_{\mu\nu}\alpha^{\mu}_{-n}b^{\nu}_{r+n}\ .
    \end{split}
\end{equation}
Once more, such expressions satisfy the centerless super-Virasoro algebra:
\begin{equation}
    \begin{split}
        \left\{L_m,L_n\right\}_{\text{DB}}&=i\left(n-m\right)L_{m+n}\ ,\\
        \left\{L_m,G_r\right\}_{\text{DB}}&=i\left(r-\frac{m}{2}\right)G_{m+r}\ ,\\
        \left\{G_r,G_s\right\}_{\text{DB}}&=-2iL_{r+s}\ .
    \end{split}
\end{equation}
This result concludes our compendium on the classical features of the supersymmetric string action \eqref{susypolyakovaction}. Therefore, we can now progress towards the analysis of the canonical quantisation of the theory.

%% file: Sections/Strings/quantum.tex
It is once more important to stress that, along the lines of their bosonic counterparts, superstrings are described by a constrained system. This peculiarity directly translates in a series of technical difficulties that have to be adequately dealt with, when attempting at quantising the theory. The following discussion does not aim at being complete nor self-sufficient. Therefore, we suggest to refer to \cite{Barcelos-Neto:1986abv,Rothe:2010dzf,Prokhorov:2011zz} for all the mathematical details. For what concerns our concise abridgement of the subject matter, it must be emphasised that constraints cannot be neglected when promoting the degrees of freedom of the theory to operators and constructing to corresponding Hilbert space. In fact, for the specific case of superstring theory, one might choose to adopt one of the following canonical quantisation approaches:
\begin{itemize}
    \item \textbf{Old covariant quantisation} The quantisation procedure is applied to the unconstrained system, producing a vast Hilbert space. Constraints are enforced by means of conditions that physical states have to satisfy, thus projecting out the unphysical sector of the Hilbert space. This method has the advantage of being explicitly covariant from the start. At the same time, it fails at being manifestly unitary and becomes so only at a critical value of the space-time dimension $D$. 
    \item \textbf{Light-cone quantisation} The constraints are enforced already at a classical level. Then, quantisation procedures are naturally applied to a constrained subset of the original family of oscillators, corresponding to $D-2$ space-time directions, and a physical Hilbert space is obtained. Unlike the previous strategy, here unitary is secured without further effort. On the other hand, covariance in achieved only at a critical value of the space-time dimension $D$, which turns out to be consistent with the one obtained by imposing unitarity to the theory quantised in the old covariant manner. 
\end{itemize}
Our review will not discuss path integral quantisation of superstring theory, which is an interesting, fundamental and highly rewarding topic in itself. A comprehensive discussion is contained in the two-volume monograph \cite{Polchinski:1998rq,Polchinski:1998rr}, together with an analysis of its major applications to superstring scattering amplitudes. 
\subsection{Old covariant quantisation}
As was briefly summarised in the previous discussion, quantising superstring theory in the old covariant approach corresponds to enforcing the constraints \eqref{nullenergymomentum} after having applied the standard quantisation prescription. This is the way we will follow in this analysis. Therefore, we can promote classical fields to quantum field operators straight away. On top of that, Poisson brackets are sent into commutators
\begin{equation}
    \{\ \ ,\ \ \}_{\text{P.B.}}\longrightarrow\frac{1}{i}[\ \ ,\ \ ]\ ,
\end{equation}
while Dirac brackets are sent in anti-commutators:
\begin{equation}
    \{\ \ ,\ \ \}_{\text{D.B.}}\longrightarrow\frac{1}{i}\{\ \ ,\ \ \}\ .
\end{equation}
Given the oscillator expansions presented in \ref{oscillatorexpansions}, positive and negative expansion modes are naturally identified with annihilation and creation operators, respectively. The non-trivial part of their algebra is: 
\begin{equation}
\begin{split}
        \left[\alpha^\mu_m,\alpha^\nu_n\right]&=\eta^{\mu\nu}m\delta_{m+n}\ ,\\
        \left\{b^\mu_r,b^\nu_s\right\}&=\eta^{\mu\nu}\delta_{r+s}\ .
\end{split}
\end{equation}
When focusing on closed strings the set of commutators has to be doubled, to account for both left and right-moving excitations. The number operator $N$ can be defined as the sum 
\begin{equation}
    N= N^{\left(\alpha\right)}+N^{\left(b\right)}
\end{equation}
of a component $N^{\left(\alpha\right)}$, given by excitations corresponding to world-sheet bosons, and a component $N^{\left(b\right)}$, coming from world-sheet fermions. Their explicit expressions are
\begin{equation}
    N^{\left(\alpha\right)}\equiv\sum_{n=1}^\infty \eta_{\mu\nu}\alpha^\mu_{-n}\alpha^\nu_n\ ,\quad N^{\left(b\right)}=\sum_{k\in\Z+\varphi>0}k\eta_{\mu\nu}b^\mu_{-k}b^\nu_k\ ,
\end{equation}
where, as usual, $\varphi=0$ for R boundary conditions and $\varphi=1/2$ for NS ones. Focusing on Virasoro generators, extracted as coefficients of the mode expansion of the energy-momentum tensor and the supercurrent, we note that they are defined through sums that might be ambiguous at the quantum level, unless we impose them to be normally ordered. In fact, this has a significant effect solely for the zeroth order energy-momentum tensor generator $L_0$. Together, they satisfy the quantum super-Virasoro algebra:
\begin{equation}\label{qsupervirasoro}
    \begin{split}
        \left[L_m,L_n\right]&=\left(m-n\right)L_{m+n}+\frac{D}{8}m\left(m^2-2\varphi\right)\delta_{m+n}\ ,\\
        \left[L_m,G_r\right]&=\left(\frac{m}{2}-r\right)G_{m+r}\ ,\\
        \left\{G_r,G_s\right\}&=2L_{r+s}+\frac{D}{2}\left(r^2-\frac{\varphi}{2}\right)\delta_{r+s}\ .
    \end{split}
\end{equation}
\subsubsection{Hilbert space vacuum construction}
We can now focus on constructing the superstring theory Hilbert space. First, we observe that the distinction between R and NS boundary conditions for world-sheet fermions directly produces a distinction between two sectors of the theory. The following discussion can be directly applied to closed strings by replicating it for left-movers. For open strings, it is instead only valid for NN and DD boundary conditions. When studying open strings with DN and ND boundary conditions, one has to take into account that the structure of the fermionic expansion modes indices is swapped between the usual NS and R sectors. Hence, the following considerations have to be swapped too. That said, we introduce an NS sector vacuum $\ket{0}_{\text{NS}}$ satisfying the conditions
\begin{equation}
    \begin{split}
        \alpha^\mu_n\ket{0}_{\text{NS}}&=0\ ,\quad \text{for}\ n=1,2,3\dots\\
        b_r^\mu\ket{0}_{\text{NS}}&=0\ ,\quad\text{for}\ r=\frac{1}{2},\frac{3}{2},\frac{5}{2}\dots
    \end{split}
\end{equation}
and an R sector vacuum $\ket{a}_{\text{R}}$ for which it holds that:
\begin{equation}
    \begin{split}
        \alpha^\mu_n\ket{a}_{\text{R}}&=0\ ,\quad \text{for}\ n=1,2,3\dots\\
        b_m^\mu\ket{a}_{\text{R}}&=0\ ,\quad\text{for}\ m=1,2,3\dots
    \end{split}
\end{equation}
In the above, the ground state dependence on the centre of mass momentum $p^\mu$ was made implicit. It can be noted that the NS sector ground state $\ket{0}_{\text{NS}}$ is unique. Thus, it is a spin zero state. In the R sector, instead, we have a family of degenerate ground states, labelled by an index $a$ and constructed from one another by acting with the fermionic zero modes $b_0^\mu$. Since it holds that 
\begin{equation}
    \left\{b_0^\mu,b_0^\nu\right\}=\eta^{\mu\nu}\ ,
\end{equation}
the ground states $\ket{a}_{\text{R}}$ form a representation of \textit{Clifford algebra} and $a$ is an $SO\left(D-1,1\right)$ spinorial index. Therefore, states in NS sector of the theory are space-time bosons, while those belonging to the R sector are space-time fermions, which are a true novelty of superstring theory with respect to the previous bosonic formulation. Once more, we stress that the contrary is true for DN and ND open strings.
\subsubsection{Imposing the constraints}
As should be clear by now, the Hilbert space $\mathcal{H}$ obtained by naively acting with the appropriate creation operators on the vacuum states $\ket{0}_{\text{NS}}$ and $\ket{a}_{\text{NS}}$ does not take the constraints \eqref{nullenergymomentum} into account. Therefore, we now introduce them as a set of conditions physical states have to fulfil, in order to reduce ourselves to the physical Hilbert space $\mathcal{H}_{\text{phys}}$. We impose
\begin{equation}\label{physcondNS}
    \begin{split}
        G_r\ket{\text{phys}}_{\text{NS}}&=0 \quad\text{for}\ r> 0\ ,\\
        L_m\ket{\text{phys}}_{\text{NS}}&=0 \quad\text{for}\ m> 0\ ,\\
        \left(L_0-a_0\right)\ket{\text{phys}}_{\text{NS}}&=0 \quad\text{for}\ m> 0
    \end{split}
\end{equation}
in the NS sector, where a normal ordering offset was extracted from $L_0$. For the R sector, instead, we do not need to do the same, since the normal ordering contributions from bosons and fermions cancel. Hence, we have:
\begin{equation}\label{phycondR}
    \begin{split}
        G_r\ket{\text{phys}}_{\text{R}}&=0 \quad\text{for}\ r\geq 0\ ,\\
        L_m\ket{\text{phys}}_{\text{R}}&=0 \quad\text{for}\ m> 0\ ,\\
        L_0\ket{\text{phys}}_{\text{R}}&=0 \quad\text{for}\ m> 0\ .
    \end{split}
\end{equation}
When working with closed strings, a \textit{level matching} condition between left and right-movers must be added in both sectors:
\begin{equation}
    \left(L_0-\Bar{L}_0\right)\ket{\text{phys}}=0\ .
\end{equation}
In conclusion, let's consider the super-Virasoro generator $L_0$ in itself. Its expansion, without the $a_0$ normal ordering constant we have previously extracted, is:
\begin{equation}
\begin{split}
        L_0&=\frac{1}{2}\sum_{p\in\Z}\normord{\alpha_{-p}\cdot\alpha_{p}}+\frac{1}{2}\sum_{r\in\Z+\varphi}\left(r+\frac{n}{2}\right)\normord{b_{-r}\cdot b_{r}}\\
        &=\frac{1}{2}\alpha_0\cdot\alpha_0+\frac{1}{2}\sum_{p\neq 0}\normord{\alpha_{-p}\cdot\alpha_{p}}+\frac{1}{2}\sum_{r\in\Z+\varphi}\left(r+\frac{n}{2}\right)\normord{b_{-r}\cdot b_{r}}\ .
\end{split}
\end{equation}
Since we have:
\begin{equation}
    \alpha_0^\mu=\frac{\pi\sqrt{2\alpha'}}{l}p^\mu\ ,
\end{equation}
the first term for closed strings ($l=2\pi$) is nothing more than
\begin{equation}
    \alpha_0\cdot\alpha_0=\frac{\alpha'}{2}p\cdot p\ ,
\end{equation}
while for open strings ($l=\pi$) it is:
\begin{equation}
    \alpha_0\cdot\alpha_0=2\alpha'p\cdot p\ .
\end{equation}
If we consider on-shell states, with $p\cdot p=-m^2$, which are physical, so that they are annihilated by $L_0+a_0$, we are left with the \textit{mass formula} for open strings
\begin{equation}
    \alpha'm^2_{\text{open}}=N+\frac{\left(\Delta X\right)^2}{4\pi^2\alpha'}+a_0 \ ,
\end{equation}
where a term dependent on the energy stored in the string being stretched appears, and that for closed ones, in which the level matching condition was employed:
\begin{equation}
\alpha'm^2_{\text{closed}}=2\left(N+\Bar{N}+a_0\right)\ .
\end{equation}
If we want the theory to be well-defined and ghost free, we must impose the number of space-time dimensions $D$ to be equal to $10$. Namely, the construction ensures manifest unitarity only in $D=10$. Reproducing the derivation of such result goes beyond the scope of our review. For a detailed discussion, we suggest to refer to \cite{Green:2012oqa,Green:2012pqa}. We limit ourselves at pointing out that, despite the expectation one might have had coming from standard quantum field theory model building, the number of space-time dimensions is not a parameter one can tune from the outside. Conversely, it is fixed by internal consistency. This is the first and most striking way in which superstrings defy our naive approach to phenomenology, imposing stronger and novel constraints on the kind of space-time models we are allowed to build. This aspect of the theory will be explored in chapter \ref{swampchap}. 

\subsection{Spectrum and GSO Projection}
In the following discussion, we will describe the first excited level in the superstring spectrum in the R and NS sectors. We will not construct it explicitly, as is done in the many references cited at the beginning of the chapter. Instead, we will simply outline the procedure and list the major results of interest for the derivation of low energy superstring effective field theories. 
\subsubsection{Construction of the spectrum}
Starting from the NS sector vacuum $\ket{0}_{\text{NS}}$ and the R sector vacuum $\ket{a}_{\text{R}}$, which is further split in two chiralities $\ket{a}_{\text{R}}$ and $\ket{\Dot{a}}_{\text{R}}$, we can produce excited states by acting with the appropriate creation operators, coming from the mode expansions of world-sheet bosons and fermions. The mass level of each state can be computed after having obtained the specific value of the normal ordering offset $a_0$. This is much easier when quantisation is performed in the light-cone scheme, as constraints are imposed at a classical level by effectively reducing the excitable directions to a subset of $D-2$ \textit{transverse} ones. It must be clear that this procedure only accounts for one set of modes and can be directly applied to the sole open string. For what concerns closed ones, we are required to take a tensor product with a second, identical set of inversely moving modes. By doing so, our spectrum seems to be affected by two major problems:
\begin{itemize}
    \item The (NS,NS) ground state appears to be \textit{tachyonic}.
    \item There is an over-abundance of space-time fermionic degrees of freedom, which do not allow to achieve supersymmetry.
\end{itemize}
Fortunately, both such issues are nothing more than artefacts. In fact, the requirement of modular invariance of the one-loop superstring partition function forces us to implement a truncation, which takes the name of \textit{GSO projection}. In order to perform it, we must introduce the \textit{fermion number} operator $\left(-1\right)^F$, assign the eigenvalue $-1$ to $\ket{0}_{\text{NS}}$ and require all states in the NS sector to be eigenstates with eigenvalue $+1$. This way, the theory is made tachyon-free. The GSO projection imposes to introduce an analogous operator for the R sector and choose physical states to have eigenvalue $+1$ or $-1$. When deriving the closed string spectrum, there are thus two inequivalent possibilities: either the same choice has been made in left and right-moving R sectors or not. We label the former case as \textit{type IIB} superstring theory, while the latter is referred to as \textit{type IIA}. Both theories are tachyon-free and can be space-time supersymmetric, since the truncation exactly matches the $128$ space-time bosonic degrees of freedom to an equal number of fermionic ones. A thorough discussion of the subject matter, together with analyses of type I and heterotic superstring theory, can be found in \cite{Blumenhagen:2013fgp}. As far as our brief summary is concerned, it is enough to list the space-time fields corresponding to the various states belonging to type IIA and the IIB low energy spectra, that come directly from the ways in which such states fit into $SO\left(8\right)$ little group representations. This leaves us us with type IIA and type IIB supergravity, respectively characterised by $\mathcal{N}=(1,1)$ and $\mathcal{N}=(2,0)$ supersymmetry. Hence, type IIB supergravity is a \textit{chiral} theory, while type IIA is not. It is now time to list the two sets of massless excitations explicitly, along with the $SO\left(8\right)$ representation in which they transform.
\newpage
\subsubsection{Type IIA}
\begin{itemize}
    \item A scalar $\Phi$, in the $\underbar{1}$ representation.
    \item Two spin $1/2$ dilatinos $\lambda_\alpha$, in the $\underbar{8}_c\oplus\underbar{8}_s$ representation.
    \item A spin $2$ graviton $g_{\mu\nu}$, in the $\underbar{35}_v$ representation.
    \item Two spin $3/2$ gravitinos $\Psi^\mu_\alpha$, in the $\underbar{56}_c\oplus\underbar{56}_s$ representation.
    \item An anti-symmetric $2$-form $B_{\mu\nu}$, in the $\underbar{28}$ representation.
    \item A vector $C_\mu$, in the $\underbar{8}_v$ representation.
    \item An anti-symmetric $3$-tensor $C^{(3)}_{\mu\nu\sigma}$, in the $\underbar{56}_v$ representation.
\end{itemize}
\subsubsection{Type IIB}
\begin{itemize}
    \item Two scalars $\Phi$ and $C$, in the $\underbar{1}$ representation.
    \item Two spin $1/2$ dilatinos $\lambda_\alpha$, in the $\underbar{8}_c\oplus\underbar{8}_c$ representation.
    \item A spin $2$ graviton $g_{\mu\nu}$, in the $\underbar{35}_v$ representation.
    \item Two spin $3/2$ gravitinos $\Psi^\mu_\alpha$, in the $\underbar{56}_c\oplus\underbar{56}_c$ representation.
    \item Two anti-symmetric $2$-forms $B_{\mu\nu}$ and $C^{(2)}_{\mu\nu}$, in the $\underbar{28}\oplus\underbar{28}$ representation.
    \item An anti-symmetric $4$-tensor $C^{(4)}_{\mu\nu\sigma\rho}$, in the $\underbar{35}_s$ representation.
\end{itemize}
After having constructed the type IIA and type IIB low energy space-time spectra, we can now move to a more in depth analysis of the ten-dimensional dynamics induced by superstring theories. This will allow us to detach from the world-sheet framework and focus on space-time phenomenology. We once more stress that a complete treatment of the subject matter should have also included type I and heterotic string theories, which are presented in the above-mentioned references.


%% file: Sections/Strings/EFT.tex
In the previous discussion, the ten-dimensional space-time fields emerging as massless states from type IIA and type IIB superstring theory were listed. Still, no amount of information was given regarding their dynamics. Namely, the spectral analysis did not provide us with the space-time actions associated to the two supergravity theories. In order to derive them, two distinct approaches can be followed.
\paragraph{Scattering amplitudes}
Scattering amplitudes among massless string states can be computed from a world-sheet perspective, employing the powerful techniques offered by path integral quantisation and $2$-dimensional conformal field theories \cite{Polchinski:1998rq,Polchinski:1998rr}. After having identified the various string scattering states with the corresponding space-time degrees of freedom, one is therefore left, at least at tree level, with a complete set of scattering amplitudes among the various space-time effective theory fields. At that point, the ten-dimensional supergravity actions can simply be reverse engineered from such expressions.
\paragraph{Absence of Weyl anomaly}
This alternative approach stems from the idea of considering the world-sheet action of a superstring propagating in a background, in which space-time fields coming from massless string excitations are present. Therefore, a non-linear $\sigma$-model is constructed and space-time degrees of freedom enter it as non-trivial couplings for the bosonic and fermionic world-sheet fields. Thereafter, a set of $\beta$-functions, associated to string scattering amplitudes, are derived, in which the running couplings are precisely the space-time fields coming from a specific superstring theory. In order for the world-sheet Weyl invariance not to be anomalous at a quantum level, all such $\beta$-functions must be set equal to zero. This directly produces a collection of equations of motion for the space-time fields. Once more, the ten-dimensional supergravity actions can then be reverse engineered.
\subsection{Type IIA supergravity}\label{IIAsugra}
Obtaining the explicit space-time form of type IIA and type IIB supergravities goes beyond the scope of the current review. Therefore, we will choose to discuss the low energy effective theory associated to type IIA superstrings, not to encounter the technical problems associated to the presence of a self-dual form field, and state the relevant results without derivations. On top of that, we will only focus on the \textit{bosonic} part of the spectrum. It goes without saying that all the details can be found in the references \cite{Blumenhagen:2013fgp,Polchinski:1998rq,Polchinski:1998rr,Green:2012oqa,Green:2012pqa}. The massless bosonic states in the type IIA spectrum are the graviton $g_{\mu\nu}$, the dilaton $\Phi$, a vector $A_\mu$, a $2$-form $B_{\mu\nu}$ and a $3$-form $C_{\mu\nu\sigma}$. In a more implicit and geometric notation, we refer to the forms as $A_1$, $B_2$ and $C_3$. The fermionic degrees of freedom, comprised of two dilatinos $\lambda_\alpha$ and two gravitinos $\Psi^\mu_\alpha$, will be neglected for the time being. Nevertheless, it can be observed that they would enter the space-time action in way which properly realises $\mathcal{N}=\left(1,1\right)$ supersymmetry. Going back to bosons, we define the field strengths:
\begin{equation}
    F_2\equiv\diff A_1\ ,\quad H_3\equiv\diff B_2\ ,\quad F_4\equiv\diff C_3\ .
\end{equation}
Moreover, we introduce a further $4$-form:
\begin{equation}
    G_4\equiv F_4-A_1\wedge F_3\ .
\end{equation}
In the so-called \textit{string frame}, in which the integral volume element is provided with an $e^{-2\Phi}$ term, the action is 
\begin{equation}\label{IIAaction}
    S_{\text{IIA}}\equiv S_{\text{NS}}+S_{\text{R}}+S_{\text{CS}}\ ,
\end{equation}
where we have identified the (NS,NS) sector, the (R,R) sector and a \textit{Chern-Simons} term. After having defined the $10$-dimensional gravitational coupling 
\begin{equation}
    2\kappa_{10}^2\equiv\left(2\pi\right)^7\alpha'^4\ ,
\end{equation}
the corresponding expressions for the action terms, on a space-time manifold $\mathcal{M}$, are given by:
\begin{equation}
    \begin{split}
      S_{\text{NS}}&=\frac{1}{2\kappa_{10}^2}\int_\mathcal{M}\diff^{10}x\sqrt{-g}e^{-2\Phi}\left(R+4\nabla_\mu\Phi\nabla^\mu\Phi-\frac{1}{2}\left|H_3\right|^2\right)\ ,\\
      S_{\text{R}}&=-\frac{1}{4\kappa_{10}^2}\int_\mathcal{M}\diff^{10}x\sqrt{-g}e^{-2\Phi}\Bigl(\left|F_2\right|^2+\left|G_4\right|^2\Bigr)\ ,\\
      S_{\text{CS}}&=-\frac{1}{4\kappa_{10}^2}\int_\mathcal{M}B_2\wedge F_4\wedge F_4\ .
    \end{split}
\end{equation}
For the sake of clarity, the Chern-Simons term has been written in a compact geometric notation, so that the volume element became implicit.

%% file: Sections/Strings/compactification.tex
After having introduced superstring theory from the classical world-sheet perspective, having quantised the theory following the old covariant prescription and having constructed the massless spectra for type IIA and type IIB superstrings via the implementation of a GSO projection, the low energy space-time effective theory action associated to the former was stated without an explicit derivation. It is clear that an action of the form \eqref{IIAaction} is far from being connected to particle physics and general relativity. It should not be forgotten, indeed, that superstring theory seems to be consistently defined in $10$ space-time dimensions, while the world we live in appears to be $4$-dimensional. How can such a drastic difference be dealt with and accounted for, when constructing viable models from superstrings? This is arguably one the most critical questions in string theory phenomenology. As far as the following discussion is concerned, we will do nothing more than exploring the simplest possible example of the technique which is most commonly employed towards such goal: \textit{compactification}. With this term, we refer to the idea that the $10$-dimensional space-time appearing in superstring theory might factorised as the product of a four dimensional Lorentzian manifold and a, perhaps non-trivially fibered, six dimensional compact geometric object, too small to be accessible at our current energy scales. This way, the six \textit{extra} dimensions would effectively disappear from any low energy theory, producing a four-dimensional universe similar to the one we perceive. Compactification has been explored for decades in all its technical and mathematical details, as a powerful model-building tool. Some rather complete references on such topic can be found in \cite{Ibanez:2012zz,Tomasiello:2022dwe,Garcia-Compean:2005thu,Anderson:2018pui,Kachru:2000tg,Lutken:1989hm,Greene:1996cy,Grana:2011ydm,Blumenhagen:2006ci}. The idea first appeared in the works \cite{Kaluza1921,Klein1926}. Here we want to investigate compactification in a controlled setting. Thus, we will take the low energy effective theory coming from type IIA superstring theory, switch off all its field content except for the metric and the dilaton and impose one of the dimensions to be a small compact circle. We will therefore be left with a $9$-dimensional effective model.
\subsection{Circle compactification}\label{circlecompactification}
As briefly outlined above, we start from the bosonic part of the low energy effective supergravity theory coming from type IIA superstrings and switch off all form fields, leaving only the metric $g_{\mu\nu}$ and the dilaton $\Phi$. Hence, we are left with:
\begin{equation}
    S=\frac{1}{2\kappa_{10}^2}\int_\mathcal{M}\diff^{10}x\sqrt{-g}e^{-2\Phi}\bigl(R_g+4\nabla_\mu\Phi\nabla^\mu\Phi\bigr)\ .
\end{equation}
Before performing the circle compactification along one spatial direction, we move from the so called string frame, defined in our discussion of space-time actions, to the \textit{Einstein frame}, in which the $\exp{-2\Phi}$ term is absorbed into a rescaling of the metric. In order to do so, we introduce $G_{\mu\nu}$, $\omega$, $\phi$ and a constant $\phi_0$, such that:
\begin{equation}
    g_{\mu\nu}=e^{2\omega}G_{\mu\nu}\ ,\quad \Phi\equiv\phi+\phi_0\ .
\end{equation}
Analysing the expressions appearing in the action one by one, we get:
\begin{equation}
    \begin{split}
        \sqrt{-g}&=e^{10\omega}\sqrt{-G}\ ,\quad\nabla_{\mu}\Phi=\nabla_\mu\phi\ ,\quad e^{-2\Phi}=e^{-2\left(\phi+\phi_0\right)}\ ,\\
        R_g&=e^{-2\omega}R_G-18e^{-2\omega}\nabla^2\omega-72e^{-2\omega}\nabla_\mu\omega\nabla^\mu\omega\\
        g^{\mu\nu}&\nabla_\mu\phi\nabla_\nu\phi=e^{-2\omega}G^{\mu\nu}\nabla_\mu\phi\nabla_\nu\phi\ .
    \end{split}
\end{equation}
Therefore, the action becomes:
\begin{equation}
\begin{split}
        S&=\frac{1}{2\kappa_{10}^2}\int_\mathcal{M}\diff^{10}x\sqrt{-G}e^{8\omega-2\left(\phi+\phi_0\right)}\Bigl(R_G-18\nabla^2\omega\\
    &\quad\quad\quad\quad\quad-72\nabla_\mu\omega\nabla^\mu\omega+4\nabla_\mu\phi\nabla^\mu\phi\Bigr)\ .
\end{split}
\end{equation}
By imposing the rescaling parameter $\omega$ to be
\begin{equation}
    \omega\equiv\frac{\phi+\phi_0}{4}\ ,
\end{equation}
we are left with:
\begin{equation}
\begin{split}
        S&=\frac{1}{2\Bar{\kappa}_{10}^2}\int_\mathcal{M}\diff^{10}x\sqrt{-G}\left(R_G-\frac{9}{2}\nabla^2\phi-\frac{1}{2}\nabla_\mu\phi\nabla^\mu\phi\right)\ .
\end{split}
\end{equation}
In the above expression, we have defined:
\begin{equation}
    \Bar{\kappa}_{10}\equiv e^{\phi_0}\kappa_{10}\ .
\end{equation}
By removing the irrelevant boundary term produced by the Laplacian of $\phi$, we simply get:
\begin{equation}
\begin{split}\label{actionsimple}
        S&=\frac{1}{2\Bar{\kappa}_{10}^2}\int_\mathcal{M}\diff^{10}x\sqrt{-G}\left(R_G-\frac{1}{2}\nabla_\mu\phi\nabla^\mu\phi\right)\ .
\end{split}
\end{equation}
This way, we have obtained the Einstein frame version of our type IIA supergravity action, reduced to the metric-dilaton sector. In order to proceed with the circle compactification, we take the space-time manifold to be (at least locally) factorised as
\begin{equation}
    \mathcal{M}\equiv\mathcal{P}\times S^1\ ,
\end{equation}
where $\mathcal{P}$ is a $9$-dimensional Lorentzian manifold and $S^1$ is a circle, whose radius might depend on the coordinates on $\mathcal{P}$. In order not to create confusion, we use:
\begin{itemize}
    \item The symbol $y$ to refer to the circular coordinate, with $y\in\left[0,2\pi\rho\right)$. 
    \item The symbol $x^M$ to refer to the coordinates $\left(x^0,\dots,x^8\right)$ on $\mathcal{P}$.
\end{itemize}
Concerning the $10$-dimensional metric $G_{\mu\nu}$, we assume it to give rise to the specific line-element
\begin{equation}
    \diff s^2_{10}\equiv h_{MN}\left(x\right)\diff x^M\diff x^N+e^{2\varphi\left(x\right)}\diff y^2\ ,
\end{equation}
where $h_M$ and $\varphi$ are the metric tensor on $\mathcal{P}$ and a scalar, respectively. For the sake of simplicity and partially sacrificing the generality of our discussion, which has nonetheless a purely illustrative aim, we have assumed all matrix entries of the form $G_{yM}$ to vanish. As can be easily observed, both $h_{MN}$ and $\varphi$ are assumed to only depend on the non-compact coordinates. The non-triviality of $\varphi$, which we will refer to as the \textit{radion} field, allows the effective radius of the compact dimension $S^1$ to vary on $\mathcal{P}$. We can directly compute:
\begin{equation}
    \sqrt{-G}=e^{\varphi}\sqrt{-h}\ .
\end{equation}
Furthermore, we have:
\begin{equation}
\begin{split}
        R_{G}&=G^{\mu\nu}R^\alpha_{\ \mu\alpha\nu}=h^{MN}R^\alpha_{\ M\alpha N}+e^{-2\varphi}R^\alpha_{\ y\alpha y}\\
        &=h^{MN}R^P_{\ MPN}+h^{MN}R^y_{\ MyN}+e^{-2\varphi}R^P_{\ yP y}\ .
\end{split}
\end{equation}
The only non-zero Christoffel symbols are those of the form:
\begin{equation}
\begin{split}
        \Gamma^M_{\ NP}&=\frac{1}{2}g^{MQ}\left(\de_Ng_{QP}+\de_Pg_{NQ}-\de_Qg_{NP}\right)\ ,\\
     \Gamma^y_{\ yP}&=\de_P\varphi\ ,\quad \Gamma^M_{\ yy}=-e^{2\varphi}\de^M\varphi\ .
\end{split}
\end{equation}
Thus, we have the simple expression:
\begin{equation}\label{scaldecomp}
    R_G=R_h-2\nabla^2\varphi-2\nabla_M\varphi\nabla^M\varphi\ .
\end{equation}
Concerning the scalar, we Fourier-expand it as:
\begin{equation}
\phi\left(x,y\right)=\sum_{n\in\Z}\phi_n\left(x\right)e^{iny/\rho}\ .
\end{equation}
The kinetic term in \eqref{actionsimple} becomes:
\begin{equation}
    \nabla_\mu\phi\nabla^\mu\phi=h^{MN}\nabla_M\phi\nabla_N\phi+e^{-2\varphi}\nabla_y\phi\nabla_y\phi
\end{equation}
After having integrated out the $y$ coordinate and having defined the $9$-dimensional gravitational coupling
\begin{equation}
    \Bar{\kappa}_{9}^2\equiv\frac{\Bar{\kappa}_{10}^2}{2\pi\rho}\ ,
\end{equation}
the action takes the dimensionally-reduced form:
\begin{equation}\label{actionfinal}
\begin{split}
        S&=\frac{1}{2\Bar{\kappa}_{9}^2}\int_\mathcal{P}\diff^{9}x\sqrt{-h}e^{\varphi}\Biggl\{R_h-2\nabla^2\varphi-2\left(\nabla\varphi\right)^2\\
        &\quad\quad\quad\quad\quad-\frac{1}{2}\sum_{n\in\Z}\left[\left(\nabla\phi_n\right)^2+e^{-2\varphi}\frac{n^2}{\rho^2}\phi_n^2\right]\Biggr\}\ .
\end{split}
\end{equation}
In order to move to Einstein frame, we impose the conformal transformation
\begin{equation}
    h_{MN}\equiv e^{2\gamma\varphi}k_{MN}\ ,
\end{equation}
rescale the radion as $\varphi=\beta\eta$ and obtain the following expression:
\begin{equation}\label{actionfinalef}
\begin{split}
        S&=\frac{1}{2\Bar{\kappa}_{9}^2}\int_\mathcal{P}\diff^{9}x\sqrt{-k}e^{\left(1+7\gamma\right)\beta\eta}\Biggl\{R_k-\left(16\gamma+2\right)\beta\nabla^2\eta-\left(56\gamma^2+2\right)\beta^2\left(\nabla\eta\right)^2\\
        &\quad\quad\quad\quad\quad-\frac{1}{2}\sum_{n\in\Z}\left[\left(\nabla\phi_n\right)^2+e^{2\left(\gamma-1\right)\beta\eta}\frac{n^2}{\rho^2}\phi_n^2\right]\Biggr\}\ .
\end{split}
\end{equation}
By imposing $7\gamma=-1$, the above expression simplifies as:
\begin{equation}\label{actionfinalef2}
\begin{split}
        S&=\frac{1}{2\Bar{\kappa}_{9}^2}\int_\mathcal{P}\diff^{9}x\sqrt{-k}\Biggl\{R_k-\beta^2\frac{22}{7}\left(\nabla\eta\right)^2-\frac{1}{2}\sum_{n\in\Z}\left[\left(\nabla\phi_n\right)^2+e^{-16\beta\eta/7}\frac{n^2}{\rho^2}\phi_n^2\right]\Biggr\}\ .
\end{split}
\end{equation}
Since the Laplacian terms account for nothing more than a boundary term, they were safely removed. We can hence select $\beta$ so that
\begin{equation}
   \frac{22}{7}\beta^2=\frac{1}{2}\Longrightarrow \beta=\sqrt{\frac{7}{44}}
\end{equation}
and obtain the properly rescale Einstein frame action:
\begin{equation}\label{actionfinalef3}
\begin{split}
        S&=\frac{1}{2\Bar{\kappa}_{9}^2}\int_\mathcal{P}\diff^{9}x\sqrt{-k}\Biggl\{R_k-\frac{1}{2}\left(\nabla\eta\right)^2-\frac{1}{2}\sum_{n\in\Z}\left[\left(\nabla\phi_n\right)^2+\frac{n^2}{\rho^2}\phi_n^2e^{-8\eta/\sqrt{77}}\right]\Biggr\}\ .
\end{split}
\end{equation}
Expanding the exponential coupling to $\eta$ as a power series, we get:
\begin{equation}
\begin{split}
        e^{-8\eta/\sqrt{77}}\frac{n^2}{\rho^2}\phi_n^2&=\frac{n^2}{\rho^2}\phi_n^2\left(1-\eta\frac{8}{\sqrt{77}}+\frac{64}{77}\eta^2-\dots\right)\ .
\end{split}
\end{equation}
Therefore, the dynamics of each one of the \textit{Kaluza-Klein} modes $\phi_n$ is controlled by a collection of masses
\begin{equation}\label{KKmass}
    m_n\equiv\frac{n}{\rho}
\end{equation}
and a family of infinitely many higher-order interactions, containing $\phi_n^2$, with the radion field $\eta$. By introducing an order $\mathcal{O}(1)$ constant $\alpha$, writing the radius parameter $\rho$ as
\begin{equation}\label{distancefirst}
    \rho\equiv \exp{\alpha\Delta} \ ,
\end{equation}
referring to $\Delta$ with the term \textit{distance}, from the value $\rho_0=1$, and promising such nomenclature to become much clearer in chapter \ref{swampchap}, we have that our low energy effective theory in $9$-dimensions features an infinite tower of species $\phi_n$ that get exponentially lighter in the large-distance regime. In particular, they become massless when $\Delta\mapsto\infty$. This behaviour appears to be a universal feature of string compactifications and will broadly generalised when discussing the \textit{swampland distance conjecture}. The instability of $\eta$ in \eqref{actionfinalef3} is not to worry about: indeed, most of the matter content has been neglected and we have performed a trivial and phenomenologically uninteresting compactification. It must be stressed that the approach we have followed in our derivation, while working perfectly well when considering a space-time quantum field theory as fundamental and studying its compactified dynamics, does not fully capture the richness of string theory compactification. Simply taking the usual effective field theory coupled to gravity emerging from type-IIA superstings, as derived \ref{IIAsugra}, and compactifying it on a circle is not enough. The reason is that the topological features of space-time already influence the theory before quantisation, when the string equations of motion are solved in a chosen background. The presence of a compact dimension in the target space-time influences the boundary conditions on the world-sheet degrees of freedom. Therefore, in order not to neglect relevant parts of the spectrum, we are demanded to start from the world-sheet formulation of superstring theory, impose space-time to have the desired topology and then obtain the appropriate string states. For the case at hand, this will be rapidly analysed in the following discussion.

\subsubsection{World-sheet perspective}\label{WSS1}
\begin{figure}[h]
    \centering
\begin{tikzpicture}[scale=1.7]
\fill [gray!40,opacity=0.5](0,0) ellipse (1.25 and 0.5);
\draw [black](0,0) ellipse (1.25 and 0.5);
\draw (-1.25,0) -- (-1.25,-3.5);
\draw (-1.25,-3.5) arc (180:360:1.25 and 0.5);
\draw (1.25,-3.5) -- (1.25,0);  
\fill [gray!20,opacity=0.5] (-1.25,0) -- (-1.25,-3.5) arc (180:360:1.25 and 0.5) -- (1.25,0) arc (0:180:1.25 and -0.5);
\draw [dashed] (-1.25,-3.5) arc (180:360:1.25 and -0.5);
\draw [blue] (-1.25,-1.2) arc (180:360:1.25 and 0.5);
\draw [blue,dashed] (-1.25,-1.2) arc (180:360:1.25 and -0.5);
\draw [red,dashed] (-1.25,-2.4) arc (180:360:1.25 and -0.5);
\draw [red] (1.24,-2.4) arc (0:155:1.3 and -0.5);
\draw [red,dashed] (-1.25,-2.6) arc (180:360:1.25 and -0.5);
\draw [red] (-1.24,-2.4) arc (180:335:1.3 and 0.5);
\draw [black] (-1.25,0) -- (-1.25,-3.5) arc (180:360:1.25 and 0.5) -- (1.25,0) arc (0:180:1.25 and -0.5);
\draw[->] (-1.7,-3.5,0) -- (-1.7,0,0) node[pos=1.05] {$\mathcal{P}$};
\draw[->] (-1.7,-3.5) arc (180:330:1.7 and 0.95) node[pos=1.07] {\text{  }$S^1$};
\end{tikzpicture}
    \caption[Schematic depiction of a circle compactification.]{Schematic depiction of a circle compactification, in which the non compact dimensions charting the reduced space-time manifold $\mathcal{P}$ are represented by a one-dimensional, vertical direction. The blue and red closed strings, included for the sake of clarity, have winding numbers $\omega_1=\pm 1$ and $\omega_2=\pm 2$, respectively.}
    \label{fig:cirle_winding}
\end{figure}
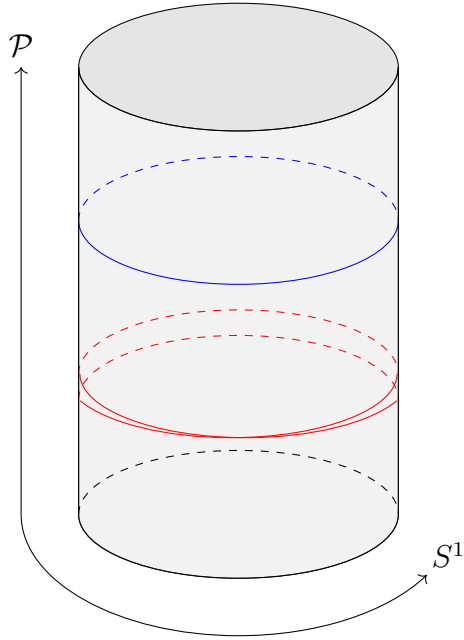

For the time being, the radon field $\eta$, whose dynamics locally controls the value of the compact dimension radius, will be imposed to be everywhere zero and removed from the effective theory. This will largely simplify our discussion, preserving its core conceptual content intact while, at the same time, making the formulas way more readable. The extension to a non-zero, fluctuating and unfrozen $\eta$ can be straightforwardly obtained by following analogous steps, while keeping track of the extra contribution from the radion. The interested reader might want to refer to \cite{Tong:2009np,Greene:1996cy,Palti:2019pca} for a more thorough discussion. That said, let's focus on the quantum mechanical properties of a closed string propagating in the compactified space-time manifold under scrutiny. The position-space representation of its state can, in general, be written as a Fourier decomposition over momentum eingenstates
\begin{equation}\label{expansion}
    \psi\left(x,y\right)=\int\diff^9p\ \diff q\ \Tilde{\psi}\left(p,q\right)e^{i\Bar{p}\cdot\Bar{x}+iqy}\ ,
\end{equation}
where all constants have been absorbed into $\Tilde{\psi}\left(p,q\right)$, the momentum variables $p_M$ are conjugate to the non-compact spatial coordinates $x^M$ and $q$, instead, corresponds to the coordinate along the circle. In order for the wave-function to be single valued, it must be periodic with period $2\pi\rho$ along the $y$ compact direction. Thus, we have:
\begin{equation}
    \psi\left(x,y\right)=\psi\left(x,y+2\pi\rho\right)\ .
\end{equation}
Imposing the above condition to \eqref{expansion}, one directly obtains:
\begin{equation}
    e^{iqy}=e^{iq\left(y+2\pi\rho\right)}\ .
\end{equation}
Therefore, purely quantum mechanical arguments force to impose the momentum $q$ along the compact direction to be quantised as:
\begin{equation}\label{quantq}
    q=\frac{n}{\rho}\ ,\quad n\in\Z\ .
\end{equation}
When solving the equations of motion for the string bosonic space-like coordinates in Minkowski, or in any other space-time manifold with no compact directions, a strict periodicity condition had to be imposed on all of them. Now, such constraint can be partially relaxed along the circle. Indeed, we have:
\begin{equation}
    \begin{split}
        X^{M}(\tau,\sigma+2\pi)=X^{M}(\tau,\sigma)\ ,\quad i=1,\dots,8,\\
        Y(\tau,\sigma+2\pi)=Y(\tau,\sigma)+2\pi \omega\rho\ ,\quad \omega\in\Z\ .
    \end{split}
\end{equation}
The capital letters have been employed for the notation to be consistent, when regarding the space-time coordinates as world-sheet bosons. The integer $\omega$ is referred to as the closed string \textit{winding number} and quantifies how many times a string wraps around the compact $S^1$ before achieving periodicity. The winding number is integer, instead of natural, since its sign allows to distinguish between clock-wise and counter-clockwise wrappings. In figure \ref{fig:cirle_winding}, a pictorial representation of the phenomenon at hand is presented. The nine spatio-temporal non-compact dimensions charting $\mathcal{P}$ have been collapsed to a single, one-dimensional direction, which takes the role of the height of the cylinder. This naturally prevents from appreciating the Lorentzian nature of the space-time metric. Nevertheless, it offers a chance to clearly represent various strings wrapping around the compact direction. The blue one has winding number $\omega_1=\pm 1$, while the red one has winding number $\omega_1=\pm 2$. The sign, as previously discussed, purely depends on the orientation of the strings.
By identifying the left-moving are right-moving contributions to $Y(\tau,\sigma)$ as 
\begin{equation}
    Y(\tau,\sigma)= Y_L(\sigma^{+})+Y_R(\sigma^{-}) \, ,
\end{equation}
where $\sigma^{\pm}\equiv\tau\pm\sigma$, the equations of motion can be solved in the usual way:
\begin{equation}
    \begin{split}
        Y_L(\sigma^+)&=\frac{y}{2}+\frac{\alpha^{'}}{2}\left(\frac{n}{\rho}+\frac{\omega\rho}{\alpha^{'}}\right)\sigma^{+}+i\sqrt{\frac{\alpha^{'}}{2}}\sum_{n\neq 0}\frac{\tilde{\alpha}^{25}_{n}}{n}e^{-in\sigma^{+}}\ ,\\
         Y_{R}(\sigma^{-})&=\frac{y}{2}+\frac{\alpha^{'}}{2}\left(\frac{n}{\rho}-\frac{\omega\rho}{\alpha^{'}}\right)\sigma^{-}+i\sqrt{\frac{\alpha^{'}}{2}}\sum_{n\neq 0}\frac{\alpha^{25}_{n}}{n}e^{-in\sigma^{-}}\ .
    \end{split}
\end{equation}
In the above, we can identify the left-moving and right-moving components of the string momentum along the circular dimension to be:
\begin{equation}
    q_L\equiv \frac{n}{\rho}+\frac{\omega\rho}{\alpha^{'}}\ ,\quad q_R\equiv \frac{n}{\rho}-\frac{\omega\rho}{\alpha^{'}}\ .
\end{equation}
It goes without saying that the total $y$-momentum $q$ is given by
\begin{equation}
    q=\frac{q_L+q_R}{2}\ ,
\end{equation}
consistently with the result obtained in \eqref{quantq}. The level-matching condition, in presence of a dimension compactified to a circle, loosens up as
\begin{equation}
    N-\Bar{N}=\omega\cdot n\ ,
\end{equation}
where $N$ and $\Bar{N}$ are the number operators associated to the two families of movers. The effective, $9$-dimensional squared mass operator on which an observer perceiving the non-compact space-time manifold $\mathcal{P}$ can perform measurements can be shown to be equal to:
\begin{equation}\label{masseffective}
    M^2_{\mathcal{P}}=\frac{ n^{2}}{\rho^{2}}+\frac{\omega^{2}\rho^{2}}{\alpha^{'2}}+\frac{2}{\alpha^{'}}\left(N+\tilde{N}-2\right)\ .
\end{equation}
The standard reference \cite{Tong:2009np} offers a clear and pedagogical derivation of the above result. Hence, string states acquire, from the perspective of the dimensionally reduced effective theory, two novel mass contributions. The first one solely depends on the quantum mechanical wave-function being well-defined and comes from the energy stored in the number of momentum quanta along the compact direction. Being it completely independent from string theory, as it would be present if the fundamental degree of freedom was taken to be a point particle, it coherently does not include $\alpha'$. It must be stressed that this is perfectly consistent with what was obtained from explicitly compactifying the space-time effective theory, from which this term was read off in \eqref{KKmass}. The second contribution, instead, emerges from the string winding modes, introducing an energy off-set associated to the stretching string tension. This is a genuine string-theoretic effect, with no direct quantum field theory analogue. The third one corresponds, as in the non-compact theory, to the energy coming from right-moving and left-moving string excitations. 
\subsubsection{T-duality and towers of states}\label{sssec:Tduality}
\begin{figure}[h]
    \centering
  \includegraphics[width=0.8\linewidth]{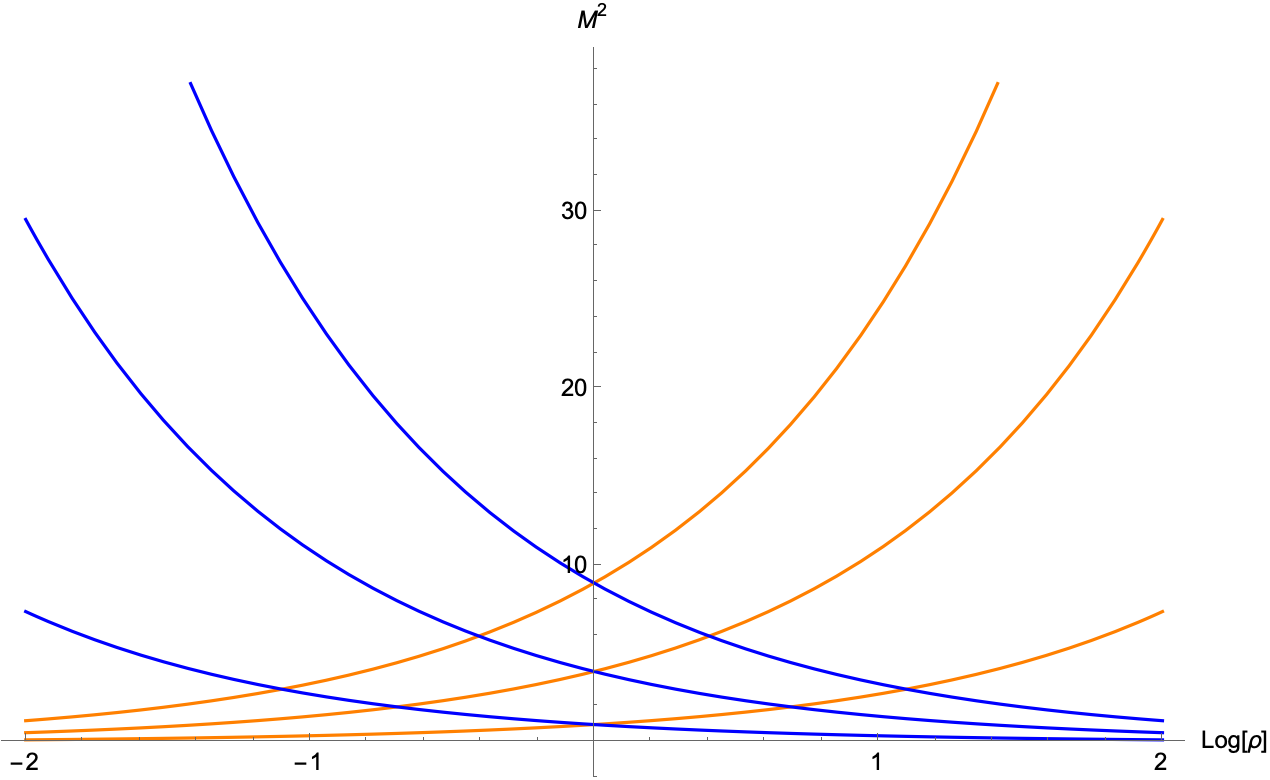}
    \caption[Masses of momentum and winding modes, expressed as functions of the compact dimension radius.]{In this figure, the dependence of the three lighter winding and momentum states on the logarithm of the compact dimension radius was depicted. As $\rho$ is sent to $\infty$, the winding modes (orange) become infinitely massive, while the momentum modes (orange) become massless. When it is, instead, sent to $0$, the opposite behaviours are achieved.}
    \label{fig:KKmasses}
\end{figure}

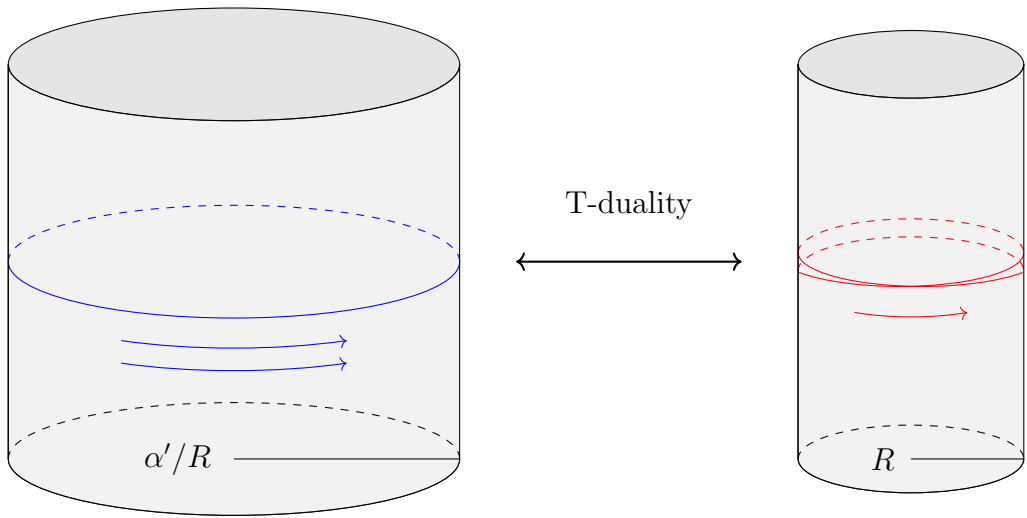
\begin{figure}[h]
    \centering
\begin{tikzpicture}[scale=1.5]
\fill [gray!40,opacity=0.5](-3,0) ellipse (2 and 0.5);
\draw [black](-3,0) ellipse (2 and 0.5);
\draw (-5,0) -- (-5,-3.5);
\draw (-5,-3.5) arc (180:360:2 and 0.5);
\draw (-5,-3.5) -- (-5,0);  
\fill [gray!20,opacity=0.5] (-5,0) -- (-5,-3.5) arc (180:360:2 and 0.5) -- (-1,0) arc (0:180:2 and -0.5);
\draw [dashed] (-5,-3.5) arc (180:360:2 and -0.5);
\draw [blue] (-5,-1.75) arc (180:360:2 and 0.5);
\draw [blue,dashed] (-5,-1.75) arc (180:360:2 and -0.5);
\draw [black] (-5,0) -- (-5,-3.5) arc (180:360:2 and 0.5) -- (-1,0) arc (0:180:2 and -0.5);
\fill [gray!40,opacity=0.5](3,0) ellipse (1 and 0.3);
\draw [black](3,0) ellipse (1 and 0.3);
\draw (2,0) -- (2,-3.5);
\draw (2,-3.5) arc (180:360:1 and 0.3);
\draw (2,-3.5) -- (2,0);  
\fill [gray!20,opacity=0.5] (2,0) -- (2,-3.5) arc (180:360:1 and 0.3) -- (4,0) arc (0:180:1 and -0.3);
\draw [dashed] (2,-3.5) arc (180:360:1 and -0.3);

\draw [red,dashed] (2,-1.67) arc (180:360:1 and -0.3);
\draw [red,dashed] (2,-1.83) arc (180:360:1 and -0.3);
\draw [red] (4,-1.67) arc (0:144:1.1 and -0.3);
\draw [red] (2,-1.67) arc (180:324:1.1 and 0.3);
\draw [black] (2,0) -- (2,-3.5) arc (180:360:1 and 0.3) -- (4,0) arc (0:180:1 and -0.3);
    \draw[<->,thick] (-0.5,-1.75) -- (1.5,-1.75);
    \node at (0.5,-1.25) {T-duality};
\draw (-3,-3.5) -- (-1,-3.5);
\node at (-3.5,-3.5) {$\alpha'/R$};
\draw (3,-3.5) -- (4,-3.5);
\node at (2.75,-3.5) {$R$};
\draw[blue,->] (-4,-2.45) arc (240:300:2 and 0.5);
\draw[blue,->] (-4,-2.65) arc (240:300:2 and 0.5);
\draw[red,->] (2.5,-2.2) arc (240:300:1 and 0.3);
\end{tikzpicture}
    \caption[Pictorial representation of T-duality.]{In this figure, the action of T-duality on the radius of the compact dimension is pictorially represented. From the perspective of a $9$-dimensional observer, only able to probe the dimensionally-reduced theory, the two configurations are indistinguishable, as long as winding and momentum modes are swapped. The number $n$ of momentum quanta is represented by the number of arrows along the compact direction, while $\omega$ is naturally depicted by the actual windings. Hence, the blue string has $\left(n_1,\omega_1\right)=\left(2,1\right)$. The red one has, subsequently, $\left(n_2,\omega_2\right)=\left(1,2\right)$.}
   \label{fig:tdual}
\end{figure}

Exploring the full circle-compactified string spectrum clearly goes beyond the scope our current analysis, if only because it would force us to decompose all those fields that we have neglected. Here, instead, we want to focus on a particular property that proves itself to be much more general, in the context of string theory model building: \textit{T-duality}. In order not to let other non-central aspects of the issue at hand to get in the way of our discussion, we will only consider string states for which:
\begin{equation}
    N+\Tilde{N}=2\ .
\end{equation}
Hence, the squared mass-formula derived in \eqref{masseffective} for Kaluza-Klein states straightforwardly reduces to the expression
\begin{equation}\label{lightmodes}
    M^2\left(n,\omega\right)=\frac{ n^{2}}{\rho^{2}}+\frac{\omega^{2}\rho^{2}}{\alpha^{'2}}\ ,
\end{equation}
which only depends on the number $n$ of momentum quanta along the compact direction and on the winding mode $\omega$, once the circle radius $\rho$ is fixed. Since both quantities only appear squared, the effective mass does not allow to tell the orientations of the string's winding and momentum. First and foremost, we should observe that:
\begin{itemize}
    \item When taking the $\rho\mapsto\infty$ limit, the mass-contribution of the winding number strongly dominates that coming from the momentum quanta.
    \item When taking the $\rho\mapsto 0$ limit, the mass-contribution of the momentum quanta strongly dominates that coming from the winding number.
\end{itemize}
It is particularly important, for reasons that will be soon made clear, to focus on two specific families of states in the $9$-dimensional effective theory. Namely, those for which either $\omega=0$ or $n=0$. In the first case, we have
\begin{equation}
    M^2_{n}=\frac{ n^{2}}{\rho^{2}}\ ,
\end{equation}
while the second provides us with:
\begin{equation}
    M^2_\omega=\frac{\omega^{2}\rho^{2}}{\alpha^{'2}}\ .
\end{equation}
We refer to such \textit{towers} of states, where the term comes from the tower-like distribution of the states' masses, as those corresponding to  \textit{momentum} and \textit{winding states}, respectively. Interestingly enough, they show opposite mass-behaviours with respect to the radius $\rho$ of the compact dimension. In figure \ref{fig:KKmasses} this was shown using $\log{\rho}$ as a variable, consistently with the notion of distance introduced in \eqref{distancefirst}.
Therefore, both the $\rho\mapsto\infty$ and the $\rho\mapsto 0$ limit are accompanied by the appearance of infinitely many massless fields, from the perspective of the $9$-dimensional theory. Regarding the circle radius $\rho$ as a \textit{modulus} of the theory, belonging to a $1$-dimensional moduli space
\begin{equation}
    \rho\in\mathrm{M}\sim\R_+\equiv\left[0,\infty\right)
\end{equation}
and endowing such manifold with the distance
\begin{equation}
    \Delta\left(\rho_2,\rho_1\right)\sim\log{\frac{\rho_2}{\rho_1}}\ ,
\end{equation}
we have that, starting from any finite value of the radius, all infinite distance limits in the moduli space are characterised by an infinite tower of massless fields. Hence, they are inconsistent. This feature will be further discussed in \ref{sec:distanceconj}, in the context of the swampland distance conjecture. Going back to the more general formula \eqref{lightmodes}, we can observe that it is left unchanged by the following substitutions:
\begin{equation}
    \rho\longrightarrow\frac{\alpha'}{\rho}\ ,\quad m\longleftrightarrow n\ .
\end{equation}
This striking property, geometrically represented in figure \ref{fig:tdual}, is typically referred to as T-duality. In practice it implies that, for a $9$-dimensional observer, a universe compactified on a circle with radius $\rho$ is indistinguishable from one compactified on a circle with radius $\alpha'/\rho$, as long as winding and momentum states are swapped. Hence, the answer to a simple question such as
\begin{center}
  \textit{What's the radius of the extra compact spatial dimension?}  
\end{center}
has two possible answers in the dimensionally-reduced effective theory, unless a specific \textit{duality frame} is chosen. Once more, we must stress that this was made possible due to the presence of strings with non-zero length, as standard quantum field theory would have not produced the winding number contribution to the masses of Kaluza-Klein modes. Superstring theory exhibits many of such dualities, connecting apparently distinct phenomenologies. Compactified type-IIA and type-IIB superstring theories themselves are, in fact, T-dual to each other. The standard references \cite{Polchinski:1998rq,Polchinski:1998rr,Greene:1996cy}, among others, cover the topic in great detail. For now, it is only significant to stress that, via T-duality, strings somehow realise a form of \textit{IR/UV mixing}, in which large and short distances -and hence small and large energy regimes- stop being decoupled and independent from each other. As will be broadly commented on in section \ref{swamplandintro}, this is expected to be a general property of quantum gravity and defies standard effective field theory reasoning.

%% file: Sections/ASwamp/swampintro.tex
After having outlined the main features of superstring theory and having described the non-trivial techniques via which space-time dynamics is obtained from the world-sheet action, the notions of compactification and dimensional reduction were introduced as ways to extract real world phenomenology from $10$-dimensional supergravity. It is now the appropriate time to discuss the most striking implications of such a framework on the shared properties of superstring low energy effective theories. The attempt at constructing viable, $4$-dimensional and predictive extensions of the standard model, coupled to a dynamical background space-time, in the context of superstring theory is by no means a new research line. It has been, on the contrary, an active and fertile field of enquiry for decades \cite{Ellis:1986qt,Ellis:1993xc,Blumenhagen:2001te,Blumenhagen:2006ci,Quevedo:1996sv,Ginsparg:1987ee,Ibanez:1992hc,Ibanez:1995tm,Witten:1985xc,Dine:1986bb,Dine:2000bf}. In fact, string theory was originally conceived as a model intended to explain the \textit{Regge slopes} appearing in hadron experiments \cite{Veneziano:1968yb,Koba:1969rw,Virasoro:1969me,Shapiro:1970gy}, and only then showed its potential as a unified theory of quantum gravity and matter. The interest in phenomenology was thus rooted in superstring theory from its birth. Nonetheless, it was with the initiation of the so-called \textit{swampland program} \cite{Vafa:2005ui}, which allowed to systematise a huge body of results in the light of a new set of organisational principles, that our understanding of the subject made its most significant leap forward. This will be the topic of the following chapter. Our discussion will be largely inspired by the standard references \cite{Palti:2019pca,Palti:2022edh,Brennan:2017rbf,Tadros:2022rkd,Casse:2022ymj,Agmon:2022thq}, but it will only cover a small portion of the available research. Namely, after a general introduction to the distinction between the string theory landscape and its complementary swampland, we will solely direct our attention to the distance conjecture and generalisations thereof. If interested in exploring the philosophical foundations of superstring phenomenology, in which the problem of non-empirical theory assessment gets central and unignorable, the reader might want to refer to \cite{Dawid1993StringTA,Dawid2016ModellingNC,Dawid2015TheNA,Cabrera2018StringTN,Polchinski2016WhyTA,rickles2014brief}.

%% file: Sections/ASwamp/eftsconstraints.tex
\begin{figure}[H]
\centering
  \includegraphics[width=0.7\linewidth]{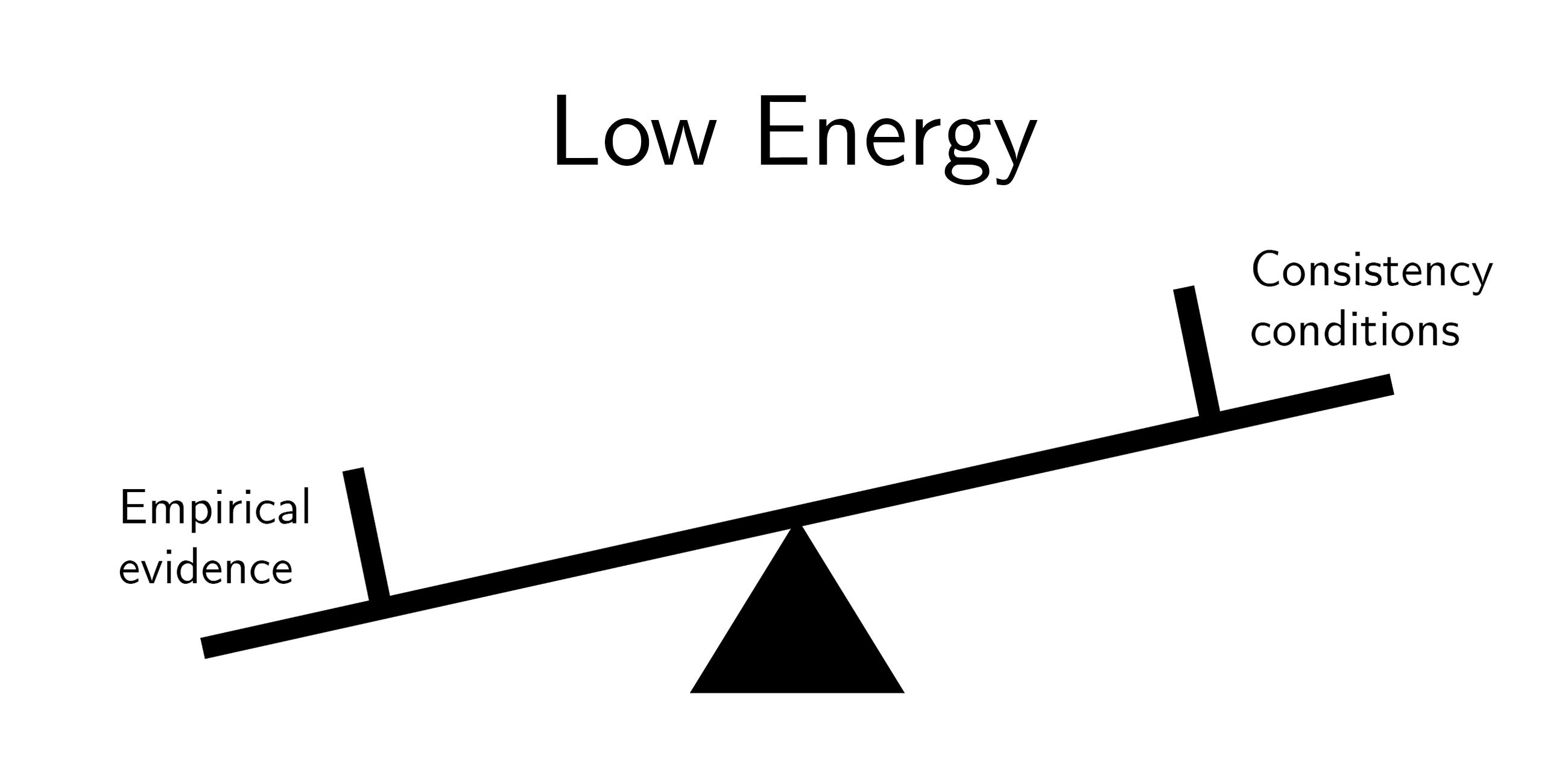}
\caption[Significance of empirical evidence in the low energy regime.]{In the low energy limit, constraints imposed by self-consistency loosen up, while experimental data becomes easier to gather. Therefore, empirical evidence \textit{weights} more than consistency conditions.}
\label{fig:lowenergy}
\end{figure}
\begin{figure}[H]
\centering
  \includegraphics[width=0.7\linewidth]{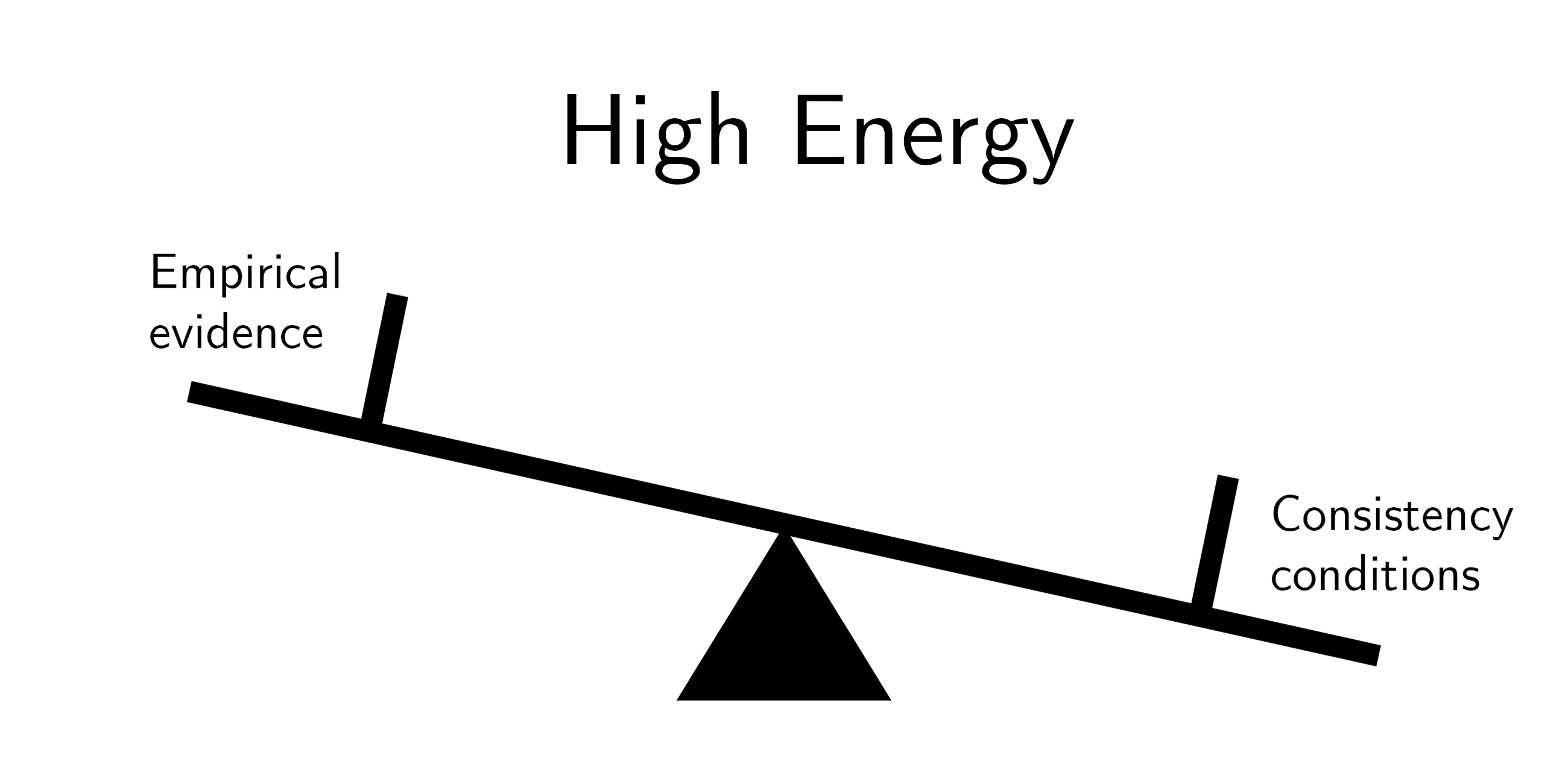}
\caption[Significance of consistency conditions in the high energy regime.]{In the high energy limit, experimental data becomes harder to gather, while constraints imposed by self-consistency tighten up. Therefore, consistency conditions \textit{weight} more than empirical evidence.}
\label{fig:highenergy}
\end{figure}
\begin{figure}[H]
\centering
  \includegraphics[width=0.7\linewidth]{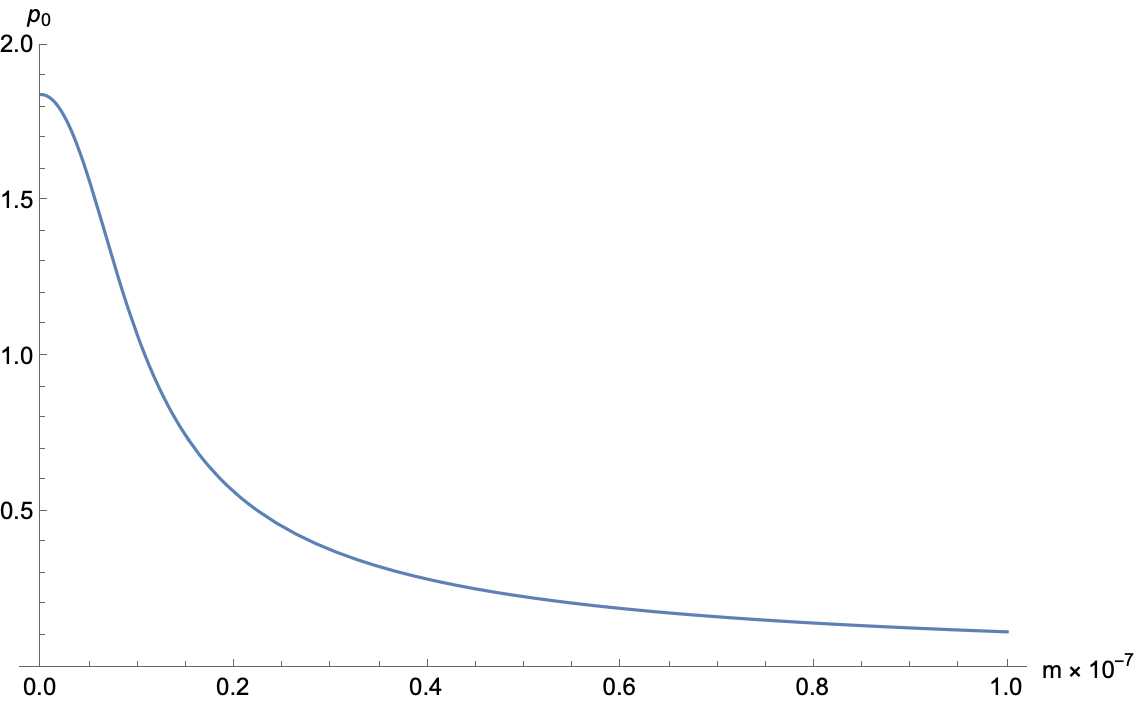}
\caption[Value of the momentum at which a particle is expected to collapse into a black hole, plotted against its mass.]{In this plot, the mass dependence of the threshold momentum value $p_0$ at which the system is expected to collapse into a black hole is shown. The momentum values are expressed in standard SI units, while the mass $m$ is re-scaled by a factor $10^{-7}$. }
\label{fig:momentummass}
\end{figure}
\begin{figure}[H]
\centering
  \includegraphics[width=0.8\linewidth]{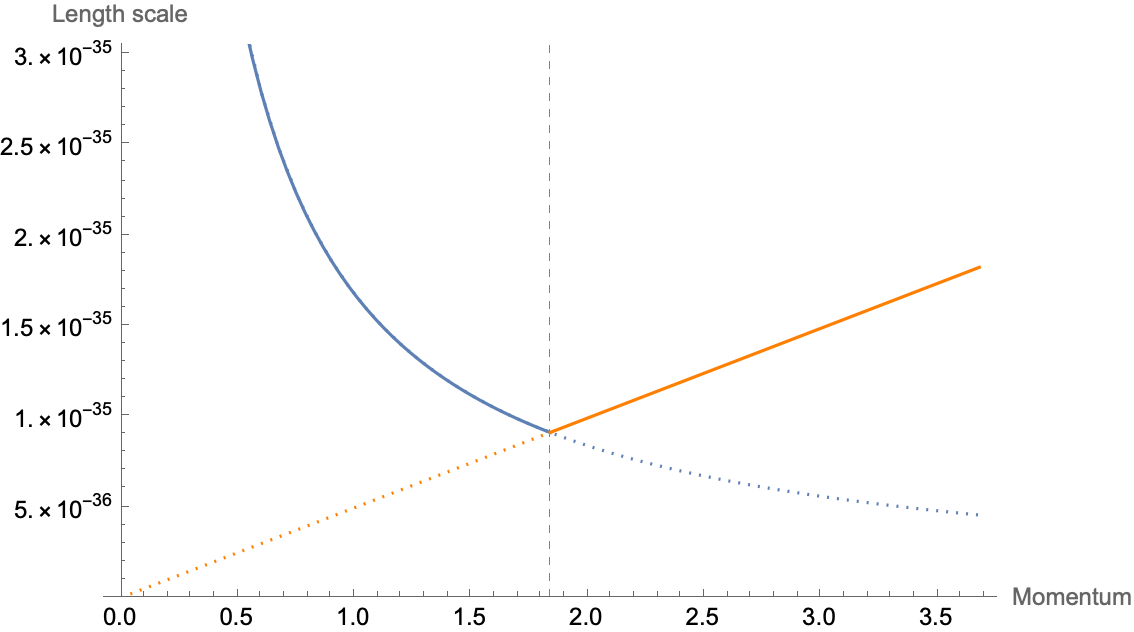}
\caption[Comparison between the wavelength associated to a particle's momentum and its Schwarzschild radius.]{In this figure, the blue plot represents the wavelength associated to a particle's momentum by formula \eqref{deb}, while the orange one depicts the linear growth of the Schwarzschild radius \eqref{schw}. The vertical dashed line corresponds to the threshold momentum value at which the latter exceeds the former, breaking the standard effective field theory reasoning.}
\label{fig:momentumplots}
\end{figure}

The Planck-Einstein equation \cite{cohen1986quantum} notoriously relates the energy $E$ of a photon to its wavelength $\lambda$, with:
\begin{equation}\label{planckeinstein}
    E=\frac{2\pi\hbar c}{\lambda}\ .
\end{equation}
In the above expression, the reduced Planck constant $\hbar$ and the speed of light in vacuum $c$ have been restored for the sake of clarity, instead of being set to one as is done in \textit{natural units}. Given the approximate values 
\begin{equation}
    \hbar\approx\SI{1.055e-34}{J\cdot s}\ ,\quad c=\SI{2.998e-8}{m\cdot s^{-1}}
\end{equation}
in \textit{SI units}, one can roughly estimate, from purely quantum mechanical reasoning, the length scale $\lambda_{\text{exp}}$ which can be resolved by a photon with energy $E_{\text{exp}}$. When considering massive particles, as the ones usually scattered in collider experiments, the formula \eqref{planckeinstein} serves as a good approximation in the relativistic limit, where the rest mass $m$ is negligible. In general, we have
\begin{equation}\label{debroglie}
    \lambda_{\text{exp}}=\frac{\hbar}{2\pi p_{\text{exp}}}=\frac{\hbar c}{2\pi}\sqrt{\frac{1}{E_{\text{exp}}^2-m^2c^4}}\ ,
\end{equation}
where $p_{\text{exp}}$ is the particle's momentum. It is clear that \eqref{debroglie} reduces to \eqref{planckeinstein} for massless particles. When computed in the centre-of-mass reference frame associated to particles colliding in high-energy accelerators, formula \eqref{debroglie} provides us with an indicator of the minimal ideal length which can be investigated by such experiments. Our chance to gather data on short-range dynamics is, in a nutshell, naturally capped by the maximum energy scale $\Lambda_{\text{EFT}}$ within the reach of our colliders. In the relativistic limit, their product is approximately constant. This is the reason, together with the issue of renormalisability, for which \textit{effective field theories} \cite{Georgi:1993mps,Burgess:2020tbq,Penco:2020kvy} gained a central importance in modern fundamental physics. They offer a powerful theoretical framework, together with valuable computational tools, to construct and study models of the relevant degrees of freedom which can be excited below a given energy. The precise features of the heavy modes lying above such scale are, in this sense, irrelevant, as their ultraviolet dynamics ends up being integrated out and absorbed into a family of operators and coupling constants governing the interactions of low energy species. From this contemporary point of view, the standard model of particle physics is thus best understood as the description of an effective dynamics, emerging as the low energy approximation of some ultraviolet theory. Superstring theory is arguably the most promising candidate to fulfil that role. But before allowing the constructions developed in chapter \ref{stringchap} to take the stage, we might devote a few more sentences to the perspective put forward by the effective field theory paradigm. Given the above discussion, the history of fundamental particle physics could be told as that of a gradual high-energy completion of our low-energy effective theories \cite{pais1986inward}. It should be clear, at this point, that the development of our comprehension of fundamental physics is strongly challenged by a lack of empirical evidence, when the energy scale under consideration is pushed towards the ultraviolet regime. Nonetheless, there is an enormous body of evidence \cite{Palti:2019pca} suggesting that, at the same time, constraints coming from the requirement of our effective theories to be consistent with quantum gravity get tighter and tighter. In a way which resembles a seesaw, it appears that as the amount of information coming from empirical evince shrinks, that originated from mathematical self-consistency expands. Such an interplay was pictorially represented in figures \ref{fig:lowenergy} and \ref{fig:highenergy}.
\newpage
This not only fuels our hope of being able to grasp the quantum properties of space-time without building a collider the size of the Milky Way, but also outlines a set of novel, general principles low energy effective theories should abide by in order to allow for an ultraviolet completion to quantum gravity. The search for such principles constitutes the bedrock of the swampland program, devoted at identifying the various ways in which gravity defies the naive effective filed theory intuition and empowers further constraints on low energy quantum field theories coupled to a dynamical space-time. Before discussing it in detail, we will outline a remarkable though experiment which, albeit being extremely simple, makes such a feature of space-time dynamics particularly evident. In order to so, we should follow a diluted version of an argument originally proposed by Matvei Petrovich Bronstein \cite{bronstein1936kvantovanie,bronstein1936quantentheorie,mead1964possible,Stachel:1999we,blum2018quantum}. We will neglect all technicalities, which can be found in the references, and stick to a purely heuristic derivation. Let's consider, moreover, the simplest, possible example we can work with. Namely, that of a spinless particle, as the one associated to a real scalar. The formula \eqref{debroglie} allows to once more estimate the wavelength associated to such a particle when its momentum modulus is set to a value $p$ as:
\begin{equation}\label{deb}
    \lambda\left(p\right)=\frac{\hbar}{2\pi p}\ .
\end{equation}
On top of that, we have that the particle energy is given by:
\begin{equation}\label{energyp}
    E\left(p\right)=\sqrt{p^2c^2+m^2c^4}\ .
\end{equation}
What the effective field theory approach suggests is that, by increasing the value of $p$, we would in principle be able to probe any length scale $l$, setting $\lambda\left(p\right)\leq l$. Nevertheless, we must remember that general relativistic gravity universally couples to the energy-momentum tensor. It is thus sourced by any kind of energy. The Schwarzschild radius associated to \eqref{energyp} is
\begin{equation}
    r_s\left(p\right)=\frac{2G}{c^4}E\left(p\right)\ ,
\end{equation}
where $G_D$ is Newton's constant. In the relativistic limit, we get:
\begin{equation}\label{schw}
    r_s\left(p\right)\approx\frac{2Gp}{c^3}\ .
\end{equation}
It can be clearly observed that, for a high enough momentum
\begin{equation}\label{critmom}
    p\geq p_{0}\equiv\sqrt{\frac{c^3\hbar}{4\pi G}}\ ,
\end{equation}
the particle's wavelength becomes smaller than its Schwarzschild radius. Hence, we should expect the system to collapse into a black hole. Such a momentum is associated to a wavelength
\begin{equation}
    \lambda\left(p\right)\leq\lambda\left(p_0\right)=\sqrt{\frac{\hbar G}{\pi c^3}}\equiv\frac{l_P}{\sqrt{\pi}}\ ,
\end{equation}
in which $l_P$ is referred to as Planck's length. On top of the fact that general relativity seems to suggest that the ultraviolet regime should be dominated by gravitational bound states, it is well know that the length scale associated to a black hole's radius grows when the system's energy is increased. Therefore, the typical correspondence between higher momenta and smaller resolved scales gets completely swapped. If gravitational degrees of freedom are included in our low energy models, as should be done in order to properly address the phenomenology of our universe, the relation \eqref{deb} suggested by standard effective field theory reasoning cannot be trusted when the momentum values characterising a system approach $p_0$. Including the corrections due to the presence of a non-zero mass, we would get:
\begin{equation}
    p\geq p_{0}\equiv\frac{1}{\sqrt{2}}\sqrt{-m^2c^2+\sqrt{m^4c^4+\frac{\hbar^2c^6}{4G^2\pi^2}}}\ .
\end{equation}
This would do nothing more than lowering the momentum threshold at which the black hole should appear, due to the presence of a non-zero rest mass energy contribution. This can be seen, for small mass values, in figure \ref{fig:momentummass}. Since the $m$-derivative of the expression for $p_0$ is always negative, no deviation is introduced for larger mass values. 
In summary, after having reached $p_0$ and having had a chance to investigate lengths comparable to $\lambda\left(p_0\right)$, we estimate that a black hole with radius $\lambda\left(p_0\right)$ would be produced. From that moment on, any increase in energy would enlarge the black hole size, pushing us back towards the infrared. Our results are nicely summarised in figure \ref{fig:momentumplots}, in which the two plots represent the momentum-dependence of the wavelength \eqref{deb} and of the Schwarzschild radius \eqref{schw}, respectively. It goes without saying that this should not be taken as a full-fledged prediction, which could only be achieved by performing the adequate computation in a quantum gravitational framework as that offered by string theory.  Nonetheless, it shows in a simple and direct way that general relativity itself, even when taken in its classical formulation, spoils our attempts at applying standard effective field theory techniques to physical systems coupled to a dynamical space-time. Further implications of this kind of phenomenon have been investigated in \cite{Dvali:2010bf,Dvali:2010jz,Dvali:2014ila,Dvali:2019ulr,Dvali:2020wqi}. 

%% file: Sections/ASwamp/swamplandconjectures.tex
\begin{figure}[htbp]
    \centering
    \begin{tikzpicture}[
            scale=0.9,
            transform shape,
            >=latex,
            node distance=2cm,
            mybox/.style={
                rectangle,
                rounded corners,
                draw,
                line width=0.7pt,
                minimum width=2.5cm,
                minimum height=1cm,
                align=center,
                outer sep=7pt,
                inner sep=7pt
            },
            mystyle/.style={->, line width=0.7pt, color=black, shorten >=1mm, shorten <=1mm}
        ]
    
        \node[mybox] (eft) {Effective theory};
        \node[mybox, right=of eft] (criteria) {Swampland criteria};
        \node[mybox, above right=of criteria] (swampland) {Swampland};
        \node[mybox, below right=of criteria] (landscape) {Landscape};
    
        \draw[mystyle] (eft) -- (criteria);
        \draw[mystyle] (criteria) -- (swampland);
        \draw[mystyle] (criteria) -- (landscape);
    
    \end{tikzpicture}
    \caption[Basic structure of swampland conjectures.]{Schematic representation of how swampland criteria are expected to work. Starting from an effective field theory, they should allow to place it either in the swampland or in the landscape only referring to its low energy features.}
    \label{fig:swampscheme}
\end{figure}
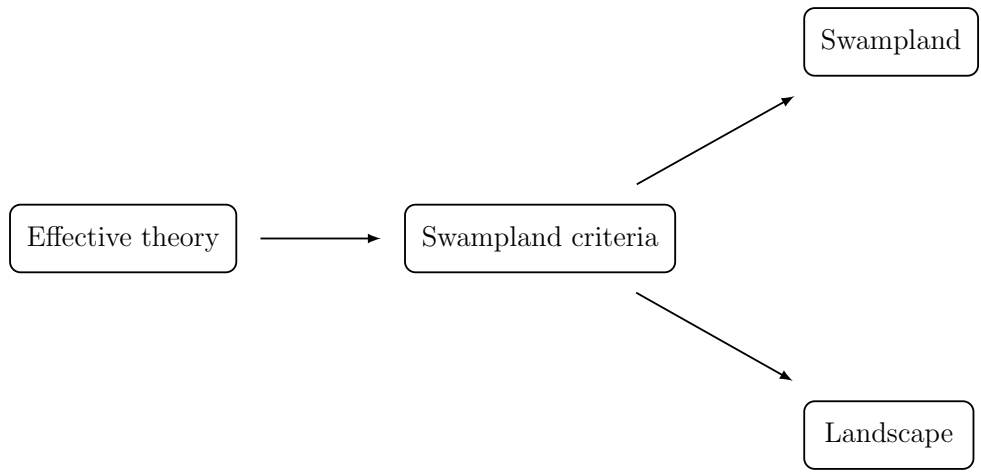

Superstrings provide us with a quantum theory of gravity. Strictly speaking, this does not rule out the fact that the actual microscopic dynamics general relativity emerges from in the real world might be different from the one described by superstring theory, nor it implies that every question concerning the nature, behaviour and non-perturbative features of space-time in such a framework has received a satisfying answer \cite{kiefer2006quantum,blau2009string,Harlow:2022qsq,stringcosmo,Horowitz:2004rn,matsubara2018spacetime,vistarini2019emergence,Huggett:2020gzt}. In contemporary research, the attempts at addressing some of the issues related to the fundamental properties of quantum gravity from the perspective of strings has produced numerous groundbreaking results, as those related to the \textit{cobordism conjecture} \cite{McNamara:2019rup,Montero:2020icj,Ooguri:2020sua,Blumenhagen:2022mqw,Andriot:2022mri,Velazquez:2022eco,Blumenhagen:2022bvh,Blumenhagen:2023abk,Debray:2023yrs}. With that being said, superstring theory, in the regimes over which we have control, can be safely assumed to be a well-behaved theory of quantum gravity. Therefore, even if it eventually turned out not to describe the specific ultraviolet dynamics realised in our world, it can serve as a theoretical laboratory in which general expectations on quantum gravity can be tested and investigated. Either it correctly describes the universe, or it offers us a useful guiding model to refine our intuition. Both options, even if they correspond to radically different scenarios, motivate the great interest in superstrings as tools to probe physics at the quantum gravity scale. For instance, when it comes to characterising black hole microstates. The inquisitive reader is encouraged to refer to \cite{Strominger:1996sh,Witten:1995ex,Maldacena:1996ky,Maldacena:1997tm,Skenderis:1999bs,Das:2000su,Lust:2018cvp,DeHaro:2019gno,David:2002wn} for a broad discussion of the subject matter. In addition, the theory possesses the promising attribute of being almost uniquely fixed by self-consistency. As broadly discussed in chapter \ref{stringchap}, the number of space-time dimensions itself is imposed by requiring the covariantly quantised theory to be unitary. Equivalently, the same result can be achieved by asking for covariance in light-cone quantisation or absence of conformal anomaly when employing path integral methods. This fact is peculiar, unexpected and should not be underestimated. It was moreover shown, in section \ref{stringefts}, how $10$-dimensional space-time dynamics is determined by forcing world-sheet scale invariance not to be broken at a quantum level, which translates into setting the $\sigma$-model $\beta$-functions to vanish. Once more, everything descended from the high degree of symmetry characterising the fundamental theory. Taken at face value, the world-sheet perturbative action \eqref{susypolyakovaction} has only one free parameter: the string length. In terms of the usual constant $\alpha'$, it can be expressed as:
\begin{equation}
    l_s=\sqrt{\alpha'}\ .
\end{equation}
These aspects precisely imply, as was previously stated, that the physics described by superstring theory is strongly constrained by consistency conditions. At least, when the energies involved in a process are high enough to resolve the full $10$-dimensional space-time theory, with both quantum corrections, parametrised by $\hbar$, and string-geometric ones, controlled by $l_s$. However, the same does not hold at any length scale. When moving towards the infrared, consistency conditions loosen up, allowing for various effective quantum field theories, coupled to a dynamical space-time metric, to be derived from the same ultraviolet degrees of freedom. Each one of them clearly comes with a cut-off $\Lambda_{EFT}$ above which it fails at providing a trustworthy approximate description of physical processes, since the states that were integrated out while moving towards low energies get too relevant not to be included in the action. This phenomenon, indeed, boils down to the existence of distinct low-energy vacua. Such a generic statement, which is expected to hold in one form or another regardless of the quantum gravity theory one might consider, becomes particularly problematic when focusing on the specific attributes of superstring theory \cite{Palti:2019pca}. In that context, the number of alternative consistent vacua has been long established to be huge, both by investigating general mechanisms via which they can be obtained and by gathering theoretical evidence from those which have been constructed explicitly. The reasons for this are multiple. Nonetheless, the major source of such degeneracy can be traced to the necessity of employing compactification techniques. The issue was addressed in \ref{seccompact}, where a simple circular dimensional reduction example was worked out, and can be summarised as follows: six of the ten space-time dimensions predicted by superstrings must be curled up into a compact manifold, in order to connect the theory to our observed phenomenology. Even if such a proposal might at first sound too imaginative, it naturally allows matter fields to emerge from the geometric properties of space-time and constitutes a legitimate scientific statement, which can be inquired via experimental means. Independently of the specific model under scrutiny, it is a well-established fact that general relativistic gravity cannot be localised on a space-time submanifold with non-zero codimension \cite{shifman2022advanced}. Therefore, from the perspective of an observer probing the $4$-dimensional effective theory, it would leak out in the extra dimensions, introducing short-scale deviations from Newton's potential. Generally speaking, one can model them \cite{extra:Floratos:1999bv,extra:Kehagias:1999my} by introducing a Yukawa-like term in Newton's gravitational potential
\begin{equation}
    V\left(r\right)=-\frac{G_4 M}{r}\Bigl(1+\alpha e^{-\beta r}\Bigr)\ ,
\end{equation}
which is exponentially suppressed as the typical length scale of an interaction grows. The constants $\alpha$ and $\beta$ depend on the specific shape and size compact dimensions are assumed to be characterised by. Direct measurements can hence bound them, setting experimental constraints on models with extra dimensions. Recent results can be found in \cite{extra:Lee:2020zjt}. After having assessed the concrete and empirical nature of our discussion, we can focus once more on superstring theory compactification. Starting from a higher-dimensional space-time action derived from it and trying to develop a particular $4$-dimensional, low-energy phenomenology, numerous decisions must be meticulously evaluated and executed. These correspond not only the choice of a precise compact geometry for the extra dimensions, but also involve the stabilisation of moduli, the analysis of background fluxes and a careful evaluation of the role played by non-perturbative extended objects, as branes and defects. Even accounting for the web of dualities under which apparently diverse theories end up inducing the same dynamics, this leads to a dramatic proliferation of admissible vacua. Let's consider, for instance, the example of $12$-dimensional F-theory. Such model, introduced in \cite{ftheory:Vafa:1996xn} as a tool to obtain realistic low energy effective theories by means of compactification on elliptically fibered Calabi-Yau four-folds \cite{ftheory:Weigand:2018rez,ftheory:Marchesano:2022qbx}, can be dimensionally-reduced on a $2$-torus $\T^2$ and mapped to a version of type-IIB superstring, in which the $SL\left(2,\Z\right)$-duality is made manifest. An extremely conservative estimate of a lower bound to the overall number of consistent flux compactifications of $4$-dimensional F-theory was set to $10^{272000}$ in \cite{ftheory:Ftheory}. In \cite{ftheory:Cvetic:2019gnh}, a family of $\mathcal{O}\left(10^{15}\right)$ vacua with the exact chiral spectrum of the standard model of particle physics and directly realising gauge coupling unification was presented. As a final example, type-IIA orbifold compactifications on $\T^6/\Z_2\times\Z_2$ were considered in \cite{vacua:12billionLoges:2022mao}, supposing the existence of $\mathcal{O}\left(100\right)$ standard models in the studied ensemble. A lot of effort was put into trying to grasp the size and features of the set of string low energy effective theories in $4$ dimensions \cite{vacua:Ashok:2003gk,vacua:Blumenhagen:2004xx,vacua:Bousso:2000xa,vacua:Buchmuller:2017wpe,vacua:Denef:2004ze,vacua:Denef:2006ad,vacua:Douglas:2003um,vacua:Douglas:2004kp,vacua:Friedmann:2012uf,vacua:Kumar:2006tn}. Recently, machine learning techniques have also been directed towards that same goal \cite{Halverson:2019tkf,Loges:2021hvn,Cole:2021nnt,Krippendorf:2021uxu}. Given all the evidence for the existence of a humongous family of viable vacua, one question naturally arises: can any conceivable quantum field theory coupled to a relativistic dynamical space-time background be achieved as a superstring effective field theory? If the answer was affirmative, the whole project of string theory phenomenology would be jeopardised, since no prediction on low energy observables could be formulated. And as there is no guarantee that humans will be able to probe arbitrarily small scales within the gap between those reached by current accelerators and Planck's length $l_P$, string theory's ultraviolet features might be practically untestable. This might simply make the theory pointless. Fortunately, things revealed themselves to be much more favourable with respect to such a gloomy scenario. Despite being relaxed, self-consistency constraints, while flowing from quantum gravity towards lower energy scales, do not disappear completely. In fact, ultraviolet degrees of freedom seem to leave a marked imprint on the general, shared features of low energy vacua. In short, not all apparently consistent quantum field theories coupled to general relativity can emerge as low energy limits of superstring theory. The qualifier \textit{apparently} is not accidental. Instead, it precisely refers to the fact that our usage of the word \textit{consistent} was the one suggested by traditional effective field theory reasoning, not necessarily allowing for an ultraviolet completion towards quantum gravity. Remarkably, superstring theory suggests that we should narrow our notion of consistency, when analysing seemingly well-behaved vacua. New principles are required. In order to make our discussion more precise, two important concept should be introduced:
\begin{itemize}
    \item We define the \textit{string theory landscape} as the family of quantum field theories coupled to a dynamical space-time metric that \textit{can} be obtained as superstring theory vacua, apparently consistent below a given energy cut-off $\Lambda$.
    \item We define the \textit{string theory swampland} as the family of quantum field theories coupled to a dynamical space-time metric that \textit{cannot} be obtained as superstring theory vacua, apparently consistent below a given energy cut-off $\Lambda$.
\end{itemize}
\begin{center}
\begin{figure}[h]
\centering
  \includegraphics[width=0.8\linewidth]{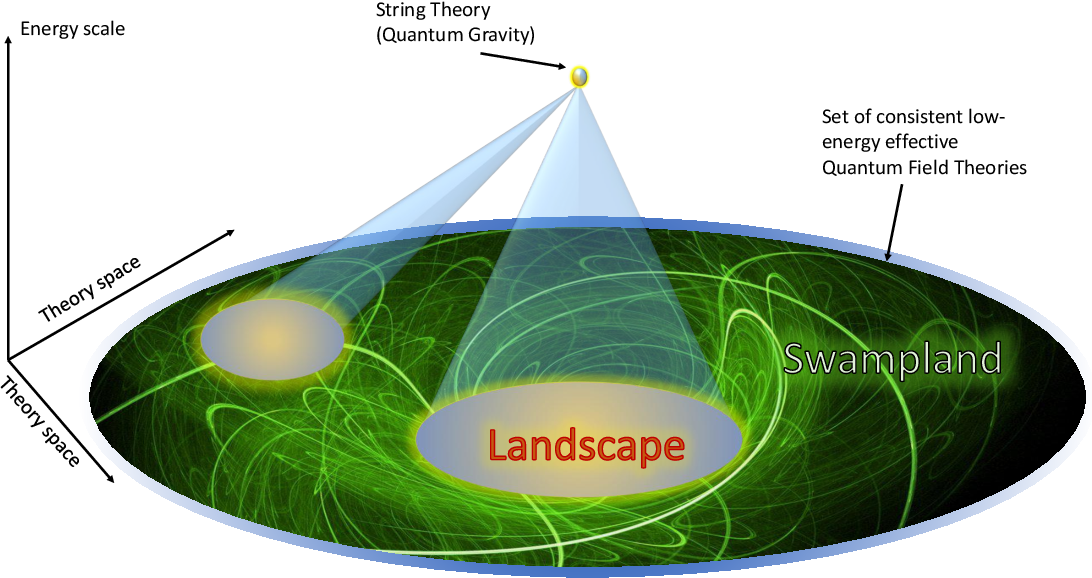}
\caption[Representation  of the superstring landscape and the swampland.]{Pictorial representation of the space of apparently consistent quantum field theories coupled to gravity at a given energy scale, from the perspective of standard effective field theory reasoning. In the ultraviolet regime, consistency constraints are expected to -almost- uniquely fix quantum gravity, while they appear to loosen up while flowing towards the infrared. The conditions which need to be satisfied in order to be part of the landscape relax accordingly. This well-known figure was taken from \cite{Palti:2019pca}.}
\label{fig:swamp}
\end{figure}
\end{center}
It goes without saying that, given the above definitions, the landscape and the swampland can be understood as complementary subsets of the encompassing family of effective field theories coupled to gravity appearing to be consistent below an energy threshold $\Lambda$. The existence of a string theory swampland was first discussed in \cite{Vafa:2005ui} and quickly became a relevant topic in superstring phenomenology \cite{swamp:Arkani-Hamed:2006emk,swamp:Huang:2007qz,swamp:Kats:2006xp,swamp:Li:2006vc,swamp:Ooguri:2006in}. Nonetheless, it was only in the last ten years that it gained the enormous traction that still characterises it. For a -largely incomplete- selection of recent, influential works on the subject matter, one might refer to \cite{Akrami:2018ylq,Alim:2021vhs,Anchordoqui:2019amx,Anchordoqui:2020sqo,Anchordoqui:2022svl,Anchordoqui:2022txe,Andriot:2020lea,Angius:2022aeq,Apers:2022zjx,Basile:2021mkd,Basile:2020mpt,Basile:2021vxh,Bedroya:2019snp,Blumenhagen:2017cxt,Blumenhagen:2018hsh,Blumenhagen:2018nts,Blumenhagen:2019vgj,Bonnefoy:2019nzv,Brown:2015iha,Bonnefoy:2020uef,Buratti:2021fiv,Buratti:2021yia,Calderon-Infante:2020dhm,Castellano:2021mmx,Castellano:2021yye,Conlon:2018eyr,Conlon:2021cjk,Cribiori:2021gbf,Das:2018rpg,Daus:2020vtf,Debray:2021vob,DallAgata:2021nnr,Dvali:2018jhn,Font:2019cxq,Garg:2018reu,Gendler:2020dfp,Gonzalo:2021fma,Gonzalo:2021zsp,Hamada:2021yxy,Heidenreich:2021xpr,Kallosh:2019axr,Katz:2020ewz,Kehagias:2019akr,Klaewer:2018yxi,Lee:2018spm,Lee:2018urn,Lee:2019skh,Lust:2019zwm,Lanza:2020qmt,Luben:2020wix,Lust:2020npd,Marchesano:2019ifh,McNamara:2020uza,Montero:2019ekk,Montero:2022prj,Obied:2018sgi,Ooguri:2016pdq,Ooguri:2018wrx,Palti:2017elp,Palti:2020qlc,Perlmutter:2020buo,Reece:2018zvv}. As already stated at the beginning of this chapter, more general introductions can be found in \cite{Palti:2019pca,Palti:2022edh,Brennan:2017rbf,Tadros:2022rkd,Casse:2022ymj,Agmon:2022thq}. Figure \ref{fig:swamp} nicely summarises the distinction between the swampland and the landscape. The central purpose of the \textit{swampland program} is to find accurate criteria of demarcation between the landscape and the swampland, formalising those consistency conditions that, while not being captured by standard effective field theory reasoning, arise from the necessity for an ultraviolet completion to quantum gravity. This is roughly summarised in figure \ref{fig:swampscheme}. Naturally, in order to be useful and bear meaning, such statements should only concern to explicit features of the low energy effective theories, without referring to the microscopic dynamics from which they are expected to emerge. 
Providing a solid proof of statements of that sort without having complete control on the fundamental, microscopic theory under scrutiny is a dramatically challenging task. Nonetheless, the body of \textit{theoretical evidence} supporting what the community refers to as \textit{swampland conjectures} has experienced a notable growth. In particular, many of such proposed criteria, suggested by evidence gathered in widely different contexts, have been shown to be linked to each other, if not absolutely equivalent. One striking example is the one offered by the refinement of the \textit{no global symmetries} conjecture \cite{Kallosh:1995hi,Harlow:2018tng} via cobordism classes, first outlined in \cite{McNamara:2019rup}. The \textit{emergence proposal} \cite{Palti:2019pca,vanBeest:2021lhn,Castellano:2022bvr,Castellano:2023qhp,Harlow:2015lma,Heidenreich:2017sim}, on the other hand, is a direct attempt at deriving multiple swampland conjecture from the same microscopic principles. What was previously grouped under the generic label of \textit{theoretical evidence} comes, in practice, from three main sources:
\begin{itemize}
    \item \textbf{Known string vacua} First and foremost, one can observe the ever-growing set of understood string vacua and hypothesise that their shared properties are, indeed, common to all effective field theories belonging to the landscape. This approach is particularly solid when the vacua are explicitly derived from superstrings, possess a full-fledged world-sheet formulation and are therefore certain to allow for an ultraviolet completion to quantum gravity. Unfortunately, this is not always the case. Many investigated models, albeit being inspired by the kind of features and objects one typically finds in string theory, are established after having made numerous assumptions. Due to their nature, such examples provide evidence that should be taken with a grain of salt. The more a vacuum is \textit{string-inspired}, rather than \textit{string-derived}, the less reliable it is when trying to assessing a swampland conjecture.
 \item \textbf{Quantum gravity arguments} While establishing the shared features of low energy effective field theories than can be successfully completed to quantum gravity, it is clear that our general expectations on the properties of such ultraviolet limit can play a crucial role. Evidence from black holes thermodynamics, holography and quantum field theory in curved space-time can substantiates our conjectural constraints on infrared dynamics by unveiling contradictions and violations of fundamental principles. Namely, by shedding light on those processes in which the effective description would break down and require either to be regarded as irredeemably inconsistent, or to be rescued by the some non-trivial quantum gravitational effects. A noteworthy example is that of the Hawking radiation emitted by black holes \cite{Bardeen:1973gs,Hartle:1976tp,Hawking:1974rv,Hawking:1975vcx,Hawking:1976de}, which is expected to break unitarity if the behaviour predicted by standard methods can be extrapolated for long evaporation times. These arguments possess the strength of being broad and almost independent from our assumptions on the specific quantum gravity dynamics realised in the ultraviolet, but lack the sharpness and precision of those assessed by analysing explicit string vacua. Moreover, they require to make a series of strong assumptions on the validity range of our effective models.
    \item \textbf{Microscopic principles} Starting from a specific and -at least partially- well understood ultraviolet theory, or from a set of microscopic principles quantum gravity is expected to satisfy, it might in principle be possible to straightforwardly derive a family of constraints consistent low energy effective field theories would have to satisfy. Such propositions would be derived in a purely top-down fashion. Hence, the associated body of evidence would be much more solid, from both a conceptual and a formal perspective, than that gathered via exploiting general quantum gravity arguments or shared properties of known string vacua. Nonetheless, it would also be heavily dependent on our comprehension of the microscopic behaviour of the space-time metric. The search for microscopic arguments should always be guided and supported by bottom-up data. 
\end{itemize}
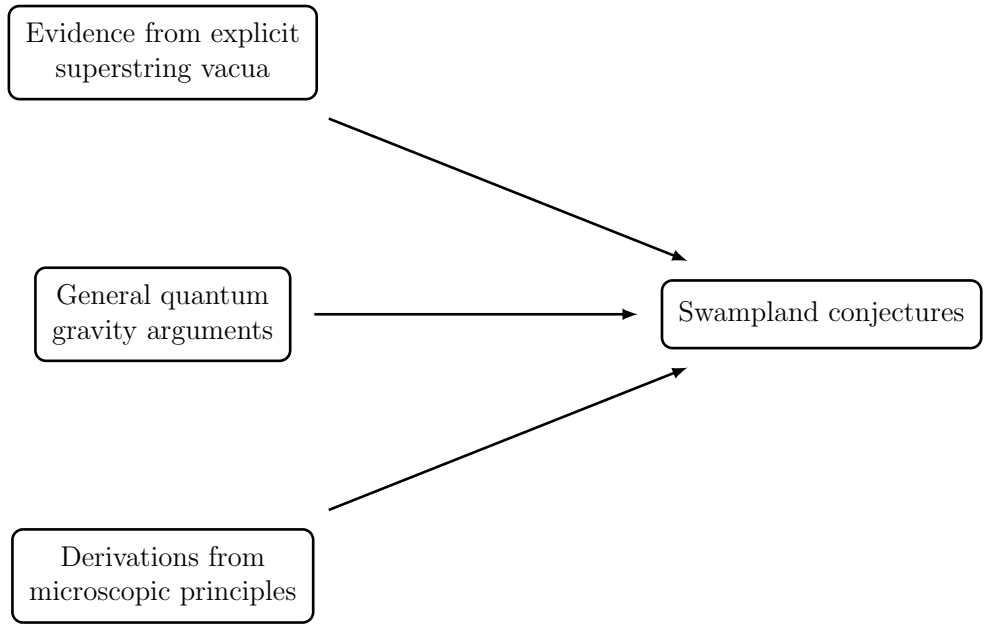
\begin{figure}[htbp]
    \centering
    \begin{tikzpicture}[
            scale=0.9,
            transform shape,
            >=latex,
            node distance=2cm,
            mybox/.style={
                rectangle,
                rounded corners,
                draw,
                line width=1pt,
                minimum width=3cm,
                minimum height=1cm,
                align=center,
                outer sep=7pt,
                inner sep=7pt
            },
            mystyle/.style={->, line width=1pt, color=black, shorten >=1mm, shorten <=1mm}
        ]
    
        \node[mybox] (evidence) {Evidence from explicit\\superstring vacua};
        \node[mybox, below=of evidence] (arguments) {General quantum\\gravity arguments};
        \node[mybox, below=of arguments] (derivations) {Derivations from\\microscopic principles};
        \node[mybox, right=5cm of arguments] (conjectures) {Swampland conjectures};
    
        \draw[mystyle] (evidence) -- (conjectures);
        \draw[mystyle] (arguments) -- (conjectures);
        \draw[mystyle] (derivations) -- (conjectures);
    
    \end{tikzpicture}
    \caption[The three sources of evidence supporting swampland conjectures.]{Schematic representation of the distinct forms of evidence backing up swampland conjectures. Usually, they tend to agree and point towards the same direction. Obviously, the amount of data supporting the various conjectures is extremely heterogeneous: some of them are at the verge of being proven, while others are still highly speculative.}
    \label{fig:swampland_connections}
\end{figure}
Given a specific swampland conjecture, the evidence backing it up is usually found to be comprised of contributions from all the three sources briefly described above. This is summarised in figure \ref{fig:swampland_connections}. A strong argument in favour of the swampland program comes, as discussed in \cite{Palti:2019pca}, from the fact that general quantum gravity arguments, inductive inferences from known string vacua and, whenever possible, derivations from microscopic principle have, in many distinct contexts, proven themselves to be extremely coherent with each other. Furthermore, it oftentimes happens that seemingly distinct conjectures, referring to different features of low energy theories and posing separate constraints, end up being unified into a common, broader perspective, highlighting novel connections between apparently independent aspects of quantum field theories coupled to gravity. As was broadly discussed at the beginning of this chapter, standard effective field theories come equipped with an energy cut-off $\Lambda_{\text{EFT}}$, above which the low energy description is expected to break down. Swampland conjecture typically pose the existence of a further scale $\Lambda_{\text{Swamp}}$, appearing when space-time dynamics is taken into account, the Planck's mass $M_{P}$ is assumed to be finite and quantum fields are, subsequently, coupled to gravity. The \textit{swampland scale} $\Lambda_{\text{Swamp}}$ must be interpreted at the energy value at which physical processes start to be heavily influenced by quantum gravity, allowing for an ultraviolet completion towards superstring theory. If the parameters of the infrared theory are such that $\Lambda_{\text{EFT}}>\Lambda_{\text{Swamp}}$, the presence of $\Lambda_{\text{Swamp}}$ concretely translates into the fact that, when flowing towards smaller length scales, non-trivial quantum gravity effects break the effective description down before it would have been predicted by standard techniques. When the gravity decoupling limit $M_{P}\rightarrow\infty$ is taken, we have that $\Lambda_{\text{Swamp}}\rightarrow\infty$, gravity decouples and the traditional conceptual framework of quantum field theory in a fixed space-time background is restored. 
\begin{figure}[h]
    \centering
    \begin{tikzpicture}[
            >=latex,
            node distance=2.5cm,
            mybox/.style={
                rectangle,
                rounded corners,
                draw,
                line width=1pt,
                minimum width=2.5cm,
                minimum height=1cm,
                align=center,
                inner sep=7pt,
                outer sep=7pt
            },
            mystyle/.style={<->, line width=1pt, color=black, shorten >=1mm, shorten <=1mm}
        ]
    
        \node[mybox] (A) at (0,4) {Completeness\\ Hypothesis};
        \node[mybox] (B) at (4.5,4) {No Global\\ Symmetries};
        \node[mybox] (C) at (9,4) {Cobordism\\ Conjecture};
        \node[mybox] (D) at (0,1) {Weak Gravity\\ Conjecture};
        \node[mybox] (E) at (4.5,1) {Distance\\ Conjecture};
        \node[mybox] (F) at (9,1) {Ricci Flow\\ Conjecture};
        \node[mybox] (H) at (4.5,-2) {de Sitter\\ Conjecture};
        \node[mybox] (I) at (9,-2) {Anti-de Sitter\\ Conjecture};
        
        \draw[mystyle] (A) -- (B);
        \draw[mystyle] (B) -- (C);
        \draw[mystyle] (A) -- (D);
        \draw[mystyle] (B) -- (D);
        \draw[mystyle] (D) -- (E);
        \draw[mystyle] (B) -- (E);
        \draw[mystyle] (C) -- (E);
        \draw[mystyle] (E) -- (F);
        \draw[mystyle] (D) -- (H);
        \draw[mystyle] (H) -- (E);
        \draw[mystyle] (I) -- (E);
        \draw[mystyle] (I) -- (F);
    \end{tikzpicture}
    \caption[Partial web of interconnected swampland conjecture.]{The diagram contains a limited number of swampland conjectures and highlights the most significant conceptual connections among them. This is by no means intended to be a complete assessment of the state of current debates, nor it is aimed at containing all pertinent information. Rather, it should be seen as a pictorial representation of the intricate web of interrelated swampland criteria, from which a deeper and general perspective is slowly emerging.}
    \label{fig:swampchartconn}
\end{figure}
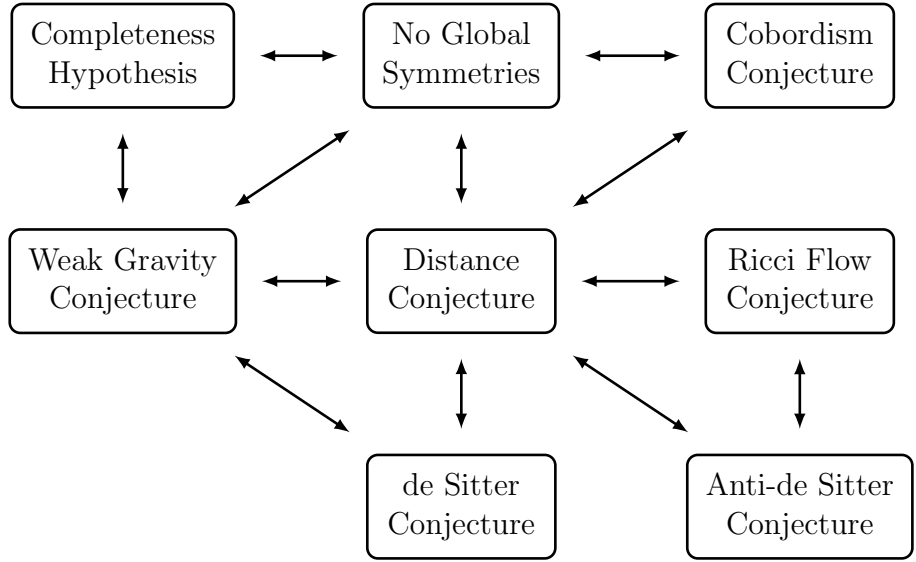
\subsection{The Swampland distance conjecture}\label{sec:distanceconj}
\begin{figure}
    \centering
    \begin{tikzpicture}
\fill[color=gray!10,rounded corners=10pt](3,0)--(2.6,0.7)--(2.6,1.5)--(1.9,1.9)--(1.5,2.6)--(0.7,2.6)--(0,3)--(-0.7,2.6)--(-1.5,2.6)--(-1.9,1.9)--(-2.6,1.5)--(-2.6,0.7)--(-3,0)--(-2.6,-0.7)--(-2.6,-1.5)--(-1.9,-1.9)--(-1.5,-2.6)--(-0.7,-2.6)--(0,-3)--(0.7,-2.6)--(1.5,-2.6)--(1.9,-1.9)--(2.6,-1.5)--(2.6,-0.7)--(3,0);
\draw[rounded corners=10pt](3,0)--(2.6,0.7)--(2.6,1.5)--(1.9,1.9)--(1.5,2.6)--(0.7,2.6)--(0,3)--(-0.7,2.6)--(-1.5,2.6)--(-1.9,1.9)--(-2.6,1.5)--(-2.6,0.7)--(-3,0)--(-2.6,-0.7)--(-2.6,-1.5)--(-1.9,-1.9)--(-1.5,-2.6)--(-0.7,-2.6)--(0,-3)--(0.7,-2.6)--(1.5,-2.6)--(1.9,-1.9)--(2.6,-1.5)--(2.6,-0.7)--(3,0);
\shade[left color=gray!10, right color=gray!110, rounded corners=6pt](2,1.2)--(2.4,0.5)--(10,0)--(2.4,-0.5)--(2,-1.2);
\draw[rounded corners=6pt](2,1.2)--(2.4,0.5)--(10,0)--(2.4,-0.5)--(2,-1.2);
\node at (-0.7,0.7) {$\mathbf{M}_{\Phi}$};
\draw[black,rounded corners=12pt](0.7,-1.2)--(1.3,-1.1)--(2.65,0.2)--(4,-0.1)--(5.5,0.06)--(7.5,-0.03)--(9,0.02)--(9.8,0);
\node at (0.4,-1.4) {$A$};
\fill[red] (0.7,-1.2) circle (2pt);
\draw[black] (0.7,-1.2) circle (2pt);
\node at (10.2,0.3) {$B$};
\fill[blue] (9.9,0) circle (2pt);
\draw[black] (9.9,0) circle (2pt);
\end{tikzpicture}
    \caption[Representation of moduli space infinite distance limits.]{Portrayal of the moduli space $\mathbf{M}_{\Phi}$, whose points unambiguously correspond to sets of vacuum expectation values for the scalar fields appearing in \eqref{msact}. Generally speaking, the moduli space can be characterised by complicated metric and topological properties. Nonetheless, infinite geodesic distances are always expected to be accompanied by infinite towers of massless states. The representation of one of such pathological limits, where theories become inconsistent, was included in the figure.}
    \label{fig:moduli space}
\end{figure}
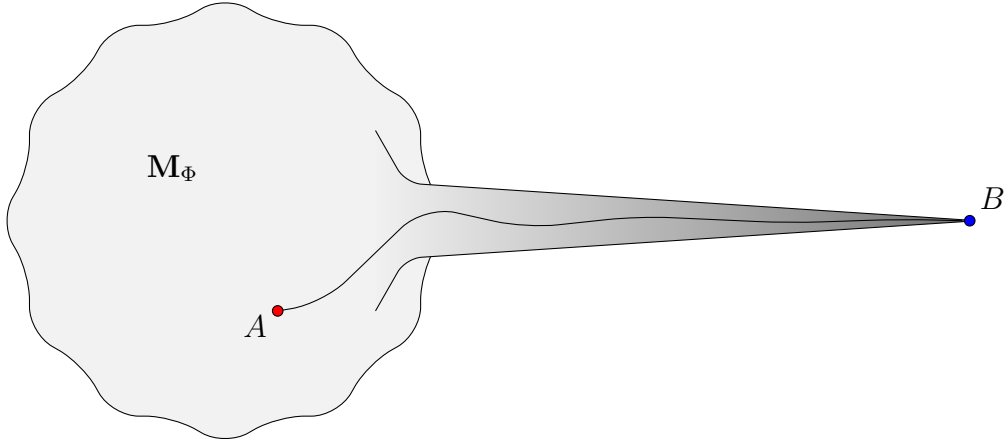

In section \ref{seccompact}, a rudimentary mass spectrum analysis of circle-compactified type-IIA superstring theory highlighted the presence of two infinite towers of states, associated to winding and momentum excitations. Their masses can be simply expressed as
\begin{equation}\label{massestowers}
    M_n=\frac{n}{\rho}\ ,\quad M_\omega=\frac{\omega\rho}{\alpha'}\ ,
\end{equation}
where $\rho$ is the circle radius in a given set of units and the integers $n$ and $\omega$ respectively count the momentum quanta and the closed string windings along the compact direction. Remarkably, as was nicely depicted in figure \ref{fig:KKmasses}, they display opposite behaviours with respect to fluctuations of the compact dimension size. Furthermore, it was argued that such length scale, up to all the details related to stability conditions, could provide a non-ambiguous way to distinguish between different circle-compactifications of superstring theory. Namely, it could be understood as a \textit{modulus} parametrising the space of the resulting dimensionally-reduced effective field theories. Nonetheless, instead of regarding the value of $\rho$ itself as a modulus, it would be natural to connect such a notion to the vacuum expectation value of some scalar field. This is indeed possible and quite direct. Considering the string-frame expression \eqref{IIAaction} for the $10$-dimensional action emerging from type-IIA superstring theory and integrating out the compactified circle direction, one can absorb the resulting factor into the dilaton volume term $\exp{-2\Phi}$. This way, the $9$-dimensional dilaton $\Phi_9$ is defined by
\begin{equation}
    e^{-2\Phi_9}\equiv 2\pi\rho M_{\text{s}}  e^{-2\Phi}\ ,
\end{equation}
where the \textit{string scale} $M_{\text{s}}$ is nothing more than:
\begin{equation}
    M_{\text{s}}\equiv\frac{1}{2\pi\sqrt{\alpha'}}\ .
\end{equation}
This directly translates into the relation:
\begin{equation}
    \Phi_9= \Phi-\frac{1}{2}\log{\left(2\pi\rho M_{\text{s}}\right) }\ .
\end{equation}
Imposing the $9$-dimensional Planck's mass, which acquired a non-trivial dependence on the compact dimension radius, to be equal to one, we must pick units so that:
\begin{equation}
    M_{\text{s}}\sim\left(2\pi\rho\right)^{1/7}\ .
\end{equation}
This was extensively discussed in \cite{Palti:2019pca} and provided, in Planck units, the following expressions:
\begin{equation}\label{emasses}
        M^{(E)}_n=\frac{n}{\rho}\left(\frac{1}{2\pi\rho}\right)^{1/7}\ ,\quad M^{(E)}_\omega=\frac{\omega\rho}{\alpha'}\left(2\pi\rho\right)^{1/7}\ .
\end{equation}
At this point, we can juxtapose low-energy effective theories characterised by different vacuum expectation values $\ev{\Omega |\Phi|\Omega}$ of the dilaton. In order for such a comparison to be meaningful from the perspective of a low-energy observer, we want to perform it while keeping the $9$-dimensional dilaton vacuum expectation value $\ev{\Omega |\Phi_9|\Omega}$ fixed. Hence, any displacement 
\begin{equation}
    \ev{\Omega |\Phi|\Omega}\longrightarrow\ev{\Omega |\Phi|\Omega}+\delta\ev{\Omega |\Phi|\Omega}
\end{equation}
should be compensated by a radius displacement:
\begin{equation}
    \rho\longrightarrow\rho+\delta\rho\ .
\end{equation}
In particular, we have that:
\begin{equation}
    \delta\ev{\Omega |\Phi_9|\Omega}=0\Longrightarrow \delta\ev{\Omega |\Phi|\Omega}=\frac{1}{2}\frac{\delta\rho }{\rho }\ .
\end{equation}
Therefore, using 
\begin{equation}
    \varphi\equiv\ev{\Omega |\Phi|\Omega}\in\mathcal{M}_{\phi}\sim\R
\end{equation}
as a modulus of the theory and introducing the natural distance
\begin{equation}\label{phidis}
    \Delta\left(\varphi_2,\varphi_1\right)\equiv \left|\varphi_2-\varphi_1\right|\ ,
\end{equation}
the mass scalings in \eqref{massestowers} get can be expressed, in Einstein-frame units, as:
\begin{equation}
    M_n\sim e^{\alpha\varphi}\ ,\quad M_\omega\sim e^{-\alpha\varphi}\ .
\end{equation}
 Given the above the discussion, the arbitrary notion of moduli space distance introduced in \eqref{distancefirst} has thus been properly and solidly justified. The constant $\alpha$ can be computed to be equal to
\begin{equation}
    \alpha=\frac{8}{\sqrt{77}}\sim\mathcal{O}\left(1\right)
\end{equation}
as it depends on the specific factors \eqref{emasses} appearing in the reduced mass operator. Rephrasing the analysis performed in \ref{seccompact} in these new and more precise terms, we have that any infinite distance limit in the $1$-dimensional moduli space $\mathcal{M}_{\Phi}\sim\R$ is accompanied by an infinite tower of states getting exponentially lighter. Namely, starting from any finite value $\varphi_1\in\mathcal{M}_{\Phi}$, considering a distinct theory characterised by $\varphi_2\in\mathcal{M}_{\Phi}$ and computing the distance between the two as in \eqref{phidis}, the infinite distance limits $\varphi_2\mapsto\pm\infty$ are characterised by infinitely many massless species appearing in the spectrum. They are hence inconsistent. In other words, the microscopic features of superstring theory, from which our models emerge as low-energy effective limits, force infinite distance limits in the moduli space $\mathcal{M}_{\Phi}$ to be in the swampland. This is a first and straightforward example of a general behaviour of string compactifications, formally described by the so-called swampland distance conjecture \cite{Ooguri:2006in,Klaewer:2016kiy,Baume:2016psm}. In the following discussion, the explicit dependence on Planck's mass will be restored for the sake of clarity. The original statement of the swampland distance conjecture, in the absence of a potential, can be expressed as follows.
\begin{distance}
Let $\Phi\equiv\left(\phi^1,\dots,\phi^N\right)$ be the fields whose dynamics is described by a low energy effective field theory coupled to a dynamical $D$-dimensional space-time $\mathcal{M}$, with $\ket{\Omega}$ being a corresponding vacuum state. Furthermore, let the $N$-dimensional manifold $\mathbf{M}_{\Phi}$ represent its moduli space, charted by the vacuum expectation values:
\begin{equation}
    \varphi^i\equiv\ev{\Omega |\phi^i|\Omega}\ ,\quad i=1,\dots,N\ .
\end{equation}
If two points $\Bar{\varphi}_A$ and $\Bar{\varphi}_B$ in $\mathbf{M}_{\Phi}$ are considered, their geodesic distance can be defined and is referred to as $\Delta_{AB}=\Delta\left(\Bar{\varphi}_A,\Bar{\varphi}_B\right)$. Therefore, there must exist an infinite tower of additional fields $\psi^j$, characterised by a moduli-dependent mass threshold 
\begin{equation}
    m:\mathbf{M}_{\Phi}\longrightarrow\R
\end{equation}
displaying an exponential drop in the geodesic distance, when approaching an infinite distance limit in the moduli space. Namely, every infinite distance limit in $\mathbf{M}_{\Phi}$ should be accompanied by an infinite tower scaling as
\begin{equation}
    m\left(\Bar{\varphi}_B\right)\sim m\left(\Bar{\varphi}_A\right)\exp{-\alpha\frac{\Delta_{AB}}{\sqrt{M^{D-2}_{\text{P}}}}}
\end{equation}
when $\Delta_{AB}\gtrsim M_{\text{P}}$, with $\alpha\sim\mathcal{O}\left(1\right)$ and positive. 
\end{distance}
While stating the above conjecture, many assumptions have been made and multiple concepts have been taken for granted. Before relaxing the former and commenting on some notable generalisations of our current discussion, the latter will be made more explicit. In particular, the notion of a moduli space \textit{geodesic distance} will be properly introduced. Acting on the presumption that $\mathbf{M}_{\Phi}$ can be accurately modelled as a topological manifold, which is straightforward in the case at hand, we are allowed to endow it with a moduli space metric $h_{ij}\left(\Bar{\varphi}\right)$. For the specific and instructive example of a family of scalar fields $\Bar{\varphi}$, the tensor can be directly read off
from the low-energy effective theory space-time action
\begin{equation}\label{msact}
    S\left[g,\Phi\right]=\frac{1}{2\kappa^2_D}\int_{\mathcal{M}}\diff^Dx\sqrt{-g}\left[R_g-h_{ij}\left(\Phi\right)\de^{\mu}\phi^i\de_\mu\phi^j\right]\ ,
\end{equation}
where the $x^\mu$ are a family of space-time coordinates, the signature of the space-time metric $g_{\mu\nu}$ is taken to be the mostly positive one $\left(-,+,\dots,+\right)$ and the moduli space metric appears in the scalar fields kinetic terms. Naturally, in order to be interpreted as such, $h_{ij}$ has to be evaluated on the set $\Bar{\varphi}$ of vacuum expectation values, which play the role of coordinates on $\mathbf{M}_\Phi$. A redefinition of the chosen basis $\phi^i$ of the fields space, which induces an analogous transformation of the vacuum expectation values $\varphi^i$ in the quantum theory, consistently translates into changes in the kinetic terms and, consequently, in the moduli space metric. Moreover, the matrix $h_{ij}$ must be positive definite, in order for the equations of motion for the fields to be well defined from an analytic perspective. The differentiable manifold $\mathbf{M}_\Phi$ is, hence, Riemannian, with signature $\left(+,\dots,+\right)$. The Levi-Civita conditions
\begin{equation}
    \nabla_k h_{ij}=0
\end{equation}
uniquely fix a connection on the moduli space, with Christoffel symbols given by:
\begin{equation}
    \Gamma^i_{jk}=\frac{h^{il}}{2}\left(\frac{\de h_{lj}}{\de \varphi^k}+\frac{\de h_{lk}}{\de \varphi^j}-\frac{\de h_{jk}}{\de \varphi^l}\right)\ .
\end{equation}
If $\Bar{\varphi}_A$ and $\Bar{\varphi}_B$ are taken to be two points in $\mathbf{M}_{\Phi}$, there is always one curve 
\begin{equation}
    \Bar{\gamma}:\left[0,1\right]\longrightarrow\mathbf{M}_\Phi
\end{equation}
connecting them, with $\Bar{\gamma}\left(0\right)=\Bar{\varphi}_A$ and $\Bar{\gamma}\left(1\right)=\Bar{\varphi}_B$, that satisfies the geodesic equation:
\begin{equation}
    \frac{\diff^2\gamma^i}{\diff s^2}+\Gamma^i_{jk}\left(\Bar{\gamma}\right)\frac{\diff\gamma^j}{\diff s}\frac{\diff\gamma^k}{\diff s}=0\ .
\end{equation}
Hence, the geodesic distance $\Delta_{AB}$ between the two points can be obtained as the length of the minimal geodesic that connects them:
\begin{equation}
    \Delta_{AB}\equiv\int_{0}^{1}\diff s\sqrt{h_{ij}\left(\Bar{\gamma}\right)\frac{\diff\gamma^i}{\diff s}\frac{\diff\gamma^j}{\diff s}}\ .
\end{equation}
The adjective \textit{minimal} was included for the sake of clarity, as there are instances in which two moduli space points can be connected by more than one geodesic. Then, the shortest one has to be chosen. With our distance definition, trivial large curve-lengths produced by non-geodesic paths are excluded from discussions related to the swampland distance conjecture. The distance between two moduli space points is defined as nothing more than the length of the shortest curve connecting them. Furthermore, when such a length is sent to infinity an infinite tower of massless states is expected to appear in the spectrum. This behaviour is schematically depicted in figure \ref{fig:moduli space}.
\subsubsection{Interpreting infinite distances}
In the example presented in \ref{circlecompactification} and further developed in the above discussion, in which a circle compactification of a $10$-dimensional space-time action was analysed, infinite distances in the moduli space either corresponded to the radius $\rho\in\R_+$ of the compact dimension going to infinity or to zero. From the perspective of an observer able to probe the $9$-dimensional effective field theory, defined on the non-compact directions, up to an energy scale $\Lambda_{\text{EFT}}$, the large-$\rho$ regime corresponds to a \textit{decompactification} limit. Namely, as $\rho$ grows, the energy required to resolve the size of the hidden dimension decreases. An estimate of the size at which such energy gets smaller than the effective field theory cut-off $\Lambda_{\text{EFT}}$ can be obtained from \eqref{debroglie}. For a massless probe, ignoring all the phenomenological and technical details of a realistic experimental setting, we have:
\begin{equation}
    \rho\sim\rho_{\text{c}}\equiv\frac{\hbar c}{\Lambda_{\text{EFT}}}\ .
\end{equation}
When the circle radius $\rho$ is comparable to the \textit{critical} radius $\rho_{\text{c}}$, the extra spatial dimension becomes accessible to the low energy observer and cannot be ignored. Therefore, the description offered by the dimensionally-reduced theory breaks down and becomes inconsistent. For $\rho>\rho_{\text{c}}$, the cut-off above which the effective theory has to be modified is lowered by the necessity of preventing the hidden dimension to be revealed. The other limit, in which the compact dimension size shrinks to zero, can be straightforwardly interpreted as being conceptually equivalent to the first one by means of T-duality \ref{sssec:Tduality}. Otherwise, it can be seen as corresponding to a regime in which the compact dimension size is commensurate to the string length. Roughly, this translates into:
\begin{equation}
    \rho\sim l_{\text{s}}\equiv\sqrt{\alpha'}\ .
\end{equation}
In such a setting, the effective field theory description in terms of local, interacting quantum fields and point-like excitations is doomed to fail. This is due to the fact that the extended nature of the fundamental degrees of freedom propagating along the circular dimension cannot be neglected anymore. Along the same line of reasoning, the practice of modelling space-time as a smooth, Lorentzian, and differentiable manifold might be jeopardised by string-theoretic corrections to the geometry. Whereas decompactification limits are already present at the level of dimensionally-reduced quantum field theories, without any peculiar feature of superstring theory being necessary for them to be realised, the low-energy effective field theory breakdown when the internal dimension radius gets small is deeply rooted into the properties of our particular quantum gravity framework. This is reflected into the former being characterised by an infinite tower of Kaluza-Klein modes getting exponentially lighter, while the latter is compromised by the appearance of light winding modes. As has been commented on in \ref{sssec:Tduality}, only the second ones are tied to the specificities of superstrings, since they emerge from closed strings winding around the extra dimension and do not figure when compactifying traditional quantum field theories. The swampland distance conjecture has been widely explored and extended \cite{Grimm:2018cpv,Grimm:2018ohb,Heidenreich:2018kpg,Lanza:2021udy,Lanza:2022zyg,Castellano:2021yye,Corvilain:2018lgw,Etheredge:2022opl,Lee:2021uuu}, fruitfully applied in the context of cosmology \cite{Agrawal:2018own,Brandenberger:2021pzy,Calderon-Infante:2022nxb,Scalisi:2018eaz} and related to either the restoration of global symmetries, which should be absent from any consistent theory of quantum gravity \cite{Zeldovich:1976vw,Zeldovich:1977mk,Abbott:1989jw,Banks:1988yz,Kallosh:1995hi,Banks:2010zn,Harlow:2018jwu,Harlow:2018tng,Harlow:2020bee,Harlow:2021lpu}, or the emergence of a critical string \cite{Lee:2019wij,Alvarez-Garcia:2021pxo,Basile:2022zee}. In a series of works \cite{Kehagias:2019akr,Anchordoqui:2019amx,Bonnefoy:2019nzv,Anchordoqui:2020sqo,Luben:2020wix,DeBiasio:2022omq,Velazquez:2022eco,DeBiasio:2022nsd,DeBiasio:2022zuh,DeBiasio:2021yoe}, some of which will constitute the backbone of part \ref{novel}, multiple generalisations of the conjecture to the space-time metric itself were put forward. Before considering the details of the results derived in \cite{DeBiasio:2022omq} and \cite{DeBiasio:2022zuh}, we will now provide a concise introduction to the motivations, guiding principles and typical tools employed when trying to construct a distance conjecture for space-time geometries.

%% file: Sections/GeometricFlows/intro.tex
The swampland distance conjecture, stated in \cite{Ooguri:2006in} and comprehensively discussed in section \ref{sec:distanceconj}, relates infinite distance limits in the moduli space of effective theories coupled to gravity to the appearance of infinite towers of asymptotically massless states. Given that a similar behaviour is observed when sending the negative cosmological constant characterising AdS space-time to zero, an appropriate extension of the conjecture was proposed in \cite{Lust:2019zwm}. This, together with the fact that many of the scalars appearing in dimensionally-reduced superstring low energy limits emerge from the $10$-dimensional metric degrees of freedom, motivated the search for a distance conjecture for the geometry, specifically referring to metric displacements. A candidate statement was presented in \cite{Kehagias:2019akr}, where the mathematics of \textit{geometric flows} allowed to frame the AdS flat limit in a more encompassing context. Further refinements \cite{Anchordoqui:2019amx,Bonnefoy:2019nzv,Anchordoqui:2020sqo,DeBiasio:2021yoe}, generalisations \cite{DeBiasio:2022omq,DeBiasio:2022zuh} and applications \cite{Luben:2020wix,Velazquez:2022eco,DeBiasio:2022nsd} thereof were subsequently explored. Geometric flow equations were originally applied in the context of string theory and condensed matter physics by Daniel Friedan, while analysing an extension of Aleksandr Markovič Poljakov's discussion \cite{Polyakov:1975rr} of the $O\left(N\right)$-invariant nonlinear $\sigma$-model renormalisation in $2+\epsilon$ dimensions \cite{friedan1985nonlinear}. Due to its analytic properties and striking connection to the $\sigma$-model graviton $\beta$-function derived from superstring theory, the most notable example of a geometric flow can be found in Richard Hamilton's \textit{Ricci flow} \cite{hamilton1982,chow2004ricci}, defined as:
\begin{equation}\label{intro:RF}
\frac{\diff g_{\mu\nu}}{\diff s}=-2R_{\mu\nu}\ .
\end{equation}
In the above expression, $R_{\mu\nu}$ is nothing more than the Ricci curvature tensor associated to the metric $g_{\mu\nu}$ and $s$ is a real \textit{flow parameter} on which all such quantities depend. Therefore, a one-parameter family of metric tensors $g_{\mu\nu}\left(s\right)$ is said to evolve according to Ricci flow, up to a flow-dependent diffeomorphism, if the equation \eqref{intro:RF} is satisfied for every value of $s$. In order for the differential problem to be solved, one naturally has to supply it with an initial condition $\Bar{g}_{\mu\nu}$. The mathematically oriented reader might want to refer to the monographs \cite{Cao2006ACP,morgan2007ricci,kleiner2008notes} for a deeper and formal discussion of the techniques employed in this chapter. Before concluding our brief introduction, it must be emphasised that geometric flows of Lorentzian manifolds are both more problematic and less understood than their Euclidean analogues. The \textit{Geometrisation conjecture}, for instance, has so far not been extended nor adapted to Lorentzian manifolds. Grigori Perelman's proof \cite{Perelman2002TheEF,Perelman2003FiniteET,Perelman2003RicciFW} of its Riemannian counterpart, which served as an establishment of the \textit{Poincaré conjecture}, specifically relied on Ricci flow, together with the tools offered by entropy functionals and surgery procedures. In the following discussion, we will start from the considerations contained in \cite{Lust:2019zwm} and gradually connect the swampland distance conjecture to such techniques, constructing a framework in which distances between nonequivalent metrics can be meaningfully defined and computed.

%% file: Sections/GeometricFlows/ads.tex
The cosmological constant $\Lambda$, which provides the general relativistic field equations 
\begin{equation}\label{cosmoefe}
    R_{\mu\nu}-\frac{1}{2}Rg_{\mu\nu}+\Lambda g_{\mu\nu}=\kappa_D T_{\mu\nu}
\end{equation}
with a term proportional to the metric tensor, was first proposed by Albert Einstein in 1917 \cite{cc:EinsteinKosmologischeBZ,cc:Kerszberg1989TheIU,cc:Schucking1991TheIO}. At the time, it served the purpose of counter-weighting the attractive nature of gravity, so that static space-time geometries could be constructed. Regardless of its philosophical appeal, the conjectural staticity of space-time was soon disproved by Edwin Hubble's experiments, suggesting that the universe was indeed expanding. Therefore, the motives for which a non-zero $\Lambda$ was considered in the first place ceased to exist \cite{Rugh:2000ji}. The necessity for a positive cosmological constant resurged only at the end of the last century, after a thorough analysis of the light emitted by a set of type Ia supernovae \cite{SupernovaCosmologyProject:1997zqe,SupernovaCosmologyProject:1998vns,SupernovaSearchTeam:1998bnz} suggested that the universal expansion rate was not diminishing, as expected, nor remaining constant. Instead, it was increasing. Postulating a small, positive cosmological constant can be arguably regarded as the most natural way to account for such evidence. Assuming $T_{\mu\nu}=0$ and $\Lambda>0$, the simplest vacuum solution to \eqref{cosmoefe} is that of \textit{de Sitter} space-time, here written in static coordinates $\left(t,r,\vartheta_1,\dots,\vartheta_{D-2}\right)$:
\begin{equation}
    \diff s^2=-\left(1-\frac{r^2}{\alpha^2}\right)\diff t^2+\left(1-\frac{r^2}{\alpha^2}\right)^{-1}\diff r^2+r^2\diff\Omega^2_{D-2}\ .
\end{equation}
The de Sitter horizon radius $\alpha$ is defined as:
\begin{equation}
    \alpha\equiv\sqrt{\frac{\left(D-1\right)\left(D-2\right)}{2\Lambda}}\ .
\end{equation}
Any candidate theory of quantum gravity would, thus, be naively expected to motivate the \textit{ad hoc} introduction of a cosmological constant from first principles, or at least not to prohibit it. 
While the endeavour to construct an explicit quantum de Sitter vacuum is a worthwhile one, which has also been undertaken in the context of superstring theory \cite{Kachru:2003aw,Kachru:2003sx,Lust:2022lfc}, there is a growing body of evidence suggesting that this might be unfeasible. It has been notably argued, without explicitly evoking any particular quantum gravity framework, that de Sitter space-time might only be realised as a temporary excited coherent state of gravitons \cite{Dvali:2018jhn,Dvali:2017eba,Berezhiani:2021zst}, with a typical quantum break-time equal to:
\begin{equation}\label{qbtdvali}
    t_{\text{Q}}=\frac{1}{H^3}\left(\frac{M_{\text{P}}}{\sqrt{N_{\text{sp}}}}\right)^2\ .
\end{equation}
The right-hand side of the formula \eqref{qbtdvali} contains the \textit{Hubble parameter} $H$ characterising a de Sitter patch, the number $N_{\text{sp}}$ of species contained in a low energy effective theory and the Planck mass $M_{\text{P}}$. The last two are, moreover, combined into the \textit{species scale}
\begin{equation}
    \Lambda_{\text{sp}}\equiv\frac{M_{\text{P}}}{\sqrt{N_{\text{sp}}}}\ ,
\end{equation}
at which the ultraviolet quantum gravity degrees of freedom are supposed to become relevant \cite{Dvali:2007hz}. Starting from the excited quantum state corresponding to de Sitter space-time and allowing it to evolve for a duration comparable to $t_{\text{Q}}$, non-perturbative effects are expected to make it depart from any semi-classical approximation. In a nutshell, a positive cosmological constant vacuum might not be admissible in quantum gravity. In order to replicate the large-scale properties of the universe, one would thus have to obtain the measured value of $\Lambda$ by temporarily exciting an appropriate space-time ground state. Assuming the argument to hold, no theory of quantum gravity should allow for eternal de Sitter states. These considerations, while not conclusive, seem to suggest that the apparent obstruction to the existence of de Sitter vacua encountered in superstring phenomenology is one of the model's strengths, rather than a shortcoming. The standard formulation of the swampland \textit{de Sitter conjecture} \cite{Obied:2018sgi,Ooguri:2018wrx}, along with its various refinements, nicely synthesises this putative feature of the theory. 
\begin{dS}
Let $\Phi\equiv\left(\phi^1,\dots,\phi^N\right)$ be a family of scalar fields whose dynamics is controlled by a potential $V\left(\Phi\right)$ and described by a low energy effective field theory coupled to a dynamical $D$-dimensional space-time $\mathcal{M}$. We have that $V\left(\Phi\right)$ has to satisfy at least one of the following conditions:
\begin{equation}\label{dsconj1}
        \left|\nabla_i V\right|\geq\frac{c_1}{M_{\text{P}}}\cdot V\ ,
\end{equation}
\begin{equation}\label{dsconj2}
            \min{\left(\nabla_i\nabla_j V\right)}\leq-\frac{c_2}{M^2_{\text{P}}}\cdot V\ .
\end{equation}
In the inequalities, the positive constants $c_1$ and $c_2$ are taken to be $\mathcal{O}\left(1\right)$ and the usual fields space metric is assumed. Furthermore, the scalar potential Hessian $\nabla_i\nabla_j V$ is computed in an orthonormal frame.
\end{dS}
As derived in the above references, this conjecture prevents the construction of stable de Sitter vacua. This can be clearly deduced by observing that \eqref{dsconj1} forbids any potential minimum to produce a positive vacuum energy, whereas \eqref{dsconj2} implies the existence of an instability. When the gravity decoupling limit $M_{\text{P}}\rightarrow\infty$ is taken, the bounds \eqref{dsconj1} and \eqref{dsconj2} are trivially satisfied. This is perfectly consistent with the fact that swampland conjectures are allegedly enforced by non-trivial gravitational effects. Regardless of how fascinating it might be, further exploring the issue at hand would go beyond the scope of this chapter. The reader is once more encouraged to refer to \cite{stringcosmo} for an up-to-date and comprehensive review of the subject matter. As far as the following discussion is concerned, it is instead crucial to emphasise how the difficulties arising when trying to obtain de Sitter metric via string compactifications do not affect $\Lambda=0$ nor $\Lambda\leq 0$ solutions. The latter, in particular, are ubiquitous and offer a way to study space-time manifolds with non-zero cosmological constant in a controlled, superstring setting. We will hence focus, for the remainder of our work, on \textit{Anti-de Sitter} space-time, which can be expressed in global coordinates $\left(t,r,\vartheta_1,\dots,\vartheta_{D-2}\right)$ as:
\begin{equation}\label{adsmetric}
    \diff s^2=-\left(1+\frac{r^2}{\beta^2}\right)\diff t^2+\left(1+\frac{r^2}{\beta^2}\right)^{-1}\diff r^2+r^2\diff\Omega^2_{D-2}\ .
\end{equation}
The Anti-de Sitter radius $\beta$ is defined by the equation:
\begin{equation}
    \beta\equiv\sqrt{\frac{\left(1-D\right)\left(D-2\right)}{2\Lambda}}\ .
\end{equation}
Albeit being less directly connected to real world phenomenology than solutions with a positive cosmological constant, Anti-de Sitter space-time proved itself to be a fertile and useful theoretical laboratory in which to test our general expectations on quantum gravity. First and foremost, as it offered an explicit realisation of the \textit{holographic principle} by means of the AdS/CFT correspondence \cite{Aharony:1999ti,Bigatti:1999dp,Harlow:2018fse,Maldacena:1997re,Maldacena:2003nj,Nastase:2007kj,Polchinski:2010hw,Witten:1998qj}. Inspired by the references \cite{Gunaydin:1984fk,Gunaydin:1985cu,Kim:1985ez,Pernici:1985ju,Schwarz:1983qr}, which offer a far more elaborate analysis, we will now describe an explicit supergravity solution that allows to embed $5$-dimensional Anti-de Sitter space-time in a supergravity background. 
\subsection{Conjecture statement}
It is a well-understood fact that a $D$-dimensional Anti-de Sitter space-time $AdS_D$, characterised by a radius $\beta$, can be isometrically embedded into a flat $\left(D+1\right)$-dimensional manifold $\mathcal{M}_{D+1}$, with signature $\left(-,-,+,\dots,+\right)$ and coordinates $\left(t_1,t_2,\eta_1,\dots,\eta_{D-1}\right)$. More specifically, it corresponds to the submanifold defined by the constraint equation:
\begin{equation}\label{adsembedd}
    \beta^2=t^2_2+t^2_1-\sum_{i=1}^{D-1}x^2_i\ .
\end{equation}
This directly allows Anti-de Sitter space-time to inherit the $SO\left(2,D-1\right)$ isometry group of $\mathcal{M}_{D+1}$, as its action leaves \eqref{adsembedd} unchanged. Such a group generally admits supersymmetric enhancements. A classification of extended supersymmetries in $AdS_D$, with $D=4,5,6,7$, was provided in \cite{adssusy:Nahm:1977tg}. Here, we are interested in the landmark example of $5$-dimensional Anti de Sitter space-time, for which we have the gauged supergravities:
\begin{equation}
    SO\left(2,2|\mathcal{N}/2\right)\ ,\quad\mathcal{N}=2,4,6,8\ .
\end{equation}
The $\mathcal{N}=8$ theory, with global symmetry $E_6$ and gauge group $SU\left(4\right)$, can be straightforwardly derived in the context of $10$-dimensional type IIB supergravity. In particular, the construction involves considering an $S^5$ compactification and assuming the \textit{Freund-Rubin} ansatz, in which the self-dual field strength 
\begin{equation}
    F\equiv\diff C_4\ ,
\end{equation}
associated to the $4$-form $C_4$, is set to be proportional to the volume element of the compact dimensions. Namely, it is assumed to take the form
\begin{equation}\label{frans}
    F^{\mu_1\dots\mu_5}=\frac{\epsilon^{\mu_1\dots\mu_5}}{\sqrt{g_{\text{S}}}}f\ ,
\end{equation}
where the $\mu_1,\dots,\mu_5$ indices refer to the compact directions, $g_{\text{S}}$ is the sphere metric determinant and $f$ is a constant. The \eqref{frans} ansatz can be shown to satisfy the equations of motion for $C_4$. Furthermore, it corresponds to the appearance of a non-trivial energy-momentum tensor in Einstein field equations \eqref{cosmoefe}. By setting the cosmological constant $\Lambda$ to zero, such equations can be solved by the $10$-dimensional Cartesian product of an $AdS_5$ and an $S^5$ manifold, both having radius:
\begin{equation}
    \varrho=\frac{2\sqrt{2}}{f}\ .
\end{equation}
From the perspective of an observer probing the non-compact $AdS_5$ manifold, the presence of $F$ sources an effective cosmological constant:
\begin{equation}
    \Lambda_{\text{eff}}=-\frac{3}{4}f^2\ .
\end{equation}
Regarding $\Lambda_{\text{eff}}$ as a coordinate on the $1$-dimensional moduli space of distinct $5$-dimensional effective theories, we have that the radius of the unobservable compact dimensions can be written as a moduli space function:
\begin{equation}
    \varrho=\sqrt{6}\left|\Lambda_{\text{eff}}\right|^{-\frac{1}{2}} \ .
\end{equation}
It can be easily shown that, as for the circle compactification discussed in \ref{circlecompactification}, the mass threshold of the Kaluza-Klein tower realised by wrappings around $S^5$ scales as:
\begin{equation}
    M_{\text{KK}}\sim\frac{1}{\varrho}\sim\left|\Lambda_{\text{eff}}\right|^{\frac{1}{2}}\ .
\end{equation}
Therefore, we have that the flat space-time limit $\Lambda_{\text{eff}}\rightarrow 0$, which corresponds to a decompactification limit of the internal dimensions, must be accompanied by an infinite tower of asymptotically massless states. This phenomenon, while concerning the displacement of a parameter entering the space-time metric, precisely resembles the distance conjecture. Hence, it led to the formulation of the \textit{AdS distance conjecture} \cite{Lust:2019zwm}.
\begin{AdS}
Let's consider a one-parameter family of $D$-dimensional Anti-de Sitter space-time metrics $g_{\mu\nu}\left(\Lambda\right)$, obtained as quantum gravity low-energy effective theories and labelled by the value of the cosmological constant $\Lambda$ which sets their curvature radius. Let $\mathbf{M}_{\Lambda}\sim\R_{-}$ be the corresponding moduli space, distinguishing between different values of $\Lambda$. Therefore, there must exist an infinite tower of fields $\psi^j$, characterised by a moduli-dependent mass threshold 
\begin{equation}
    m:\mathbf{M}_{\Lambda}\longrightarrow\R
\end{equation}
displaying a power-law drop
\begin{equation}\label{adsmassas}
    m\sim\left|\Lambda\right|^{\alpha}
\end{equation}
in the cosmological constant, when approaching the flat space-time $\Lambda\rightarrow 0$ limit in the moduli space. In the above equation, expressed in Planck units, the number $\alpha$ is taken to be $\sim\mathcal{O}\left(1\right)$ and positive. 
\end{AdS}
A further refinement is offered by the so-called \textit{Strong AdS distance conjecture}, which requires $\alpha=1/2$ for supersymmetric AdS vacua. It was subsequently shown \cite{Bonnefoy:2020uef} that the above discussion can be generalised to an arbitrary number of space-time dimensions.

%% file: Sections/GeometricFlows/moduli.tex
In the previous derivation, it was shown how dimensionally-reduced superstring $AdS_5$ low energy effective theories can give rise to the appearance of infinite towers of asymptotically massless states. More precisely, it was argued that such states are expected to become light in the flat space-time limit $\Lambda\rightarrow 0$. This sticking analogy with the analysis presented in \ref{sec:distanceconj} motivated the formulation of the Anti-de Sitter distance conjecture, together with its refinement. Cosmological constant displacements, which translate into displacements of the space-time metric, seem to obey a particular form of the swampland distance conjecture, traditionally applied to vacuum expectation values of scalar fields. Alongside the observation that many of the scalars populating superstring low energy effective theories emerge, via compactification, from degrees of freedom originally pertaining to the $10$-dimensional metric tensor, the interplay between cosmological constant limits and towers of light states strongly hints at the need for a more general theoretical framework. Namely, it appears to suggest that any space-time metric displacement should be accompanied by the emergence of an infinite tower of asymptotically massless states. Another argument in favour of this perspective might come from the very origin of geometric and matter degrees of freedom in the context of superstring theory, which are unified as excitations of the same fundamental objects. Our discussion of metric displacements, so far, lacks the precision of the standard distance conjecture, mainly due to need for a generalised notion of moduli space able to account for distinct geometries, which should itself be endowed with a suitable metric structure. The aim of the following discussion is to make such an intuition precise. Hence, distances between different solutions for the space-time metric will be defined in a meaningful and formal way. Along the lines of \cite{Lust:2019zwm}, this endeavour requires us to start by recalling that, when dealing with scalar fields, the moduli space metric $h_{ij}\left(\Phi\right)$ was obtained from the Lagrangian kinetic term
\begin{equation}
    \mathscr{L}_{\text{kin}}\left(\Phi\right)\sim-h_{ij}\left(\Phi\right)\de^\mu\phi^i\de_\mu\phi^j
\end{equation}
appearing in the action \eqref{msact}. From that, the length of a one-parameter curve 
\begin{equation}
    \Bar{\gamma}:\left[s_0,s_1\right]\subset\R\longrightarrow\mathbf{M}_\Phi
\end{equation}
in the moduli space $\mathbf{M}_\Phi$, with $\Bar{\gamma}\equiv\left(\gamma^1,\dots,\gamma^N\right)$, can be computed as:
\begin{equation}\label{ddis}
    \mathbf{L}\left[\Bar{\gamma}\right]=\int_{s_0}^{s_1}\diff s \left|h_{ij}\left(\Bar{\gamma}\right)\frac{\diff\gamma^i}{\diff s}\frac{\diff\gamma^j}{\diff s}\right|^{\frac{1}{2}}\ .
\end{equation}
More broadly, the Lagrangian kinetic term for a space-time field $\psi$ with components $\psi_{\hat{\mu}}$, with $\hat{\mu}\equiv\mu_1\dots\mu_n$ being a family of indices, is supposed to take the form 
\begin{equation}
    \mathscr{L}_{\text{kin}}\left(\psi\right)\sim-K^{\hat{\mu}\hat{\nu}}\left(\psi\right)\mathcal{D}\psi_{\hat{\mu}}\cdot\mathcal{D}\psi_{\hat{\nu}}\ ,
\end{equation}
where $\mathcal{D}$ is a suitable derivative operator. The field $\psi$ is generally expected, unlike typical moduli or the cosmological constant, to have a space-time dependence. Any consideration on the moduli space distance between two distinct low energy solution should, nonetheless, be independent from the choice of a space-time point. Hence, the space-time average
\begin{equation}\label{staverage}
    \ev{F}\equiv\frac{1}{V_{\mathcal{M}}}\int_{\mathcal{M}}\diff^Dx\sqrt{\left|g\right|} F\left(x\right)
\end{equation}
of a function $F:\mathcal{M}\rightarrow C_F$ over the $D$-dimensional space-time manifold $\mathcal{M}$, with metric tensor $g_{\mu\nu}$, must be introduced. In the above expression, the symbol $V_{\mathcal{M}}$ serves as a normalisation and refers to the space-time volume:
\begin{equation}\label{vol}
    V_{\mathcal{M}}\equiv\int_{\mathcal{M}}\diff^Dx\sqrt{\left|g\right|}\ .
\end{equation}
It must be stressed that $V_{\mathcal{M}}$, as defined in \eqref{vol}, might not be finite. In that case, both the integral with which $V_{\mathcal{M}}$ is computed and that appearing explicitly in formula \eqref{staverage} must be properly regularised. For the sake of simplicity and unless faced with an explicit computation, we will from now on keep such regularisations implicit. For a space-time constant function $F_0$, we obviously have:
\begin{equation}
    \ev{F_0}=F_0\ .
\end{equation}
If the $D$-dimensional space-time manifold is assumed to be the trivially-fibered Cartesian product $\mathcal{M}=\mathcal{N}\times\mathcal{C}$ of two manifolds with dimensions $d$ and $D-d$, charted by coordinates $z\equiv\left(z^1,\dots,z^{d}\right)$ and $y\equiv\left(y^1,\dots,y^{D-d}\right)$, the space-time volume factorises as
\begin{equation}
    V_{\mathcal{M}}=V_{\mathcal{N}}\cdot V_{\mathcal{C}}
\end{equation}
and the average of a product function $F\left(z,y\right)\equiv F_1\left(z\right)F_2\left(y\right)$ is simply given by:
\begin{equation}\label{factstaverage}
    \ev{F}_{\mathcal{M}}=\ev{F_1}_{\mathcal{N}}\cdot\ev{F_2}_{\mathcal{C}}
\end{equation}
In the above, the specific manifold on which an average is taken was made explicit for clarity. In the case in which $\mathcal{C}$ models a set of internal compact dimension and in which we are working with the degrees of freedom $A\left(z\right)$ of the dimensionally-reduced low energy effective theory, that can only depend on the non-compact directions, we simply have:
\begin{equation}
    \ev{A}_{\mathcal{M}}=\ev{A}_{\mathcal{N}}\ .
\end{equation}
Hence, the average can be equivalently taken on the full $D$-dimensional space-time or on its non-compact $d$-dimensional factor. That being said, let's now consider a generalised moduli space $\mathbf{M}_\psi$, in which every point corresponds to a space-time configuration of $\psi$ solving the appropriate field equations. Moreover, let 
\begin{equation}
    \sigma:\left[s_0,s_1\right]\longrightarrow\mathbf{M}_\psi
\end{equation}
be a curve in such moduli space with components $\sigma_{\hat{\mu}}$, corresponding to a one-parameter family of field configurations. The most natural way of extending \eqref{ddis} to the case at hand and defining a measure of the length of $\sigma$ is:
\begin{equation}
    \mathbf{L}\left[\sigma\right]\equiv\int_{s_0}^{s_1}\diff s\left|\ev{K^{\hat{\mu}\hat{\nu}}\left(\sigma\right)\frac{\diff\sigma_{\hat{\mu}}}{\diff s}\frac{\diff\sigma_{\hat{\nu}}}{\diff s}}\right|^{\frac{1}{2}}
\end{equation}
Further expanding the discussion to variations of multiple fields $\Psi\equiv\left(\psi^1,\dots,\psi^N\right)$, labelled by $i=1,\dots,N$ and with components $\psi^i_{\hat{\mu}(i)}\equiv\psi^i_{\mu_1\dots\mu_{n(i)}}$, the distance along a curve $\Sigma$ in the moduli space $\mathbf{M}_\Psi$ is generalised as follows:
\begin{equation}\label{generalstdistance}
    \mathbf{L}\left[\Sigma\right]\equiv\int_{s_0}^{s_1}\diff s\left|\ev{K^{\hat{\mu}(i)\hat{\nu}(j)}_{\ ij}\left(\Sigma\right)\frac{\diff\Sigma^i_{\hat{\mu}(i)}}{\diff s}\frac{\diff\Sigma^j_{\hat{\nu}(j)}}{\diff s}}\right|^{\frac{1}{2}}\ .
\end{equation}
As before, the moduli space metric should be read off from the Lagrangian kinetic term:
\begin{equation}
    \mathscr{L}_{\text{kin}}\left(\Psi\right)\sim-K^{\hat{\mu}(i)\hat{\nu}(j)}_{\ ij}\left(\Psi\right)\mathcal{D}\Psi^i_{\hat{\mu}(i)}\cdot\mathcal{D}\Psi^j_{\hat{\nu}(j)}\ .
\end{equation}
When either imposing $\Sigma$ to be space-time constant or to have no $\Bar{\mu}$ indices, the previously analysed cases can be consistently recovered from \eqref{generalstdistance}.
\subsubsection{Smooth metric variations}\label{subsub:metvar}
Now that the extended formula \eqref{generalstdistance} for the length of a curve in the generalised moduli space of all space-time fields has been defined, we can focus on the particular case of geometry displacements
\begin{equation}
    g_{\mu\nu}\longrightarrow g_{\mu\nu}+a_{\mu\nu}\ ,
\end{equation}
where $a_{\mu\nu}$ is a symmetric and traceless tensor field, so that the Levi-Civita condition is preserved. Introducing a smooth one-parameter dependence in $a_{\mu\nu}\left(s\right)$, with $s\in\left[s_0,s_1\right]$, and imposing the initial variation $a_{\mu\nu}\left(s_0\right)$ to vanish, a curve
\begin{equation}
    G_{\mu\nu}\left(s\right)\equiv g_{\mu\nu}+a_{\mu\nu}\left(s\right)
\end{equation}
in the generalised moduli space of space-time metrics is defined, with:
\begin{equation}
    G_{\mu\nu}\left(s_0\right)= g_{\mu\nu}\ .
\end{equation}
From the $D$-dimensional Einstein-Hilbert action
\begin{equation}
    S_{\text{EH}}=\frac{1}{2\kappa^2_D}\int_{\mathcal{M}}\diff^Dx\sqrt{-h}R_h
\end{equation}
for the generic Lorentzian metric $h_{\mu\nu}$, the kinetic terms matrix can be read off as:
\begin{equation}\label{DeWitt}
    K^{\mu\nu\alpha\beta}=\frac{1}{2}\left(h^{\mu\alpha}h^{\nu\beta}+h^{\mu\beta}h^{\nu\alpha}-2h^{\mu\nu}h^{\alpha\beta}\right)\ .
\end{equation}
The above expression, up to a $\sqrt{-g}$ factor which we have chosen to move to the definition \eqref{staverage} of the space-time average, is referred to as the \textit{DeWitt metric} \cite{DW:DeWitt:1967yk,DW:Giulini:1994dx,DW:Giulini:2006xi,DW:Pekonen:1987bc}. Such an object is usually derived and discussed in the context of canonical general relativity, analysing the Hamiltonian structure of the theory. Furthermore, a parameter $\tau$ is typically introduced as follows
\begin{equation}
    K^{\mu\nu\alpha\beta}_{\tau}=\frac{1}{2}\left(h^{\mu\alpha}h^{\nu\beta}+h^{\mu\beta}h^{\nu\alpha}-2\tau h^{\mu\nu}h^{\alpha\beta}\right)\ ,
\end{equation}
producing a whole $\tau$-dependent family of DeWitt metrics. Therefore, assuming to work in this more general setting, the curve length of $G_{\mu\nu}\left(s\right)$ in the space of metric deformations can be computed as:
\begin{equation}\label{ttdistance}
\begin{split}
        \mathbf{L}_{\tau}\left[G\right]&=\int_{s_0}^{s_1}\diff s\left|\ev{K^{\mu\nu\alpha\beta}_{\tau}\frac{\diff G_{\mu\alpha}}{\diff s}\frac{\diff G_{\nu\beta}}{\diff s}}\right|^{\frac{1}{2}}\\
        &=\int_{s_0}^{s_1}\diff s\left|\frac{1}{V_{\mathcal{M}}}\int_{\mathcal{M}}\diff^Dx\sqrt{-G}\left(G^{\mu\nu}G^{\alpha\beta}-\tau G^{\mu\alpha}G^{\nu\beta}\right)\frac{\diff G_{\mu\alpha}}{\diff s}\frac{\diff G_{\nu\beta}}{\diff s}\right|^{\frac{1}{2}}\ .
\end{split}
\end{equation}
Therefore, the above formula seems to provide us with a general notion of path length in the generalised moduli space of space-time metrics, which requires to be further specified. Before doing so, we will briefly discuss the geodesics associated to \eqref{ttdistance} and consider a simple example of geometry displacements.
\paragraph{Geodesic equation}
The functional \eqref{ttdistance} assigns a length to any specific path in the generalised moduli space of space-time metrics. Still, selecting two moduli space points, corresponding to distinct geometries $g_{\mu\nu}^{\left(0\right)}$ and $g_{\mu\nu}^{\left(1\right)}$, and measuring the length of a curve connecting them is not enough to properly quantify the distance between them. The reason is that the choice of a moduli space path, without further restriction, is ambiguous. In order for such ambiguity to be removed, we impose the distance between two points to be equal to the length of the minimal geodesic curve connecting them. This choice perfectly analogous to the one taken in the context of the swampland distance conjecture. Focusing on the $\tau=0$ case, which will anyway be the most relevant one in the following sections, we can derive the associated geodesic equation. Infinitesimally, we have:
\begin{equation}
    \begin{split}
    \mathbf{L}_{0}\left[G+\delta G\right]&\sim \mathbf{L}_{0}\left[G\right]+\delta \mathbf{L}_{0}\left[G\right]\ .
    \end{split}
\end{equation}
In the above expression, the length variation can be implicitly expressed as:
\begin{equation}
    \begin{split}
        \delta \mathbf{L}_{0}\left[G\right]&=\frac{1}{2}\int_{s_0}^{s_1}\diff s\Biggl\{\delta\left(\frac{1}{V_{\mathcal{M}}}\int_{\mathcal{M}}  \text{vol}\left(G\right)\cdot\text{Tr}\left[\left(G^{-1}\frac{\diff G}{\diff s}\right)^2\right]\right)\\
    &\quad\quad\quad\quad\quad\quad\quad\cdot\left|\frac{1}{V_{\mathcal{M}}}\int_{\mathcal{M}} \text{vol}\left(G\right)\cdot\text{Tr}\left[\left(G^{-1}\frac{\diff G}{\diff s}\right)^2\right]\right|^{-\frac{1}{2}}\Biggr\}\ .
    \end{split}
\end{equation}
We refer to the infinitesimal variation appearing in the integral as:
\begin{equation}
    \delta\mathcal{A}\equiv \delta\left(\frac{1}{V_{\mathcal{M}}}\int_{\mathcal{M}} \text{vol}\left(G\right)\cdot\text{Tr}\left[\left(G^{-1}\frac{\diff G}{\diff s}\right)^2\right]\right)\ .
\end{equation}
Therefore, we have:
\begin{equation}
    \begin{split}
        \delta\mathcal{A}&=\frac{1}{V_{\mathcal{M}}}\int_{\mathcal{M}} \text{vol}\left(G\right)\frac{1}{2}\left\{\text{Tr}\left[\left(G^{-1}\frac{\diff G}{\diff s}\right)^2\right]-\ev{\text{Tr}\left[\left(G^{-1}\frac{\diff G}{\diff s}\right)^2\right]}\right\}G^{-1}\delta G\\
        &\quad+\frac{1}{V_{\mathcal{M}}}\int_{\mathcal{M}} \text{vol}\left(G\right)\cdot\delta\text{Tr}\left[\left(G^{-1}\frac{\diff G}{\diff s}\right)^2\right]
    \end{split}
\end{equation}
The last term, which we label as $\delta\mathcal{A}_{0}$, can be computed as:
\begin{equation}
    \begin{split}
        \delta\mathcal{A}_0&=\frac{2}{V_{\mathcal{M}}}\int_{\mathcal{M}} \text{vol}\left(G\right)\cdot \frac{\diff G}{\diff s}G^{-1}\frac{\diff G}{\diff s}\delta G^{-1}+\frac{2}{V_{\mathcal{M}}}\int_{\mathcal{M}} \text{vol}\left(G\right)\cdot G^{-1}\frac{\diff G}{\diff s}G^{-1}\delta\frac{\diff G}{\diff s}\\
        &=\frac{2}{V_{\mathcal{M}}}\int_{\mathcal{M}} \text{vol}\left(G\right)\cdot \frac{\diff G}{\diff s}G^{-1}\frac{\diff G}{\diff s}\delta G^{-1}-\frac{4}{V_{\mathcal{M}}}\int_{\mathcal{M}} \text{vol}\left(G\right)\cdot \frac{\diff{G^{-1}}}{\diff s}\frac{\diff G}{\diff s}G^{-1}\delta G\\
        &\quad+\frac{1}{V_{\mathcal{M}}}\int_{\mathcal{M}} \text{vol}\left(G\right)\left\{\ev{\text{Tr}\left[G^{-1}\frac{\diff G}{\diff s}\right]}-\text{Tr}\left[G^{-1}\frac{\diff G}{\diff s}\right]\right\}G^{-1}\frac{\diff G}{\diff s}G^{-1}\delta G\\
        &\quad-\frac{2}{V_{\mathcal{M}}}\int_{\mathcal{M}} \text{vol}\left(G\right)\cdot G^{-1}\frac{\diff^2 G}{\diff s^2}G^{-1}\delta G\ .
    \end{split}
\end{equation}
Putting everything together and expressing each term with respect to the variation $\delta G$ of the metric, we obtain:
\begin{equation}\label{varfinmet}
    \begin{split}
        \delta\mathcal{A}&=\frac{1}{V_{\mathcal{M}}}\int_{\mathcal{M}} \text{vol}\left(G\right)\frac{1}{2}\left\{\text{Tr}\left[\left(G^{-1}\frac{\diff G}{\diff s}\right)^2\right]-\ev{\text{Tr}\left[\left(G^{-1}\frac{\diff G}{\diff s}\right)^2\right]}\right\}G^{-1}\delta G\\
        &\quad-\frac{2}{V_{\mathcal{M}}}\int_{\mathcal{M}} \text{vol}\left(G\right)\cdot \frac{\diff G^{-1}}{\diff s}G\frac{\diff G^{-1}}{\diff s}\delta G-\frac{4}{V_{\mathcal{M}}}\int_{\mathcal{M}} \text{vol}\left(G\right)\cdot \frac{\diff{G^{-1}}}{\diff s}\frac{\diff G}{\diff s}G^{-1}\delta G\\
        &\quad+\frac{1}{V_{\mathcal{M}}}\int_{\mathcal{M}} \text{vol}\left(G\right)\left\{\ev{\text{Tr}\left[G^{-1}\frac{\diff G}{\diff s}\right]}-\text{Tr}\left[G^{-1}\frac{\diff G}{\diff s}\right]\right\}G^{-1}\frac{\diff G}{\diff s}G^{-1}\delta G\\
        &\quad-\frac{2}{V_{\mathcal{M}}}\int_{\mathcal{M}} \text{vol}\left(G\right)\cdot G^{-1}\frac{\diff^2 G}{\diff s^2}G^{-1}\delta G\ .
    \end{split}
\end{equation}
Imposing the overall variation to vanish, so that $G\left(s\right)$ is a geodesic trajectory, one has:
\begin{equation}
    \begin{split}
    \frac{\diff^2 G}{\diff s^2}&=\frac{\diff G}{\diff s}G^{-1}\frac{\diff G}{\diff s}+\frac{1}{2}\left\{\ev{\text{Tr}\left[G^{-1}\frac{\diff G}{\diff s}\right]}-\text{Tr}\left[G^{-1}\frac{\diff G}{\diff s}\right]\right\}\frac{\diff G}{\diff s}\\
        &\quad+\frac{1}{4}\left\{\text{Tr}\left[\left(G^{-1}\frac{\diff G}{\diff s}\right)^2\right]-\ev{\text{Tr}\left[\left(G^{-1}\frac{\diff G}{\diff s}\right)^2\right]}\right\}G\ .
    \end{split}
\end{equation}
Hence, the geodesic equation \cite{Bonnefoy:2019nzv} associated to the moduli space metric induced by \eqref{ttdistance} takes the following form:
\begin{equation}\label{metgeod1}
\begin{split}
\Ddot{G}_{\mu\nu}&=G^{\alpha\beta}\Dot{G}_{\mu\alpha}\Dot{G}_{\nu\beta}+\frac{1}{4}G^{\alpha\beta}G^{\gamma\delta}\Dot{G}_{\alpha\gamma}\Dot{G}_{\beta\delta}G_{\mu\nu}-\frac{1}{2}G^{\alpha\beta} \Dot{G}_{\alpha\beta}\Dot{G}_{\mu\nu}\\
       &\quad-\frac{1}{4}\ev{G^{\alpha\beta}G^{\gamma\delta}\Dot{G}_{\alpha\gamma}\Dot{G}_{\beta\delta}}G_{\mu\nu}+\frac{1}{2}\ev{G^{\alpha\beta} \Dot{G}_{\alpha\beta}}\Dot{G}_{\mu\nu}\ .
\end{split}
\end{equation}
In the above, the derivative indicated with the dot notation $\Dot{A}$ is taken with respect to the proper time $t$, infinitesimally defined by:
\begin{equation}
    \diff t\equiv\diff s\left|\frac{1}{V_{\mathcal{M}}}\int_{\mathcal{M}}\diff^Dx\sqrt{-G}G^{\mu\nu}G^{\alpha\beta}\frac{\diff G_{\mu\alpha}}{\diff s}\frac{\diff G_{\nu\beta}}{\diff s}\right|^{\frac{1}{2}}\ .
\end{equation}
In order to simplify the above expression, a new variable
\begin{equation}\label{defH}
    H^{\mu}{}_{\nu}\equiv G^{\mu\alpha}\Dot{G}_{\alpha\nu}
\end{equation}
can be introduced. By doing so, the geodesic equation \eqref{metgeod1} takes the following form:
\begin{equation}\label{metgeod2}
\begin{split}
        \Dot{H}^{\mu}{}_{\nu}=\frac{1}{4}H^{\alpha}{}_{\beta}H^{\beta}{}_{\alpha}\delta^{\mu}{}_{\nu}-\frac{1}{2}H^{\alpha}{}_{\alpha}H^{\mu}{}_{\nu}-\frac{1}{4}\ev{H^{\alpha}{}_{\beta}H^{\beta}{}_{\alpha}}\delta^{\mu}{}_{\nu}+\frac{1}{2}\ev{H^{\alpha}{}_{\alpha}}H^{\mu}{}_{\nu}\ .
\end{split}
\end{equation}
\paragraph{Anti-de Sitter space-time}
Now, we will apply such result to the interesting example of Anti-de Sitter space-time. In Poincaré coordinates \cite{Bayona:2005nq}, the metric of $D$-dimensional AdS space-time \eqref{adsmetric} can be simply written as
\begin{equation}\label{poincar}
    \diff s^2=\frac{\left(1-D\right)\left(D-2\right)}{2\Lambda}\left(\frac{\diff u^2}{u^2}+u^2\eta_{\mu\nu}\diff x^\mu\diff x^\nu\right)\equiv\Lambda^{-1}\diff \sigma^2\ ,
\end{equation}
where $\eta_{\mu\nu}$ is the $\left(D-1\right)$-dimensional Minkowski metric and $\diff\sigma$ is the differential unit length in Anti-de Sitter space-time when $\Lambda=1$. Assuming to work with a metric deformation which can be completely moved to the cosmological constant, leaving $\diff\sigma$ untouched, we can promote $\Lambda$ to an $s$-dependent function and compute:
\begin{equation}
    H^{\mu}{}_{\nu}= -\frac{\Dot{\Lambda}}{\Lambda}\delta^{\mu}{}_{\nu}\ .
\end{equation}
Hence, the geodesic equation \eqref{metgeod2} becomes:
\begin{equation}
\begin{split}
        \Ddot{\Lambda}\Lambda-\Dot{\Lambda}^2=0\ .
\end{split}
\end{equation}
The above equation can be solved, after having introduced two constants $C_0$ and $C_1$, by the family of smooth solutions:
\begin{equation}
    \Lambda\left(t\right)=C_0e^{C_1t}\ .
\end{equation}
Therefore, by choosing an initial and a final moduli space point, respectively characterised by $\Lambda_0$ and $\Lambda_1$, and imposing $t_0=0$ and $t_1=1$, the geodesic connecting the two can be obtained by fixing $C_0$ and $C_1$ as:
\begin{equation}
       \Lambda\left(t\right)=\Lambda_0\left(\frac{\Lambda_1}{\Lambda_0}\right)^t\ ,
\end{equation}
where $\Lambda\left(0\right)=\Lambda_0$ and $\Lambda\left(1\right)=\Lambda_1$.

\paragraph{Weyl deformations}
In this example, we consider Weyl metric variations
\begin{equation}\label{weylrescaling}
    W_{\mu\nu}\left(s\right)=g_{\mu\nu}+w_{\mu\nu}\left(s\right)\ ,
\end{equation}
where we have imposed $s\in\left[s_0,s_1\right]$, $\omega\left(s_0\right)=0$ and:
\begin{equation}
    w_{\mu\nu}\left(s\right)=\left[e^{\omega\left(s\right)}-1\right]g_{\mu\nu}\ .
\end{equation}
The metric variation \eqref{weylrescaling} clearly induces the inverse metric variation
\begin{equation}\label{invweylrescaling}
    W^{\mu\nu}\left(s\right)=g^{\mu\nu}-v^{\mu\nu}\left(s\right)\ ,
\end{equation}
where we have introduced:
\begin{equation}
    v^{\mu\nu}\left(s\right)=\left[e^{-\omega\left(s\right)}-1\right]g^{\mu\nu}\ .
\end{equation}
Thus, we have preserved the condition:
\begin{equation}
    W_{\mu\nu}\left(s\right)W^{\mu\nu}\left(s\right)=\mathbb{1}\ .
\end{equation}
Under such assumptions, the curve length formula \eqref{ttdistance} simplifies to:
\begin{equation}
        \mathbf{L}_{\tau}\left[W\right]=\left|D\left(1-\tau D\right)\omega\left(s_1\right)\right|\ .
\end{equation}
Interestingly, the path length for Weyl rescalings only depends on the path end points: it could have been either computed on a geodesic or non-geodesic moduli space curve, leading to the same result. Hence, it directly provides us with a measure of the geodesic distance, without the need of explicitly solving any kind of geodesic equation. Focusing on the case of Anti-de Sitter space-time and once more exploiting Poincaré coordinates \eqref{poincar}, it is clear that any deformation of the cosmological constant from an the value $\Lambda_0$ to a final value $\Lambda_1$ can be encoded in a Weyl rescaling of the form presented in \eqref{weylrescaling}, with:
\begin{equation}\label{bcomega}
    \omega\left(s_1\right)=\log{\frac{\Lambda_0}{\Lambda_1}}\ .
\end{equation}
It must be stressed that, as long as \eqref{bcomega} is satisfied, the specific dependence of the rescaling on $s$ does not matter. This reflects the fact that $\Lambda$ can be tuned, towards a final value $\Lambda_1$, in many different ways, without affecting the overall path length. As discussed above, the length formula
\begin{equation}
    \Delta\left(\Lambda_1,\Lambda_0\right)\equiv\mathbf{L}_{\tau}\left[W\right]=\left|D\left(1-\tau D\right)\log{\frac{\Lambda_0}{\Lambda_1}}\right|\ .
\end{equation}
correctly measures the moduli space geodesic distance between $\Lambda_0$ and $\Lambda_1$. From this we can see that, for $\tau\neq D^{-1}$, the flat space-time limit $\Lambda_1\rightarrow 0$ discussed in the Anti-de Sitter distance conjecture sits at an infinite distance in the cosmological constant moduli space:
\begin{equation}
        \lim_{\Lambda_1\to 0}\Delta\left(\Lambda_1,\Lambda_0\right)=\infty\ .
\end{equation}
Moreover, the cosmological constant dependence in the formula \eqref{adsmassas} for the infinite tower mass scale gets rephrased as:
\begin{equation}
    m\sim\left|\Lambda\right|^\alpha\sim e^{-\alpha\Delta}\ .
\end{equation}
The mass threshold of the infinite tower of light states precisely drops with an exponential dependence on the moduli space geodesic distance. Thus, the deep connection between the Anti-de Sitter distance conjecture and the standard swampland distance conjecture, within the framework offered by our generalised notion of moduli space path length \eqref{generalstdistance}, is finally established in its full strength. 
\subsubsection{The Riemannian manifold of Riemannian metrics}\label{sss:modulisp}
As previously mentioned, we will now introduce a more precise notion \textit{generalised moduli space} for space-time geometries. In order to do so, we will follow the analysis developed in \cite{1992math1259G}, using \cite{michor2011manifolds} as a more general reference, and direct our attention to Riemannian manifolds. Namely, we will temporarily disregard Lorentzian metrics in favour of Euclidean ones, for which the mathematical aspects of the problems at hand are better understood and explored. The discussion of the specific difficulties introduced by Lorentzian signature, together with a series of measures which should be enforced in order to avoid them, will be postponed to \ref{sec:ric}. That being said, let $\mathcal{M}$ be a smooth and finite dimensional differentiable manifold. Let's furthermore consider the set $\mathbf{G}_{\mathcal{M}}$ of all Riemannian metrics with which $\mathcal{M}$ can be endowed. Such set can, as widely discussed in the above mentioned reference \cite{1992math1259G}, be itself provided with a topological and differential structure. Starting from $\mathcal{M}$, one can naturally construct the vector bundle $S^{2}T^{*}\mathcal{M}$ of symmetric $\left(0,2\right)$-tensor fields on $\mathcal{M}$. Among those, the subset $S_+^{2}T^{*}\mathcal{M}$ of all the positive-definite ones can be extracted. The manifold $\mathbf{G}_{\mathcal{M}}$ of all Riemannian metrics on $\mathcal{M}$ is, hence, nothing more that the space of smooth sections $\Sigma_{\infty}\left(S_+^{2}T^{*}\mathcal{M}\right)$. Each element 
\begin{equation}
    g\in\mathbf{G}_{\mathcal{M}}\equiv\Sigma_{\infty}\left(S_+^{2}T^{*}\mathcal{M}\right)
\end{equation}
corresponds to a smooth assignment of a positive-definite symmetric $\left(0,2\right)$-tensor to every point $p\in\mathcal{M}$. Generally speaking, $\mathbf{G}_{\mathcal{M}}$ can be expected to be an infinite dimensional smooth manifold. Its tangent bundle can be obtained as
\begin{equation}
T\mathbf{G}_{\mathcal{M}}=\Sigma_{\infty}\left(S^{2}_+T^{*}\mathcal{M}\right)\times\mathcal{D}\left(S^{2}T^{*}\mathcal{M}\right)\ ,
\end{equation}
where $\mathcal{D}\left(S^{2}T^{*}\mathcal{M}\right)$ is the space of sections of the vector bundle $S^{2}T^{*}\mathcal{M}$ with compact support on $\mathcal{M}$. Given a point $g\in\mathbf{G}_{\mathcal{M}}$, which corresponds to a Riemannian metric on $\mathcal{M}$, and two elements $\left(g,a\right)$ and $\left(g,b\right)$ in $T\mathbf{G}_{\mathcal{M}}$, which can be analogously though of as vectors $a$ and $b$ in the tangent space $T_g\mathbf{G}_{\mathcal{M}}$ at $g$, the canonical Riemannian metric on $\mathbf{G}_{\mathcal{M}}$ is typically defined as:
\begin{equation}\label{canonicalmetric}
    G_{g}(h_{1},h_{2})\equiv\int_{\mathcal{M}}vol\left(g\right)tr\left(g^{-1}ag^{-1}b\right)\ .
\end{equation}
Charting $\mathcal{M}$ with an appropriate set of coordinates $x^{\mu}$, \eqref{canonicalmetric} can be written as:
\begin{equation}\label{canonicalmetriccomponent}
    G_{g}(h_{1},h_{2})=\int_{\mathcal{M}}\diff^Dx\sqrt{g} g^{\mu\nu}g^{\alpha\beta}a_{\mu\alpha}b_{\nu\beta} \ .
\end{equation}
Considering a one parameter path 
\begin{equation}
    G:\left[s_0,s_1\right]\subset\R\longrightarrow\mathbf{G}_\mathcal{M}\ ,
\end{equation}
choosing a specific value $\Bar{s}\in\left[s_0,s_1\right]$ and regarding the first derivatives of the components of $G$ at $\Bar{s}$ as elements of $T_{G\left(s\right)}\mathbf{G}_{\mathcal{M}}$, the canonical metric \eqref{canonicalmetriccomponent} assigns to $G$ the following path length:
\begin{equation}
    \Tilde{\mathbf{L}}\left[G\right]=\int_{s_0}^{s_1}\diff s\left|\int_{\mathcal{M}}\diff^Dx\sqrt{G}G^{\mu\nu}G^{\alpha\beta}\frac{\diff G_{\mu\alpha}}{\diff s}\frac{\diff G_{\nu\beta}}{\diff s}\right|^{\frac{1}{2}}\ .
\end{equation}
Except from the fact that we are working with a Riemannian metric a for the absence of a volume normalisation, which is useful in many applications, this is equivalent to the result that can be obtained from the formula $\mathbf{L}_\tau\left[G\right]$ presented in \eqref{ttdistance} by setting $\tau=0$. Hence, we have now obtained, with $\mathbf{G}_{\mathcal{M}}$ and up to the discussed caveats, a proper formalisation of the intuitive notion of a generalised moduli space for space-time geometries. Before proceeding, it is crucial to stress that \textit{physical} metrics are distinguished up to diffeomorphisms:
\begin{equation}
    \varphi:\mathcal{M}\longrightarrow\mathcal{M}\ ,\quad\varphi\in\text{Diff}\left(\mathcal{M}\right)\ .
\end{equation}
The generalised moduli space of physically distinguishable geometries on a given space-time manifold $\mathcal{M}$ should, therefore, be obtained as the quotient:
\begin{equation}
    \mathbf{P}_{\mathcal{M}}\equiv\quotient{\mathbf{G}_{\mathcal{M}}}{\text{Diff}\left(\mathcal{M}\right)}\ .
\end{equation}
This redundancy under diffeomorphisms will not affect the remainder of our analysis in any particular manner. Nonetheless, it will be reflected, in the context of \textit{geometric flows}, in the addition of a flow-dependent diffeomorphism term. 
\subsubsection{Metric-scalar systems}\label{sss:metscalsysgeod}
Before introducing the particular class of moduli space trajectories defined by \textit{geometric flows}, which will be the main focus of the next sections, we will now generalise the previous discussion to a setting in which both a scalar field and a dynamical space-time geometry are present. In particular, we will derive the geodesic equation in the corresponding extended moduli space. Therefore, let $\mathbf{G}_{\mathcal{M}}$ be the generalised moduli space of Riemannian metrics over a $D$-dimensional Riemannian manifold $\mathcal{M}$. Moreover, let 
\begin{equation}
    \phi:\mathcal{M}\longrightarrow C_{\phi}
\end{equation}
be a field defined on $\mathcal{M}$, with values in $C_{\phi}$. The extended moduli space, accounting for both the space-time configuration of the geometry degrees of freedom described by $g$ and that of the field $\phi$, is obtained as: 
\begin{equation}\label{metscalmodulispace}
    \mathbf{\Gamma}_{\mathcal{M}}\equiv \mathbf{G}_{\mathcal{M}}\times\mathbf{\Phi}_{\mathcal{M}}\ .
\end{equation}
In the above expression, the generalised moduli space $\mathbf{\Phi}_{\mathcal{M}}$ of all space-time configurations of $\phi$ should not be confused with the moduli space $\mathbf{M}_\Phi$ of its vacuum expectation values. Any one-parameter curve in $\mathbf{\Gamma}_{\mathcal{M}}$ can be defined as
\begin{equation}
    \Gamma\left(s\right)\equiv\bigl[G\left(s\right),\Phi\left(s\right)\bigr]\ ,\quad\forall s\in\left[s_0,s_1\right]\ ,
\end{equation}
where $G$ and $\Phi$ are one-parameter curves in in $\mathbf{G}_{\mathcal{M}}$ and $\mathbf{\Phi}_{\mathcal{M}}$, respectively. We assume the theory on $\mathcal{M}$ to be controlled by an action of the form:
\begin{equation}
    S\left[g,\phi\right]=\frac{1}{2\kappa_D^2}\int_{\mathcal{M}}\diff^Dx\sqrt{-g}\left(R_g-\frac{1}{2}\nabla^\mu\phi\nabla_\mu\phi\right)\ .
\end{equation}
Consistently with \eqref{ttdistance}, the path length of $\Gamma$ can be thus computed employing the natural notion of path length on $\mathbf{\Gamma}_{\mathcal{M}}$:
\begin{equation}\label{ttdistance7}
        \mathbf{L}_{\tau}\left[\Gamma\right]=\int_{s_0}^{s_1}\diff s\left|\ev{\left(G^{\mu\nu}G^{\alpha\beta}-\tau G^{\mu\alpha}G^{\nu\beta}\right)\frac{\diff G_{\mu\alpha}}{\diff s}\frac{\diff G_{\nu\beta}}{\diff s}+\frac{1}{2}\left(\frac{\diff\Phi}{\diff s}\right)^2}\right|^{\frac{1}{2}}\ .
\end{equation}
Fixing two points $\gamma_1\equiv\left(g_1,\phi_1\right)$ and $\gamma_2\equiv\left(g_2,\phi_2\right)$ in $\mathbf{\Gamma}_{\mathcal{M}}$ and extremising the length $\mathbf{L}_{\tau}$ of the path connecting them, one can obtain a geodesic equation, solve it and properly define the geodesic distance between $\gamma_1$ and $\gamma_2$. We will derive the geodesic equation associated with \eqref{ttdistance7} for a general value of $\tau$ and a scalar moduli space path satisfying
\begin{equation}\label{volanspre}
    \frac{\diff\Phi}{\diff s}=\sqrt{2}\lambda G^{\mu\nu}\frac{\diff G_{\mu\nu}}{\diff s}\ ,
\end{equation}
with $\lambda\in\R$. The reason for such an ansatz will become evident in the following sections, when discussing Perelman's combined flow. Therefore, \eqref{ttdistance7} reduces to
\begin{equation}\label{ttdistance8}
        \mathbf{L}_{\vartheta}\left[\Gamma\right]=\int_{s_0}^{s_1}\diff s\left|\ev{\left(G^{\mu\nu}G^{\alpha\beta}+\vartheta G^{\mu\alpha}G^{\nu\beta}\right)\frac{\diff G_{\mu\alpha}}{\diff s}\frac{\diff G_{\nu\beta}}{\diff s}}\right|^{\frac{1}{2}}\ ,
\end{equation}
in which $\vartheta\equiv\lambda^2-\tau$. In a more implicit and clear notation, we have:
\begin{equation}\label{lengthgen2}
    \mathbf{L}_{\vartheta}\left[\Gamma\right]=\int_{s_0}^{s_1}\diff s\left|\ev{\text{Tr}\left[\left(G^{-1}\frac{\diff G}{\diff s}\right)^2\right]}+\vartheta\ev{\text{Tr}\left[G^{-1}\frac{\diff G}{\diff s}\right]^2}\right|^{\frac{1}{2}}\ .
\end{equation}
Now, we will generalise the computation performed in \ref{subsub:metvar} and derive the geodesic equation associated to \eqref{lengthgen2}. Namely, we variate a path $\Gamma\equiv\left(G,\Phi\right)$ and impose the induced first order variation of the curve length to vanish. We obtain
\begin{equation}
    \delta\mathbf{L}_{\vartheta}\left[\Gamma\right]=\frac{1}{2}\int_{s_0}^{s_1}\diff s\left|\ev{\text{Tr}\left[\left(G^{-1}\frac{\diff G}{\diff s}\right)^2\right]}+\vartheta\ev{\text{Tr}\left[G^{-1}\frac{\diff G}{\diff s}\right]^2}\right|^{-\frac{1}{2}}\delta\mathcal{A}\ ,
\end{equation}
where we have defined the local variation
\begin{equation}
    \delta\mathcal{A}=\delta\mathcal{A}_{1}+\vartheta\cdot\delta\mathcal{A}_{2}\ ,
\end{equation}
expressed as the sum of two terms:
\begin{equation}
    \delta\mathcal{A}_{1}\equiv\delta\ev{\text{Tr}\left[\left(G^{-1}\frac{\diff G}{\diff s}\right)^2\right]}\ ,\quad \delta\mathcal{A}_{2}\equiv\ev{\text{Tr}\left[G^{-1}\frac{\diff G}{\diff s}\right]^2}\ .
\end{equation}
From the result expressed in \eqref{varfinmet}, we have that
\begin{equation}
    \begin{split}
        \delta\mathcal{A}_1&=\frac{1}{V_{\mathcal{M}}}\int_{\mathcal{M}} \text{vol}\left(G\right)\mathcal{B}_1\cdot\delta G\ ,
    \end{split}
\end{equation}
where we have defined:
\begin{equation}
    \begin{split}
        \mathcal{B}_1&\equiv G^{-1}\Biggl\{\frac{1}{2}\left(\text{Tr}\left[\left(G^{-1}\frac{\diff G}{\diff s}\right)^2\right]-\ev{\text{Tr}\left[\left(G^{-1}\frac{\diff G}{\diff s}\right)^2\right]}\right)G+2\frac{\diff G}{\diff s}G^{-1}\frac{\diff G}{\diff s}\\
        &\quad\quad\quad\quad-2\frac{\diff^2 G}{\diff s^2}+\left(\ev{\text{Tr}\left[G^{-1}\frac{\diff G}{\diff s}\right]}-\text{Tr}\left[G^{-1}\frac{\diff G}{\diff s}\right]\right)\frac{\diff G}{\diff s}\Biggr\}G^{-1}\ .
    \end{split}
\end{equation}
The second term, instead, requires us to perform an explicit computation. We obtain
\begin{equation}
    \begin{split}
        \delta\mathcal{A}_2&=\frac{1}{V_{\mathcal{M}}}\int_{\mathcal{M}} \text{vol}\left(G\right)\mathcal{B}_2\cdot\delta G\ ,
    \end{split}
\end{equation}
in which we have defined:
\begin{equation}
    \begin{split}
        \mathcal{B}_2&\equiv \frac{1}{2}\left(\text{Tr}\left[G^{-1}\frac{\diff G}{\diff s}\right]^2-\ev{\text{Tr}\left[G^{-1}\frac{\diff G}{\diff s}\right]^2}\right)G^{-1}+2\text{Tr}\left[G^{-1}\frac{\diff G}{\diff s}\right]\frac{\diff G^{-1}}{\diff s}\\
        &\quad+\Biggl(\ev{\text{Tr}\left[G^{-1}\frac{\diff G}{\diff s}\right]}-\text{Tr}\left[G^{-1}\frac{\diff G}{\diff s}\right]\Biggr)\text{Tr}\left[G^{-1}\frac{\diff G}{\diff s}\right]G^{-1}\\
        &\quad-2\Biggl(\text{Tr}\left[\frac{\diff G^{-1}}{\diff s}\frac{\diff G}{\diff s}\right]G^{-1}+\text{Tr}\left[G^{-1}\frac{\diff^2 G}{\diff s^2}\right]G^{-1}+\text{Tr}\left[G^{-1}\frac{\diff G}{\diff s}\right]\frac{\diff G^{-1}}{\diff s}\Biggr)\ .
    \end{split}
\end{equation}
Therefore, imposing the overall variation
\begin{equation}
    \delta\mathcal{A}=\frac{1}{V_{\mathcal{M}}}\int_{\mathcal{M}} \text{vol}\left(G\right)\left(\mathcal{B}_{1}+\vartheta\cdot\mathcal{B}_{2}\right)\delta G\ ,
\end{equation}
to vanish, we are left with the geodesic equation:
\begin{equation}
    \begin{split}
        \frac{\diff^2 G}{\diff s^2}&=\frac{1}{4}\left(\text{Tr}\left[\left(G^{-1}\frac{\diff G}{\diff s}\right)^2\right]-\ev{\text{Tr}\left[\left(G^{-1}\frac{\diff G}{\diff s}\right)^2\right]}\right)G+\frac{\diff G}{\diff s}G^{-1}\frac{\diff G}{\diff s}\\
        &\quad+\frac{1}{2}\left(\ev{\text{Tr}\left[G^{-1}\frac{\diff G}{\diff s}\right]}-\text{Tr}\left[G^{-1}\frac{\diff G}{\diff s}\right]\right)\frac{\diff G}{\diff s}\\
        &\quad+\frac{\vartheta}{4}\left(\text{Tr}\left[G^{-1}\frac{\diff G}{\diff s}\right]^2-\ev{\text{Tr}\left[G^{-1}\frac{\diff G}{\diff s}\right]^2}\right)G\\
        &\quad+\frac{\vartheta}{2}\Biggl(\ev{\text{Tr}\left[G^{-1}\frac{\diff G}{\diff s}\right]}-\text{Tr}\left[G^{-1}\frac{\diff G}{\diff s}\right]\Biggr)\text{Tr}\left[G^{-1}\frac{\diff G}{\diff s}\right]G\\
        &\quad-\vartheta\Biggl(\text{Tr}\left[\frac{\diff G^{-1}}{\diff s}\frac{\diff G}{\diff s}\right]G+\text{Tr}\left[G^{-1}\frac{\diff^2 G}{\diff s^2}\right]G\Biggr)\ .
    \end{split}
\end{equation}
Once more introducing $H^{\mu}{}_\nu$ as in \eqref{defH}, the above expression simplifies to:
\begin{equation}
    \begin{split}
        \frac{\diff H}{\diff s}+\vartheta\text{Tr}\left[\frac{\diff H}{\diff s}\right]\mathbb{1}&=\frac{1}{4}\Bigl(\text{Tr}\left[H^2\right]-\ev{\text{Tr}\left[H^2\right]}\Bigr)\mathbb{1}+\frac{1}{2}\Bigl(\ev{\text{Tr}\left[H\right]}-\text{Tr}\left[H\right]\Bigr)H\\
        &\quad-\frac{\vartheta}{4}\Bigl(\text{Tr}\left[H\right]^2-2\text{Tr}\left[H\right]\ev{\text{Tr}\left[H\right]}+\ev{\text{Tr}\left[H\right]^2}\Bigr)\mathbb{1}\ .
    \end{split}
\end{equation}
\paragraph{Non-normalised distance}
Focusing once more on the generalised moduli space $\mathbf{\Gamma}_{\mathcal{M}}$, whose points correspond to distinct configurations of a dynamical space-time metric and a scalar field, an alternative approach can be studied. Namely, the volume normalisation appearing in the path length formulas \eqref{ttdistance7} and \eqref{ttdistance8} can be removed, in order to stick to the convention typically employed in mathematical literature. Assuming to work with a manifold that allows such non-normalised distances to be properly behaved and once more enforcing the ansatz \eqref{volanspre}, we are left with the alternative definition:
\begin{equation}\label{lengthgen3}
    \mathbf{\Tilde{L}}_{\vartheta}\left[\Gamma\right]=\int_{s_0}^{s_1}\diff s\left|\int_{\mathcal{M}}\text{vol}\left(G\right)\left\{\text{Tr}\left[\left(G^{-1}\frac{\diff G}{\diff s}\right)^2\right]+\vartheta\text{Tr}\left[G^{-1}\frac{\diff G}{\diff s}\right]^2\right\}\right|^{\frac{1}{2}}\ .
\end{equation}
Imposing the path length variation associated to a displacement 
\begin{equation}
    \Gamma\left(s\right)\longrightarrow\Gamma\left(s\right)+\delta\Gamma\left(s\right)
\end{equation}
to vanish, we obtain the following geodesic equation:
\begin{equation}
    \begin{split}
        \frac{\diff^2 G}{\diff s^2}&=\frac{1}{4}\text{Tr}\left[\left(G^{-1}\frac{\diff G}{\diff s}\right)^2\right]G+\frac{\diff G}{\diff s}G^{-1}\frac{\diff G}{\diff s}\\
        &\quad-\frac{1}{2}\text{Tr}\left[G^{-1}\frac{\diff G}{\diff s}\right]\frac{\diff G}{\diff s}+\frac{\vartheta}{4}\text{Tr}\left[G^{-1}\frac{\diff G}{\diff s}\right]^2G\\
        &\quad-\frac{\vartheta}{2}\text{Tr}\left[G^{-1}\frac{\diff G}{\diff s}\right]\text{Tr}\left[G^{-1}\frac{\diff G}{\diff s}\right]G\\
        &\quad-\vartheta\Biggl(\text{Tr}\left[\frac{\diff G^{-1}}{\diff s}\frac{\diff G}{\diff s}\right]G+\text{Tr}\left[G^{-1}\frac{\diff^2 G}{\diff s^2}\right]G\Biggr)\ .
    \end{split}
\end{equation}
Introducing, once again, $H^{\mu}{}_\nu$ as in \eqref{defH}, the above equation takes the form:
\begin{equation}
    \begin{split}
        \frac{\diff H}{\diff s}+\vartheta\cdot\text{Tr}\left[\frac{\diff H}{\diff s}\right]\mathbb{1}&=\Bigl[\text{Tr}\left[H^2\right]+\vartheta\left(2\text{Tr}\left[H\right]-\text{Tr}\left[H\right]^2\right)\Bigr]\frac{\mathbb{1}}{4}-\text{Tr}\left[H\right]\frac{H}{2}\ .
    \end{split}
\end{equation}
For $\vartheta=0$, we find ourselves working with the extremely simple formula:
\begin{equation}
    \begin{split}
        \frac{\diff H}{\diff s}&=\text{Tr}\left[H^2\right]\frac{\mathbb{1}}{4}-\text{Tr}\left[H\right]\frac{H}{2}\ .
    \end{split}
\end{equation}

%% file: Sections/GeometricFlows/ricci.tex
While outlining the properties of the generalised moduli space $\mathbf{G}_{\mathcal{M}}$ of space-time geometries which $\mathcal{M}$ can be endowed with, particular attention was devoted to constructing an explicit procedure to measure the length of any moduli space path $g\left(s\right)$, corresponding to a one-parameter family of metric tensors on $\mathcal{M}$. Nonetheless, the only meaningful example of a moduli space path appearing in the previous discussion was that represented by geodesic curves \eqref{metgeod2}. Albeit being of great help when computing distances, the geodesic equation seems to bear little direct physical significance and is often extremely hard to solve, at least without making strong assumptions on system's symmetries and general features. Therefore, we now intend to explore an alternative class of moduli space paths, employing the tools offered by \textit{geometric flows}. Taking Ricci flow as an obvious and natural starting point, both due to its useful mathematical properties and to its direct connection to the superstring theory graviton $\beta$-function \cite{Polchinski:1998rq,Polchinski:1998rr}, we will then generalise it and integrate it within the context of the swampland distance conjecture. Most of our discussion will follow the research line initiated in \cite{Kehagias:2019akr}. Still, we will approach the subject without the degree of rigour any proper, formal treatment would require. The interested reader is again suggested to refer to \cite{hamilton1982,Perelman:2006un,chow2004ricci,Cao2006ACP,morgan2007ricci,kleiner2008notes} for more detailed expositions.
\subsection{Ricci flow}
Let's consider a Lorentzian manifold $\mathcal{M}$ with signature $\left(-,+,\dots,+\right)$ and the space $\mathbf{G}_{\mathcal{M}}$ of all possible metrics on $\mathcal{M}$, itself endowed with a manifold structure. Even if we are now working with a non-Riemannian manifold, the construction of $\mathbf{G}_{\mathcal{M}}$ is analogous to the one outlined in \ref{sss:modulisp}. Let's furthermore define
\begin{equation}
    g:\left[s_0,s_1\right]\subset\R\longrightarrow\mathbf{G}_{\mathcal{M}}
\end{equation}
as a one-parameter smooth path in $\mathbf{G}_{\mathcal{M}}$, with initial point $g\left(s_0\right)=\Bar{g}$. The evolution of $g\left(s\right)$ in $s$ is said to be dictated by Ricci flow if it satisfies:
\begin{equation}\label{implRicci}
    \frac{\diff g}{\diff s}=-2\text{Ric}\left(g\right)\ ,\quad\forall s\in\left(s_0,s_1\right)\ .
\end{equation}
In the above equation, $\text{Ric}\left(g\right)$ is nothing more than the Ricci tensor associated to $g\left(s\right)$ at a specific value of $s$, which is usually referred to as the \textit{flow parameter}. Charting $\mathcal{M}$ with an appropriate set of coordinates, the explicit version of \eqref{implRicci} can be written down as:
\begin{equation}\label{explRicci}
        \frac{\diff g_{\mu\nu}}{\diff s}=-2R_{\mu\nu}\ .
\end{equation}
It can be easily observed that the above equations implies the scalar curvature flow:
\begin{equation}\label{scalRF}
    \frac{\diff R}{\diff s}=\nabla^2 R+2R_{\mu\nu}R^{\mu\nu}\ .
\end{equation}
From \eqref{explRicci}, it is clear that any flat geometry for which
\begin{equation}
    R_{\mu\nu}=0
\end{equation}
is a Ricci flow fixed point. As commented on at the end of \ref{sss:modulisp}, distinguishable physical metrics are uniquely defined up to diffeomorphisms. Taking such an aspect into account, Ricci flow can be generalised to the so-called \textit{Hamilton-DeTurk-Ricci flow}
\begin{equation}\label{HamDet}
    \frac{\diff g_{\mu\nu}}{\diff s}=-2R_{\mu\nu}+\mathcal{L}_\xi g_{\mu\nu}\ ,
\end{equation}
in which a flow-dependent diffeomorphism, induced by a one-parameter family of vector fields $\xi\left(s\right)$ on $\mathcal{M}$, was included. Therefore, a path $g\left(s\right)$ in $\mathbf{G}_{\mathcal{M}}$ satisfies Hamilton-DeTurk-Ricci flow equations if there exists a one-parameter family of vector fields $\xi\left(s\right)$ such that \eqref{HamDet} is fulfilled for every value of the flow parameter. From now on, we will always assume to work with \eqref{HamDet} and simply name it Ricci flow, in order for our discussion to be consistent with the existing literature. Thanks to the introduction of the addition of the diffeomorphism term, a new class of significant solutions to our flow equations can be defined: that of \textit{Ricci solitons}. A one-parameter family of geometries $g\left(s\right)$ is said to describe a Ricci soliton if there is a family of vector fields $\xi\left(s\right)$ such that:
\begin{equation}\label{condsol}
    2R_{\mu\nu}=\mathcal{L}_{\xi}g_{\mu\nu}\ ,\quad\forall s\in\left(s_0,s_1\right)\ .
\end{equation}
Hence, Ricci solitons are characterised by the fact that the evolution induced by the Ricci curvature tensor term can be undone by a flow-dependent diffeomorphism. Ricci flow fixed points, for which $R_{\mu\nu}$ vanishes, represent the most trivial subset of solitons. In order to make \eqref{condsol} more explicit, the Lie derivative of the metric components with respect to $\xi$ can be expressed as:
\begin{equation}
    \mathcal{L}_{\xi}g_{\mu\nu}=\nabla_\mu\xi_\nu+\nabla_\nu\xi_\mu\ .
\end{equation}
A simple $2$-dimensional Ricci soliton example will be derived in the following. Subsequently, the main properties and most natural generalisations of Ricci flow will be discussed.
\subsubsection{Ricci soliton example}
Let the space-time Lorentzian manifold $\mathcal{M}$ be $2$-dimensional and charted by coordinates $\Bar{x}\equiv\left(t,x\right)$. Moreover, let the diagonal tensor
\begin{equation}\label{2dinitial}
    \Bar{g}_{\mu\nu}\equiv\begin{bmatrix}
A\left(r\right) & 0 \\
0 & B\left(r\right) 
\end{bmatrix}
\end{equation}
be a metric on $\mathcal{M}$, which will serve as the Ricci flow initial point. The Ricci curvature scalar associated to \eqref{2dinitial} is
\begin{equation}
    R=\frac{B\left(A^{'}\right)^2-2AA^{''}B+AA^{'}B^{'}}{2 A^2
   B^2}\ ,
\end{equation}
where the $F^{'}$ notation refers to $r$-derivatives. Since every $2$-dimensional manifold is an Einstein manifold, we have:
\begin{equation}
    R_{\mu\nu}=\frac{R}{2}g_{\mu\nu}\ .
\end{equation}
Promoting $A$ and $B$ to flow-dependent quantities, the corresponding flow equations can be written down, at every value of the flow parameter, as:
\begin{equation}\label{compbycomp}
    \frac{\diff A}{\diff s}=-R\cdot A+\mathcal{L}_\xi A\ ,\quad \frac{\diff B}{\diff s}=-R\cdot B+\mathcal{L}_\xi B\ .
\end{equation}
Assuming $\xi$ to be $t$-independent and computing its covariant derivatives, we get:
\begin{equation}
    \begin{split}
        \nabla_{t}\xi_{t}&=\frac{A^{'}}{2B}\xi_r\ ,\quad\nabla_{r}\xi_{r}=-\frac{B^{'}}{2B}\xi_r\ ,\\   
        \nabla_{t}\xi_{r}&=-\frac{A^{'}}{2A}\xi_t\ ,\quad\nabla_{r}\xi_{t}=\de_{r}\xi_{t}-\frac{A^{'}}{2A}\xi_t\ .
    \end{split}
\end{equation}
First of all, we assume $\xi_t=0$ in order for the diffeomorphism not to affect the off-diagonal zero metric components. Imposing the equations \eqref{compbycomp} to identically vanish along the flow, we have:
\begin{equation}\label{solcond}
    R\cdot A\cdot B=A^{'}\xi_r\ ,\quad R\cdot B^2=-B^{'}\xi_r\ .
\end{equation}
For the above equation to admit solutions, we restrict the initial metric ansatz and impose
\begin{equation}
    \frac{A^{'}}{A}=-\frac{B^{'}}{B}\Longrightarrow B\left(r\right)=-\frac{C_{0}}{A\left(r\right)}\ ,
\end{equation}
with $C_0\in\R_{+}$ being an arbitrary positive constant, which allows to preserve the correct signature. Under such assumption, the Ricci tensor simplifies to
\begin{equation}
    R=\frac{A^{''}}{C_0}
\end{equation}
and the Ricci soliton conditions \eqref{solcond} leave us with a single equation:
\begin{equation}\label{solcond2}
    \xi_r=-\frac{A^{''}}{A^{'}}\ .
\end{equation}
Raising the index, the vector field components are:
\begin{equation}\label{vectorcomp}
\xi^t=0\ ,\quad \xi^r=\frac{A\cdot A^{''}}{C_0\cdot A^{'}}\ .
\end{equation}
Since the diffeomorphism defined by $\xi^\mu$ is able to undo the evolution imposed by the flow equations \eqref{compbycomp}, we have successfully constructed a $2$-dimensional Ricci soliton for any choice of $A\left(r\right)$ and $C_0$.
\subsubsection{Well-posedness of the flow}
\label{wellposed}
In order to investigate whether a differential equation of the form \eqref{HamDet} is well-posed or not, at least in the simplest non-trivial setting, we will consider small perturbations
\begin{equation}\label{linearmet}
    g_{\mu\nu}\left(s\right)\sim\delta_{\mu\nu}+\varepsilon h_{\mu\nu}\left(s\right)
\end{equation}
around flat space, where $\left|\varepsilon\right|\ll 1$ and $h_{\mu\nu}\sim\mathcal{O}\left(1\right)$ component by component. For the time being, we will assume to work with a Euclidean metric and postpone the discussion of the specificities of Lorentzian geometries. It must be noted that, in \eqref{linearmet}, the flow dependence was completely moved to the metric perturbations. Defining 
\begin{equation}
    h^{\mu\nu}\equiv\delta^{\mu\alpha}\delta^{\nu\beta}h_{\alpha\beta}\ ,\quad h^{\mu}{}_{\nu}\equiv\delta^{\mu\alpha}h_{\alpha\nu}\ ,\quad h\equiv h^{\mu}{}_{\mu}\ ,\quad\nabla^2 h_{\mu\nu}\equiv \delta^{\alpha\beta}\de_\alpha\de_\beta h_{\mu\nu}\ ,
\end{equation}
the leading-order $\varepsilon$-expansion of the Ricci curvature tensor can be written as:
\begin{equation}\label{linrt}
    R_{\mu\nu}\sim\frac{\varepsilon}{2}(\partial_\sigma\partial_\mu h^\sigma{}_\nu + \partial_\sigma\partial_\nu h^\sigma{}_\mu - \partial_\mu\partial_\nu h - \nabla^2 h_{\mu\nu})\ .
\end{equation}
By plugging both \eqref{linearmet} and \eqref{linrt} into \eqref{HamDet}, one is left with the leading-order linearized Ricci flow equation
\begin{equation}
    \frac{\diff h_{\mu\nu}}{\diff s}=\nabla^2 h_{\mu\nu}+\partial_\mu\partial_\nu h-\partial_\sigma\partial_\mu h^\sigma{}_\nu - \partial_\sigma\partial_\nu h^\sigma{}_\mu+\de_\mu\xi_\nu+\de_\nu\xi_\mu  \ ,
\end{equation}
in which the flow-dependent diffeomorphism has been written down explicitly. By introducing a flow-dependent vector field
\begin{equation}
    \omega^{\mu}\equiv\frac{\de^{\mu}h}{2}-\de_\sigma h^{\sigma\mu}\ ,
\end{equation}
whose index can be lowered with the flat space metric, the flow equation gets to be:
\begin{equation}
    \frac{\diff h_{\mu\nu}}{\diff s}=\nabla^2 h_{\mu\nu}+\de_\mu\left(\omega_\nu+\xi_\nu\right)+\de_\nu\left(\omega_\mu+\xi_\mu\right)\ .
\end{equation}
By exploiting our flow-dependent diffeomorphism and setting $\xi^\mu=-\omega^\mu$, we are left with a \textit{heat equation} for each of the metric components:
\begin{equation}\label{heatmet}
    \frac{\diff h_{\mu\nu}}{\diff s}=\nabla^2 h_{\mu\nu}\ .
\end{equation}
Therefore, the well-posedness of \eqref{heatmet}, at least in Euclidean signature, is inherited from that of the heat equation. A more thorough discussion, also concerning perturbations of non-trivial Ricci flat fixed points of the flow, can be found in \cite{Headrick:2006ti}.

\subsubsection{Lorentzian signature}
We have so far assumed to apply Ricci flow together with the additional diffeomorphism term, to Euclidean metrics. This framework allowed us to temporarily simplify our discussion and investigate some significant properties of such system of differential equations. Nonetheless, it is evidently inadequate for the phenomenological applications we will eventually try to address. Euclidean gravity, albeit often representing a useful intermediate step towards the understanding of standard Lorentzian problems, does not offer an appropriate model to make contact with observations \cite{BarberoG:1995tgc}. Nature, as far as our current experimental evidence is concerned, appears to be Lorentzian. After having commented on the well-posedness of the flow equation \eqref{HamDet} when applied to Euclidean metrics, it is therefore necessary to discuss the supplementary pathologies that might arise if, instead, Lorentzian geometries are taken into account. Hence, we will now focus on geometries characterised by a $\left(-,+,\dots,+\right)$ signature. The main problem one might encounter when applying Ricci flow equations, and generalisations thereof, to such geometries is that modes with timelike momentum can introduce infinitely many instabilities \cite{Headrick:2006ti}. Let's focus, for instance, on the specific example of a perturbation 
\begin{equation}
    g_{\mu\nu}\left(s\right)\sim\eta_{\mu\nu}+\varepsilon a_{\mu\nu}\left(s\right)
\end{equation}
around flat Minkowski space-time, where $\left|\varepsilon\right|\ll 1$ and $a_{\mu\nu}\sim\mathcal{O}\left(1\right)$ component by component. Consistently with what was done in Euclidean space, the whole flow dependence was moved to the perturbation. An appropriate flow-dependence diffeomorphism can simplify the linearized flow equations as
\begin{equation}\label{linlor}
    \frac{\diff a_{\mu\nu}}{\diff s}=\nabla^2 a_{\mu\nu}\ ,
\end{equation}
with the flat space-time Laplacian being defined as $\nabla^2\equiv\eta^{\mu\nu}\de_\mu\de_\nu$. For the sake of simplicity, we will furthermore restrict ourselves to the case in which the space-time manifold is $2$-dimensional. Thus, by employing coordinates $\left(t,x\right)$, the flow equations \eqref{linlor} become:
\begin{equation}\label{2dlinlor}
        \frac{\diff a_{\mu\nu}}{\diff s}=\frac{\de^2 a_{\mu\nu}}{\de x^2}-\frac{\de^2 a_{\mu\nu}}{\de t^2}\ .
\end{equation}
Such an equation, when applied to a generic initial metric perturbation depending on both $t$ and $x$, might reveal itself to be extremely problematic. The problem lies in the fact that, due to the Lorentzian space-time signature, the two differential operators on the right-hand side of \eqref{2dlinlor} appear with an opposite sign. More specifically, the minus sign in front of the second derivative in time corresponds to a negative diffusion constant. It hence induces a reversed heat flow along the temporal direction. Differential equations displaying such a feature are troublesome, either due to their solutions being unbounded in finite flow parameter intervals or not existing at all. It is crucial to highlight that the issue analogously appears in any dimension and would have not been solved by choosing the opposite signature convention $\left(+,-,\dots,-\right)$. Similarly, it cannot be worked out by reversing the sign of the flow parameter. The obstacle really lies in the mixed nature of the metric signature. The most natural way of avoiding such difficulty would be to only consider static, $t$-independent initial configurations, reducing \eqref{2dlinlor} to a family of well-behaved diffusion problems. For a more detailed analysis of the subject, together with a more detailed analysis of the features Lorentzian manifolds must have in order to admit a well-defined Ricci flow, the reader is encouraged to explore the references \cite{garcia2014homogeneity,Dhumuntarao:2018jup,Woolgar:2007vz}. From now on, we will assume to work with initial data such that the problems coming from working in Lorentzian signature can be kept under control.
\subsubsection{Natural generalisations}
The flow equations presented in \eqref{HamDet} allow for some natural generalisations. In order to motivate them, we first have to observe that the volume $V_\mathcal{M}$ of $\mathcal{M}$, as defined in \ref{vol}, displays the following Ricci flow evolution
\begin{equation}
    \frac{\diff V_\mathcal{M}}{\diff s}=-\int_{\mathcal{M}}\diff^Dx\sqrt{\left|g\right|}R=-V_\mathcal{M}\cdot\ev{R}\ ,
\end{equation}
in which the diffeomorphism term has been neglected for simplicity. The volume of a manifold is hence not invariant under Ricci flow. This can compromise many of the most interesting mathematical applications of Ricci flow, in which it is used to smoothen out the geometry of a manifold with the aim of better understanding its general properties. In the \textit{normalised Ricci flow} 
\begin{equation}\label{normalisedRF}
    \frac{\diff g_{\mu\nu}}{\diff s}=-2R_{\mu\nu}+\frac{2\ev{R}}{D}g_{\mu\nu}+\mathcal{L}_\xi g_{\mu\nu}\ ,
\end{equation}
a term proportional to the scalar curvature space-time average $\ev{R}$ was introduced as a way to overcome the issue at hand. In particular, it can be easily observed that, once more neglecting the diffeomorphism, we have:
\begin{equation}
    \frac{\diff V_\mathcal{M}}{\diff s}=0\ .
\end{equation}
While having the nice feature of preserving the volume along the flow, \eqref{normalisedRF} has the drawback of being highly non-local. By restricting ourselves to local geometric flows, it is natural to consider \textit{Ricci-Bourguignon flow}
\begin{equation}\label{RBF}
    \frac{\diff g_{\mu\nu}}{\diff s}=-2R_{\mu\nu}+\vartheta Rg_{\mu\nu}+\mathcal{L}_\xi g_{\mu\nu}\ ,
\end{equation}
in which a real parameter $\vartheta$ was introduced. Assuming to work in Euclidean signature and linearizing \eqref{RBF} around flat space, as was previously done for \eqref{HamDet}, we have:
\begin{equation}
\begin{split}
        \frac{\diff h_{\mu\nu}}{\diff s}&=\nabla^2 h_{\mu\nu}+\partial_\mu\partial_\nu h-\partial_\sigma\partial_\mu h^\sigma{}_\nu - \partial_\sigma\partial_\nu h^\sigma{}_\mu\\
&\quad+\vartheta\left(\de_{\alpha}\de_{\beta}h^{\alpha\beta}-\nabla^2 h\right)\delta_{\mu\nu}+\de_\mu\xi_\nu+\de_\nu\xi_\mu  \ .
\end{split}
\end{equation}
By once more performing the appropriate diffeomorphism, we are left with:
\begin{equation}\label{linrbf}
        \frac{\diff h_{\mu\nu}}{\diff s}=\nabla^2 h_{\mu\nu}+\vartheta\left(\de_{\alpha}\de_{\beta}h^{\alpha\beta}-\nabla^2 h\right)\delta_{\mu\nu} \ .
\end{equation}
Thus, we decompose the perturbation $h_{\mu\nu}$ as
\begin{equation}
    h_{\mu\nu}\equiv p\delta_{\mu\nu}+q_{\mu\nu}\ ,
\end{equation}
where we have taken the second term to satisfy:
\begin{equation}
    \nabla^2 q^{\mu\nu}\delta_{\mu\nu}=\de_{\mu}\de_{\nu}q^{\mu\nu}\ .
\end{equation}
The trace-less part is characterised by $\delta^{\mu\nu}k_{\mu\nu}=0$. The two components respectively flow, under \eqref{linrbf}, as:
\begin{equation}
        \frac{\diff p}{\diff s}=\left[1-\vartheta\left(D-1\right)\right]\nabla^2 p \ ,\quad
        \frac{\diff q_{\mu\nu}}{\diff s}=\nabla^2 q_{\mu\nu}\ .
\end{equation}
Therefore, in order for the evolution to be well-posed, we must impose:
\begin{equation}
    \vartheta\leq\frac{1}{D-1}\ .
\end{equation}
It is interesting to notice that, as we move towards $D\rightarrow\infty$, the $\vartheta$ parameter gets confined to being less or equal to zero. The behaviour of Swampland conjectures at a large number of space-time dimensions was recently explored in \cite{Bonnefoy:2020uef}. For a more general examination of \eqref{RBF}, the interested reader is encouraged to refer to \cite{Headrick:2006ti,Catino2015}. It must be noted that Ricci-Bourguignon flow reduces to Ricci flow if $\vartheta$ is set equal to zero. Furthermore, it can be easily observed that \eqref{RBF} induces the following volume evolution:
\begin{equation}
    \frac{\diff V_\mathcal{M}}{\diff s}=\left(\frac{D\vartheta}{2}-1\right)V_\mathcal{M}\cdot\ev{R}\ .
\end{equation}
Similarly, it forces the scalar curvature $R$ to flow according to:
\begin{equation}
    \frac{\diff R}{\diff s}=2R_{\mu\nu}R^{\mu\nu}-\vartheta R^2+\left[1-\left(D-1\right)\vartheta\right]\nabla^2 R\ .
\end{equation}
The Ricci-Bourguignon flow equation \eqref{RBF} can be further generalised to 
\begin{equation}\label{moregen}
    \frac{\diff g_{\mu\nu}}{\diff s}=-2R_{\mu\nu}+\vartheta Rg_{\mu\nu}+\varphi g_{\mu\nu}+\mathcal{L}_\xi g_{\mu\nu}\ ,
\end{equation}
in which $\varphi$ is nothing more than a constant. While not affecting the local well-posedness of the linearized flow, the presence of $\varphi$ produces the following volume evolution
\begin{equation}
    \frac{\diff V_\mathcal{M}}{\diff s}=2\bigl[\left(D\vartheta-2\right)\ev{R}+D\varphi\bigr]V_\mathcal{M}\ .
\end{equation}
and allows for fixed points with non-vanishing curvature. This can be clearly seen by taking an Einstein manifold, for which
\begin{equation}
    R_{\mu\nu}=\frac{R}{D}g_{\mu\nu}\ ,
\end{equation}
and imposing the standard condition for fixed points:
\begin{equation}
    -2R_{\mu\nu}+\vartheta Rg_{\mu\nu}+\varphi g_{\mu\nu}=0\Longrightarrow R=\frac{D\varphi}{2-D\vartheta}\ .
\end{equation}
Hence, for a fixed point to be a solution with cosmological constant $\Lambda$, we must take:
\begin{equation}
    \varphi=\frac{2\left(2-D\vartheta\right)}{D-2}\Lambda\ .
\end{equation}
After having discussed the properties and most direct generalisations of Ricci flow, it is time to apply it to a physical setting and connect it to the swampland distance conjecture.
\subsection{Conjecture statement}\label{sec:rfconj}
\begin{center}
\begin{figure}[h]
\centering
  \includegraphics[width=0.7\linewidth]{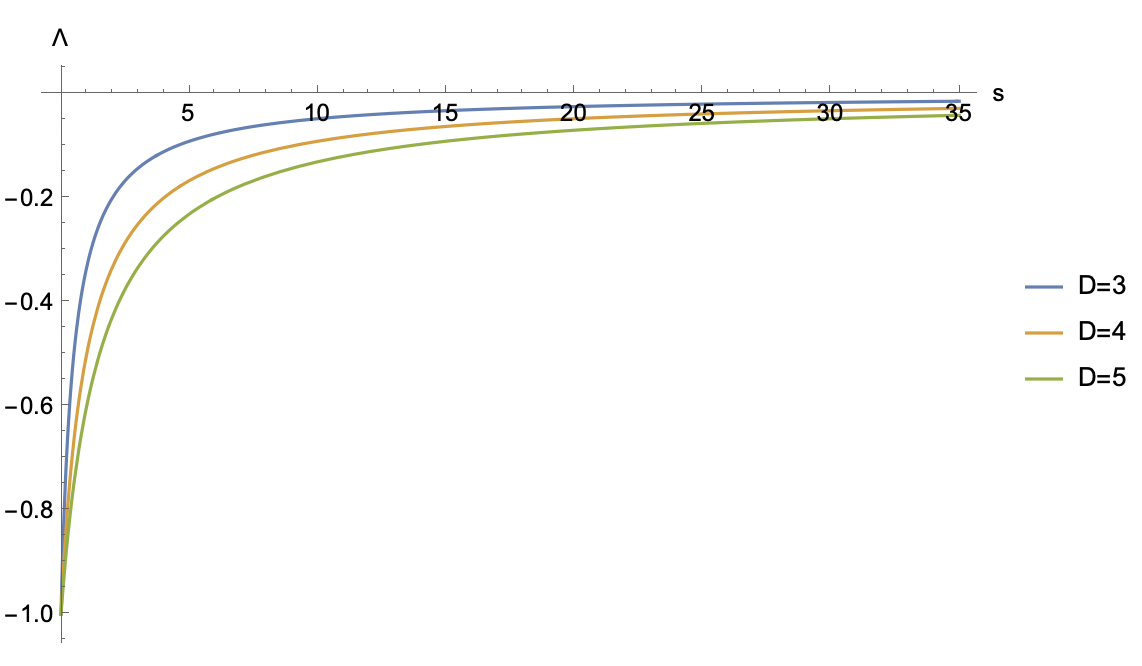}
\caption[Ricci flow behaviour of the Anti-de Sitter cosmological constant.]{Ricci flow behaviour of the Anti-de Sitter cosmological constant, in different numbers of space-time dimensions. The initial value was set $\Lambda_0=-1$, while the three coloured curves, as reported in the legend, correspond to $D=3,4,5$.}
\label{fig:adsflow}
\end{figure}
\end{center}

We will now once more focus on the example of $D$-dimensional Anti-de Sitter space-time \eqref{adsmetric} and compute its Ricci flow. Stressing the facts that we are dealing with an Einstein manifold, for which
\begin{equation}
    R=\frac{2D}{D-2}\Lambda\ ,
\end{equation}
and moving the whole flow dependence to the cosmological constant $\Lambda$, it is enough to solve the induced flow equation \eqref{scalRF} for the scalar curvature. Rephrasing it in terms of the cosmological constant, we get:
\begin{equation}
    \frac{\diff \Lambda}{\diff s}=\frac{2\Lambda^2}{D-2}\ .
\end{equation}
Therefore, by choosing the initial value $\Lambda_0$, the flow behaviour is given by:
\begin{equation}
    \Lambda\left(s\right)=\Lambda_0\left(1-\frac{2\Lambda_0}{D-2}s\right)^{-1}\ .
\end{equation}
In figure \ref{fig:adsflow}, the flow behaviour of $\Lambda$ in three, four and five spatial dimensions was plotted, starting from an initial value $\Lambda_0=-1$. Whatever the initial, negative value of the Anti de-Sitter cosmological constant might be, we always have:
\begin{equation}
    \lim_{s\to\infty}\Lambda\left(s\right)=0\ .
\end{equation}
Therefore, Ricci flow \eqref{HamDet} forces Anti-de Sitter space-time to approach, in the $s\to\infty$ limit, flat Minkowski space-time, which is a flow fixed point characterised by a vanishing cosmological constant. As widely discussed in section \ref{sec:adscon}, this limit, when quantum gravity effects are appropriately considered, is expected to be accompanied by an infinite tower of asymptotically massless states. The above observation led to a further refinement of the swampland distance conjecture, when generalised in order to account for moduli space displacements in the space-time geometry: the \textit{Ricci flow conjecture}. First proposed in \cite{Kehagias:2019akr}, the Ricci flow conjecture can be stated as follows.
\begin{ricci}\label{ricciconjecture}
Let's consider a $D$-dimensional space-time manifold $\mathcal{M}$ and the generalised moduli space $\mathbf{G}_{\mathcal{M}}$ of metric tensors it can be endowed with.
Moreover, let $g$ be a one-parameter curve
\begin{equation}
g:\left[s_0,s_1\right)\longrightarrow\mathbf{G}_{\mathcal{M}}\ ,
\end{equation}
in ${G}_{\mathcal{M}}$, with $s_1$ either finite infinite, such that:
\begin{itemize}
    \item The initial point $g_0\equiv g\left(s_0\right)$ is the geometry corresponding to a consistent quantum gravity low energy effective theory.
    \item The $s$-dependence of $g$ is dictated by Ricci flow
    \begin{equation}
        \frac{\diff g}{\diff s}=-2\text{Ric}\left(g\right)+\mathcal{L}_\xi g\ ,
    \end{equation}
    up to a diffeomorphism induced by the one-parameter family of vector fields $\xi$.
    \item The final point 
    \begin{equation}
        g_1\equiv\lim_{s\to s_1}g\left(s\right)
    \end{equation}
    is a Ricci flow fixed point.
    \item The geodesic distance between the initial and the final point of the curve is infinite:
    \begin{equation}
        \lim_{s\to s_1}\Delta\left(s,s_0\right)=\infty\ .
    \end{equation}
\end{itemize}
Therefore, there must be an infinite tower of additional fields $\psi^j$, characterised by a flow-dependent mass threshold 
\begin{equation}
    m:\left[s_0,s_1\right)\longrightarrow\R
\end{equation}
displaying an exponential drop in the geodesic distance, when following the flow towards $g_1$. Namely, any Ricci flow fixed point at infinite distance should be accompanied by an infinite tower scaling as:
\begin{equation}
    m\left(s\right)\sim m\left(s_0\right)\exp{-\alpha\frac{\Delta\left(s,s_0\right)}{\sqrt{M^{D-2}_{\text{P}}}}}\ .
\end{equation}
 In the above equation, the number $\alpha$ is taken to be $\sim\mathcal{O}\left(1\right)$ and positive, while $\Delta\left(s,s_0\right)$ is the geodesic distance between $\Bar{g}$ and $g\left(s\right)$.
\end{ricci}
Up to this point, geometric flow equations have been postulated and taken as given, without being derived from more general principles. In the subsequent discussion, that will constitute the conclusion of the current chapter, we will address such topic by introducing moduli space \textit{entropy functionals}. In doing so, we will adhere to \cite{Kleiner2006NotesOP,Perelman2002TheEF,Perelman2003FiniteET,Perelman2003RicciFW} and \cite{Kehagias:2019akr}.

%% file: Sections/GeometricFlows/perelman.tex
The geometric evolution induced by Ricci flow, as defined in \eqref{HamDet}, poses a particularly well-behaved differential problem, admits careful extensions to Lorentzian metric tensors and correctly reproduces the expected behaviour for Anti-de Sitter space-time. It therefore provides us with a theoretical framework suited for generalising the swampland distance conjecture to displacements in the space-time geometry. Nevertheless, except for pointing out a direct connection to the superstring theory graviton $\beta$-function, we have so far not provided any derivation of Ricci flow from a more fundamental object. We will now precisely do so, introducing the notion of an entropy functional and postponing the problem of grounding it in physics to the next chapters.
\subsection{The entropy functional}\label{sec:perconj}
Let $\mathcal{M}$ be a $D$-dimensional Riemannian manifold, with $\mathbf{G}_{\mathcal{M}}$ being the generalised moduli space of Lorentzian metrics on $\mathcal{M}$, and 
\begin{equation}
    \phi:\mathcal{M}\longrightarrow C_{\phi}
\end{equation}
be a field defined on $\mathcal{M}$, with values in $C_{\phi}$. We can construct the generalised moduli space $\mathbf{\Phi}_{\mathcal{M}}$ of all space-time configurations of $\phi$, which should not be confused with the moduli space $\mathbf{M}_\Phi$ of its vacuum expectation values, previously introduced in the context of fields not being subject to a non-trivial potential. We assume $C_{\phi}\subset\R$ for the sake of simplicity and take the theory on $\mathcal{M}$ to be controlled by a general action of the form:
\begin{equation}
    S\left[g,\phi\right]=\frac{1}{2\kappa_D^2}\int_{\mathcal{M}}\diff^Dx\sqrt{-g}\left(R_g-\frac{1}{2}\nabla^\mu\phi\nabla_\mu\phi\right)\ .
\end{equation}
The extended moduli space, which accounts for both the space-time configuration of the metric $g$ and that of the field $\phi$, is simply obtained as: 
\begin{equation}
    \mathbf{\Gamma}_{\mathcal{M}}\equiv \mathbf{G}_{\mathcal{M}}\times\mathbf{\Phi}_{\mathcal{M}}\ .
\end{equation}
Starting from any couple of one-parameter curves
\begin{equation}\label{curveF}
        G:\left[s_0,s_1\right]\subset\R\longrightarrow\mathbf{G}_{\mathcal{M}}\ ,\quad \Phi:\left[s_0,s_1\right]\subset\R\longrightarrow\mathbf{\Phi}_{\mathcal{M}}\ ,
\end{equation}
respectively in $\mathbf{G}_{\mathcal{M}}$ and $\mathbf{\Phi}_{\mathcal{M}}$, a one-parameter curve in $\mathbf{\Gamma}_{\mathcal{M}}$ can be constructed as:
\begin{equation}
    \Gamma\left(s\right)\equiv\bigl[G\left(s\right),\Phi\left(s\right)\bigr]\ ,\quad\forall s\in\left[s_0,s_1\right]\ .
\end{equation}
Once more following the steps that led to \eqref{ttdistance}, the length of $\Gamma$ can be naturally computed employing an extended notion of path length on $\mathbf{\Gamma}_{\mathcal{M}}$:
\begin{equation}\label{ttdistance2}
        \mathbf{L}_{\tau}\left[\Gamma\right]=\int_{s_0}^{s_1}\diff s\left|\ev{\left(G^{\mu\nu}G^{\alpha\beta}-\tau G^{\mu\alpha}G^{\nu\beta}\right)\frac{\diff G_{\mu\alpha}}{\diff s}\frac{\diff G_{\nu\beta}}{\diff s}+\frac{1}{2}\left(\frac{\diff\Phi}{\diff s}\right)^2}\right|^{\frac{1}{2}}\ .
\end{equation}
Extremising the length $\mathbf{L}_{\tau}$ of the path connecting two points $\gamma_1\equiv\left(g_1,\phi_1\right)$ and $\gamma_2\equiv\left(g_2,\phi_2\right)$ in $\mathbf{\Gamma}_{\mathcal{M}}$, a geodesic equation can be obtained and solved. Therefore, the distance between $\gamma_1$ and $\gamma_2$ can be consistently defined as the length of such geodesic. While this is perfectly coherent with the discussion developed in \ref{sss:metscalsysgeod} for $\mathbf{G}_{\mathcal{M}}$, we will now define an alternative notion of distance, more directly connected to geometric flows. Eventually, we will accordingly restate the Ricci flow conjecture. In order to so, we must introduce a Lorentzian version of Perelman's $\mathcal{F}$-entropy functional on $\mathbf{\Gamma}_{\mathcal{M}}$:
\begin{equation}\label{Perelentropy}
    \mathcal{F}\left[g,\phi\right]\equiv\int_{\mathcal{M}}\diff^Dx\sqrt{-g}e^{-\phi}\bigl(R_g+\nabla^\mu\phi\nabla_\mu\phi\bigr)\ .
\end{equation}
It must be first of all stressed that, up to some minor differences \cite{Kehagias:2019akr}, \eqref{Perelentropy} resembles the string frame action for the metric and a rescaled dilaton. This identification has remarkable physical implications in the context of string theory. In fact, it allows us to connect $\phi$ with the string coupling $g_s$ via the simple equation:
\begin{equation}
    g_s^2=e^\phi\ .
\end{equation}
Therefore, a geometric flow in $\phi$ could in principle be translated into a non-trivial evolution of the string coupling. For reasons that will be soon made clear, we introduce the string frame volume functional:
\begin{equation}\label{sfvol}
    \Tilde{V}_\mathcal{M}\equiv\int_{\mathcal{M}}\diff^Dx\sqrt{-g}e^{-\phi}\ .
\end{equation}
We now want to obtain Ricci flow, as long as a similar flow equation for $\phi$, as a volume-preserving gradient flow of \eqref{Perelentropy}. Namely, we perform field variations
\begin{equation}
    g_{\mu\nu}\rightarrow g_{\mu\nu}+v_{\mu\nu}\ ,\quad\phi\rightarrow\phi+h
\end{equation}
such that the variation of the string-frame volume functional is set to vanish:
\begin{equation}\label{vanishvol}
    \delta\Tilde{V}_\mathcal{M}=0\ .
\end{equation}
The variation of the metric forces the inverse metric to variate as:
\begin{equation}
    g^{\mu\nu}\rightarrow g^{\mu\nu}-v^{\mu\nu}\ .
\end{equation}
Moreover, a straightforward computation shows that \eqref{vanishvol} implies:
\begin{equation}\label{volpres}
2h=g^{\mu\nu}v_{\mu\nu}\ .
\end{equation}
By computing the induced variation of \eqref{Perelentropy} and always neglecting terms defined on the boundary, where variations are set to vanish, we get:
\begin{equation}
        \delta\mathcal{F}=-\int_{\mathcal{M}}\diff^Dx\sqrt{-g}e^{-\phi}\left(R_{\mu\nu}+\nabla_\mu\nabla_\nu\phi\right)v^{\mu\nu}\ .
\end{equation}
By once more introducing a curve $\gamma\left(s\right)\equiv\left[g\left(s\right),\phi\left(s\right)\right]$ in $\mathbf{\Gamma}_{\mathcal{M}}$, imposing the $s$-evolution of $g$ and $\phi$ to be fixed by the volume-preserving variation of the $\mathcal{F}$-entropy functional and fixing the flow of $\phi$ from that of $g$ thanks to \eqref{volpres}, we get:
\begin{equation}\label{flow1}
    \begin{split}
        \frac{\diff g_{\mu\nu}}{\diff s}&=-2\left(R_{\mu\nu}+\nabla_\mu\nabla_\nu\phi\right)+\mathcal{L}_\xi g_{\mu\nu}\ ,\\
        \frac{\diff\phi}{\diff s}&=-R-\nabla^2\phi+\mathcal{L}_\xi\phi\ .
    \end{split}
\end{equation}
In the above equations, a factor of two was introduced for normalisation purposes and the standard diffeomorphism dependent term was included, due to the corresponding redundancy of physical solutions. After having performed a diffeomorphism set by
\begin{equation}
    \Bar{\xi}^\mu\equiv \nabla^\mu\phi\ ,
\end{equation}
we finally obtain the following system of flow equations:
\begin{equation}\label{flow2}
    \begin{split}
        \frac{\diff g_{\mu\nu}}{\diff s}&=-2R_{\mu\nu}+\mathcal{L}_\xi g_{\mu\nu}\ ,\\
        \frac{\diff\phi}{\diff s}&=-R-\nabla^2\phi+\left(\nabla\phi\right)^2+\mathcal{L}_\xi\phi\ .
    \end{split}
\end{equation}
Hence, the volume preserving gradient flow of \eqref{Perelentropy} forces the metric to evolve according to Ricci flow, while the scalar transforms in such a way that the variation of the string frame volume vanishes. This combined system of geometric flow equations is typically referred to as \textit{Perelman's combined flow} \cite{Perelman2002TheEF}. It can be easily shown \cite{Kleiner2006NotesOP} that the $\mathcal{F}$-entropy functional is monotonic along the flow \eqref{flow2}. It therefore provides us with a natural and alternative way to define distances along flow trajectories. By considering a curve $\gamma\left(s\right)\equiv\left[g\left(s\right),\phi\left(s\right)\right]$ in $\mathbf{\Gamma}_{\mathcal{M}}$, with $s\in\left[s_0,s_1\right)$ and $s_1$ either finite or infinite, whose $s$-evolution is a solution to \eqref{flow2} and a specific value $s$ of the flow parameter, the $\mathcal{F}$-distance between $\gamma\left(s_0\right)$ and $\gamma\left(s\right)$ is defined as
\begin{equation}
    \Delta_{\mathcal{F}}\left(s,s_0\right)\equiv\log{\frac{\mathcal{F}\left(s_0\right)}{\mathcal{F}\left(s\right)}}\ ,
\end{equation}
where $\mathcal{F}\left(s_0\right)\equiv\mathcal{F}\left[\gamma\left(s_0\right)\right]$ and $\mathcal{F}\left(s\right)\equiv\mathcal{F}\left[\gamma\left(s\right)\right]$. In general, the $\mathcal{F}$-distance between two flow trajectory points should not be expected to be equivalent to their geodesic distance. Given the above definitions, the Ricci flow conjecture can be refined and slightly altered as follows.
\begin{dilmet}\label{perelconj}
Let's consider a $D$-dimensional space-time manifold $\mathcal{M}$, the generalised moduli space $\mathbf{G}_{\mathcal{M}}$ of metric tensors it can be endowed with and the moduli space $\mathbf{\Phi}_{\mathcal{M}}$ of space-time configurations of a scalar field $\phi$ defined on $\mathcal{M}$. Moreover, let $\gamma\equiv\left(g,\phi\right)$ be a one-parameter curve
\begin{equation}
\gamma:\left[s_0,s_1\right)\longrightarrow\mathbf{\Gamma}_{\mathcal{M}}\equiv\mathbf{G}_{\mathcal{M}}\times\mathbf{\Phi}_{\mathcal{M}}\ ,
\end{equation}
in ${\Gamma}_{\mathcal{M}}$, with $s_1$ either finite infinite, such that:
\begin{itemize}
    \item The initial point $\gamma_0\equiv \gamma\left(s_0\right)$ is the combination of geometry and a scalar field configuration corresponding to a consistent quantum gravity low energy effective theory.
    \item The $s$-dependence of $\gamma$ is dictated by Perelman's combined flow
    \begin{equation}
    \begin{split}
        \frac{\diff g_{\mu\nu}}{\diff s}&=-2R_{\mu\nu}+\mathcal{L}_\xi g_{\mu\nu}\ ,\\
        \frac{\diff\phi}{\diff s}&=-R-\nabla^2\phi+\left(\nabla\phi\right)^2+\mathcal{L}_\xi\phi\ ,
    \end{split}
    \end{equation}
    up to a diffeomorphism induced by the one-parameter family of vector fields $\xi$.
    \item The final point 
    \begin{equation}
        \gamma_1\equiv\lim_{s\to s_1}\gamma\left(s\right)
    \end{equation}
    is a Perelman's combined flow fixed point.
    \item The $\mathcal{F}$-distance between the initial and the final point of the curve is infinite:
    \begin{equation}
        \lim_{s\to s_1}\Delta_{\mathcal{F}}\left(s,s_0\right)=\infty\ .
    \end{equation}
\end{itemize}
Therefore, there must be an infinite tower of additional fields $\psi^j$, characterised by a flow-dependent mass threshold 
\begin{equation}
    m:\left[s_0,s_1\right)\longrightarrow\R
\end{equation}
displaying an exponential drop in the $\mathcal{F}$-distance, when following the flow towards $\gamma_1$. Namely, any Perelman's combined flow fixed point at infinite distance should be accompanied by an infinite tower scaling as:
\begin{equation}
    m\left(s\right)\sim m\left(s_0\right)\exp{-\alpha\frac{\Delta_{\mathcal{F}}\left(s,s_0\right)}{\sqrt{M^{D-2}_{\text{P}}}}}\ .
\end{equation}
 In the above equation, the number $\alpha$ is taken to be $\sim\mathcal{O}\left(1\right)$ and positive, while $\Delta_{\mathcal{F}}\left(s,s_0\right)$ is the $\mathcal{F}$-distance
 \begin{equation}\label{pereldistanceconj}
         \Delta_{\mathcal{F}}\left(s,s_0\right)\equiv \Delta_0\log{\frac{\mathcal{F}\left(s_0\right)}{\mathcal{F}\left(s\right)}}\ ,
 \end{equation}
 defined from the $\mathcal{F}$-entropy functional
 \begin{equation}
    \mathcal{F}\left[g,\phi\right]\equiv\int_{\mathcal{M}}\diff^Dx\sqrt{-g}e^{-\phi}\bigl(R_g+\nabla^\mu\phi\nabla_\mu\phi\bigr)
\end{equation}
and from the dimensionful model-dependent constant $\Lambda_0$.
\end{dilmet}

%% file: Sections/Bubbles/intro.tex
Throughout the following chapter, largely based on \cite{DeBiasio:2022omq}, we will once more assume to work with a smooth $D$-dimensional manifold $\mathcal{M}$, endowed with a dynamical Lorentzian metric $g$ and a real scalar field $\phi$. The generalised moduli space of space-time configurations $\gamma\equiv\left(g,\phi\right)$ of such fields will be $\mathbf{\Gamma}_{\mathcal{M}}$, as introduced in \eqref{metscalmodulispace}. Moreover, for the sake of phenomenological interest, the number of space-time dimensions will be set to $D=4$. In the context of such a conceptual framework, a class of moduli space paths produced by specific geometric flow equations will be carefully analysed. In particular, we will direct our attention towards scalar bubble solutions, embedded in Minkowski, de Sitter, Anti-de Sitter and Schwarzschild space-time backgrounds, and negative cosmological constant bubble solutions, coupled to a space-time constant scalar field. While the former will be imposed to evolve according to Perelman's combined flow equations \eqref{flow2}, which will translate to flows in the background geometry and in the bubble radius within the thin-wall approximate regime, the latter will be studied under a novel family of geometric flow equations, specifically designed in order for the cosmological to be fixed far from the bubble and derived from a suitable entropy functional.

%% file: Sections/Bubbles/bubblesflow.tex
\label{bubbs}
 This section is aimed at studying the evolution of \textit{scalar bubble} configurations of $\phi$ under Perelman's combined flow \eqref{flow2}. We will therefore construct a family of simplified toy models, designed to capture the main features of scalar bubbles coupled to general relativistic background geometries. Such an objective will be achieved by relaxing the equations of motion in the non-trivial portions of the bubbles, where they interpolate among different vacua. Thus, we will allow our scalar profiles and metrics not to solve them in those limited space-time shells and derive simpler mathematical descriptions. Deviations from proper on-shell solutions will be kept under control by imposing to work in the \textit{thin-wall} approximation. Hence, we will take such settings as initial conditions for our geometric flow equations and investigate their induced moduli space trajectories and asymptotic behaviour in the flow parameter. Having developed an intuitive picture of the properties of bubbles under geometric flows, less trivial constructions will be thereafter discussed. 
\subsection{Bubble construction}\label{subs:bubblescal}
Before considering specific background geometries and scalar field configurations, we will describe the class of bubble toy models of our interest in a general fashion. The detailed form of the metric will now be discussed, followed by a quick investigation of the scalar profile. Thus, all subsequent analyses will be performed after having fixed the values of some solution parameters. Spherical symmetry will always be assumed, together with staticity. We will hence find ourselves with a set of purely radial functions of one variable.
\subsubsection{Background Metric}\label{metbubbleconstruct}
For what concerns the space-time geometry, we will consider a family of spherically symmetric, static and uncharged black holes with no angular momentum, embedded in a $4$-dimensional cosmological constant background:
\begin{equation}\label{genmet}
\begin{split}
    \ds^2=&-F\left(r,M,\Lambda\right)\diff t^2+\frac{\diff r^2}{F\left(r,M,\Lambda\right)}+r^2\left(\diff\theta^2+\sin^2{\theta}\diff\varphi^2\right)\ .
\end{split}
\end{equation}
In the above formula, expressed in standard spherical coordinates $\left(t,r,\theta,\varphi\right)$, the radial function $F\left(r,M,\Lambda\right)$ was defined in terms of the black hole mass $M$ and the cosmological constant $\Lambda$ as:
\begin{equation}\label{metricfunctionmlambda}
    F\left(r,M,\Lambda\right)\equiv 1-\frac{2M}{r}-\frac{\Lambda r^{2}}{3}\ .
\end{equation}
It goes without saying that, for the remainder of our analysis, we will work in units such that both the $4$-dimensional Newton constant $G_N$ and the speed of light in vacuum $c$ are chosen to be dimensionless and set to one. From the explicit expression \eqref{metricfunctionmlambda}, it is clear that the family of geometries we are working with is a subset $\mathbf{A}$ of the generalised moduli space $\mathbf{G}_{\mathcal{M}}$ of metrics on $\mathcal{M}$, charted by two parameters:
\begin{equation}
    \left(M,\Lambda\right)\in\R_+\times\R\ .
\end{equation}
At this stage, it is not clear whether, starting for a specific initial condition $\left(M_0,\Lambda_0\right)\in\mathbf{A}$, Perelman's combined flow \eqref{flow2} trajectories would be constrained to such subset of $\mathbf{G}_{\mathcal{M}}$. This issue will be briefly addressed in the following. For a more detailed assessment of the case in which $\Lambda<0$, the reader is encouraged to refer to \cite{DeBiasio:2020xkv}. The $\Lambda>0$ one was, instead, widely analysed in \cite{Luben:2020wix}. Since the metric described by \eqref{genmet} is that of an Einstein manifold, we have that the Ricci scalar and tensor are related by the formula
\begin{equation}\label{curvspe}
    R_{\mu\nu}=\frac{R}{4}g_{\mu\nu}\ ,
\end{equation}
with the scalar curvature being equal to:
\begin{equation}\label{curvspe2}
    R=4\Lambda\ .
\end{equation}
When considering a geometric flow equation for the metric, as previously commented on, there is no guarantee that an initial point in $\mathbf{A}$ will produce a flow trajectory completely contained in $\mathbf{A}$, rather than any other general path in $\mathbf{G}_{\mathcal{M}}$. It is, nevertheless, straightforward to show \cite{DeBiasio:2020xkv} that any generalised Ricci-Bourguignon flow \eqref{moregen} defined by
\begin{equation}\label{generalflowRBF}
    \frac{\diff g_{\mu\nu}}{\diff s}=-2R_{\mu\nu}+\vartheta Rg_{\mu\nu}+\lambda g_{\mu\nu}+\mathcal{L}_\xi g_{\mu\nu}\ ,
\end{equation}
with $\vartheta,\lambda\in\R$, does, indeed, satisfy such requirement. Hence, the identities \eqref{curvspe} and \eqref{curvspe2} can be taken to hold along their induced geometric evolutions. Thus, it is much easier to focus, after having neglected the diffeomorphism term, on the flow defined by \eqref{generalflowRBF} for the curvature scalar:
\begin{equation}
    \frac{\diff R}{\diff s}=2R^{\mu\nu}R_{\mu\nu}-\left(\vartheta R+\lambda\right) R+\left[1-\left(D-1\right)\vartheta\right]\nabla^2R
\end{equation}
Imposing both \eqref{curvspe} and \eqref{curvspe2}, the above formula can be expressed in the form of a flow equation for the cosmological constant:
\begin{equation}
    \frac{\diff \Lambda}{\diff s}=2\left(1-2\vartheta\right)\Lambda^2-\lambda\Lambda\ .
\end{equation}
For $\lambda\neq 0$ and $\Lambda\left(0\right)=\Lambda_0$, the equation can be solved by:
\begin{equation}
    \Lambda\left(s\right)=\lambda\Lambda_0\bigl\{2\left(1-2\vartheta\right)\Lambda_0+e^{\lambda\cdot s}\left[\lambda-2\left(1-2\vartheta\right)\Lambda_0\right]\bigr\}^{-1}\ .
\end{equation}
For $\lambda=0$, the differential equation for the cosmological constant flow is solved by the $\lambda\to 0$ limit of the above expression:
\begin{equation}
    \Lambda\left(s\right)=\frac{\Lambda_0}{1-2\Lambda_0\left(1-2\vartheta\right)s}\ .
\end{equation}
Further reducing ourselves to Ricci flow, which is the metric flow equation appearing in \eqref{flow2} and the one we will consider in our analysis, we simply have:
\begin{equation}\label{ccflowrf}
    \Lambda\left(s\right)=\frac{\Lambda_0}{1-2\Lambda_0 s}\ .
\end{equation}
Therefore, we observe two distinct behaviours:
\begin{itemize}
    \item For a negative initial cosmological constant $\Lambda_0$, corresponding to a black hole embedded in an Anti-de Sitter background, the cosmological constant flows to zero as the parameter $s$ is sent to infinity. 
    \item For a positive initial cosmological constant $\Lambda_0$, corresponding to a black hole embedded in a de Sitter background, the cosmological constant blows to infinity at the finite flow parameter critical value:
    \begin{equation}
        s_{1}\equiv\frac{1}{2\Lambda_0}\ .
    \end{equation}
    \item If the initial value of the cosmological constant is taken to be zero, the space-time geometry does not change with the parameter $s$.
\end{itemize}
Particular care is required when dealing with either flat space-time limits or flow singularities, as the ones described above. The reason is that whether or not they lie at finite or infinite distance is not related to them corresponding a finite or infinite, respectively, value of the flow parameter. Instead, it must once more be stressed that moduli space distances should be either computed by employing the path length formula \eqref{ttdistance2} along a geodesic or by comparing the values of Perelman's entropy functional \eqref{Perelentropy} at the two moduli space points that are taken into account, as done in \eqref{pereldistanceconj}. In our case, such two points are the initial condition and either the flat space-time limit, for the $\Lambda_0<0$ case, or the flow singularity, when $\Lambda_0>0$. Regardless of the choice of one of the two notions of distance, it is clear that no definitive answer can be provided before analysing the scalar field flow behaviour. This will be done, for a variety of distinct scenarios, in the following discussion. Plugging the cosmological constant flow dependence described in \eqref{ccflowrf} into the Ricci metric part of Perelman's combined flow, one can derive the flow behaviour of the black hole mass $M$. Once more, explicit and detailed analyses of such general problems are included in \cite{DeBiasio:2020xkv,Luben:2020wix}. As far as our current discussion is concerned, it is simply important to stress that, when the initial value of the cosmological constant is set to zero, the space-time geometry reduces to that of a flat Schawrzschild black hole. Therefore, it corresponds to a flow fixed point. Having described the family of geometrical backgrounds we are interested in, having studied their Ricci flow behaviour and having assessed that the corresponding moduli space trajectories are constrained to $\mathbf{A}\subset\mathbf{G}_{\mathcal{M}}$, we can move our focus to the scalar field radial profile.
\subsubsection{Scalar Profile}
\begin{center}
\begin{figure}[h]
\centering
  \includegraphics[width=0.7\linewidth]{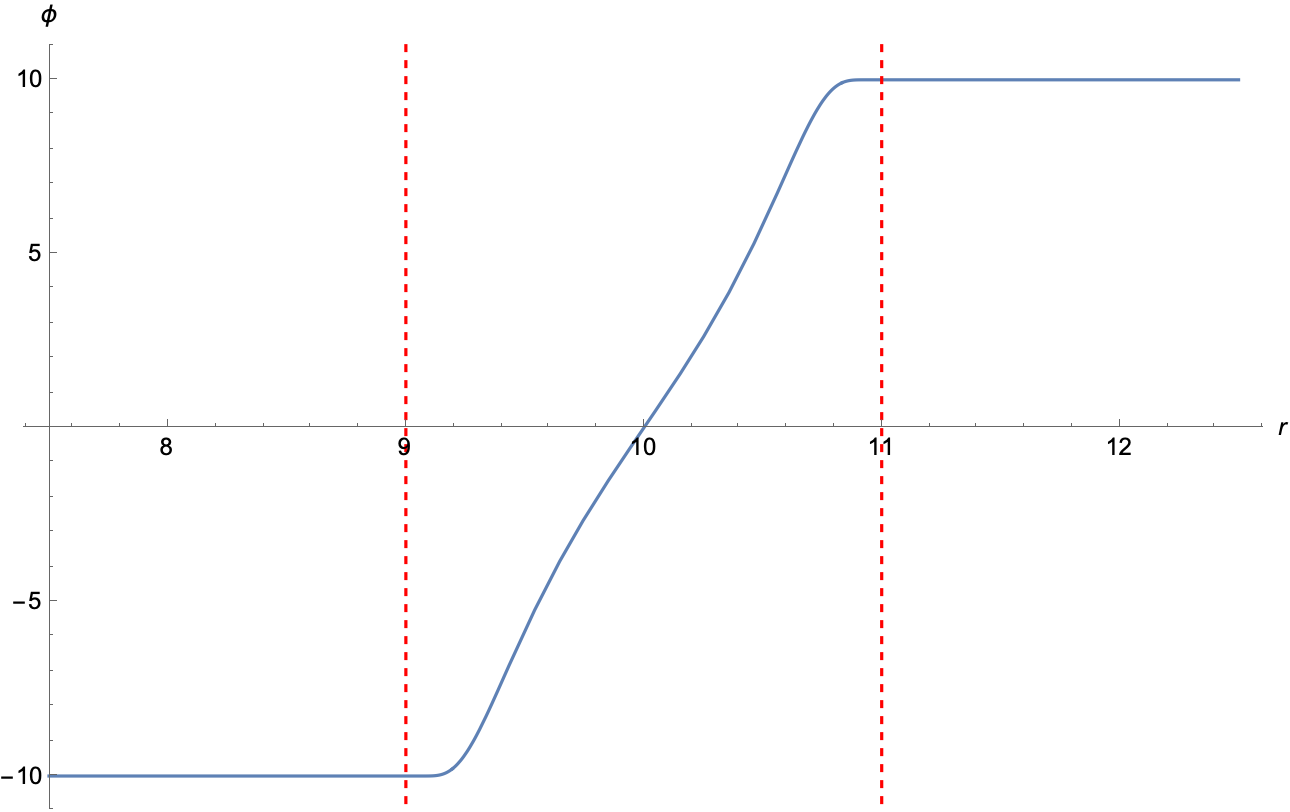}
\caption[Radial profile of a scalar field bubble.]{Radial profile of the scalar field $\phi$, when the constant values are set to $\phi_1=-10$ and $\phi_2=10$, while the spherical shell is defined by $\varrho=10$ and $\varepsilon=1$. The vertical dashed lines correspond to spherical shell boundaries.}
\label{fig:bubbleprofile}
\end{figure}
\end{center}

In order to construct the scalar profile, we start by assuming it to be static and spherically symmetric. Namely, we take $\phi=\phi\left(r\right)$ in spherical coordinates $\left(t,r,\theta,\varphi\right)$. Thereafter, we introduce the \textit{bubble radius} $\varrho$ and consider a spherical shell
\begin{equation}\label{shelldef}
    B_{\varepsilon}\left(\varrho\right)\equiv\left\{\left(t,r,\theta,\varphi\right)\in\mathcal{M}\ |\ t=\tau,\ r\in\left(\varrho-\varepsilon,\varrho+\varepsilon\right)\right\}
\end{equation}
around the origin, with thickness $2\varepsilon$. The $\tau$-dependence in $B_{\varepsilon}\left(\varrho\right)$ was purposely dropped. The reason is that, being both the metric and the scalar time-independent, $\tau$ can always be rescaled by performing an isometry in the time direction. Thus, we will from now on set it to one without loss of generality. Concerning the scalar field, we assume it to take two constant values $\phi_1$ and $\phi_2$, respectively inside and outside the spherical shell. Inside $B_{\varepsilon}\left(\varrho\right)$, which corresponds to the wall of the scalar bubble, we instead assume $\phi$ to smoothly interpolate between $\phi_1$ and $\phi_2$. In order for such a goal to be achieved, we take
\begin{equation}\label{dilaton}
    \phi\left(r\right)\equiv\begin{cases}
        \begin{split}
        \phi_{1}&\text{ for }r\leq\varrho-\varepsilon\\
        I\left(r\right)&\text{ for }\varrho-\varepsilon\leq r\leq\varrho+\varepsilon\\
        \phi_{2}&\text{ for }r\geq\varrho+\varepsilon
        \end{split}
    \end{cases}\ ,
\end{equation}
where the interpolating radial function $I\left(r\right)$ was defined as:
\begin{equation}
    I\left(r\right)\equiv\frac{\phi_{2}+\phi_{1}}{2}+\frac{\phi_{2}-\phi_{1}}{2}\tanh{\frac{r-\varrho}{\left(\varrho+\varepsilon-r\right)\left(r-\varrho+\varepsilon\right)}}\ .
\end{equation}
In the following examples, the radial profile described by \eqref{dilaton} and depicted in figure \ref{fig:bubbleprofile} will represent, for a specific choice of the parameters $\phi_2$, $\phi_1$, $\varrho$ and $\varepsilon$, the Perelman's combined flow initial condition for the scalar field. While the presence of non trivial field gradients in the region described by the spherical shell $B_{\varepsilon}\left(\varrho\right)$ would, strictly speaking, induce a back-reaction on the space-time geometry \eqref{genmet}, the thin-wall approximation $\varrho\gg\varepsilon>0$ will allow us to neglect it without introducing strong deviations from the full-fledged on-shell solution. 
\subsection{Minkowski background}\label{Minkowski}
\begin{figure}[H]
    \centering
    \includegraphics[width=0.8\linewidth]{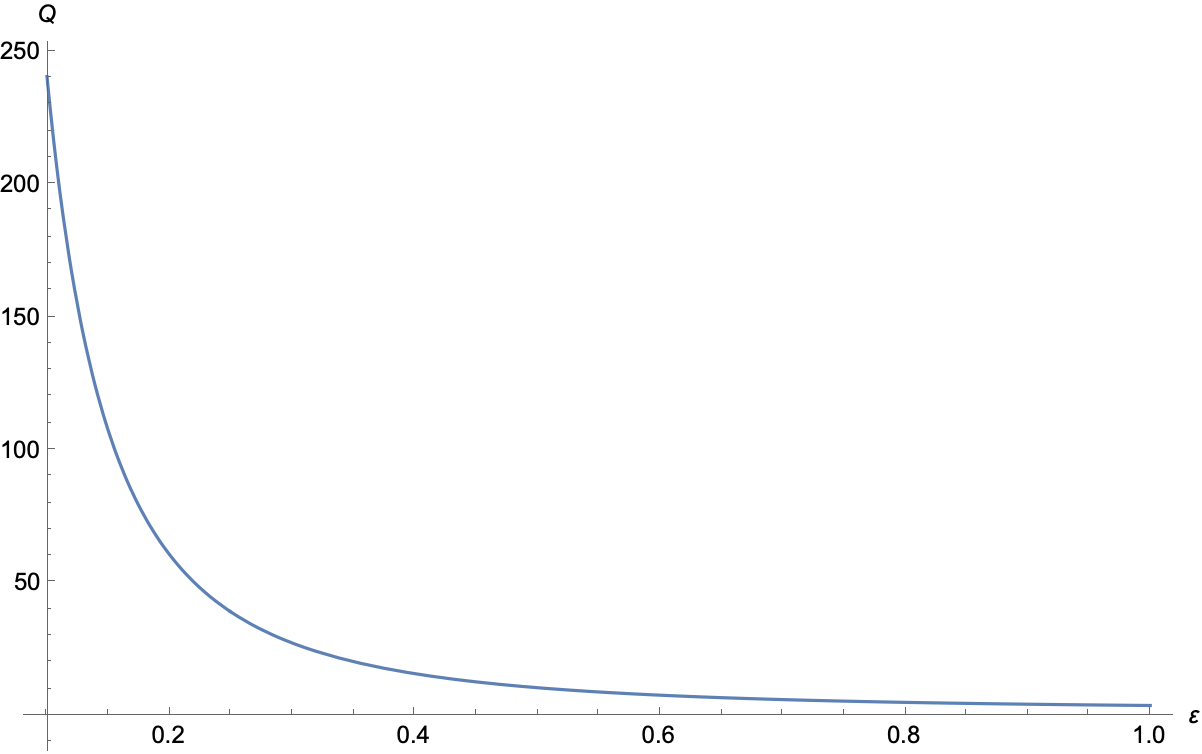}
    \caption[Plot of the function $Q\left(\varepsilon\right)$, governing the flow behaviour of a scalar field bubble.]{Plot of $Q\left(\varepsilon\right)$ with respect to the spherical shell halved thickness. At least within the thin-wall approximation, which is the regime we are interested in, such a function is clearly restricted to positive values.}
    \label{fig:QMinkowski}
\end{figure}

The first example we focus on is, arguably, the simplest possible. Namely, we take both $M=0$ and $\Lambda=0$, so that the background metric \eqref{genmet} reduces to that of Minkowski space-time. First of all, we observe that such a space-time geometry is characterised by a vanishing Ricci curvature tensor:
\begin{equation}\label{ricciflatmink}
    R_{\mu\nu}=0\ .
\end{equation}
Therefore, the metric part of the system is constant along Perelman's combined flow \eqref{flow2}. This directly translates in $M$ and $\Lambda$ having no evolution in the flow parameter $s$. Since \eqref{ricciflatmink} implies the curvature scalar to vanish too, the flow equation for the scalar field $\phi$ reduces to:
\begin{equation}\label{flsc1}
    \frac{\diff\phi}{\diff s}=\left(\nabla \phi\right)^2-\nabla^2\phi+\mathcal{L}_\xi\phi\ .
\end{equation}
Working in Minkowksi space-time, we simply have:
\begin{equation}\label{flsc}
    \frac{\diff\phi}{\diff s}=\left(\de \phi\right)^2-\de^2\phi+\mathcal{L}_\xi\phi\ .
\end{equation}
From this point on, we will neglect the diffeomorphism term. It can be straightforwardly observed that the scalar field derivative vanishes for:
\begin{equation}
    r\notin\left(\varrho-\varepsilon,\varrho+\varepsilon\right)\Longrightarrow\de_\mu\phi=0\ .
\end{equation}
The right-hand side of \eqref{flsc} is hence zero outside the spherical shell \eqref{shelldef}, implying that both $\phi_{1}$ and $\phi_{2}$ are constant along the flow.
Since both sides of \eqref{flsc} solely depend on the radial coordinate, it can be safely deduced that Perelman's flow does not spoil the spherical symmetry of $\phi$. Starting from the previous observation and from the constancy of $\phi_1$ and $\phi_2$, we now introduce an appropriate flow ansatz, assuming that the functional form \eqref{dilaton} properly describes $\phi$ at any value of the flow parameter $s$. The flow is therefore regarded as a deformation of the wall-like interface between the two regions in which the scalar field is taken as constant. Moreover, we take $\varepsilon$ to be constant along the flow. This last hypothesis is backed-up by the so-called thin-wall approximation, which we will assume to well describe our solution for the remainder of the following discussion. Namely, we impose
\begin{equation}
    \varrho\gg\varepsilon>0
\end{equation}
and assume it to hold along the flow, so that any variation of $\varepsilon$ is negligible with respect to the overall evolution. Hence, we can fully model the geometric flow as an evolution in the wall position $\varrho$. If bubbles will be found to shrink along the flow, the approximation of constant thickness can be expected to break down roughly after a critical flow time $s_c$. Some extra care will thus be required. That said, we now choose to work within a regime in which $\varrho\gg\varepsilon$, define
\begin{equation}\label{bubb}
    \phi(r)\equiv\varphi\left(r-\varrho\right)
\end{equation}
and move the whole flow dependence to $\varrho$, with $r-\varrho\equiv x\in\left(-\varrho,+\infty\right)$. Concerning the flow equation \eqref{flsc} for the scalar profile in terms of $\varrho$ and $\varphi$, we have:
\begin{equation}
    \frac{\diff^2\varphi}{\diff x^2}=\left(\frac{\diff\varrho}{\diff s}-\frac{2}{x+\varrho}\right)\frac{\diff\varphi}{\diff x}+\left(\frac{\diff\varphi}{\diff x}\right)^2\ .
\end{equation}
By multiplying both sides by the $x$-derivative of $\varphi$, we obtain:
\begin{equation}\label{wall1flow}
\frac{1}{2}\frac{\diff}{\diff x}\left(\frac{\diff\varphi}{\diff x}\right)^2=\left(\frac{\diff\varrho}{\diff s}-\frac{2}{x+\varrho}\right)\left(\frac{\diff\varphi}{\diff x}\right)^2+\left(\frac{\diff\varphi}{\diff x}\right)^3\ .
\end{equation}
In particular, we have
\begin{equation}\label{derphi}
    \frac{\diff\varphi}{\diff x}=\begin{cases}
        0\ \text{ for }-\varrho\leq x\leq-\varepsilon\\
        K(x)\text{ for }-\varepsilon\leq x\leq\varepsilon\\
        0\ \text{ for }x\geq\varepsilon
    \end{cases}\ ,
\end{equation}
in which we have defined the function $K\left(x\right)$ as:
\begin{equation}
    K(x)=\frac{\phi_{2}-\phi_{1}}{2}\frac{\left(x^{2}+\varepsilon^2\right)}{(x-\varepsilon)^2(x+\varepsilon)^2}\cosh^{-2}{\frac{x}{\varepsilon^2-x^2}}\equiv\frac{\phi_{2}-\phi_{1}}{2}G(x)\ .
\end{equation}
At this point, we perform an integration of \eqref{wall1flow} in the variable $x$, taken to go from $0$ to $+\infty$, and read-off $\diff\varrho/\diff s$ from the resulting expression. By doing so, we are left with the following equation:
\begin{equation}\label{rhoeq}
    \frac{\diff\varrho}{\diff s}\int_{-\varepsilon}^{\varepsilon}G(x)^{2}\diff x=2\int_{-\varepsilon}^{\varepsilon}\frac{G(x)^{2}}{x+\varrho}\diff x+\frac{\phi_{1}-\phi_{2}}{2}\int_{-\varepsilon}^{\varepsilon}G(x)^{3}\diff x\ .
\end{equation}
Evidently, \eqref{rhoeq} is way too complicated to be solved analytically. Anyway, we can easily observe that $G(x)$ is always positive in $(-\varepsilon,\varepsilon)$. Hence, the flow behaviour of the bubble radius $\varrho$ is characterised by three positive functions:
\begin{equation}\label{functions}
    \begin{split}
        A\left(\varepsilon\right)\equiv\int_{-\varepsilon}^{\varepsilon}G(x)^{2}\diff x\ ,\quad
        B\left(\varepsilon,\varrho\right)\equiv\int_{-\varepsilon}^{\varepsilon}\frac{G(x)^{2}}{x+\varrho}\diff x\ ,\quad
        C\left(\varepsilon\right)\equiv\int_{-\varepsilon}^{\varepsilon}G(x)^{3}\diff x\ .
    \end{split}
\end{equation}
Therefore, the flow equation for $\varrho$ can be written in the compact form:
\begin{equation}\label{rhof}
    \frac{\diff\varrho}{\diff s}=2\frac{B\left(\varepsilon,\varrho\right)}{A\left(\varepsilon\right)}+\frac{\phi_{1}-\phi_{2}}{2}\frac{C\left(\varepsilon\right)}{A\left(\varepsilon\right)}\ .
\end{equation}
Since we work in the $\varepsilon\ll\varrho$ regime, we have:
\begin{equation}
    B\left(\varepsilon,\varrho\right)\sim\int_{-\varepsilon}^{\varepsilon}\frac{G(x)^{2}}{\varrho}\diff x=\frac{A\left(\varepsilon\right)}{\varrho}\ .
\end{equation}
Given the above approximate relation, the geometric flow equation \eqref{rhof} reduces to
\begin{equation}\label{equaa}
        \frac{\diff\varrho}{\diff s}=\frac{2}{\varrho}+\frac{\phi_{1}-\phi_{2}}{2}Q\left(\varepsilon\right)\ ,
\end{equation}
where the function $Q\left(\varepsilon\right)$ of the halved spherical shell thickness $\varepsilon$ is defined as the ratio:
\begin{equation}\label{eq:Qdef}
   Q\left(\varepsilon\right)\equiv\frac{C\left(\varepsilon\right)}{A\left(\varepsilon\right)}\ .
\end{equation}
As displayed in figure \ref{fig:QMinkowski}, the function $Q\left(\varepsilon\right)$ is strictly restricted to positive values, at least for thin enough interfaces. Therefore, whether the second contribution appearing on the right-hand side of \eqref{equaa} will tend to inflate or deflate the bubble, within which the scalar takes the value $\phi_1$, will depend on the magnitudes of the constants $\phi_1$ and $\phi_2$. We hence consider the cases $\phi_2>\phi_1$ and $\phi_2<\phi_1$ separately. Before doing so, it is important to stress that the first term in the right-hand side \eqref{equaa} will always contribute to the scalar field bubble getting inflated. Nonetheless, the significance of such effect will linearly decrease with the size of the aforesaid bubble.
\subsubsection{Bubble with a smaller scalar value}
\begin{figure}[H]
    \centering
    \includegraphics[width=0.9\linewidth]{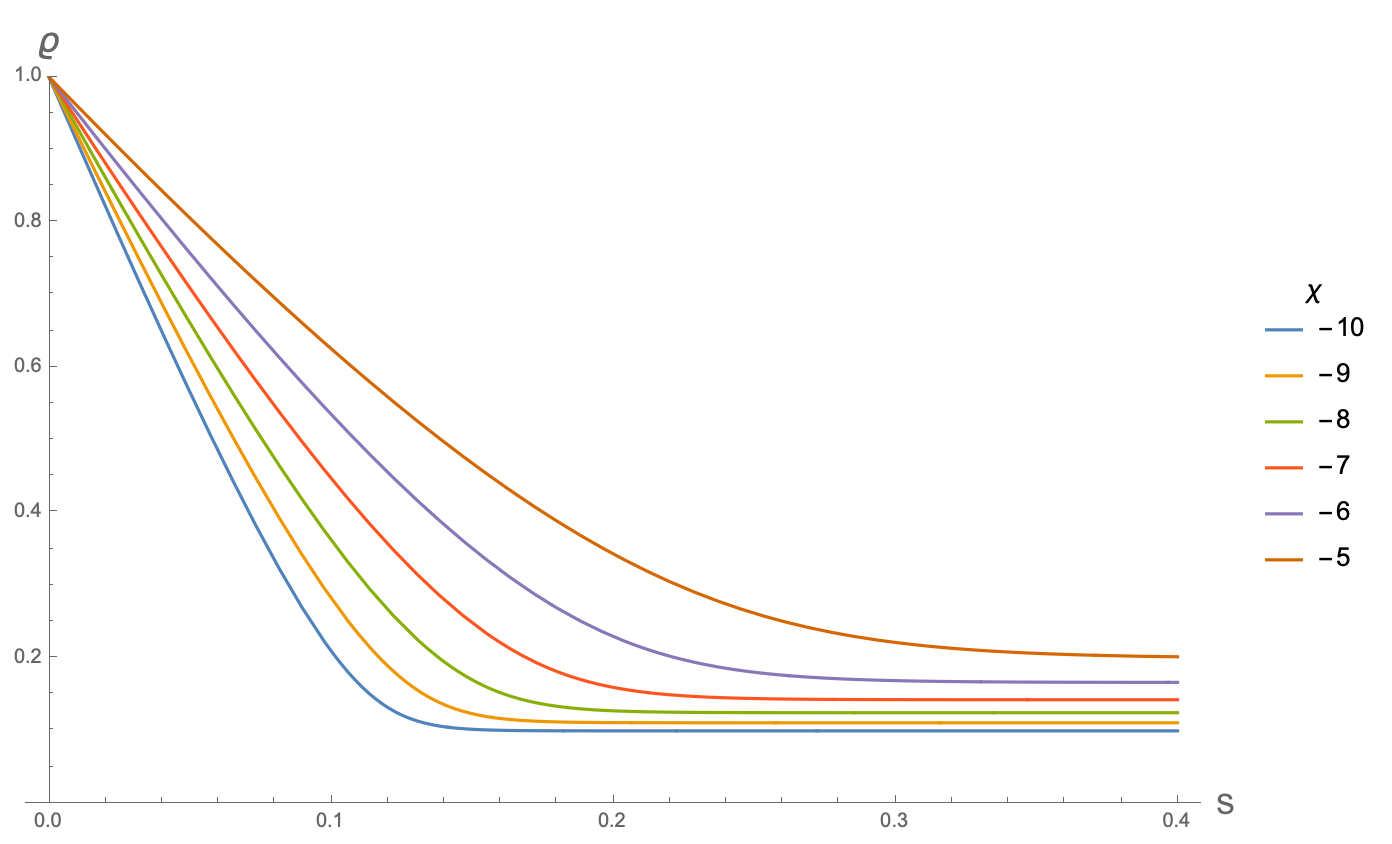}
    \caption[Flow behaviour of the bubble radius, when the value of the scalar inside it is smaller than the one outside.]{Flow behaviour of $\varrho$ for different values of $\chi$, with $\varrho_{0}=1$ and assuming to be considering a setting in which $\phi_2>\phi_1$.}
    \label{fig:smallbubb}
\end{figure}

As an initial condition, we assume $\phi_{2}>\phi_{1}$ and take the spherical shell radius to have a big enough value $\varrho_{0}$, when compared to its thickness. This way the thin-wall approximation can be safely employed. Furthermore, we define the function
\begin{equation}\label{chii}
    \chi\left(\varepsilon\right)\equiv\frac{\phi_{1}-\phi_{2}}{2}Q\left(\varepsilon\right)\ ,
\end{equation}
depending both on the constant $\phi_2$ and $\phi_1$ and on the previously introduced function $Q\left(\varepsilon\right)$ of the halved shell thickness. By doing so, the flow equation \eqref{equaa} for the radius of the bubble can be solved by
\begin{equation}\label{rhosol}
    \varrho\left(s\right)=-\frac{2}{\chi\left(\varepsilon\right)}\left\{1+W\left[-\exp{-1-2\chi^2\left(\varepsilon\right)s-\frac{\varrho_{0}}{2}\chi\left(\varepsilon\right)}\left(1+\frac{\varrho_{0}}{2}\chi\left(\varepsilon\right)\right)\right]\right\}\ ,
\end{equation}
where $W$ refers to the positive branch of Lambert's function, which only admits arguments in $\left(-e^{-1},+\infty\right)$. In order for such an object to be well-defined, we hence take:
\begin{equation}\label{initial}
    \varrho_{0}\geq\frac{1}{\phi_{2}-\phi_{1}}\frac{4}{Q\left(\varepsilon\right)}\ .
\end{equation}
The regime in which the value of the dilaton inside the bubble is smaller than the one outside, namely when $\phi_{2}>\phi_{1}$, produces shrinking bubbles. In particular, starting from an initial value $\varrho_{0}$ satisfying \eqref{initial}, the bubbles shrink approaching the asymptotic value
\begin{equation}
    \varrho_{\infty}\equiv\frac{1}{\phi_{2}-\phi_{1}}\frac{4}{Q\left(\varepsilon\right)}\ ,
\end{equation}
for which $\diff\varrho/\diff s=0$ and the above bound gets saturated. In figure \ref{fig:smallbubb}, the flow behaviour of $\varrho$, for different values of $\chi$ and starting from an initial condition $\varrho_0=1$, is depicted.

\subsubsection{Bubble with a larger scalar value}
We now assume the bubble to be characterise by a larger scalar field value, which translates into a choice of parameters $\phi_{2}<\phi_{1}$, and take a big enough value of $\varrho_{0}$ as the flow initial condition, so that the thin-wall approximation can be safely employed. Defining $\chi\left(\varepsilon\right)$ as in \eqref{chii}, it can be directly noticed that the solution for $\varrho$ presented in \eqref{rhosol} does not apply to the current scenario, since it would imply dealing with negative arguments in Lambert's function. Nevertheless, we can observe that right-hand side of \eqref{equaa} is always positive. Hence, $\varrho$ is forced to grow, asymptotically approaching a linear behaviour
\begin{equation}\label{asy}
    \Tilde{\varrho}\left(s\right)\approx\varrho_{0}+\chi\left(\varepsilon\right)\cdot s\ ,
\end{equation}
for which the $2/\varrho$ in \eqref{equaa} can be neglected. Therefore, when the scalar field value inside the bubble is larger than the one outside, the bubble grows indefinitely. For large enough values of the flow parameter $s$, moreover, the simple expression \eqref{asy} properly approximates the flow behaviour of the bubble radius.
\subsection{Cosmological constant background}
After having discussed the Minkowski background case in detail, we will now move to a slightly more complicated example. More specifically, we will still assume $M=0$, while the initial value $\Lambda_0$ of the cosmological constant will be taken not to vanish. The distinct scenarios associated to de Sitter and Anti-de Sitter space-time will therefore be analysed separately. As far as the metric flow is concerned, the evolution of the cosmological constant in both such instances was described in \ref{metbubbleconstruct} as:
\begin{equation}\label{eq:flowcc}
    \Lambda\left(s\right)=\frac{\Lambda_0}{1-2\Lambda_0\cdot s}\ .
\end{equation}
\subsubsection{Anti-de Sitter background}
As observed in \ref{metbubbleconstruct}, if the initial metric configuration is taken to be that of Anti-de Sitter space-time, we have that the flow asymptotically approaches Minkowski as $s\rightarrow\infty$. That being said, we now study how such an evolution of the geometry affects the flow of the scalar bubble. As far as large values of the flow parameter are concerned, we can expect the evolution to asymptotically approach the one studied in section \ref{Minkowski} as $\Lambda\left(s\right)$ goes to zero. Regarding the flow behaviour of the constants $\phi_2$ and $\phi_1$, respectively characterising the scalar field $\phi$ inside and outside the bubble, we get:
\begin{equation}\label{bbf}
    \frac{\diff\phi_2}{\diff s}=\frac{\diff\phi_1}{\diff s}=\frac{\Lambda_{0}}{1-2\Lambda_{0}\cdot s}\ .
\end{equation}
Therefore, such constants are not preserved along the flow in $\lambda$, as they were in section \ref{Minkowski}. In particular, we have:
\begin{equation}
\begin{split}
\phi_{2}\left(s\right)&=\phi_{2}\left(0\right)+2\log{\left(1-2\Lambda_{0}\cdot s\right)}\ ,\\
    \phi_{1}\left(s\right)&=\phi_{1}\left(0\right)+2\log{\left(1-2\Lambda_{0}\cdot s\right)}\ .
\end{split}
\end{equation}
The logarithmic behaviour is precisely produced by the fact that we asymptotically approach Minkowski space-time, for which the source term on the right-hand side of \eqref{bbf} progressively weakens. It can be observed that both $\phi_{1}$ and $\phi_{2}$ grow towards infinity. Fortunately, as we will see below, the flow equation for the bubble size will only depend on their difference, so the $s$-dependent part will be factored out. Plugging \eqref{bubb}, together with the metric, into the geometric flow equation \eqref{flow2} for $\phi$ and moving the $s$-dependence to the bubble radius $\varrho$, we get:
\begin{equation}
    \frac{\diff^2\varphi}{\diff x^2}=-4\Lambda+\left(\frac{\diff\varrho}{\diff s}-\frac{2}{x+\varrho}\right)\frac{\diff\varphi}{\diff x}+\left[1-\frac{\Lambda}{3}\left(x+\varrho\right)^2\right]\left(\frac{\diff\varphi}{\diff x}\right)^2\ .
\end{equation}
By multiplying both sides by the $x$-derivative of $\varphi$, we obtain:
\begin{equation}\label{walladsflow}
\begin{split}
\frac{1}{2}\frac{\diff}{\diff x}\left(\frac{\diff\varphi}{\diff x}\right)^2=&-4\Lambda\frac{\diff\varphi}{\diff x}+\left(\frac{\diff\varrho}{\diff s}-\frac{2}{x+\varrho}\right)\left(\frac{\diff\varphi}{\diff x}\right)^2+\\
&+\left[1-\frac{\Lambda}{3}\left(x+\varrho\right)^2\right]\left(\frac{\diff\varphi}{\diff x}\right)^3\ .
\end{split}
\end{equation}
The $x$-derivative of $\varphi$ takes the form illustrated in \eqref{derphi}. By integrating the above equation and getting rid of boundary terms, we get:
\begin{equation}
\begin{split}
    \frac{\diff\varrho}{\diff s}\int_{-\varepsilon}^{\varepsilon}G(x)^{2}\diff x=&2\int_{-\varepsilon}^{\varepsilon}\frac{G(x)^{2}}{x+\varrho}\diff x+\frac{\phi_{1}-\phi_{2}}{2}\int_{-\varepsilon}^{\varepsilon}G(x)^{3}\diff x+\\
    &-\frac{\Lambda}{3}\frac{\phi_{1}-\phi_{2}}{2}\int_{-\varepsilon}^{\varepsilon}G(x)^{3}\left(x+\varrho\right)^2\diff x\ .
\end{split}
\end{equation}
The flow in characterised by the three positive functions defined in \eqref{functions}, together with a further positive function:
\begin{equation}
    D\left(\varepsilon,\varrho\right)\equiv\int_{-\varepsilon}^{\varepsilon}G(x)^{3}\left(x+\varrho\right)^2\diff x\ .
\end{equation}
Hence, we can write the flow equation for the bubble radius as:
\begin{equation}\label{rhof2}
    \frac{\diff\varrho}{\diff s}=2\frac{B\left(\varepsilon,\varrho\right)}{A\left(\varepsilon\right)}+\frac{\phi_{1}-\phi_{2}}{2}\frac{C\left(\varepsilon\right)}{A\left(\varepsilon\right)}-\frac{\Lambda}{3}\frac{\phi_{1}-\phi_{2}}{2}\frac{D\left(\varepsilon,\varrho\right)}{A\left(\varepsilon\right)}\ .
\end{equation}
Given the thin-wall approximation, we get the approximate relation
\begin{equation}
    D\left(\varepsilon,\varrho\right)\approx\int_{-\varepsilon}^{\varepsilon}G(x)^{3}\varrho^2\diff x=C\left(\varepsilon\right)\varrho^2
\end{equation}
and write the above evolution in the following, simple, form
\begin{equation}\label{eqbb}
    \frac{\diff\varrho}{\diff s}=\frac{2}{\varrho}+\frac{\phi_1-\phi_2}{2}\left(1-\frac{\Lambda\varrho^2}{3}\right)Q\left(\varepsilon\right)\ ,
\end{equation}
in which the function $Q\left(\varepsilon\right)$ is defined in \eqref{eq:Qdef}.
Introducing, as was done in \eqref{chii}, the function $\chi\left(\varepsilon\right)$, we finally obtain:
\begin{equation}\label{eqeq}
       \frac{\diff\varrho}{\diff s}=\frac{2}{\varrho}+\left(1-\frac{\Lambda\varrho^2}{3}\right)\chi\left(\varepsilon\right)\ .
\end{equation}
We observe that, for Anti de Sitter space-time, $\Lambda$ does nothing more than enhancing the contribution of $\chi$ to the source term on the right-hand side of \eqref{eqeq}, particularly for early flow times and large values of $\varrho$. Therefore, there is no qualitative difference with the analysis developed in section \eqref{Minkowski}. Since $\Lambda\rightarrow 0$, even the asymptotic radius of shrinking bubbles in the $\chi\left(\varepsilon\right)<0$ regime is unchanged from the discussion developed in the case of a Minkowski background.
\subsubsection{de Sitter background}
Since its formal derivation coincides with the one developed for Anti de Sitter space-time, we find ourselves with the same thin-wall regime flow equation:
\begin{equation}
       \frac{\diff\varrho}{\diff s}=\frac{2}{\varrho}+\left(1-\frac{\Lambda\varrho^2}{3}\right)\chi\left(\varepsilon\right)\ .
\end{equation}
As previously discussed, the evolution in the flow parameter $s$ reaches a singularity at
\begin{equation}
    s_{1}=\frac{1}{2\Lambda_{0}}\ ,
\end{equation}
where $\Lambda\rightarrow\infty$ and $\phi_{i}\rightarrow-\infty$, for $i=1,2$. The function $\chi\left(\varepsilon\right)$ is still constant along the flow. Nevertheless, it can be easily observed that this setting is way richer and more subtle that the ones studied before. Therefore, different regions in the space of parameters $\left(\varrho_{0},\varepsilon,\Lambda_{0}\right)$ produce a wide variety of flow behaviours and must be analysed separately. Depending on the values of the parameters, the bubbles can steadily grow, shrink or follow non-monotonic behaviours.
\begin{figure}[H]
    \centering
    \includegraphics[width=0.9\linewidth]{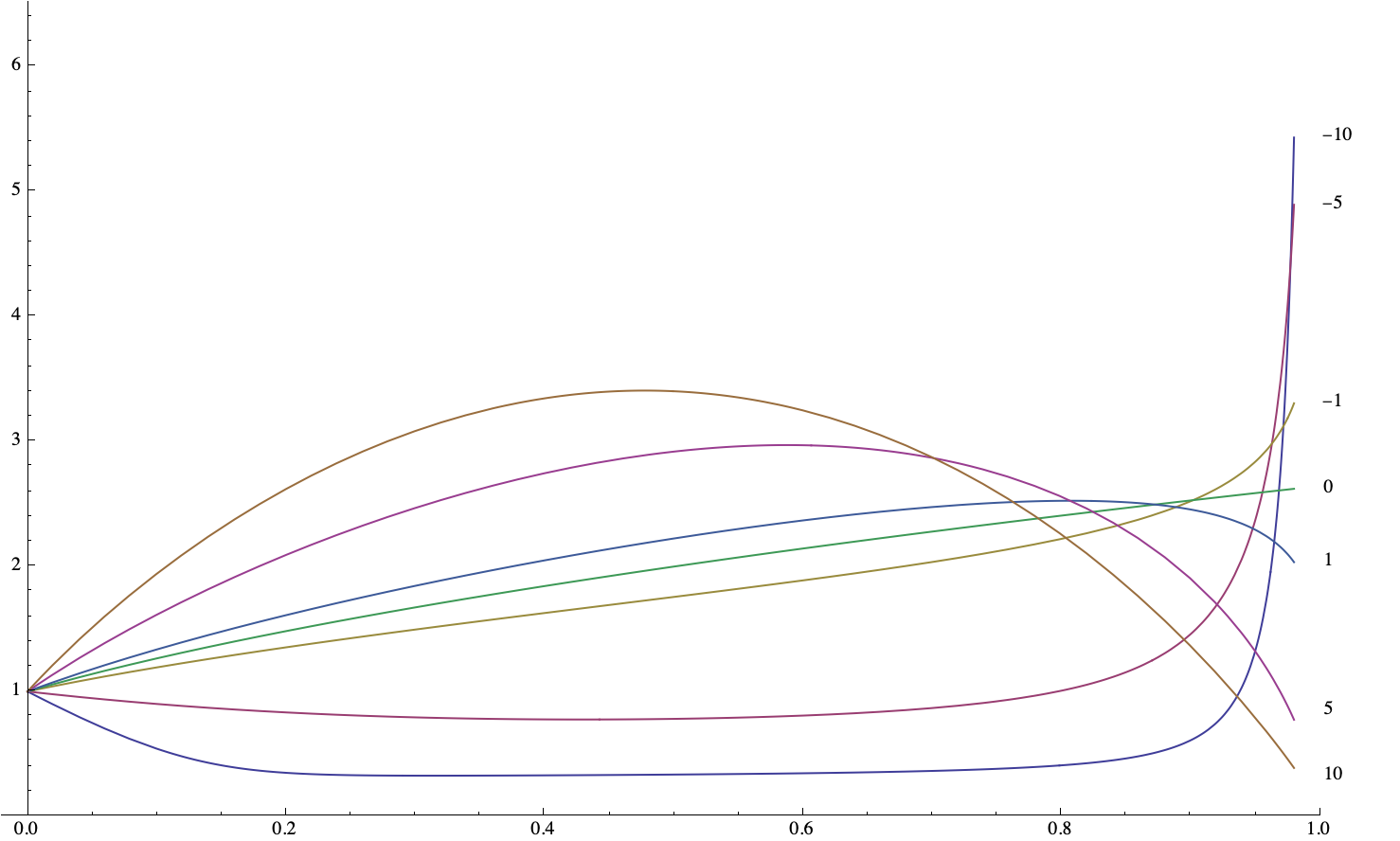}
    \caption[Flow behaviour of a bubble radius in de Sitter background.]{Flow behaviour $\varrho\left(s\right)$ of the scalar bubble radius, with respect to the flow parameter $s$ and for different values of the function $\chi\left(\varepsilon\right)$, depending on the bubble halved thickness. The initial condition is chosen to be $\varrho_{0}=1$.}
    \label{fig:sdbackgroundflowchi}
    \end{figure} 
In figure \ref{fig:sdbackgroundflowchi}, the flow behaviour $\varrho\left(s\right)$ of the scalar bubble radius, with respect to the flow parameter $s$ and for different values of the function $\chi\left(\varepsilon\right)$, is depicted. There, the initial condition was chosen to be $\varrho_0=1$. Different choices would have led to qualitatively analogous, albeit quantitatively different, evolutions. The figure clearly highlight how different flow behaviours can be produced by appropriately choosing the initial values of the solution parameters. In particular, there are regions of the parameters space for which the bubble radius exceeds that of de Sitter space-time along the evolution. This clearly generates an inconsistency, as the bubble would pass through the cosmological horizon. This feature might signal the fact that the only allowed choices for the initial parameters are the ones that do not lead us into such a pathological situation.

\subsection{Schwarzschild Background}
\begin{figure}[H]
    \centering
    \includegraphics[width=0.8\linewidth]{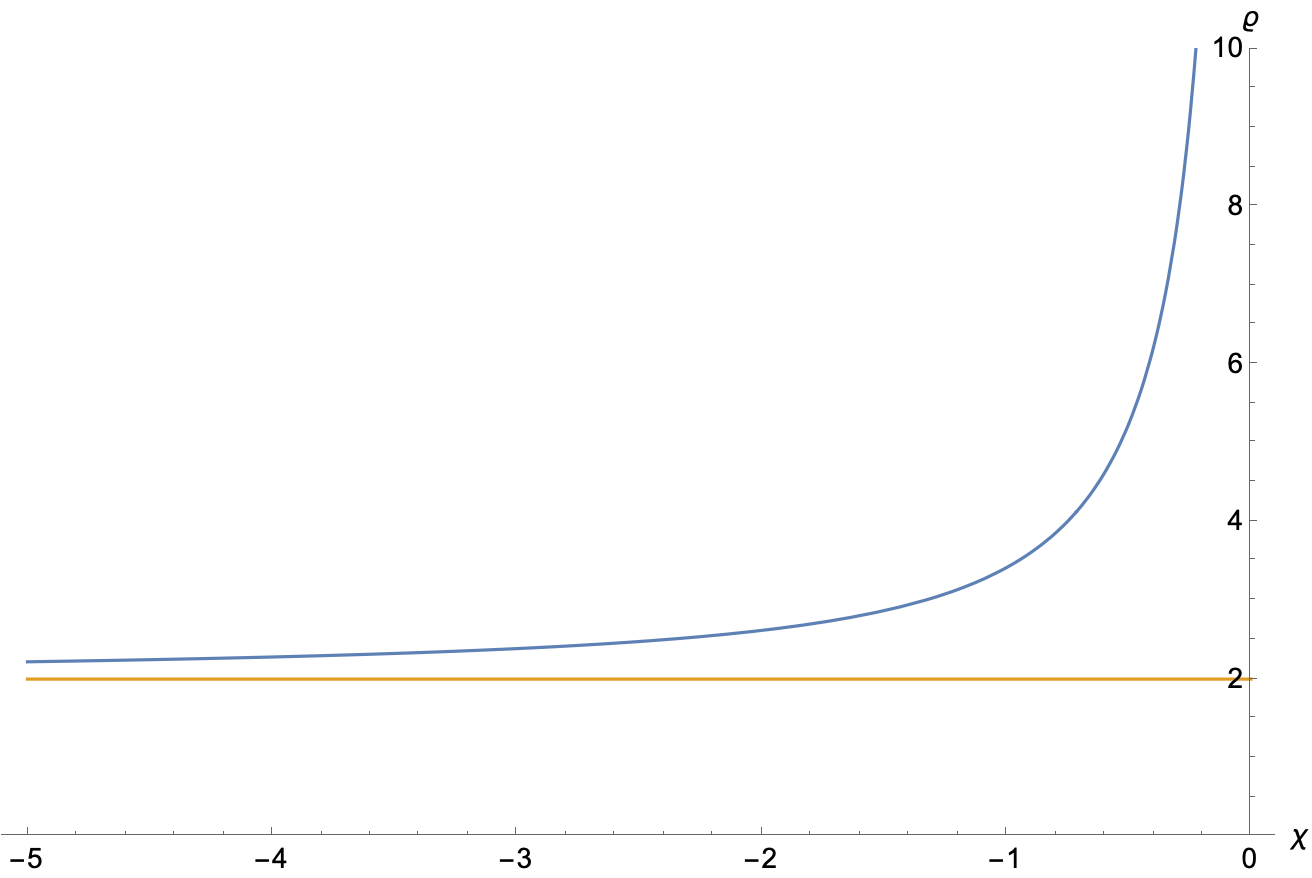}
    \caption[Flow behaviour of a bubble radius in Schwarzschild background.]{Plot of $\Bar{\varrho}_{+}$ (blue line), for $M=1$, against negative values of $\chi\left(\varepsilon\right)$. As can be clearly observed, it is bigger than the horizon radius (orange line).}
    \label{fig:rhominrad}
\end{figure}

Having got to this point, it is almost natural to focus on the case in which the initial value of the black hole mass $M_0$ is taken to be strictly positive, while setting the cosmological constant to zero. Analysing the much more complicated case in which the metric \eqref{genmet} is considered in its full generality goes beyond the scope of our current work. Hence, we simply consider the standard coordinates expression of the Schwarzschild metric:
\begin{equation}\label{schw}
    \ds^2=-\left(1-\frac{2M}{r}\right)\diff t^2+\left(1-\frac{2M}{r}\right)^{-1}\diff r^2+r^2\diff\theta^2+r^2\sin{\theta}^2\diff\phi^2\ .
\end{equation}
Since \eqref{schw} describes a flat geometry, the metric remains constant along the flow defined by \eqref{flow2}. Studying the region outside the black hole event horizon, we only consider bubbles larger than the horizon radius $r_{h}=2M$. For what concerns the scalar profile, we stick to the form presented in \eqref{dilaton} and take the ansatz \eqref{bubb}. Furthermore, we assume the thin wall approximation $\varrho\gg\varepsilon$. Being it radial, its Laplacian is:
\begin{equation}
    \Delta\phi=\frac{1}{\sqrt{-g}}\de_{\mu}\left(\sqrt{-g}g^{\mu\nu}\de_{\nu}\phi\right)=\left(1-\frac{2M}{r}\right)\frac{\diff^2\phi}{\diff r^2}+2\frac{r-M}{r^2}\frac{\diff\phi}{\diff r}\ .
\end{equation}
Hence, the flow equation in \eqref{flow2} for $\phi$ takes the form:
\begin{equation}\label{equasc}
    \frac{\diff\phi}{\diff s}=-\left(1-\frac{2M}{r}\right)\frac{\diff^2\phi}{\diff r^2}-2\frac{r-M}{r^2}\frac{\diff\phi}{\diff r}+\left(1-\frac{2M}{r}\right)\left(\frac{\diff\phi}{\diff r}\right)^2\ .
\end{equation}
By plugging \eqref{bubb} in \eqref{equasc}, we have:
\begin{equation}
     \frac{\diff^2\varphi}{\diff x^2}=\left(1-\frac{2M}{x+\varrho}\right)^{-1}\left[\frac{\diff\varrho}{\diff s}-2\frac{x+\varrho-M}{\left(x+\varrho\right)^2}\right]\frac{\diff\varphi}{\diff x}+\left(\frac{\diff\varphi}{\diff x}\right)^2\ .
\end{equation}
Since $x\in\left(-\varepsilon,\varepsilon\right)$, we have $x+\varrho\approx\varrho$. Moreover, we work with $\varrho>2M$. Hence, the flow equation can be approximated as:
\begin{equation}
    \frac{\diff^2\varphi}{\diff x^2}=\frac{\varrho}{\varrho-2M}\left(\frac{\diff\varrho}{\diff s}+2\frac{M-\varrho}{\varrho^2}\right)\frac{\diff\varphi}{\diff x}+\left(\frac{\diff\varphi}{\diff x}\right)^2\ .
\end{equation}
By multiplying both sides by $\diff\varphi/\diff x$ and integrating as was done in section \ref{Minkowski}, we have:
\begin{equation}
   \frac{\diff\varrho}{\diff s}\int_{-\varepsilon}^{\varepsilon}\left(\frac{\diff\varphi}{\diff x}\right)^2\diff x=2\frac{\varrho-M}{\varrho^2}\int_{-\varepsilon}^{\varepsilon}\left(\frac{\diff\varphi}{\diff x}\right)^2\diff x-\frac{\varrho-2M}{\varrho}\int_{-\varepsilon}^{\varepsilon}\left(\frac{\diff\varphi}{\diff x}\right)^3\diff x\ .
\end{equation}
Therefore, we obtain the flow equation for the scalar bubble radius as:
\begin{equation}
    \frac{\diff\varrho}{\diff s}=2\frac{\varrho-M}{\varrho^2}+\chi\left(\varepsilon\right)\frac{\varrho-2M}{\varrho}\ .
\end{equation}
It can be observed that, when $M=0$, the above reduces to \eqref{equaa}. Moreover, by studying the near-horizon behaviour of the flow, it can be shown that the $\varrho>2M$ assumption is conserved along the flow. Namely, by taking $\varrho=2M+\mu$, with $\mu\ll 2M$, we have:
\begin{equation}
    \frac{\diff\varrho}{\diff s}\sim\frac{1}{2M}>0\ .
\end{equation}
Hence, starting from $\varrho_{0}>2M$ we are forced to stay in the $\varrho>2M$ regime. Namely, if taken to be outside the black hole, the bubble wall cannot cross the horizon along the geometric evolution in $s$. Concerning fixed points of the flow, we observe:
\begin{equation}
    \frac{\diff\varrho}{\diff s}=0\quad\Longrightarrow\quad 2\frac{\varrho-M}{\varrho^2}+\frac{\varrho-2M}{\varrho}\chi\left(\varepsilon\right)=0
\end{equation}
Thus, it seems like we have fixed points $\Bar{\varrho}_{\pm}$ of the flow solving
\begin{equation}
    \Bar{\varrho}^2+2\frac{1-M\cdot\chi\left(\varepsilon\right)}{\chi\left(\varepsilon\right)}\Bar{\varrho}-\frac{2M}{\chi\left(\varepsilon\right)}=0\ ,
\end{equation}
but we still have to discuss whether any of the two lies outside the black hole event horizon, which is the region we are interested in. We have:
\begin{equation}
    \Bar{\varrho}_{\pm}=M-\frac{1}{\chi\left(\varepsilon\right)}\pm\sqrt{M^2+\frac{1}{\chi\left(\varepsilon\right)^2}}\ .
\end{equation}
We observe that, for $M>0$, only $\Bar{\varrho}_{+}$ can be bigger than $2M$. Moreover, we can only achieve this with $\chi\left(\varepsilon\right)<0$. Therefore:
\begin{itemize}
    \item $\chi\left(\varepsilon\right)>0$: There is no fixed point of the flow.
    \item $\chi\left(\varepsilon\right)<0$: The bubble radius $\Bar{\varrho}_{+}$ is a fixed point of the flow.
\end{itemize}
We now study the $\chi\left(\varepsilon\right)<0$ case in detail. As discussed above, we have a fixed point at:
\begin{equation}
       \Bar{\varrho}_{+}=M-\frac{1}{\chi\left(\varepsilon\right)}-\frac{\sqrt{M^2\chi\left(\varepsilon\right)^2+1}}{\chi\left(\varepsilon\right)}\ .
\end{equation}
In figure \ref{fig:rhominrad}, $\Bar{\varrho}_{+}$ is plotted against negative values of the function $\chi\left(\varepsilon\right)$. It can be easily observed that $\diff\varrho/\diff s>0$ for $2M<\varrho<\Bar{\varrho}_{+}$ and $\diff\varrho/\diff s<0$ for $\Bar{\varrho}_{+}<\varrho$. Namely, we always tend towards the fixed point at $\Bar{\varrho}_{+}$. For $\chi\left(\varepsilon\right)>0$, instead, we simply have that $\diff\varrho/\diff s>0$ for every value of $\varrho$. Therefore, the bubble is forced to grow indefinitely.

\section{Geometric flow of space-time bubbles}\label{norms}
In \ref{subs:bubblescal}, a spherically symmetric scalar field bubble was embedded in a variety of distinct space-time metrics. After having properly described such configuration and having assumed the thin-wall approximation, the evolution induced by Perelman's combined flow was assessed and discussed. In this section, instead of doing so for the scalar field, we will consider $4$-dimensional bubble configurations of the space-time geometry itself. In particular, we will construct space-time kinematical states with negative cosmological constant bubbles, connecting distinct Anti-de Sitter solutions. This will imply taking two negative values $\Lambda_1$ and $\Lambda_2$ for the cosmological constant, respectively characterising the inside and the outside of the bubble, and interpolating between them in a narrow spherical shell. Thereafter, we will study the systems evolution under a particular set of geometric flow equations, specifically designed to keep $\Lambda_1$ and $\Lambda_2$ fixed along the flow. Obtaining them will constitute the first step of the following derivation.
\subsection{Geometric flow equations}
Before constructing the space-time bubble interpolating between distinct Anti-de Sitter vacua, we will derive a new set of geometric flow equations, allowing us to easily force the cosmological constant values far from the bubble wall not to change along the flow. In order to do so, let's consider a $D$-dimensional space-time manifold $\mathcal{M}$, on which a Lorentzian metric $g_{\mu\nu}$ and a scalar field $\phi$ are defined. Let's moreover consider a slightly altered version of Perelman's entropy $\mathcal{F}$-functional, defined by 
\begin{equation}\label{FNC}
    \mathcal{F}\left[g,\phi\right]=\int\diff^{D}x\sqrt{-g}e^{-\phi}R\cdot\phi
\end{equation}
and on which we perform volume-preserving variations of the fields, such that:
\begin{equation}
    \delta\left(\sqrt{-g}e^{-\phi}\right)=0\ .
\end{equation}
Namely, starting from the usual variations
\begin{equation}
    g_{\mu\nu}\longrightarrow g_{\mu\nu}+v_{\mu\nu}\ ,\quad \phi\longrightarrow\phi+h\ ,
\end{equation}
we impose the identity:
\begin{equation}
    2h=g^{\mu\nu}v_{\mu\nu}\ .
\end{equation}
The overall variation of the entropy functional can be computed as:
\begin{equation}
    \begin{split}
        \delta\mathcal{F}&=\int\diff^{D}x\sqrt{-g}e^{-\phi}\left[\frac{R}{2}g_{\mu\nu}-\phi R_{\mu\nu} +e^{\phi}\nabla_{\mu}\nabla_\nu\left(e^{-\phi}\phi \right) -g_{\mu\nu}e^{\phi}\nabla^2\left(e^{-\phi}\phi\right) \right]v^{\mu\nu}\ .
    \end{split}
\end{equation}
Therefore, the flow equations can be written, in an extremely compact form and introducing a suitable normalisation of the flow parameter, as:
\begin{equation}
    \begin{split}
        \frac{\diff g_{\mu\nu}}{\diff s}&=-\frac{\phi}{2} R_{\mu\nu}+\frac{R}{4}g_{\mu\nu} +\frac{1}{2}\left[\nabla_{\mu}\nabla_\nu\left(e^{-\phi}\phi \right) -g_{\mu\nu}\nabla^2\left(e^{-\phi}\phi\right) \right]e^{\phi}\ ,\\
        \frac{\diff\phi}{\diff s}&=\frac{R}{4}\left(\frac{D}{2}-\phi\right)+\frac{1-D}{4}\nabla^2\left(e^{-\phi}\phi \right)e^{\phi} \ .
    \end{split}
\end{equation}
In $D=4$, which is the case we will consider, the above equations become:
\begin{equation}\label{eq:flownorm}
    \begin{split}
        \frac{\diff g_{\mu\nu}}{\diff s}&=-\frac{\phi}{2} R_{\mu\nu}+\frac{R}{4}g_{\mu\nu} +\frac{1}{2}\left[\nabla_{\mu}\nabla_\nu\left(e^{-\phi}\phi \right) -g_{\mu\nu}\nabla^2\left(e^{-\phi}\phi\right) \right]e^{\phi}\ ,\\
        \frac{\diff\phi}{\diff s}&=\frac{R}{4}\left(2-\phi\right)-\frac{3}{4}\nabla^2\left(e^{-\phi}\phi \right)e^{\phi} \ .
    \end{split}
\end{equation}
In \cite{DeBiasio:2022omq}, an alternative set of flow equations was derived and applied to cosmological constant bubble example. Nonetheless, being the entropy functional employed in such work not invariant under diffeomorphism, which is a feature \eqref{FNC} instead possesses, we choose to stick to the flow equations \eqref{eq:flownorm} for the remainder of the section. Even though they are not generally equivalent to the geometric flows presented in \cite{DeBiasio:2022omq}, they reduce to the same expressions when considering the particular geometric configuration at hand and for a specific scalar field profile. 
\subsection{Bubble construction}
Let the $4$-dimensional space-time manifold $\mathcal{M}$, endowed with a scalar field $\phi$ and with the Lorentzian metric tensor $g_{\mu\nu}$, be charted by coordinates $\left(t,r,x^i\right)$. 
Hence, we once again consider a spherical shell \eqref{shelldef} around the origin, with thickness $2\varepsilon$ and radius $\varrho$, and work with a metric tensor of the form
\begin{equation}\label{ADS}
    \ds^{2}=\frac{r^2}{\alpha^2\left(r\right)}\diff t^2+\frac{\alpha^2\left(r\right)}{r^2}\diff r^2+\frac{r^2}{\alpha^2\left(r\right)}\diff\Bar{x}^2\ ,
\end{equation}
where the radial function $\alpha(r)$ is defined as
\begin{equation}
    \alpha(r)\equiv\begin{cases}
        \begin{split}
        \alpha_{0}&\text{ for }r\leq\varrho-\varepsilon\\
        G(r)&\text{ for }\varrho-\varepsilon\leq r\leq \varrho+\varepsilon\\
        \alpha_{2}&\text{ for }r_1\geq \varrho+\varepsilon
        \end{split}
    \end{cases}\ ,
\end{equation}
and the interpolating function is:
\begin{equation}
    G(r)\equiv\frac{\alpha_2+\alpha_1}{2}+\frac{\alpha_2-\alpha_1}{2}\tanh{\frac{r-\varrho}{\left(\varepsilon+\varrho-r\right)\left(r-\varrho+\varepsilon\right)}}\ .
\end{equation}
In the above expressions, the constants $\alpha_2$ and $\alpha_1$ correspond to:
\begin{equation}
    \alpha_1\equiv\sqrt{-\frac{\Lambda_1}{3}}\ ,\quad \alpha_2\equiv\sqrt{-\frac{\Lambda_2}{3}}\ .
\end{equation}
By forcing the scalar field to be constant and equal to
\begin{equation}
    \phi_0\equiv 2\ ,
\end{equation}
the geometric flow equations reduce to:
\begin{equation}\label{eq:flownorm2}
    \begin{split}
        \frac{\diff g_{\mu\nu}}{\diff s}=-R_{\mu\nu}+\frac{R}{4}g_{\mu\nu}\ ,\quad \frac{\diff\phi}{\diff s}=0 \ .
    \end{split}
\end{equation}
Moving the full flow dependence to the function $\alpha(r)$, we can translate \eqref{eq:flownorm2} into a single equation for the scalar curvature and simplify the problem. Indeed, we get:
\begin{equation}\label{SRF}
    \frac{\de R}{\de\lambda}=R^{\mu\nu}R_{\mu\nu}-\frac{\nabla^2 R}{4}-\frac{R^2}{4}\ .
    \end{equation}
Considering the flow away from $\varrho$, we see that $\alpha_1$ and $\alpha_2$ are fixed in $s$. Hence, the full flow dependence can be moved to $\varepsilon$ and $\varrho$. At this point, we move to the thin-wall approximation, assume that the whole flow dependence can be pushed to $\varrho$, introduce a new radial variable $x\equiv r-\varrho$ and take:
\begin{equation}
   \alpha\left(r\right)\equiv\psi\left(r-\varrho\right)=\psi\left(x\right)\ .
\end{equation}
Referring with $\Dot{f}$ to $s$-derivatives and with $f'$ to $x$-derivatives, we have:
\begin{equation}
    \frac{\de\alpha}{\de r}=\psi'\ ,\quad\Dot{\alpha}=-\frac{\diff\varrho}{\diff s}\cdot\psi'\ .
\end{equation}
From the above expression for the metric, we get:
\begin{equation}
    R=\frac{6}{\psi^4} \left\{\psi\left(x+\varrho\right)\left[6 \psi'+\left(x+\varrho\right)\psi''\right]-2 \psi^2-4 \left(x+\varrho\right)^2 \psi'^2\right\}\ .
\end{equation}
By taking the $\lambda$-derivative, we obtain:
\begin{equation}
\begin{split}
    \Dot{R}&=\frac{\diff\varrho}{\diff s}\frac{6}{\psi^4} \biggl\{4\psi'\left(x+\varrho\right)6 \psi'+4\psi'\left(x+\varrho\right)^2\psi''-8\psi' \psi-16 \left(x+\varrho\right)^2 \frac{\psi'^3}{\psi}+\\
    &\quad-\psi'^2\left(x+\varrho\right)6 -\psi'\left(x+\varrho\right)^2\psi''+\psi6 \psi'+\psi\left(x+\varrho\right)\psi''+\\
    &\quad-6\psi''\psi\left(x+\varrho\right)-\psi\left(x+\varrho\right)^2\psi'''+\psi''\psi\left(x+\varrho\right)+4 \psi\psi'+\\
    &\quad-8 \left(x+\varrho\right) \psi'^2+8 \left(x+\varrho\right)^2 \psi'\psi''\biggr\}=\\
    &=\frac{\diff\varrho}{\diff s}\frac{6}{\psi^4} \biggl\{10\psi'\left(x+\varrho\right) \psi' +2 \psi\psi'-4\psi''\psi\left(x+\varrho\right)-\psi\left(x+\varrho\right)^2\psi'''+\\
    &\quad+11\left(x+\varrho\right)^2 \psi'\psi''-16 \left(x+\varrho\right)^2 \frac{\psi'^3}{\psi}\biggr\}=\\
    &=\frac{\diff\varrho}{\diff s}\left[C_{2}\left(x\right)\left(x+\varrho\right)^2+C_{1}\left(x\right)\left(x+\varrho\right)+C_{0}\left(x\right)\right]\ .
\end{split}
\end{equation}
In the above, we have introduced:
\begin{equation}
    \begin{split}
        C_{2}\left(x\right)&\equiv\frac{6}{\psi^5}\left[11\psi\psi'\psi''-\psi^2\psi'''-16\psi'^3\right]\ ,\\
        C_{1}\left(x\right)&\equiv\frac{12}{\psi^4}\left[5\psi'^2-2\psi\psi''\right]\ ,\\
        C_{0}\left(x\right)&\equiv\frac{12}{\psi^3}\psi'\ .
    \end{split}
\end{equation}
By collecting $\varrho$ terms in $\Dot{R}$, we are left with
\begin{equation}
    \Dot{R}=\frac{\diff\varrho}{\diff s}\left[D_{2}\left(x\right)\varrho^2+D_{1}\left(x\right)\varrho+D_0\left(x\right)\right]\ ,
\end{equation}
where:
\begin{equation}
    \begin{split}
        D_{2}\left(x\right)&\equiv C_{2}\left(x\right)\ ,\\
        D_{1}\left(x\right)&\equiv 2xC_{2}\left(x\right)+C_{1}\left(x\right)\ ,\\
        D_{0}\left(x\right)&\equiv x^2C_{2}\left(x\right)+xC_{1}\left(x\right)+C_0\left(x\right)\ .
    \end{split}
\end{equation}
Now, we want to write in a similar fashion the right-hand side
\begin{equation}
    K\equiv R^{\mu\nu}R_{\mu\nu}-\frac{\Delta R}{4}-\frac{R^2}{4}
\end{equation}
of the scalar curvature flow equation. We get:
\begin{equation}
\begin{split}
    K&=-\frac{3 \left(x+\varrho\right)}{2 \psi^8} \biggl\{16 \left(x+\varrho\right)^2 \psi'^2 \psi \left[20 \psi'+9 \left(x+\varrho\right)
   \psi''\right]-\\
   &\quad+\left(x+\varrho\right) \psi^2 \bigl[13 \left(x+\varrho\right)^2 \psi''^2+2 \left(x+\varrho\right) \psi' \bigl(91
   \psi''+\\
   &\quad+9 \left(x+\varrho\right) \psi'''\bigr)+208 \psi'^2\bigr]+\psi^3 \bigl[\left(x+\varrho\right)\bigl(\left(x+\varrho\right)^2 \psi''''+\\
   &\quad+50 \psi''+14 \left(x+\varrho\right) \psi'''\bigr)+40
   \psi'\bigr]-152 \left(x+\varrho\right)^3 \psi'^4\biggr\}=\\
   &=\sum_{k=1}^{4}G_{k}\left(x\right)\left(x+\varrho\right)^k\ .
\end{split}
\end{equation}
In the above expression, we've introduced:
\begin{equation}
    \begin{split}
        G_{4}\left(x\right)&\equiv-\frac{3}{2\psi^8}\left\{144 \psi'^2 \psi  \psi''-\psi^2 13\psi''^2+\psi^2 18  \psi'   \psi'''-152  \psi'^4+ \psi^3\psi''''\right\}\ ,\\
        G_{3}\left(x\right)&\equiv-\frac{3}{2\psi^8}\left\{\psi^3 14 \psi'''+320 \psi'^3\psi+182\psi^2   \psi' 
   \psi''\right\}\ ,\\
        G_{2}\left(x\right)&\equiv-\frac{3}{2\psi^8}\left\{\psi^2 208 \psi'^2+\psi^3 50 \psi''\right\}\ ,\\
        G_{1}\left(x\right)&\equiv-\frac{60}{\psi^8}\psi^3 
   \psi'\ .
    \end{split}
\end{equation}
At this point, we collect powers of $\varrho$ and get
\begin{equation}
    K\equiv N_{4}\left(x\right)\varrho^4+N_{3}\left(x\right)\varrho^3+N_{2}\left(x\right)\varrho^2+N_{1}\left(x\right)\varrho+N_{0}\left(x\right)\ ,
\end{equation}
where we have defined the functions:
\begin{equation}
    \begin{split}
        N_{4}\left(x\right)&\equiv G_{4}\left(x\right)\ ,\\
        N_{3}\left(x\right)&\equiv 4xG_{4}\left(x\right)+G_{3}\left(x\right)\ ,\\
        N_{2}\left(x\right)&\equiv 6x^2 G_{4}\left(x\right)+3x G_{3}\left(x\right)+G_{2}\left(x\right)\ ,\\
        N_{1}\left(x\right)&\equiv 4x^3 G_{4}\left(x\right)+3x^2 G_{3}\left(x\right)+2xG_{2}\left(x\right)+G_{1}\left(x\right)\ ,\\
        N_{0}\left(x\right)&\equiv x^4 G_{4}\left(x\right)+x^3 G_{3}\left(x\right)+x^2 G_{2}\left(x\right)+x G_{1}\left(x\right)\ .
    \end{split}
\end{equation}
Observing that all $D_{i}\left(x\right)$ and $N_{i}\left(x\right)$ are zero for $x\not\in\left[-\varepsilon,+\varepsilon\right]$, we can introduce the integrated constants:
\begin{equation}
    \mathcal{D}_{i}\equiv\int_{-\varepsilon}^{+\varepsilon}D_{i}\left(x\right)\diff x\ ,\quad \mathcal{N}_{i}\equiv\int_{-\varepsilon}^{+\varepsilon}N_{i}\left(x\right)\diff x\ .
\end{equation}
Therefore, the flow equation can simply be written as:
\begin{equation}
    \frac{\diff\varrho}{\diff s}\sum_{i=0}^{2}\mathcal{D}_i\varrho^i=\sum_{j=0}^{4}\mathcal{N}_j\varrho^j\ .
\end{equation}
By defining
\begin{equation}
    \mathcal{S}\equiv\frac{\mathcal{N}_0}{\mathcal{D}_0}\ ,\quad\mathcal{L}\equiv\frac{\mathcal{N}_4}{\mathcal{D}_2}\ ,
\end{equation}
we have that the \textit{large} $\varrho$ behaviour is controlled by sign of $\mathcal{L}$, while the \textit{small} $\varrho$ behaviour is controlled by sign of $\mathcal{S}$. In order to investigate the general flow behaviour, we fix $\alpha_2\equiv 1$ without loss of generality, as it only corresponds to setting a scale, and study the sign of $\Dot{\varrho}$ as a function of $\varrho$ and $\alpha_1$. It can be straightforwardly observed that the value of $\varrho$ doesn't really affect the sign of $\Dot{\varrho}$. Therefore, given the values of $\alpha_1$ and $\alpha_2$, the bubble follows either a monotonic growing or monotonic shrinking behaviour, regardless of the initial value $\varrho_0$ of its radius. Thus, the full qualitative dynamics can be captured by a plot of the sign of $\Dot{\varrho}$ against $\alpha_1$.
    \begin{figure}[H]
    \centering
    \includegraphics[width=0.9\linewidth]{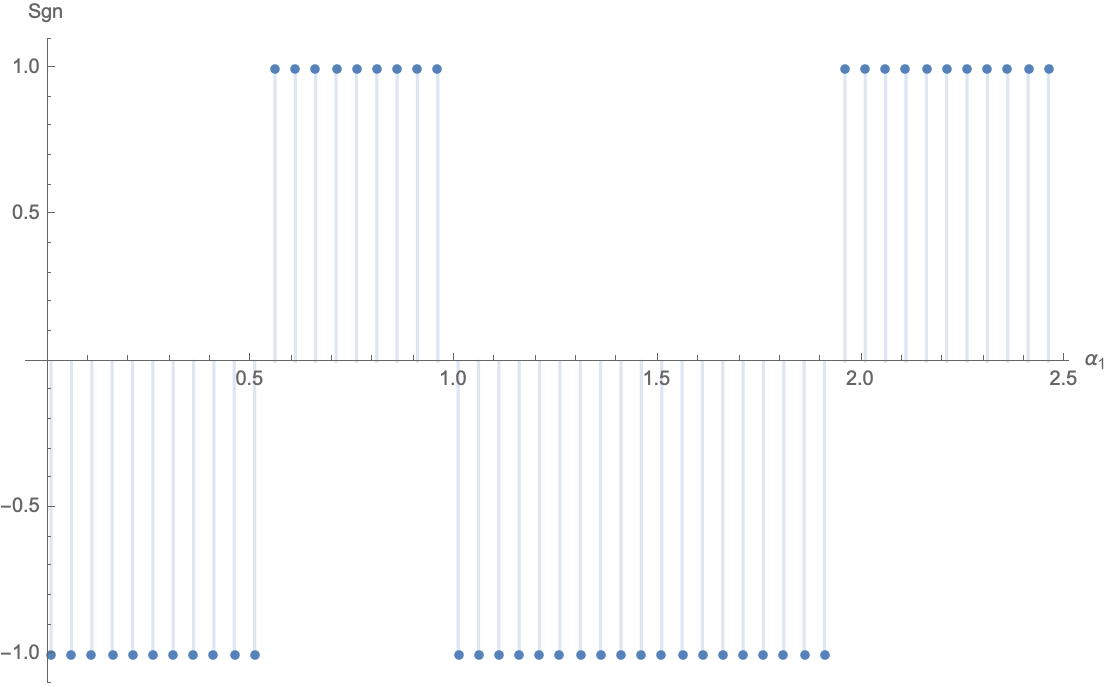}
    \caption[Plot of the sign of the bubble radius derivative in the flow parameter, with respect to $\alpha_1$.]{Plot of the sign of $\Dot{\varrho}$ against the value of $\alpha_1$, when setting $\alpha_2\equiv 1$. Regardless of the initial radius $\varrho_0$, the bubble will either grow or shrink monotonically.}
\end{figure}
\subsubsection{Distances along the Flow}
Concerning the evaluation of distances along the flow, at $D=4$, we can compute them with the formula
\begin{equation}
    \Delta\left(\varrho,\varrho_0\right)=\log{\frac{\mathcal{F}\left(\varrho\right)}{\mathcal{F}\left(\varrho_0\right)}}\ ,
\end{equation}
where $\mathcal{F}\left(\varrho\right)$ is the entropy functional computed for the metric characterised by radius $\varrho$.
Introducing an upper radial cut-off $\Omega$ and extracting as $\mathcal{F}_0$ the non-radial part of the entropy functional, which will anyway simplify in $\Delta$, we get:
\begin{equation}
    \mathcal{F}_{\Omega}=\mathcal{F}_0\int_{0}^{\Omega}\diff r\ \frac{6r}{\alpha^5}\left[\alpha r\left(6\alpha'+r\alpha''\right)-2\alpha^2-4r^2\alpha'^2\right]\equiv\mathcal{F}_0\cdot\mathcal{I}_{\Omega}\ .
\end{equation}
The above integral can be split as:
\begin{equation}
\begin{split}
    \mathcal{I}_{\Omega}&=\int_{\varrho-\varepsilon}^{\varrho+\varepsilon}\diff r\ \frac{6r}{\alpha^5}\left[\alpha r\left(6\alpha'+r\alpha''\right)-2\alpha^2-4r^2\alpha'^2\right]+\\
    &\quad -\int_{0}^{\varrho-\varepsilon}\diff r\ \frac{12r}{\alpha_2^3}-\int_{\varrho+\varepsilon}^{\Omega}\diff r\ \frac{12r}{\alpha_2^3}=\\
    &=\int_{\varrho-\varepsilon}^{\varrho+\varepsilon}\diff r\ \frac{6r}{\alpha^5}\left[\alpha r\left(6\alpha'+r\alpha''\right)-2\alpha^2-4r^2\alpha'^2\right]+\\
    &\quad -\frac{6}{\alpha_2^3}\left(\varrho-\varepsilon\right)^2-\frac{6}{\alpha_2^3}\left[\Omega^2-\left(\varrho+\varepsilon\right)^2\right]\ .
\end{split}
\end{equation}
By working in the \textit{thin-wall} approximation, we have $\varrho+\varepsilon\sim\varrho-\varepsilon\sim\varrho$. Hence, we get:
\begin{equation}
    \mathcal{I}_{\Omega}\left(\varrho\right)\sim-\frac{6}{\alpha_2^3}\varrho^2-\frac{6}{\alpha_2^3}\left[\Omega^2-\varrho^2\right]\ .
\end{equation}
We can therefore give a rough estimate of $\Delta_{\Omega}$, where $\Omega$ must be sent to $\infty$ \textit{after} having studied the limits in $\varrho_f$, as:
\begin{equation}
    \Delta_{\Omega}\sim\log{\frac{\mathcal{I}_{\Omega}\left(\varrho\right)}{\mathcal{I}_{\Omega}\left(\varrho_0\right)}}\ .
\end{equation}
We only focus on the $\varrho_f\mapsto\infty$ limit, since the shrinking behaviour leads to a breakdown of the thin-wall approximation. Indeed, the infinite radius bubble limit sits at infinite distance even before sending $\Omega$ to infinity, with:
\begin{equation}
    \Delta_{\Omega}\propto\log{\varrho}\ .
\end{equation}
By keeping the dependence on the radial cut-off $\Omega$, we are trying to discuss the \textit{intensive} distance in the same spirit as the one motivating the quantum field theoretic version of the \textit{information metric} presented in \cite{stout2021infinite}.
\newpage


%% file: Sections/On-Shellflow/intro.tex
In the following chapter, a collection of results contained in \cite{DeBiasio:2022omq} and \cite{DeBiasio:2022zuh} will be presented, broadened and organised in a coherent fashion. First of all, we will extend the results obtained in \ref{sec:perel} for the $\mathcal{F}$-entropy functional. In order to do so, we will generalise it, by introducing a family of supplementary terms, and derive the induced geometric flows for a system comprised of a metric tensor and a scalar field. The resulting differential equations will reduce to \eqref{flow2} for a specific choice of the constants up to which the functional will be defined. After having done so, we will start from an Einstein frame space-time action, move to string frame and use the resulting expression as an entropy functional, from which geometric flow equations will be deduced. We will therefore describe a natural path in the moduli space, straightforwardly fixed by the dynamics without introducing any additional, unphysical object. We will thereafter address an issue that has been postponed up to this point. Namely, we will consider the fact that, once a set of flow equations for both the metric and the scalar field is imposed, together with an on-shell initial configuration for the flow, there is no guarantee that the $s$-evolution will produce on-shell configurations, even for very small values of the flow parameter. In general, instead, a deviation from the equations of motion of the theory can be observed as soon as the flow is initiated. Hence, we will try to construct on-shell flows, preserving at least part of the equations of motion of the theories under consideration. Even if this will only be achieved for the metric, relaxing the on-shell condition on the scalar field, it will nevertheless be fruitful to project the path produced by the flow in $\mathbf{\Gamma}_{\mathcal{M}}$ to the generalised moduli space directions related to the geometry. By doing so, a partially on-shell flow will be obtained, probing that portion of $\mathbf{G}_{\mathcal{M}}$ consistent with Einstein field equations. 

%% file: Sections/On-Shellflow/generalflow.tex
\label{gfaaa}
In the following discussion, we will once more consider a $D$-dimensional Lorentzian space-time manifold $\mathcal{M}$, a scalar field $\phi$ and the generalised moduli space $\mathbf{\Gamma}_{\mathcal{M}}$ of geometries and scalar field configurations $\mathcal{M}$ can be provided with. The entropy functional \eqref{Perelentropy} on $\mathbf{\Gamma}_{\mathcal{M}}$ can be generalised, via the introduction of a Laplacian term and an infinite family of self interactions, as:
\begin{equation}\label{funk}
    \mathcal{F}_{\left(\alpha,\beta,\gamma\right)}\left[g,\phi\right]=\int_{\mathcal{M}}\diff^{D}x\sqrt{-g}e^{-\phi}\left[R_g+\alpha\left(\nabla\phi\right)^2+\gamma\nabla^2\phi+\sum_{n=0}^{\infty}\beta_{n}\phi^n\right]\ .
\end{equation}
We can avoid introducing an overall constant in front of $\mathcal{F}_{\left(\alpha,\beta,\gamma\right)}\left[g,\phi\right]$ without loss of generality, as it could anyway be reabsorbed into the flow parameter and therefore generate the same geometric flow equations. Hence, we set it to one. From this point on, we define:
\begin{equation}
\Bar{\mu}\equiv\left(\alpha,\gamma,\beta_{0}\dots\right)\ .
\end{equation}
It must be noted that the $\alpha$ parameter measures the significance of domain walls for $\phi$, as the term usually included in the \textit{Ginzburg-Landau} theory free energy functional. The standard $\mathcal{F}$-entropy functional \eqref{Perelentropy} can be obtained, from \eqref{funk}, by setting:
\begin{equation}
    \Bar{\mu}=\left(1,0,0\dots\right)\ .
\end{equation}
In order to derive the flow equations associated to \eqref{funk}, we follow the procedure outlined in \ref{sec:perel} and perform variations in $\phi$ and $g_{\mu\nu}$ so that:
\begin{equation}\label{volpres}
    \delta\int_{\mathcal{M}}\diff^Dx\sqrt{-g}e^{-\phi}=0\ .
\end{equation}
Namely, we impose the string frame volume functional $\Tilde{V}_{\mathcal{M}}$, defined in \eqref{sfvol}, not to change along the flow. By taking the first order variations
\begin{equation}
    g_{\mu\nu}\rightarrow g_{\mu\nu}+v_{\mu\nu}\ ,\quad\phi\rightarrow\phi+h\ ,
\end{equation}
we have the inverse metric variation:
\begin{equation}
    g^{\mu\nu}\rightarrow g^{\mu\nu}-v^{\mu\nu}\ .
\end{equation}
The requirement \eqref{volpres} imposes, as in \ref{sec:perel}, that:
\begin{equation}\label{presvol}
    2h=g^{\mu\nu}v_{\mu\nu}\ .
\end{equation}
In order to render the following computations more manageable, we start by listing the induced variations of the different terms appearing in the entropy functional \eqref{funk}. Indeed, after having define $v\equiv g^{\mu\nu}v_{\mu\nu}$, we have:
\begin{equation}
\begin{split}
        \delta\left(\nabla\phi\right)^2&=-v^{\mu\nu}\nabla_{\mu}\phi\nabla_{\nu}\phi +\nabla^{\mu}\phi\nabla_{\mu}v\ ,\quad     \delta\phi^n=\frac{n}{2}\phi^{n-1}v \ ,\\
    \delta\nabla^{2}\phi&=-v^{\mu\nu}\nabla_{\mu}\nabla_{\nu}\phi-\left(\nabla_{\mu}v^{\mu\nu}-\frac{1}{2}\nabla^{\nu}v\right)\nabla_{\nu}\phi+\frac{1}{2}\nabla^{2}v\ ,\\
    \delta R&=\nabla_{\mu}\nabla_{\nu}v^{\mu\nu}-\nabla^{2}v-R_{\mu\nu} v^{\mu\nu}\ .
\end{split}
\end{equation}
At this point, we can start analysing all the functional terms separately, with the aim of singling the field variations out. We will always neglect terms which are purely defined on the boundary, as there variations are taken to vanish. Starting from the polynomial term, it can be easily seen that:
\begin{equation}
    \delta\int_{\mathcal{M}}\diff^{D}x\sqrt{-g}e^{-\phi}\sum_{n=0}^{\infty}\beta_{n}\phi^n=\int_{\mathcal{M}}\diff^{D}x\sqrt{-g}e^{-\phi}\sum_{n=1}^{\infty}\frac{n\beta_{n}}{2}\phi^{n-1} v\ .
\end{equation}
Studying the kinetic term for the scalar and getting rid of boundary terms, we obtain:
\begin{equation}
    \delta\int_{\mathcal{M}}\diff^{D}x\sqrt{-g}e^{-\phi}\left(\nabla\phi\right)^2=\int_{\mathcal{M}}\diff^{D}x\sqrt{-g}e^{-\phi}\bigl\{\left[\left(\nabla\phi\right)^2-\nabla^2\phi\right]g_{\mu\nu}-\nabla_\mu\phi\nabla_\nu\phi\bigr\}v^{\mu\nu}\ .
\end{equation}
Moving to the scalar curvature term, we get:
\begin{equation}
    \begin{split}
        \delta\int_{\mathcal{M}}\diff^{D}x\sqrt{-g}e^{-\phi}R&=\int_{\mathcal{M}}\diff^{D}x\sqrt{-g}e^{-\phi}\bigl\{\nabla_\mu\phi\nabla_\nu\phi-\nabla_\mu\nabla_\nu\phi\\
        &\quad\quad\quad\quad\quad\quad\quad\quad\quad\quad+\left[\nabla^2\phi-\left(\nabla\phi\right)^2\right]g_{\mu\nu}-R_{\mu\nu}\bigr\}v^{\mu\nu}\ .
    \end{split}
\end{equation}
At this point, we analyse the two derivatives term:
\begin{equation}
    \begin{split}
        \delta\int_{\mathcal{M}}\diff^{D}x\sqrt{-g}e^{-\phi}\nabla^{2}\phi&=\int_{\mathcal{M}}\diff^{D}x\sqrt{-g}e^{-\phi}\bigl\{\left[\left(\nabla\phi\right)^2-\nabla^2\phi\right]g_{\mu\nu}-\nabla_\mu\phi\nabla_\nu\phi\bigr\}v^{\mu\nu}\ .
    \end{split}
\end{equation}
Combining all the above contribution into the overall variation of the entropy functional \eqref{funk}, we are left with the following expression:
\begin{equation}\label{varfunk}
    \begin{split}
    \delta\mathcal{F}_{\left(\alpha,\beta,\gamma\right)}&=\int_{\mathcal{M}}\diff^{D}x\sqrt{-g}e^{-\phi}\Bigl\{\left(1-\alpha-\gamma\right)\nabla_\mu\phi\nabla_\nu\phi-\nabla_\mu\nabla_\nu\phi+g_{\mu\nu}\sum_{n=1}^{\infty}\frac{n\beta_{n}}{2}\phi^{n-1}\\
        &\quad\quad\quad+\left(1-\alpha-\gamma\right)\left[\nabla^2\phi-\left(\nabla\phi\right)^2\right]g_{\mu\nu}-R_{\mu\nu}\Bigr\}v^{\mu\nu}\ .
    \end{split}
\end{equation}
Introducing the usual factor of two for normalisation purposes, including a diffeomorphism term, deriving the induced flow equation for the metric from \eqref{varfunk} and the one for the scalar from \eqref{presvol}, we get the following system of differential equations:
\begin{equation}
\begin{split}
        \frac{\diff g_{\mu\nu}}{\diff s}&=-2R_{\mu\nu}+2\left(1-\alpha-\gamma\right)\nabla_\mu\phi\nabla_\nu\phi-2\nabla_\mu\nabla_\nu\phi+\sum_{n=1}^{\infty}n\beta_{n}\phi^{n-1}g_{\mu\nu}\\
        &\quad-2\left(1-\alpha-\gamma\right)\bigl[\left(\nabla\phi\right)^2-\nabla^2\phi\bigr]g_{\mu\nu}+\mathcal{L}_\xi g_{\mu\nu}\ ,\\
        \frac{\diff\phi}{\diff s}&=-R+\left(1-D\right)\left(1-\alpha-\gamma\right)\left(\nabla\phi\right)^2+\frac{D}{2}\sum_{n=1}^{\infty}n\beta_{n}\phi^{n-1}\\
        &\quad+\left[D\left(1-\alpha-\gamma\right)-1\right]\nabla^2\phi+\mathcal{L}_\xi\phi\ .
\end{split}
\end{equation}
By performing the same diffeomorphism that deformed \eqref{flow1} into \eqref{flow2} for the sake of consistency, we are finally left with:
\begin{equation}\label{geneq}
\begin{split}
        \frac{\diff g_{\mu\nu}}{\diff s}&=-2R_{\mu\nu}+2\left(1-\alpha-\gamma\right)\nabla_\mu\phi\nabla_\nu\phi+\sum_{n=1}^{\infty}n\beta_{n}\phi^{n-1}g_{\mu\nu}\\
        &\quad-2\left(1-\alpha-\gamma\right)\bigl[\left(\nabla\phi\right)^2-\nabla^2\phi\bigr]g_{\mu\nu}+\mathcal{L}_\xi g_{\mu\nu}\ ,\\
        \frac{\diff\phi}{\diff s}&=-R+\left[\left(1-D\right)\left(1-\alpha-\gamma\right)+1\right]\left(\nabla\phi\right)^2+\frac{D}{2}\sum_{n=1}^{\infty}n\beta_{n}\phi^{n-1}\\
        &\quad+\left[D\left(1-\alpha-\gamma\right)-1\right]\nabla^2\phi+\mathcal{L}_\xi\phi\ .
\end{split}
\end{equation}
It can be clearly observed that, by imposing $\Bar{\mu}=\left(1,0,0\dots\right)$, the standard form of Perelman's combined flow \eqref{flow2} can be achieved. The coefficients of the various terms appearing in the above equations depend both on the parameters $\Bar{\mu}$ introduced in the functional \eqref{funk} and on the dimension $D$ of the manifold on which the fields are defined. By choosing a specific value of $D$ and suitably adjusting the parameters, some terms can be set to zero and removed. However, one might be interested in grounding the choice of the parameters in a more fundamental physical principle, instead of simply imposing them by hand. In \eqref{gfa}, a proposal to address such a necessity is discussed in detail.
\subsection{Flow equations from an action}\label{gfa}

The introduction of an entropy functional, at first in the standard form considered in \eqref{Perelentropy} and then with the more general expression \eqref{funk}, allowed us to derive a set geometric flow equations for a system comprised of a dynamical metric tensor and a scalar field, instead of simply postulating them. While representing a significant achievement, this new conceptual framework might arguably not be regarded as satisfying from the perspective of physics. At a close inspection, part of the arbitrariness which was previously exerted in the specification of a geometric flow was simply pushed to the choice of an entropy functional. Therefore, we now want to start from a particular space-time theory for a $D$-dimensional metric and a scalar, provided with an appropriate dynamics, and use it to completely determine a set of geometric flow equations. By doing so, we will get rid of the above mentioned arbitrariness and exploit the one functional that physics already grants us with: the action. In practice, this will translate into taking the Einstein frame action for a theory, moving to string frame and treating the resulting expression for the action as an entropy functional, from which a set of geometric flow equations will be obtained. We hence start from the Lorentzian dynamics governed by:
\begin{equation}\label{aact}
    \tilde{S}\left[g,\phi\right]=\frac{1}{2\kappa^2_D}\int_{\mathcal{M}}\diff^{D}x\sqrt{-g}\left[R_{g}-\frac{1}{2}\left(\nabla \phi\right)^{2}-\sum_{n=0}^{\infty}\frac{g_{n}}{n!}\phi^{n}\right]\ .
\end{equation}
It can be clearly observed that the equations of motion associated to above action are
\begin{equation}\label{aact:eom}
    R_{\mu\nu}-\frac{1}{2}Rg_{\mu\nu}= \kappa^2_D T_{\mu\nu}\ ,\quad\nabla^2\phi=\sum_{n=0}^{\infty}\frac{g_{n+1}}{n!}\phi^{n}\ ,
\end{equation}
where the energy momentum tensor is defined as:
\begin{equation}\label{eq:emt}
    \kappa^2_D T_{\mu\nu}\equiv\frac{1}{2}\nabla_\mu \phi\nabla_\nu \phi - \left[\frac{1}{4}\left(\nabla \phi\right)^{2}+\sum_{n=0}^{\infty}\frac{g_{n}}{n!}\cdot\frac{\phi^{n}}{2}\right]g_{\mu\nu}\ . 
\end{equation}
We now rescale both the metric $g$ and the scalar $\phi$, in order to obtain the string frame expression for \eqref{aact}. We thus introduce the fields $g$ and $\phi$, associated to the previous ones by:
\begin{equation}
    g_{\mu\nu}\rightarrow e^{2\varphi}g_{\mu\nu},\quad \phi\rightarrow\sigma\phi\ .
\end{equation}
Under such redefinition of the dynamical degrees of freedom, the action \eqref{aact} becomes:
\begin{equation}\label{sfaact}
\begin{split}
        S\left[g,\phi\right]&\equiv\frac{1}{2\kappa^2_D}\int_{\mathcal{M}}\diff^{D}x\sqrt{-g}e^{\left(D-2\right)\varphi}\biggl[R_g-\left(D-2\right)\left(D-1\right)\left(\nabla\varphi\right)^2\\
     &\quad\quad\quad\quad\quad\quad-2\left(D-1\right)\nabla^2\varphi-\frac{\sigma^{2}}{2}\left(\nabla\phi\right)^{2}-e^{2\varphi}\sum_{n=0}^{\infty}\frac{\sigma^{n}g_{n}}{n!}\phi^{n}\biggr]\ .
\end{split}
\end{equation}
For the volume element to take the appropriate string frame form, we take:
\begin{equation}
    \varphi\equiv\frac{1}{2-D}\phi\ .
\end{equation}
Therefore, we obtain:
\begin{equation}\label{sfact2}
\begin{split}
        S\left[g,\phi\right]&\equiv\frac{1}{2\kappa^2_D}\int_{\mathcal{M}}\diff^{D}x\sqrt{-g}e^{-\phi}\biggl[R_g-\left(\frac{\sigma^{2}}{2}+\frac{D-1}{D-2}\right)\left(\nabla\phi\right)^2\\
     &\quad\quad\quad\quad\quad\quad+2\frac{D-1}{D-2}\nabla^2\phi-\exp{\frac{2\phi}{2-D}}\sum_{n=0}^{\infty}\frac{\sigma^{n}g_{n}}{n!}\phi^{n}\biggr]\ .
\end{split}
\end{equation}
The constant $\sigma$ can be fixed by imposing, without loss of generality, that:
\begin{equation}
    \frac{\sigma^{2}}{2}+\frac{D-1}{D-2}=2\Longrightarrow \sigma=\sqrt{\frac{2D-6}{D-2}}\ .
\end{equation}
We are therefore left with:
\begin{equation}\label{sfact2}
\begin{split}
        S\left[g,\phi\right]&\equiv\frac{1}{2\kappa^2_D}\int_{\mathcal{M}}\diff^{D}x\sqrt{-g}e^{-\phi}\Biggl[R_g-2\left(\nabla\phi\right)^2+2\frac{D-1}{D-2}\nabla^2\phi\\
     &\quad\quad\quad\quad\quad\quad-\exp{\frac{2\phi}{2-D}}\sum_{n=0}^{\infty}\frac{g_{n}}{n!}\phi^{n}\left(\frac{2D-6}{D-2}\right)^{\frac{n}{2}}\Biggr]\ .
\end{split}
\end{equation}
The above expression can be made simpler by defining the constants
\begin{equation}
    p\equiv\frac{2}{2-D},\quad q_{n}\equiv\frac{g_{n}}{n!}\left(\frac{2D-6}{D-2}\right)^{n/2}
\end{equation}
and observing that:
\begin{equation}
    e^{p\phi}\sum_{n=0}^{\infty}q_n\phi^n=\sum_{n=0}^{\infty}\phi^n\sum_{k=0}^{n}\frac{g_{k}}{\left(n-k\right)!k!}\left(\frac{2D-6}{D-2}\right)^{\frac{k}{2}}\left(\frac{2}{2-D}\right)^{n-k} \ .
\end{equation}
Therefore, by defining
\begin{equation}\label{consteqai}
    s_{\ n}^{(D)}\equiv\sum_{k=0}^{n}\frac{g_{k}}{\left(n-k\right)!k!}\left(\frac{2D-6}{D-2}\right)^{\frac{k}{2}}\left(\frac{2}{2-D}\right)^{n-k}
\end{equation}
for $n\geq 0$, the string frame action takes the following form:
\begin{equation}\label{entract}
     S\left[g,\phi\right]=\frac{1}{2\kappa^2_D}\int_{\mathcal{M}}\diff^{D}x\sqrt{-g}e^{-\phi}\Biggl[R_g-2\left(\nabla \phi\right)^2+2\frac{D-1}{D-2}\nabla^2\phi-\sum_{n=0}^{\infty}s_{\ n}^{(D)}\phi^{n}\Biggr]\ .
\end{equation}
Hence, starting from the space-time action \eqref{aact} and moving to string frame, we have obtained a functional over the moduli space $\mathbf{\Gamma}_{\mathcal{M}}$ that falls into the general class described by \eqref{funk}. More specifically, it corresponds to the choices:
\begin{equation}
    \alpha=-2,\quad\gamma=2\frac{D-1}{D-2},\quad\beta_{n}=-s_{\ n}^{(D)}\ .
\end{equation}
The set of geometric flow equations induced by \eqref{entract} for $g$ and $\phi$, which can be obtained by plugging the above constant into the general formulas \eqref{geneq}, are:
\begin{equation}\label{acteq}
\begin{split}
        \frac{\diff g_{\mu\nu}}{\diff s}&=-2R_{\mu\nu}+2\frac{D-4}{D-2}\nabla_\mu\phi\nabla_\nu\phi-\sum_{n=1}^{\infty}ns_{\ n}^{(D)}\phi^{n-1}g_{\mu\nu}\\
        &\quad-2\frac{D-4}{D-2}\bigl[\left(\nabla\phi\right)^2-\nabla^2\phi\bigr]g_{\mu\nu}+\mathcal{L}_\xi g_{\mu\nu}\ ,\\
        \frac{\diff\phi}{\diff s}&=-R-\frac{D^2-6D+6}{D-2}\left(\nabla\phi\right)^2-\frac{D}{2}\sum_{n=1}^{\infty}ns_{\ n}^{(D)}\phi^{n-1}\\
        &\quad+\frac{D^2-5D+2}{D-2}\nabla^2\phi+\mathcal{L}_\xi\phi\ .
\end{split}
\end{equation}
Starting from the $D$-dimensional Lorentzian action for a system comprised of dynamical metric and a scalar field, in which the latter is subject to arbitrary polynomial self interactions, and treating its string frame expression as an entropy functional of the form presented in \eqref{funk}, we have derived a set of geometric flow equations without the necessity of postulating any additional, arbitrary structure. It is remarkable that, in $D=4$, the equations take the simple form:
\begin{equation}\label{4acteq}
\begin{split}
        \frac{\diff g_{\mu\nu}}{\diff s}&=-2R_{\mu\nu}-\sum_{n=1}^{\infty}ns_{\ n}^{(4)}\phi^{n-1}g_{\mu\nu}+\mathcal{L}_\xi g_{\mu\nu}\ ,\\
        \frac{\diff\phi}{\diff s}&=-R+\left(\nabla\phi\right)^2-\nabla^2\phi-2\sum_{n=1}^{\infty}ns_{\ n}^{(4)}\phi^{n-1}+\mathcal{L}_\xi\phi\ ,
\end{split}
\end{equation}
where the constants appearing the polynomial terms are just:
\begin{equation}
    s_{\ n}^{(4)}=\sum_{k=0}^{n}\frac{g_{k}\left(-1\right)^{n-k}}{\left(n-k\right)!k!}=\frac{1}{n!}\sum_{k=0}^{n}\binom{n}{k}\left(-1\right)^{n-k}g_{k}\ .
\end{equation}
Since, except for the standard Ricci flow contribution, there is no other term on the right hand side of the metric flow equation containing its derivatives, we are sure that the well-posedness of the differential problem is not affected by our generalisation. The only necessary requirement obviously concerns the number of space-time dimensions, which must be set to be bigger than two.
\subsubsection{Circle-compactified theory}
In the present discussion, we will follow a procedure analogous to the one outlined in \ref{circlecompactification}, start from a free general relativistic gravity theory in $\left(D+1\right)$-dimensions and compactify it on a circle. Under a simple ansatz, in which the vector degrees of freedom of the geometry will be set to zero, the dimensionally-reduced theory will be that of a dynamical metric and a scalar radion field. A set of geometric flow equations will be then naturally obtained according to the results presented in \ref{gfa}. Starting, thus, from the Einstein-Hilbert action
\begin{equation}\label{grmet}
    S_{\text{EH}}\left[G\right]\equiv\frac{1}{2\kappa^2_{D+1}}\int_{\mathcal{M}}\diff^{D+1}x\sqrt{-G}R_{G}
\end{equation}
for a $\left(D+1\right)$-dimensional metric tensor $G_{MN}$ over a Lorentzian manifold $\mathcal{M}$, we can decompose the degrees of freedom of the geometry in a $D$-dimensional metric tensor $h_{\mu\nu}$, a $D$-dimensional vector field $A_\mu$ and a scalar field $\sigma$. In particular, we impose:
\begin{equation}
   G_{MN} \equiv \begin{pmatrix}
 h_{\mu\nu} + e^{2\varphi} A_\mu A_\nu & e^{2\varphi} A_\mu \\
 e^{2\varphi} A_\nu & e^{2\varphi}\end{pmatrix}.
\end{equation}
From the perspective of the $\left(D+1\right)$-dimensional line element, we obtain
\begin{equation}
    ds^2_{D+1}= h_{\mu\nu}dx^\mu dx^\nu + e^{2\varphi} \bigl(A_\mu dx^\mu + dx^D\bigr)^2.
\end{equation}
with the greek indices run from $0$ to $D-1$. At this point, for the sake of simplicity and focusing on a particular example, we assume $\varphi$ and $h_{\mu\nu}$ not to depend on $x^D$ and the vector field $A_\mu$ to vanish:
\begin{equation}
    A_{\mu}=0\ .
\end{equation}
Furthermore, we take the space-time manifold to be expressed as a Cartesian product
\begin{equation}
    \mathcal{M}\equiv\mathcal{N}\times S^{1}\ ,
\end{equation}
where $\mathcal{N}$ is a $D$-dimensional space-time manifold, with coordinates $x^\mu$ and metric tensor $h_{\mu\nu}$, and $S^{1}$ is a $1$-dimensional compact circle parametrised by $x^D\in\left[0,2\pi\rho\right)$. It goes without saying that, taken at face value, the radion field $\varphi$ fixes the variation of the compact dimension radius along the non-compact manifold directions. The constant $\rho$ hence corresponds to the size of the compact dimension where $\varphi=0$. Similarly to what was done in \eqref{scaldecomp}, we can derive:
\begin{equation}
    R_G=R_h-2\nabla^2\varphi-2\nabla^\mu\varphi\nabla_\mu\varphi\ .
\end{equation}
By plugging such result in \eqref{grmet} and integrating the compact direction out, the action governing the $D$-dimensional dynamics on $\mathcal{N}$ can be obtained as
\begin{equation}
S_{0}\left[h,\varphi\right]\equiv\frac{1}{\kappa^2_{D}}\int_{\mathcal{N}}\diff^Dx\sqrt{-h}e^{\varphi}\left(R_h-2\nabla^2\varphi-2\nabla^\mu\varphi\nabla_\mu\varphi\right)\ ,
\end{equation}
in which we have defined:
\begin{equation}
    \kappa^2_{D}\equiv\frac{\kappa^2_{D+1}}{2\pi\rho}\ .
\end{equation}
In order to make contact with the action presented in \eqref{aact}, where Einstein-frame was employed, we introduce the following rescalings:
\begin{equation}
    h_{\mu\nu}=e^{2\omega}g_{\mu\nu}\ ,\quad \varphi\equiv\alpha\phi\ .
\end{equation}
Analysing the expressions appearing in the action one by one, we get:
\begin{equation}
    \begin{split}
        \sqrt{-h}&=e^{D\omega}\sqrt{-g}\ ,\quad\nabla_{\mu}\varphi=\alpha\nabla_\mu\phi\ ,\quad e^{\varphi}=e^{\alpha\phi}\ ,\\
        R_h&=e^{-2\omega}R_g-2(D-1)e^{-2\omega}g^{\mu\nu}\nabla_\mu\nabla_\nu\omega-(D-2)(D-1)e^{-2\omega}g^{\mu\nu}\nabla_\mu\omega\nabla_\nu\omega\ ,\\
h^{\mu\nu}&\nabla_\mu\varphi\nabla_\nu\varphi=\alpha^2e^{-2\omega}g^{\mu\nu}\nabla_\mu\phi\nabla_\nu\phi\ ,\\
h^{\mu\nu}&\nabla_\mu\nabla_\nu\varphi=\alpha e^{-2\omega}\left[g^{\mu\nu}\nabla_\mu\nabla_\nu\phi+\left(D-2\right)g^{\mu\nu}\nabla_\mu\phi\nabla_\nu\omega\right]\ .
    \end{split}
\end{equation}
Therefore, the $D$-dimensional action for $g$ and $\phi$ becomes:
\begin{equation}
\begin{split}
    S_{1}\left[g,\phi\right]\equiv\frac{1}{\kappa^2_{D}}\int_{\mathcal{N}}\diff^Dx\sqrt{-g}e^{\alpha\phi+\left(D-2\right)\omega}&\Bigl\{R_g-2(D-1)\nabla^2\omega-2\alpha^2\nabla^\mu\phi\nabla_\mu\phi\\
        &-(D-2)(D-1)\nabla^\mu\omega\nabla_\mu\omega\\
        &-2\alpha \left[\nabla^2\phi+\left(D-2\right)\nabla^\mu\phi\nabla_\mu\omega\right]\Bigr\}\ .
\end{split}
\end{equation}
In order to properly fix the metric rescaling, we impose
\begin{equation}
    \omega=\frac{\alpha}{2-D}\phi
\end{equation}
and reduce the action, from which we remove the Laplacian boundary terms, to:
\begin{equation}
\begin{split}
    S_{1}\left[g,\phi\right]\equiv\frac{1}{\kappa^2_{D}}\int_{\mathcal{N}}\diff^Dx\sqrt{-g}&\Bigl\{R_g-\alpha^2\frac{D-1}{D-2}\nabla^\mu\phi\nabla_\mu\phi\Bigr\}\ .
\end{split}
\end{equation}
Requiring the free rescaling parameter $\alpha$ to satisfy
\begin{equation}
    \alpha\equiv\sqrt{\frac{D-2}{D-1}}\ ,
\end{equation}
we have nothing more than the Einstein frame action
\begin{equation}
\begin{split}
    S\left[g,\phi\right]\equiv\frac{1}{\kappa^2_{D}}\int_{\mathcal{N}}\diff^Dx\sqrt{-g}&\left(R_g-\frac{1}{2}\nabla^\mu\phi\nabla_\mu\phi\right)\ .
\end{split}
\end{equation}
for a free real scalar field in a dynamical Lorentzian background space-time. Exploiting the string-frame action as an entropy functional as described in the previous discussion, the flow equations induced by our circle-compactified gravitational theory can be read-off from \eqref{acteq}, by imposing all the $g_k$ self-coupling constants in \eqref{aact} to vanish, as:
\begin{equation}\label{acteqcircle}
\begin{split}
        \frac{\diff g_{\mu\nu}}{\diff s}&=-2R_{\mu\nu}+2\frac{D-4}{D-2}\nabla_\mu\phi\nabla_\nu\phi-2\frac{D-4}{D-2}\bigl[\left(\nabla\phi\right)^2-\nabla^2\phi\bigr]g_{\mu\nu}+\mathcal{L}_\xi g_{\mu\nu}\ ,\\
        \frac{\diff\phi}{\diff s}&=-R-\frac{D^2-6D+6}{D-2}\left(\nabla\phi\right)^2+\frac{D^2-5D+2}{D-2}\nabla^2\phi+\mathcal{L}_\xi\phi\ .
\end{split}
\end{equation}
Taking the effective theory to be in $4$ dimensions, the flow remarkably reduces to:
\begin{equation}\label{acteqcircleD4}
\begin{split}
        \frac{\diff g_{\mu\nu}}{\diff s}&=-2R_{\mu\nu}+\mathcal{L}_\xi g_{\mu\nu}\ ,\\
        \frac{\diff\phi}{\diff s}&=-R+\left(\nabla\phi\right)^2-\nabla^2\phi+\mathcal{L}_\xi\phi\ .
\end{split}
\end{equation}
We once more find ourselves with Perelman's combined flow. Hence, such flow equations can be physically interpreted, in the sense specified in \ref{gfa} and setting the vector part of the geometry to vanish, as being induced by the $4$-dimensional circle compactification of a $5$-dimensional purely gravitational theory.
\subsection{Scalar field with a sextic potential}
\begin{figure}[H]
    \centering
    \includegraphics[width=0.8\linewidth]{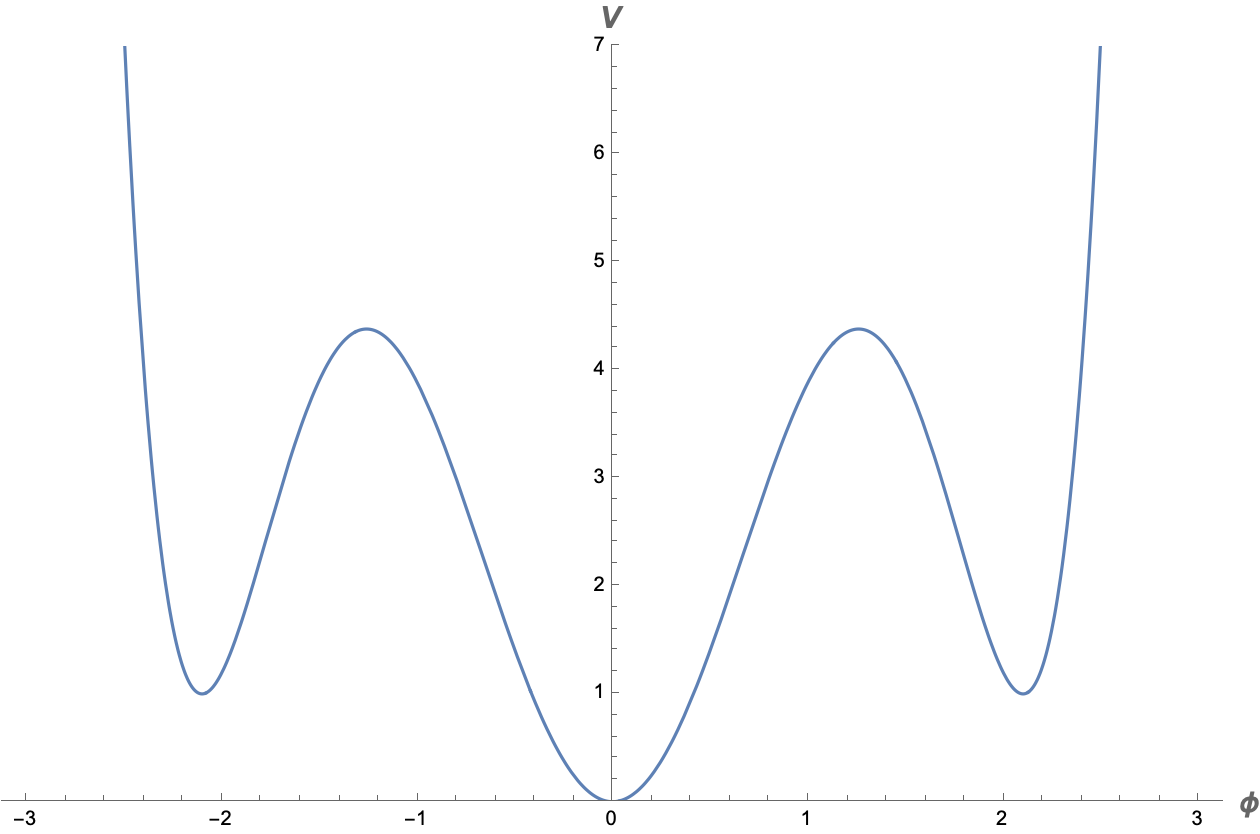}
    \caption{Plot of the behaviour of $V\left(\phi\right)$ around $\phi=0$, when $V_0=1$.}
    \label{pottV}
\end{figure}

\begin{figure}[H]
    \centering
    \includegraphics[width=0.7\linewidth]{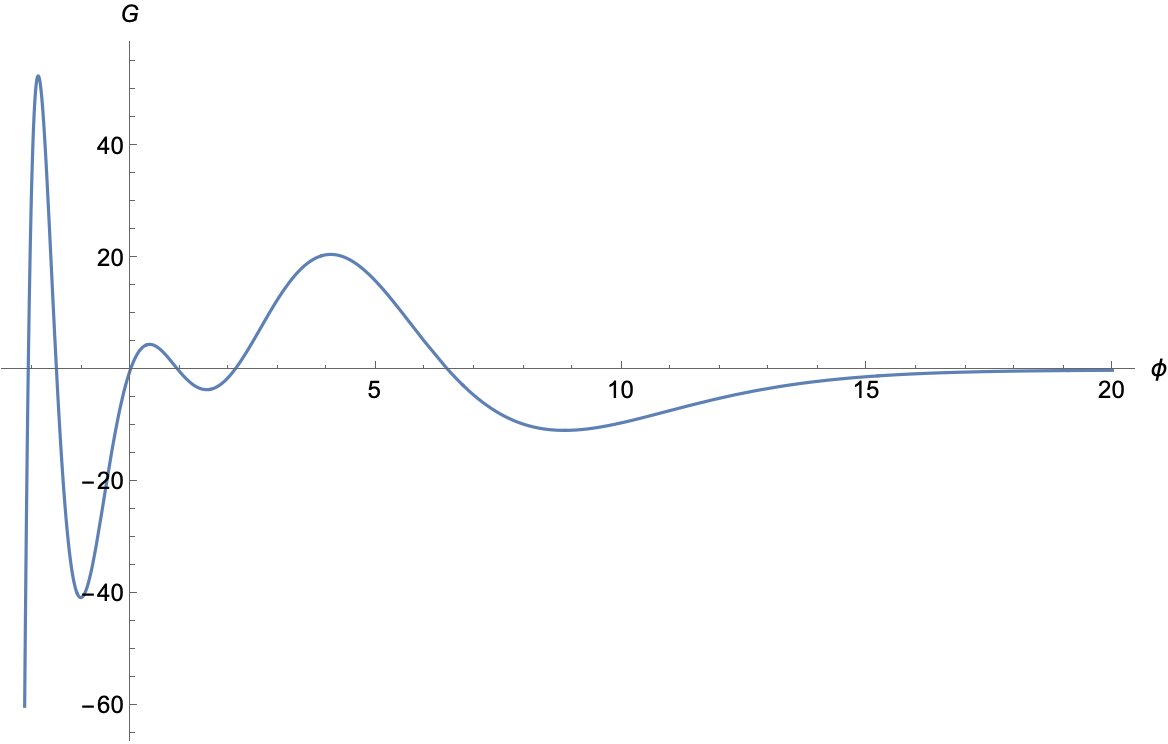}
    \caption{Plot of $G_4\left(\phi\right)$ with respect to the scalar field value $\phi$, in the case in which the value of the potential at the non-trivial minima is $V_0=1$.}
    \label{fig:G_4}
\end{figure}

\begin{figure}[H]
    \centering
    \includegraphics[width=0.7\linewidth]{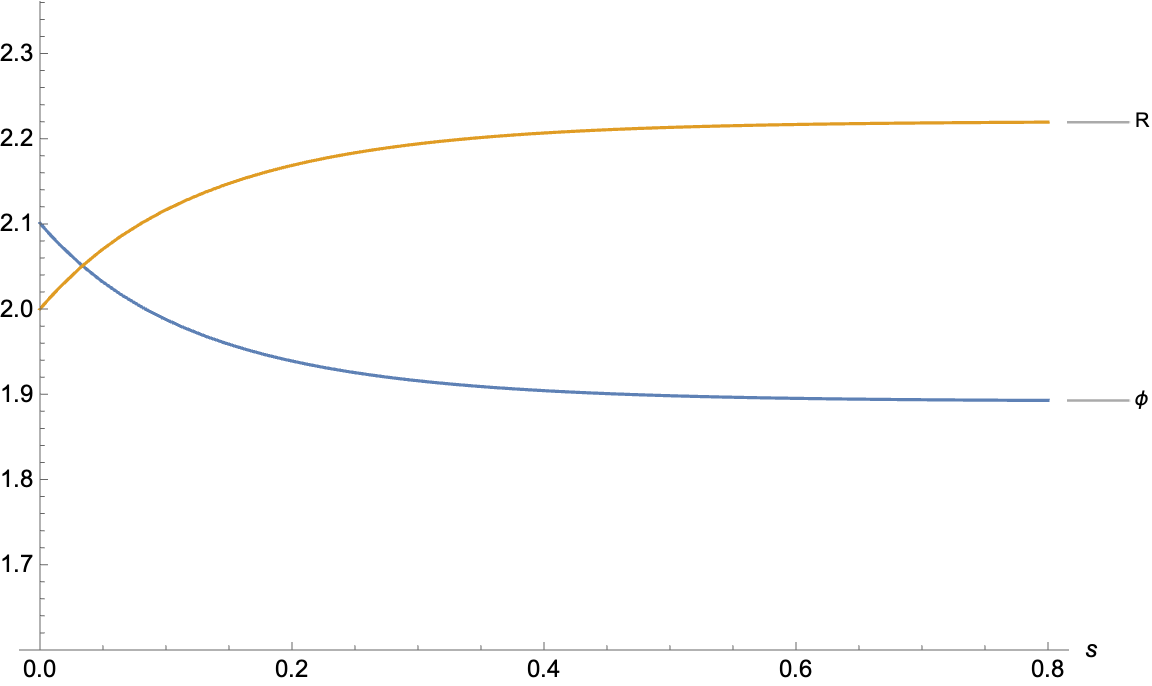}
    \caption{Plot of $\phi$ and $R$ with respect to the flow parameter $s$, when starting from the initial on-shell configuration $\left(\phi_3,R_3\right)$.}
    \label{fig:first}
\end{figure}

In the following discussion, we assume to work with a single scalar field $\phi$, subject to a potential of the form
\be \label{kpot}
V\left(\phi\right)=\frac{V_0}{\alpha_0}\phi^2\left(\phi^4-9\phi^2+21\right)\ ,
\ee
with $\alpha_0=9-4\sqrt{2}$, so that the values of the potential minima of our interest are normalised to $V_0$, which is taken to be a positive parameter. The potential is shown in figure \ref{pottV}.
Before considering a specific solution, we derive the action-induced flow equations \eqref{acteq} associated to a potential of the form described in \eqref{kpot}. In particular, we observe that:
\begin{equation}
    g_{2}=42\frac{V_0}{\alpha_0}\ ,\quad g_{4}=-9\cdot 4!\frac{V_0}{\alpha_0}\ ,\quad g_{6}=6!\frac{V_0}{\alpha_0}\ .
\end{equation}
From the above expressions, we can directly compute the $s_{\ n}^{(D)}$ constants \eqref{consteqai} appearing in the action-induced flow equations. By doing so, we get:
\begin{equation}
    \begin{split}
        s_{\ 0}^{(D)}&=0\ ,\quad s_{\ 1}^{(D)}=0\ ,\quad s_{\ 2}^{(D)}=21\frac{V_0}{\alpha_0}\cdot\frac{2D-6}{D-2}\ ,\\
        s_{\ 3}^{(D)}&=-84\frac{V_0}{\alpha_0}\cdot\frac{D-3}{\left(D-2\right)^2}\ ,\quad  s_{\ 4}^{(D)}=\frac{V_0}{\alpha_0}\left[84\frac{D-3}{\left(D-2\right)^3}-9\left(\frac{2D-6}{D-2}\right)^2\right]\ ,\\
        s_{\ 5}^{(D)}&=8\frac{V_0}{\alpha_0}\left[9\left(D-3\right)-\frac{7}{D-2}\right]\frac{D-3}{\left(D-2\right)^3}\ ,\\
        s_{\ n\geq 6}^{(D)}&=8\frac{V_0}{\alpha_0}\Biggl[\frac{84}{\left(n-2\right)!}\left(\frac{1}{2-D}\right)^{4}-\frac{18}{\left(n-4\right)!}\left(\frac{D-3}{D-2}\right)\left(\frac{1}{2-D}\right)^{2}\\
        &\quad\quad\quad\quad\quad\quad\quad\quad\quad\quad\quad+\frac{1}{\left(n-6\right)!}\left(\frac{D-3}{D-2}\right)^{2}\Biggr]\frac{D-3}{D-2}\left(\frac{2}{2-D}\right)^{n-6}\ .
        \end{split}
\end{equation}
Imposing to work in $D=4$, the above formulas simplify to:
\begin{equation}
    \begin{split}
        s_{\ 0}^{(4)}&=0\ ,\quad s_{\ 1}^{(4)}=0\ ,\quad s_{\ 2}^{(4)}=21\frac{V_0}{\alpha_0}\ ,\quad
        s_{\ 3}^{(4)}=-21\frac{V_0}{\alpha_0}\ ,\quad  s_{\ 4}^{(4)}=\frac{3}{2}\cdot\frac{V_0}{\alpha_0}\ ,\\
        s_{\ 5}^{(4)}&=\frac{11}{2}\cdot\frac{V_0}{\alpha_0}\ ,\quad
        s_{\ n\geq 6}^{(4)}=\left(-1\right)^{n-6}\frac{V_0}{\alpha_0}\Biggl[\frac{21}{\left(n-2\right)!}-\frac{9}{\left(n-4\right)!}+\frac{1}{\left(n-6\right)!}\Biggr]\ .
        \end{split}
\end{equation}
Focusing once more on the potential \eqref{kpot}, it can be observed that it possesses three minima, at the field values
\begin{equation}
    \phi_1\equiv-\sqrt{3+\sqrt{2}}\ ,\quad\phi_2\equiv 0\ ,\quad \phi_{3}\equiv \sqrt{3+\sqrt{2}}\ ,
\end{equation}
corresponding to the potential values:
\begin{equation}
    V_1\equiv V\left(\phi_1\right)=V\left(\phi_3\right)=V_0\ ,\quad V_2\equiv V\left(\phi_2\right)=0\ .
\end{equation}
Therefore, assuming to work with a geometry characterised by a constant scalar curvature, we have three, distinct on-shell configurations that can be used as initial conditions for the flow. We have two solutions with de Sitter background metric and constant scalar, with
\begin{equation}
    \left(\phi_1,R_1\right)=\left(-\sqrt{3+\sqrt{2}},2V_0\right)\ ,\quad \left(\phi_3,R_3\right)=\left(\sqrt{3+\sqrt{2}},2V_0\right)\ ,
\end{equation}
and a Minkowski background solution with everywhere vanishing scalar:
\begin{equation}
    \left(\phi_2,R_2\right)=\left(0,0\right)\ .
\end{equation}
Whatever of those we choose as an initial condition, it can be easily assessed \cite{DeBiasio:2022zuh} that both $R$ and $\phi$ remain constant along the flow. The action-induced flow equations, thus, reduce to the simple form
\begin{equation}\label{acteq88}
\begin{split}
        \frac{\diff R}{\diff s}&=\frac{R}{2}\left[R+G_4\left(\phi\right)\right]\ ,\quad
        \frac{\diff\phi}{\diff s}=-\left[R+G_4\left(\phi\right)\right]\ ,
\end{split}
\end{equation}
where we have introduced the function:
\begin{equation}
    G_4\left(\phi\right)\equiv 2\sum_{n=1}^{\infty}ns_{\ n}^{(4)}\phi^{n-1}\ .
\end{equation}
The form of $G_4\left(\phi\right)$ is represented, for $V_0=1$, in figure \ref{fig:G_4} and can be made more explicit as follows:
\begin{equation}
    G_4\left(\phi\right)=-\frac{2V_0}{\alpha_0} \left(\phi^6-6\phi^5-9\phi^4+36\phi^3+21\phi^2-42\phi\right)e^{-\phi}\ .
\end{equation}
It can be straightforwardly observed, by substituting in the right-hand side of the flow equations the initial values for the scalar field and the curvature, that the $\left(\phi_2,R_2\right)$ solution is a fixed point of the flow. For the other two, of which we will only consider the second one as an example, the flow equations must be solved numerically. From now on, we will set $V_0=1$ for the sake of simplicity. Considering the $\left(\phi_3,R_3\right)$ configuration, in which the scalar field starts from a positive value, we obtain the flow behaviour depicted in figure \eqref{fig:first}. As can be clearly observed, both $\phi$ and $R$ approach some asymptotic values. 

\subsection{Scalar field with quartic potential}\label{quart}
\begin{figure}[H]
    \centering
    \includegraphics[width=0.8\linewidth]{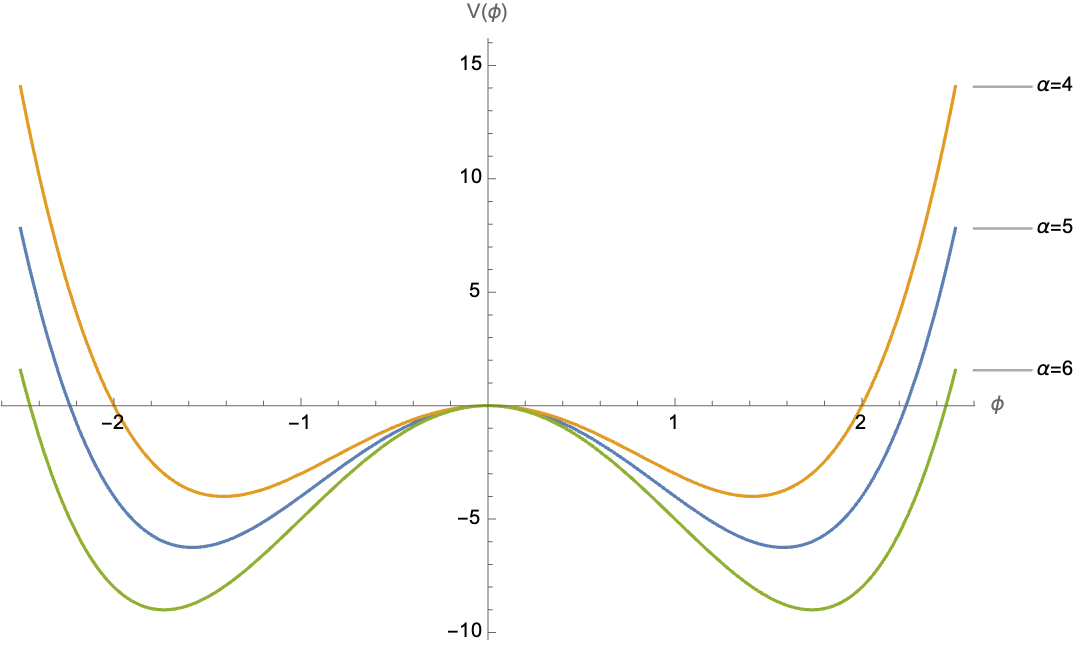}
    \caption{Plot of the behaviour of $V\left(\phi\right)$ around $\phi=0$, for three different values of $\alpha$.}
    \label{pottVg}
\end{figure}

\begin{figure}[H]
    \centering
    \includegraphics[width=0.9\linewidth]{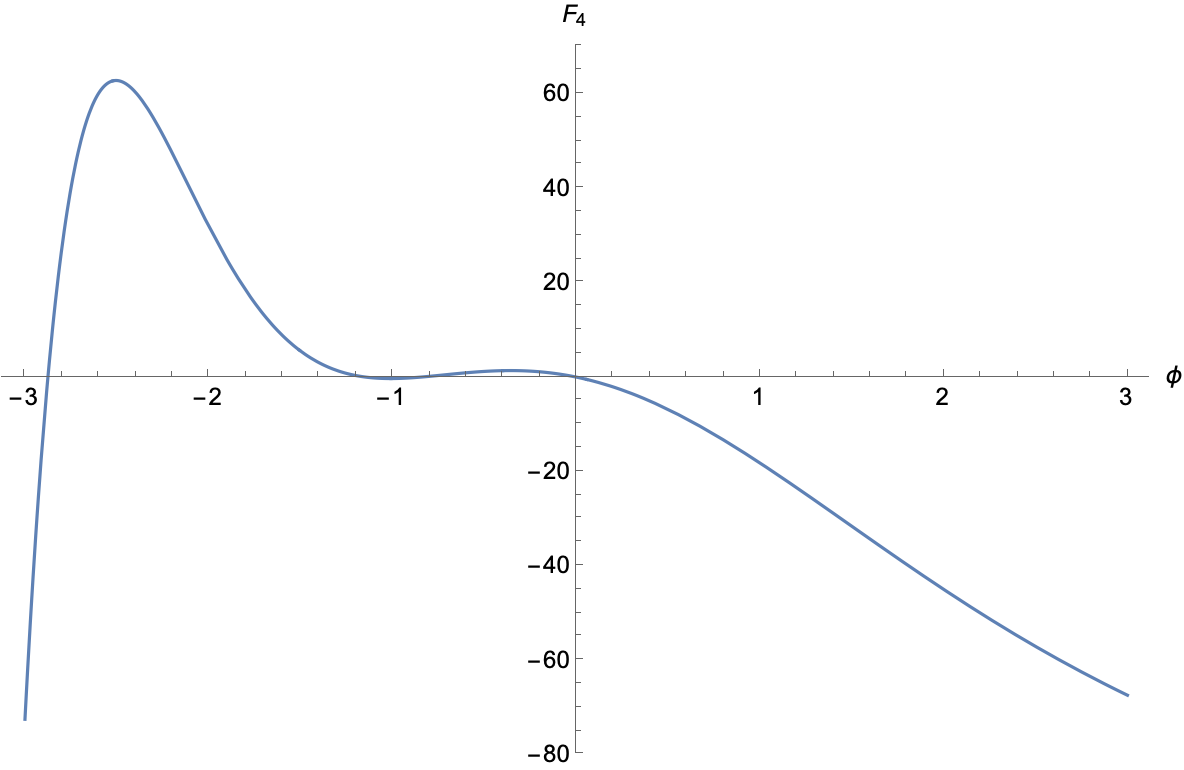}
    \caption{Plot of $F_4\left(\phi\right)$ with respect to the scalar field value $\phi$.}
    \label{fig:F_4}
\end{figure}

\begin{figure}[H]
    \centering
    \includegraphics[width=0.8\linewidth]{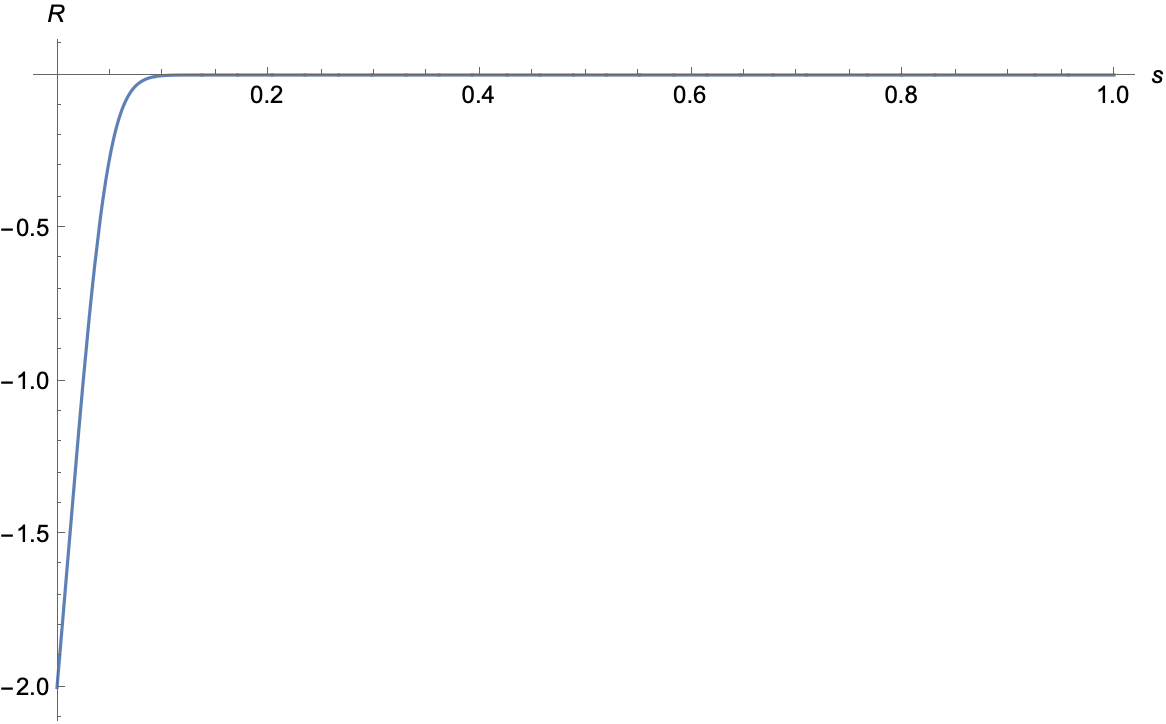}
    \caption{Plot of $R\left(s\right)$ with respect to the flow parameter $s$, which clearly displays how the curvature tends to vanish.}
    \label{fig:Fplot}
\end{figure}

\begin{figure}[H]
    \centering
    \includegraphics[width=0.8\linewidth]{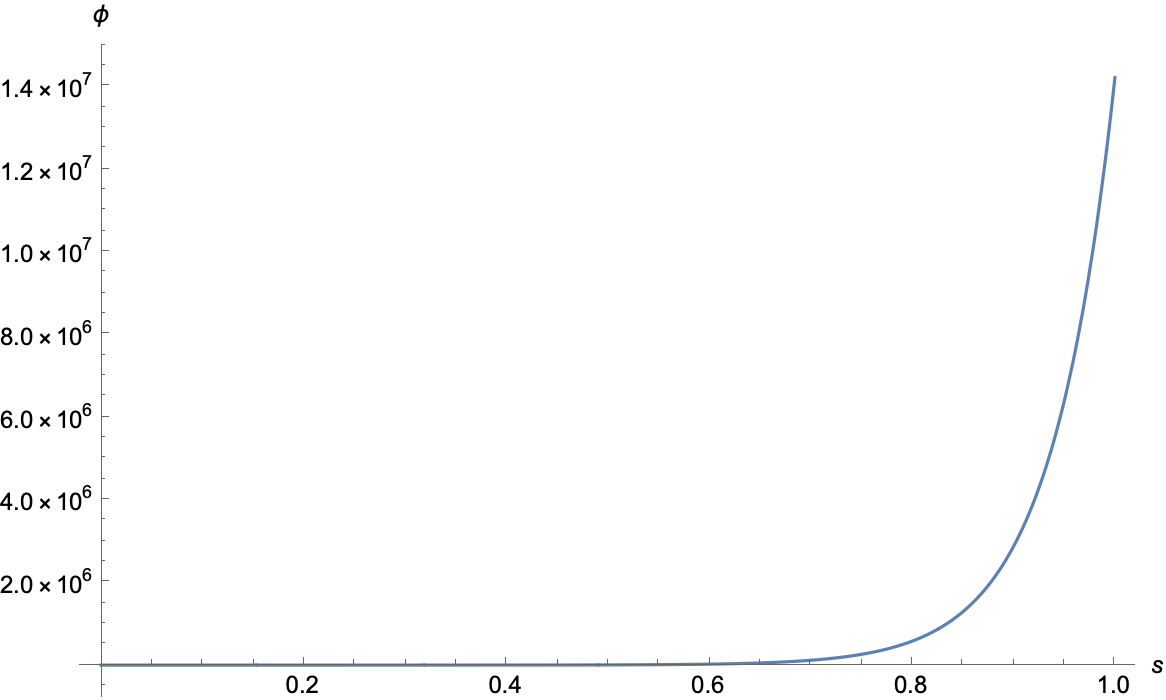}
    \caption{Plot of $\phi\left(s\right)$ with respect to the flow parameter $s$, which clearly displays how the scalar field value tends to blow up.}
    \label{fig:phiplot}
\end{figure}

In the following discussion, we will apply the action-induced flow equations to a particular example.
In order to do so, we will assume the scalar field $\phi$ to be subject to a potential
\be \label{6pot}
V\left(\phi\right)=\phi^4-\alpha\phi^2\ ,
\ee
with $\alpha$ being a positive constant to be fixed by the desired space-time curvature initial value, which will be a function of $\alpha$ in our specific initial point configuration. The potential, for different values of $\alpha$, is shown in figure \ref{pottVg}. At the minima corresponding to the scalar field values
\begin{equation}
    \phi_{\pm}=\pm\sqrt{\frac{\alpha}{2}}\ ,
\end{equation}
the potential takes the $\alpha$-dependent value:
\begin{equation}
    V\left(\phi_\pm\right)=-\frac{\alpha^2}{4}\ .
\end{equation}
Assuming the scalar field to either take the constant value $\phi_+$ or the constant value $\phi_-$ at the beginning of the flow, the potential produces an effective cosmological constant term
\be
\Lambda_{\text{eff}}=-\frac{\alpha^2}{8}\ .
\ee
Therefore, the field equations allow our initial metric to be that of $D$-dimensional Anti-de Sitter spacetime with scalar curvature:
\be 
R_0=-\frac{\alpha^2D}{4\left(D-2\right)}\ .
\ee
Without loss of generality, we assume to work with the initial flow condition:
\begin{equation}
    R_0=-\frac{\alpha^2D}{4\left(D-2\right)}\ ,\quad \phi_+=\sqrt{\frac{\alpha}{2}}\ .
\end{equation}
Being both the space-time curvature and the scalar field constancy preserved along the flow \cite{DeBiasio:2022zuh}, the action-induced flow equations \eqref{acteq} greatly simplify and assume the following form
\begin{equation}\label{acteqquartic}
\begin{split}
        \frac{\diff g_{\mu\nu}}{\diff s}&=-\frac{2}{D}Rg_{\mu\nu}-\sum_{n=1}^{\infty}ns_{\ n}^{(D)}\phi^{n-1}g_{\mu\nu}+\mathcal{L}_\xi g_{\mu\nu}\ ,\\
        \frac{\diff\phi}{\diff s}&=-R-\frac{D}{2}\sum_{n=1}^{\infty}ns_{\ n}^{(D)}\phi^{n-1}+\mathcal{L}_\xi\phi\ ,
\end{split}
\end{equation}
where the constants $s_{\ n}^{(D)}$ are defined by the usual equation:
\begin{equation}
    s_{\ n}^{(D)}\equiv\sum_{k=0}^{n}\frac{g_{k}}{\left(n-k\right)!k!}\left(\frac{2D-6}{D-2}\right)^{\frac{k}{2}}\left(\frac{2}{2-D}\right)^{n-k}\ .
\end{equation}
Rephrasing the metric flow equation in \eqref{acteqquartic} in terms of the scalar curvature and neglecting the diffeomorphism term, we are left with the simple system:
\begin{equation}\label{acteqquartic2}
\begin{split}
        \frac{\diff R}{\diff s}&=\frac{2R}{D}\left[R+\frac{D}{2}\sum_{n=1}^{\infty}ns_{\ n}^{(D)}\phi^{n-1}\right]\ ,\quad\frac{\diff\phi}{\diff s}=-\left[R+\frac{D}{2}\sum_{n=1}^{\infty}ns_{\ n}^{(D)}\phi^{n-1}\right]\ .
\end{split}
\end{equation}
By defining the function
\begin{equation}
    F_D\left(\phi\right)\equiv \frac{D}{2}\sum_{n=1}^{\infty}ns_{\ n}^{(D)}\phi^{n-1}\ ,
\end{equation}
the above formulas are reduced to:
\begin{equation}
    \frac{\diff R}{\diff s}=\frac{2}{D}R\cdot \bigl[R+F_D\left(\phi\right)\bigr]\ ,\quad \frac{\diff\phi}{\diff s}=-\bigl[R+F_D\left(\phi\right)\bigr]\ .
\end{equation}
In order to solve the flow equations explicitly, on top of specifying the number of space-time dimensions, we have to compute the values of the non vanishing $s_{\ n}^{(D)}$ constants and plug them into the definition of $F_D\left(\phi\right)$. We start by reading off the non-zero couplings
\begin{equation}
    g_4=4!\ ,\quad g_2=-2\alpha\ .
\end{equation}
from the potential formula \eqref{6pot}. Therefore, we have:
\begin{equation}
    \begin{split}
        s_{\ 1}^{(D)}&=0\ ,\quad s_{\ 2}^{(D)}=-\alpha\left(\frac{2D-6}{D-2}\right)\ ,\\
        s_{\ 3}^{(D)}&=-\alpha\left(\frac{2D-6}{D-2}\right)\left(\frac{2}{2-D}\right)\ ,\\
        s_{\ n\geq 4}^{(D)}&=-\frac{\alpha}{\left(n-2\right)!}\left(\frac{2D-6}{D-2}\right)\left(\frac{2}{2-D}\right)^{n-2}\\
        &\quad+\frac{1}{\left(n-4\right)!}\left(\frac{2D-6}{D-2}\right)^{2}\left(\frac{2}{2-D}\right)^{n-4}\ .
    \end{split}
\end{equation}
For the remainder of our discussion, we will assume to work with $D=4$. Hence, the above expressions reduce to:
\begin{equation}
    \begin{split}
       s_{\ 1}^{(4)}&=0\ ,\quad s_{\ 2}^{(4)}=-\alpha\ ,\quad s_{\ 3}^{(4)}=\alpha\ ,\\
        s_{\ n\geq 4}^{(4)}&=\frac{\left(-1\right)^{n}}{\left(n-2\right)!}\bigl[\left(n-2\right)\left(n-3\right)-\alpha\bigr]\ .
    \end{split}
\end{equation}
By further taking $\alpha=2$, the initial condition becomes
\begin{equation}\label{initial4D}
    R_0=-2\ ,\quad \phi=1\ ,
\end{equation}
while the constants get to be nothing more than:
\begin{equation}
    \begin{split}
        s_{\ 1}^{(4)}&=0\ ,\quad s_{\ 2}^{(4)}=-2\ ,\quad s_{\ 3}^{(4)}=2\ ,\\
        s_{\ n\geq 4}^{(4)}&=\left(-1\right)^{n}\frac{\left(n-1\right)\left(n-4\right)}{\left(n-2\right)!}\ .
    \end{split}
\end{equation}
Therefore, we also obtain the expression
\begin{equation}
\begin{split}
    F_4\left(\phi\right)&= -8\phi+12\phi^{2}+2\sum_{n\geq 4}^{\infty}\left(-1\right)^{n}\frac{n\left(n-1\right)\left(n-4\right)}{\left(n-2\right)!}\phi^{n-1}\\
    &=-16\phi-2\phi\left(\phi^3+4\phi^2+2\phi-4\right)e^{-\phi}\ ,
\end{split}
\end{equation}
which is plotted in figure \ref{fig:F_4}.
The flow equations for $R$ and $\phi$, since we are working in $D=4$, take the form:
\begin{equation}\label{acteqquartic4D}
\begin{split}
        \frac{\diff R}{\diff s}&=\frac{R}{2}\left[R+F_4\left(\phi\right)\right]\ ,\quad\frac{\diff\phi}{\diff s}=-\left[R+F_4\left(\phi\right)\right]\ .
\end{split}
\end{equation}
Starting from the initial condition \eqref{initial4D}, the evolution can therefore be solved numerically and plotted, as was done in figures \ref{fig:Fplot} and \ref{fig:phiplot}. There, it can be clearly observed that the flow forces the system towards a region of the moduli space in which $R$, albeit negative, gets really small, while $\phi$ evolves to large positive values. In such a context, we have 
\begin{equation}
    F_4\left(\phi\right)\sim -16\phi\ ,
\end{equation}
with the flow equations being nicely approximated by:
\begin{equation}
    \frac{\diff R}{\diff s}\sim-8R\cdot\phi\ ,\quad\frac{\diff\phi}{\diff s}\sim 16\phi\ .
\end{equation}
Therefore, for large enough values of the flow parameter $s$, we can approximate the evolution of the curvature and the scalar field space-time constant value by
\begin{equation}\label{approxosflow}
    R\left(s\right)\sim -C_0\cdot \exp{-\frac{C_1}{2}e^{16s}}\ ,\quad \phi\left(s\right)\sim C_1\cdot e^{16s}\ ,
\end{equation}
where $C_0$ and $C_1$ are positive constants.
From the expressions \eqref{approxosflow}, which properly provides us with the flow behaviour of $R$ and $\phi$ for large values of the flow parameter, we get:
\begin{equation}
    \lim_{s\to\infty}R\left(s\right)=0\ ,\quad\lim_{s\to\infty}\phi\left(s\right)=\infty\ .
\end{equation}
Therefore, the action-induced flow equations push our initial condition towards a Minkowski space-time limit, in which the scalar field value blows up. Distances along the flow can, as usual, be computed by exploiting the formula
\begin{equation}\label{distonf}
    \Delta\left(s,s_0\right)\sim\log{\frac{\mathcal{F}\left(s_0\right)}{\mathcal{F}\left(s\right)}}\ ,
\end{equation}
in which $\mathcal{F}\left(s\right)$ is nothing more than the entropy functional from which our flow equations descend, here represented by the string frame action associated to \eqref{6pot}, computed at the flow parameter values $s$. More specifically, we have:
\begin{equation}
\begin{split}
    \mathcal{F}\left(s\right)&=\frac{1}{2\kappa^2_D}\int_{\mathcal{M}}\diff^{D}x\sqrt{-g\left(s\right)}e^{-\phi\left(s\right)}\Biggl\{R\left(s\right)-2\bigl[\nabla \phi\left(s\right)\bigr]^2\\
    &\quad\quad\quad\quad\quad\quad\quad\quad\quad\quad\quad\quad\quad\quad\quad+3\nabla^2\phi\left(s\right)-\sum_{n=0}^{\infty}s_{\ n}^{(4)}\phi^{n}\left(s\right)\Biggr\}\ .
    \end{split}
\end{equation}
Plugging the explicit expression for the constants $s_{\ n}^{(4)}$ into the above formula, exploiting the fact that both the scalar field and space-time curvature are constant and noticing that, in standard coordinates, the determinant $g$ of Anti-de Sitter metric is does not depend on the cosmological constant and is hence fixed in $s$, we obtain:
\begin{equation}
\begin{split}
    \mathcal{F}\left(s\right)&=\frac{1}{2\kappa^2_D}V_{\mathcal{M}}\cdot\mathcal{K}\left(s\right)\ .
    \end{split}
\end{equation}
In the above expression, we have introduced
\begin{equation}
    \mathcal{K}\left(s\right)\equiv e^{-\phi\left(s\right)}\Biggl[R\left(s\right)+2\phi^{2}\left(s\right)-2\phi^{3}\left(s\right)-\sum_{n=4}^{\infty}\left(-1\right)^{n}\frac{\left(n-1\right)\left(n-4\right)}{\left(n-2\right)!}\phi^{n}\left(s\right)\Biggr]
\end{equation}
and defined the volume via the usual formula:
\begin{equation}
    V_{\mathcal{M}}\equiv\int_{\mathcal{M}}\diff^{D}x\sqrt{-g}\ .
\end{equation}
In the case of Anti-de Sitter space-time, such a quantity is not finite. This is not problematic for the case at hand, since in appears in both $\mathcal{F}\left(s_0\right)$ and $\mathcal{F}\left(s\right)$, factoring out from the distance formula \eqref{distonf}. Strictly speaking, this can be rendered precise by introducing a volume regulator and removing it after having computed the distance. By doing so, we get
\begin{equation}\label{distonf2}
    \Delta\left(s,s_0\right)\sim\log{\frac{\mathcal{K}\left(s_0\right)}{\mathcal{K}\left(s\right)}}\ ,
\end{equation}
in which the flow-dependent quantity $\mathcal{K}\left(s\right)$ can be shown to be equal to:
\begin{equation}
    \mathcal{K}\left(s\right)=e^{-\phi\left(s\right)}\bigl\{R\left(s\right)+e^{-\phi\left(s\right)}\left[2-\phi^2\left(s\right)\right]\phi^2\left(s\right) \bigr\}\ .
\end{equation}
For large values of the flow parameter $s$, the above expression can be approximated by:
\begin{equation}
\begin{split}
        \mathcal{K}\left(s\right)&\sim -\exp{-\frac{3C_1}{2}e^{16s}}\cdot\left\{C_0 +\exp{-\frac{C_1}{2}e^{16s}}\left[C_1^2\cdot e^{32s}-2\right]C_1^2\cdot e^{32s} \right\}\ .
\end{split}
\end{equation}
Further removing sub-leading terms, we obtain:
\begin{equation}
\begin{split}
        \mathcal{K}\left(s\right)&\sim -C_0\exp{-\frac{3C_1}{2}e^{16s}} \ .
\end{split}
\end{equation}
Therefore, the corresponding behaviour of the distance is given by:
\begin{equation}\label{distonf2}
    \Delta\left(s,s_0\right)\sim \frac{3C_1}{2}\left(e^{16s}-e^{16s_0}\right)\ .
\end{equation}
It can be clearly seen that the limit in which $R\to 0$ and $\phi\to\infty$ lies at infinite distance:
\begin{equation}
    \lim_{s\to\infty}\Delta\left(s,s_0\right)=\infty\ .
\end{equation}
Hence, the swampland distance conjecture would suggest it to be accompanied by an infinite tower of asymptotically massless supplementary states.

%% file: Sections/On-Shellflow/onshell.tex
Starting from a $D$-dimensional theory of the form presented in \ref{gfa}, with a dynamical metric tensor and a self-interacting scalar field, a set of action-induced geometric flow equations \eqref{acteq} can be straightforwardly derived. If the flow initial condition, comprised of a geometry and a space-time configuration of the scalar field, is taken to satisfy the equations of motion \eqref{aact:eom} associated to \eqref{aact}, there is nonetheless no reason to believe that such conditions would be preserved along the flow trajectory in $\mathbf{\Gamma}_{\mathcal{M}}$. We now want to develop a set of theoretical tools allowing us to reconcile the geometric flow equations \eqref{acteq} with the equations of motion \eqref{aact:eom}. This would allow to probe the actual generalised moduli space of \textit{physical} solutions, defined as the subset $\mathbf{D}_{\mathcal{M}}\subset\mathbf{\Gamma}_{\mathcal{M}}$ of on-shell space-time configurations of the fields. In order to do so, we will draw inspiration from the Swampland distance conjecture, introduced and discussed in \ref{sec:distanceconj} and further refined in \ref{sec:rfconj} and \ref{sec:perconj}, and from its towers of supplementary light states, which are expected to appear along the flow. At this point, it must be stressed that the idea of providing flow equations that both keep the scalar and the metric on shell might be too ambitious. The most natural choice will thus be to relax the on-shell conditions to the metric tensor alone. Temporarily focusing on a simpler setting, we will therefore consider a generic flow equation for the space-time geometry and compute an induced evolution for the energy-momentum tensor, fixed by requiring the metric equations of motion not to be violated. In this first approach the energy-momentum tensor will be considered as a generic tensor on its own, without explicitly constructing it in terms of the scalar field. We will thereafter make contact with the theory illustrated in \ref{gfa}, reintroducing the scalar field explicitly and considering the action-induced flow equations. The metric field equations will be preserved by allowing new energy-momentum contributions to appear along the flow, resembling the towers postulated by the distance conjecture.
\subsection{Generic flow equations}
Let's consider a $D$-dimensional theory with a dynamical spacetime metric, a cosmological constant and a set of matter fields, satisfying appropriate equations of motion. Moreover, let's introduce a real flow parameter $s\in\left[0,s_{1}\right)$, where $s_{1}$ is positive and can either be finite or infinite, depending on whether the flow ends up hitting a singularity. At this point, we introduce a one-parameter family of metric tensors $g\left(s\right)$, a one-parameter family of cosmological constants $\Lambda\left(s\right)$ and a one-parameter family of energy-momentum tensors $T_{\mu\nu}\left(s\right)$, so that the initial conditions $g_{\mu\nu}\left(0\right)$, $\Lambda\left(0\right)$ and $T_{\mu\nu}\left(0\right)$ satisfy the equations of motion and the $s$-evolutions of the families are induced by the flow equations:
\be \label{floweq}
        \frac{\diff g_{\mu\nu}}{\diff s}=A_{\mu\nu}\ ,\quad \frac{\diff\Lambda}{\diff s}=C\ ,\quad \frac{\diff T_{\mu\nu}}{\diff s}=\frac{1}{\kappa_D^2}B_{\mu\nu}\ .
\ee 
In the last formula of \eqref{floweq}, the $\kappa_D^{-2}$ factor is introduced for the sake of simplicity. Moreover, the flow-sources $A_{\mu\nu}$ and $B_{\mu\nu}$ are taken to be symmetric, $s$-dependent tensors. Having defined such general flow equations, we can now move to constraining $B_{\mu\nu}$ in terms of $A_{\mu\nu}$ and $C$ in a way that preserves the equations of motion
\begin{equation}\label{aact:eom2}
    R_{\mu\nu}-\frac{1}{2}Rg_{\mu\nu}+\Lambda g_{\mu\nu}= \kappa^2_D T_{\mu\nu}
\end{equation}
for $g_{\mu\nu}$ at any value of the flow parameter $s$.
\subsection{On-shell geometric flow}
In order for the metric equations of motion to be conserved along the flow, together with taking an on-shell initial condition, we must impose
\be\label{condition}
\frac{\diff}{\diff s}\left(G_{\mu\nu}+g_{\mu\nu}\Lambda\right)=\kappa_D^2\frac{\diff T_{\mu\nu}}{\diff s}
\ee
for any value of the flow parameter $ s$. To make the discussion more concrete, we will now compute the $ s$-evolution of the Einstein tensor $G_{\mu\nu}$ induced by \eqref{floweq}. First of all, we observe that we must have
\be 
\frac{ \diff g^{\mu\nu}}{ \diff s}=-A^{\mu\nu}\ ,
\ee
so that the condition $g^{\mu\nu}g_{\nu\alpha}=\delta^\mu{}_\alpha$ is preserved along the flow. Furthermore, one can straightforwardly derive  the flow equation for the Ricci tensor to be
\be\label{rtflow}
\begin{split}
  2\frac{ \diff R_{\mu\nu}}{ \diff s}&=\nabla^\sigma\nabla_\nu A_{\mu\sigma}+\nabla_\mu\nabla^\sigma A_{\sigma\nu} -\nabla_\mu\nabla_\nu A\\
  &\quad-\nabla^2  A_{\mu\nu }+R_\mu{}^\sigma A_{\sigma\nu }-R_\mu{}^\sigma{}_\nu {}^\theta A_{\sigma\theta}\ ,  
\end{split}
\ee
where we have defined $A\equiv g^{\mu\nu}A_{\mu\nu}$. From \eqref{rtflow}, one can obtain the following:
\be\label{rsflow}
\frac{ \diff R}{ \diff  s}=\nabla^\sigma\nabla^\theta A_{\sigma\theta}-\nabla^2  A-A^{\sigma\theta}R_{\sigma\theta}\ .
\ee
By combining \eqref{rtflow}, \eqref{rsflow} and the flow equation for the metric, we can derive the flow equation for the Einstein tensor:
\be\label{etflow}
\begin{split}
    2\frac{ \diff G_{\mu\nu}}{ \diff s}&=\nabla^\sigma\nabla_\nu A_{\mu\sigma}+\nabla_\mu\nabla^\sigma A_{\sigma\nu} -\nabla_\mu\nabla_\nu A-\nabla^2  A_{\mu\nu }\\
  &\quad 
  -g_{\mu\nu}\nabla^\sigma\nabla^\theta A_{\sigma\theta}+g_{\mu\nu}\nabla^2  A+g_{\mu\nu}A^{\sigma\theta}R_{\sigma\theta}\\
    &\quad+R_\mu{}^\sigma A_{\sigma\nu }-R_\mu{}^\sigma{}_\nu {}^\theta A_{\sigma\theta}-A_{\mu\nu}R\ .
\end{split}
\ee
With the expression derived in \eqref{etflow}, together with the flow equations for the cosmological constant and the energy-momentum tensor, we finally obtain the on-shell condition:
\be\label{onshellB}
\begin{split}
   2B_{\mu\nu}&=\nabla^\sigma\nabla_\nu A_{\mu\sigma}+\nabla_\mu\nabla^\sigma A_{\sigma\nu} -\nabla_\mu\nabla_\nu A-R_\mu{}^\sigma{}_\nu {}^\theta A_{\sigma\theta}\\
       &\quad+2C g_{\mu\nu}-\nabla^2  A_{\mu\nu }+R_\mu{}^\sigma A_{\sigma\nu }-A_{\mu\nu}R+2\Lambda A_{\mu\nu}\\
  &\quad 
  -g_{\mu\nu}\nabla^\sigma\nabla^\theta A_{\sigma\theta}+g_{\mu\nu}\nabla^2  A+g_{\mu\nu}A^{\sigma\theta}R_{\sigma\theta}\ .
\end{split}
\ee
If $A_{\mu\nu}$ and $C$ are chosen freely, the overall flow can be guaranteed to conserve the equations of motion \eqref{aact:eom2} for the metric tensor by imposing $B_{\mu\nu}$ to take the form illustrated in \eqref{onshellB}. Since we are generically treating the energy-momentum tensor as an appropriate source for space-time curvature, this has nothing to do with the matter content of a specific theory. If we study, for instance, the case in which the flow source $A_{\mu\nu}$ for the metric is
\be\label{rbflow}
A_{\mu\nu}=-2R_{\mu\nu}+2\omega Rg_{\mu\nu}\ ,
\ee
with $\omega$ being defined as half the constant $\vartheta$ appearing in \eqref{RBF} for the sake of the following derivations, we can get the formula $A=2\left(D\omega-1\right)R$ and the relation:
\be\label{rbosB}
\begin{split}
   B_{\mu\nu}&=-\nabla^\sigma\nabla_\nu R_{\mu\sigma}+\left[\left(2-D\right)\omega+\frac{1}{2}\right]\nabla_\mu\nabla_\nu R+R_\mu{}^\sigma{}_\nu {}^\theta R_{\sigma\theta}+\nabla^2  R_{\mu\nu }-R_\mu{}^\sigma R_{\sigma\nu }\\
       &\quad+\left(R-2\Lambda\right) R_{\mu\nu}-\left\{C+\left[\left(2-D\right)\omega +\frac{1}{2}\right]\nabla^2  R+2\omega\Lambda R-R^{\sigma\theta}R_{\sigma\theta}\right\}g_{\mu\nu}\ .
\end{split}
\ee
This way, we have explicitly derived a relation between $B_{\mu\nu}$ and $C$, in terms of $\omega$, for the case in which the metric follows Ricci-Bourguignon flow. 
\subsubsection{Einstein manifolds}
In the following discussion, we consider the particular case in which the geometry corresponds to that of a $D$-dimensional Einstein manifold. Indeed, we assume the initial conditions for the energy-momentum tensor and the cosmological constant to be
\begin{equation}
    T_{\mu\nu}\left(0\right)\equiv K_{0}\cdot g_{\mu\nu}\left(0\right)\ ,\quad \Lambda\left(0\right)\equiv\Lambda_0\ ,
\end{equation}
where $K_0$ is nothing more than a constant. Hence, the field equations \eqref{aact:eom2} allow the initial condition to satisfy:
\be\label{curv}
R_{\mu\nu}\left(0\right)=\frac{R_0}{D}g_{\mu\nu}\left(0\right)\ ,\quad R_0=\left(K_0-\Lambda_0\right)\frac{2D}{2-D}\ .
\ee
It can be easily shown \cite{DeBiasio:2020xkv} that a flow of the form presented in \eqref{rbflow} preserves the Einstein manifold condition. In particular, the flow equation for $R$ is: 
\be
\frac{ \diff R}{ \diff  s}=\frac{2\left(1-\omega\right)}{D}R^2\ .
\ee
The above can be simply solved as:
\be
R\left( s\right)=\frac{DR_0}{D-2\left(1-\omega\right)R_0 s}\ .
\ee
Furthermore, given the flow equation \eqref{floweq} for $\Lambda$, we can explicitly write its behaviour:
\be 
\Lambda\left( s\right)=\Lambda_0+\int^{ s}_0 C\left(\tau\right)\diff\tau\ .
\ee
Concerning the right-hand side of the energy-momentum tensor flow equation, we have:
\be
\begin{split}
   2B_{\mu\nu}&=\left[\left(2-D\right)\omega+\frac{D-2}{2D}\right]\nabla_\mu\nabla_\nu R-g_{\mu\nu}\left[\left(2-D\right)\omega +\frac{D-2}{2D}\right]\nabla^2  R\\
  &\quad+2\Lambda \frac{D\omega-1}{D}Rg_{\mu\nu}+C g_{\mu\nu} \ .
\end{split}
\ee
The assumption of constant curvature directly implies:
\be
\begin{split}
   B_{\mu\nu}&=\left(2\Lambda \frac{D\omega-1}{D}R+C \right)g_{\mu\nu} \ .
\end{split}
\ee
Anyway, we can once more read-off the flow behaviour of the energy momentum tensor directly from \eqref{curv}, which gives us:
\be 
K\left( s\right)=\frac{R_0\left(2-D\right)}{2D-4\left(1-\omega\right)R_0 s}+\Lambda_0+\int^{ s}_0 C\left(\tau\right)\diff\tau\ .
\ee 
Therefore, starting with an on-shell Einstein manifold, imposing the metric to evolve according to \eqref{rbflow} and freely choosing the flow behaviour of the cosmological constant, we are left with the explicit flow behaviour for the constant $K$ appearing in the energy-momentum tensor. 
\section{On-shell conditions and the action-induced flow}\label{act-inon}
After having treated the energy-momentum tensor as a generic symmetric tensorial source for Einstein field equations, will now move back to the theory defined in \ref{gfa} and characterise $T_{\mu\nu}$ in terms of the scalar field $\phi$ appearing in \eqref{aact}. Practically, we choose to work with an energy-momentum tensor defined by the formula \eqref{eq:emt}, to incorporate the cosmological constant term in the sum appearing in the action and not to make it flow, so that the only evolving objects are dynamical fields. Furthermore, we impose the geometry and scalar field to evolve according to the action-induced flow equations \eqref{acteq}, from which, neglecting the diffeomorphism term, we obtain:
\begin{equation}\label{acteqrep}
\begin{split}
        A_{\mu\nu}&=-2R_{\mu\nu}+2\frac{D-4}{D-2}\nabla_\mu\phi\nabla_\nu\phi-\sum_{n=1}^{\infty}ns_{\ n}^{(D)}\phi^{n-1}g_{\mu\nu}\\
        &\quad-2\frac{D-4}{D-2}\bigl[\left(\nabla\phi\right)^2-\nabla^2\phi\bigr]g_{\mu\nu}\ .
\end{split}
\end{equation}
Therefore, the energy-momentum flow source required to preserve the on-shell condition on the geometry along the generalised moduli space path can be directly written as:
\be
\begin{split}
   2B_{\mu\nu}&=\nabla^\sigma\nabla_\nu A_{\mu\sigma}+\nabla_\mu\nabla^\sigma A_{\sigma\nu} -\nabla_\mu\nabla_\nu A-R_\mu{}^\sigma{}_\nu {}^\theta A_{\sigma\theta}-g_{\mu\nu}\nabla^\sigma\nabla^\theta A_{\sigma\theta}\\
       &\quad-\nabla^2 A_{\mu\nu }+R_\mu{}^\sigma A_{\sigma\nu }-A_{\mu\nu}R+g_{\mu\nu}\nabla^2  A+g_{\mu\nu}A^{\sigma\theta}R_{\sigma\theta}\ .
\end{split}
\ee
Nevertheless, by combining the action-induced flow equation
\begin{equation}
\begin{split}
        \frac{\diff\phi}{\diff s}&=-R-\frac{D^2-6D+6}{D-2}\left(\nabla\phi\right)^2-\frac{D}{2}\sum_{n=1}^{\infty}ns_{\ n}^{(D)}\phi^{n-1}+\frac{D^2-5D+2}{D-2}\nabla^2\phi
\end{split}
\end{equation}
for the scalar field and the expression \eqref{eq:emt} for the energy-momentum tensor in terms of $\phi$, we would find an action-induced energy-momentum tensor flow equation 
\begin{equation}\label{emtactioninduced}
    \frac{\diff T_{\mu\nu}}{\diff s}=E_{\mu\nu}
\end{equation}
for the components of $T_{\mu\nu}$, where we have introduced the tensor $E_{\mu\nu}$ by combining the action-induced flow equations for the geometry and the scalar as:
\begin{equation}
    \begin{split}
         E_{\mu\nu}&\equiv\frac{1}{\kappa^2_D}\left(\frac{\delta T_{\mu\nu}}{\delta g_{\mu\nu}}\frac{\diff g_{\mu\nu}}{\diff s}+\frac{\delta T_{\mu\nu}}{\delta \phi}\frac{\diff \phi}{\diff s}\right)\ .
    \end{split}
\end{equation}
Here, we find ourselves addressing a crucial issue. In fact, one has no reason to believe the above two flow-sources for the energy-momentum tensor match. In general, we have:
\be 
S_{\mu\nu}\equiv E_{\mu\nu}-B_{\mu\nu}\neq 0\ .
\ee 
Therefore, since we both want the metric tensor to be on-shell and the flow to be realised as the action-induced geometric flow associated to \eqref{aact}, we assume the difference between $B_{\mu\nu}$ and $E_{\mu\nu}$ to be accounted for by the appearance of further matter fields, which might be even read off directly from the explicit expression of their energy-momentum contribution. This intuition is obviously motivated by the swampland distance conjecture, that claims we should expect towers of light states to emerge for large fields displacements in the moduli space. In order to follow the outlined procedure, one must introduce the symmetric tensor $S_{\mu\nu}$ and derive its explicit form. Once $S_{\mu\nu}$ is derived, one can write the energy momentum tensor as
\be 
T_{\mu\nu}\left(s\right)\equiv \Bar{T}_{\mu\nu}\left(s\right)+\Hat{T}_{\mu\nu}\left(s\right)\ ,
\ee 
with $\Hat{T}_{\mu\nu}\left(0\right)=0$ and:
\be 
\frac{\diff \Bar{T}_{\mu\nu}}{\diff s}=E_{\mu\nu}\ ,\quad\frac{\diff \Hat{T}_{\mu\nu}}{\diff s}=S_{\mu\nu}\ .
\ee
This way, we have an energy-momentum tensor component $\Bar{T}_{\mu\nu}$ produced by the scalar $\phi$ and evolving according to the appropriate action-induced geometric flow equations, together with an extra $\Hat{T}_{\mu\nu}$ term appearing along the flow accounting for the emergence of new matter fields, which can flow themselves, so that the metric remains on-shell. 

%% file: Sections/On-Shellflow/scalars.tex
Assuming to work with the most general form of the action \eqref{aact} in $D$ dimensions, with arbitrarily high powers appearing in the self-interaction potential for the scalar field, would produce extremely complicated flow sources $E_{\mu\nu}$ and $S_{\mu\nu}$ for the energy-momentum component associated to $\phi$ and the one introduced along the flow, respectively. Therefore, we will now focus on a particular example, characterised by specific set of polynomial self-interactions. We will construct an on-shell configuration, evolve it according to the relevant action-induced flow equations and derive the flow dependence of the additional energy-momentum term, required for the metric on-shell conditions not to be broken.
\subsection{Parabolic potential}\label{parabolicpot}
In the following discussion, we assume to work with a single scalar field $\phi$, subject to a potential of the form
\be \label{2pot}
V\left(\phi\right)=\frac{g_2}{2}\phi^2-g_1\phi-g_0\ ,
\ee
where $g_2$, $g_1$ and $g_0$ are three parameters, with $g_2$ strictly positive. As can be clearly inferred, such a potential possess one global minimum at:
\begin{equation}\label{minimqua}
    \phi_0=\frac{g_1}{g_2}\Longrightarrow  V\left(\phi_0\right)=-\left(g_0+\frac{g^2_1}{2g_2}\right)\ .
\end{equation}
Before focusing on a specific solution, which will be characterised by space-time constant values of scalar field and the curvature, we derive the action-induced flow equations \eqref{acteq} associated to the potential described in \eqref{2pot}. From its expression, we can directly compute the $s_{\ n}^{(D)}$ constants \eqref{consteqai} appearing in the flow equations. By doing so, we get:
\begin{equation}
\begin{split}
       s_{\ 0}^{(D)}&=g_{0}\ ,\quad s_{\ 1}^{(D)}=\frac{2g_{0}}{2-D}+g_{1}\sqrt{\frac{2D-6}{D-2}}\ ,\\
       s_{\ n\geq 2}^{(D)}&=\frac{g_{0}}{n!}\left(\frac{2}{2-D}\right)^{n}+\frac{g_{1}}{\left(n-1\right)!}\sqrt{\frac{2D-6}{D-2}}\left(\frac{2}{2-D}\right)^{n-1}\\
       &\quad+\frac{g_{2}}{\left(n-2\right)!2!}\frac{2D-6}{D-2}\left(\frac{2}{2-D}\right)^{n-2}\ .
\end{split}
\end{equation}
To simplify the following computations, we start by defining the function
\begin{equation}
    F_D\left(\phi\right)\equiv \sum_{n=0}^{\infty}s_{\ n}^{(D)}\phi^n
\end{equation}
and express it, in an explicit way, as follows:
\begin{equation}
    \begin{split}
        F_D\left(\phi\right)&=\exp{\frac{2\phi}{2-D}}\cdot\left(g_{0}+g_{1}\phi\sqrt{\frac{2D-6}{D-2}}+g_{2}\phi^2\frac{D-3}{D-2}\right)\ .
    \end{split}
\end{equation}
Furthermore, we introduce another function
\begin{equation}\label{Gfunctionquad}
    G_D\left(\phi\right)\equiv \frac{D}{2}\frac{\diff}{\diff\phi}F_D\left(\phi\right)
\end{equation}
and compute it, from the above formula, as:
\begin{equation}
\begin{split}
        G_D\left(\phi\right)&=\frac{D}{2}\exp{\frac{2\phi}{2-D}}\Biggl\{\frac{1}{D-2}\left[g_1\left(D-2\right)\sqrt{\frac{2D-6}{D-2}}-2g_0\right] \\
        &\quad\quad\quad-\frac{2}{D-2}\left[g_1\sqrt{\frac{2D-6}{D-2}}-\left(D-3\right)g_2\right]\phi-g_{2}\frac{2\left(D-3\right)}{\left(D-2\right)^2}\phi^2\Biggr\}\ .
\end{split}
\end{equation}
In $D=4$, the formulas for $F_D\left(\phi\right)$ and $G_D\left(\phi\right)$ reduce to:
\begin{equation}
    \begin{split}
        F_4\left(\phi\right)&=e^{-\phi}\left(g_{0}+g_{1}\phi+\frac{g_{2}}{2}\phi^2\right)\\
        G_4\left(\phi\right)&=2e^{-\phi}\left[g_1-g_0+\left(g_2-g_1\right)\phi-\frac{g_{2}}{2}\phi^2\right]\ .
    \end{split}
\end{equation}
Focusing, once more, on a generic number $D$ of dimensions, we assume the scalar field to take the space-time constant value $\phi_0$, as introduced in \eqref{minimqua}. By doing so, the energy-momentum tensor associated to $\phi$ takes the value:
\begin{equation}\label{eq:emtquad}
    \kappa^2_D T_{\mu\nu}=-\frac{1}{2}V\left(\phi_0\right)g_{\mu\nu}\ . 
\end{equation}
Hence, it produced an effective cosmological constant term, with:
\begin{equation}
    \Lambda_{\text{eff}}=-\frac{1}{2}\left(g_0+\frac{g^2_1}{2g_2}\right)\ .
\end{equation}
From the perspective of the field equations for the geometry, we are allowed to consider a metric with constant scalar curvature:
\begin{equation}
    R_0=\frac{2D}{D-2}\Lambda_{\text{eff}}=-\frac{D}{D-2}\left(g_0+\frac{g^2_1}{2g_2}\right)\ .
\end{equation}
The pair $\left(\phi_0,R_0\right)$ will subsequently serve as an initial condition for our geometric flow equations. As in the previous examples, it can be shown that the constancy of $R$ and $\phi$ is preserved along the action-induced flow equations. Writing them down explicitly and neglecting the diffeomorphism terms, we get
\begin{equation}\label{acteqGGGG}
\begin{split}
        \frac{\diff R}{\diff s}&=\frac{2}{D}R\left[R+G_D\left(\phi\right)\right]\ ,\quad
        \frac{\diff\phi}{\diff s}=-\left[R+G_D\left(\phi\right)\right]\ ,
\end{split}
\end{equation}
where $G_D\left(\phi\right)$ was defined in \eqref{Gfunctionquad}. In the following discussion, we will focus on a specific choice for the number of space-time dimensions.
\subsubsection{Four-dimensional example}
\begin{figure}[H]
    \centering
    \includegraphics[width=0.87\linewidth]{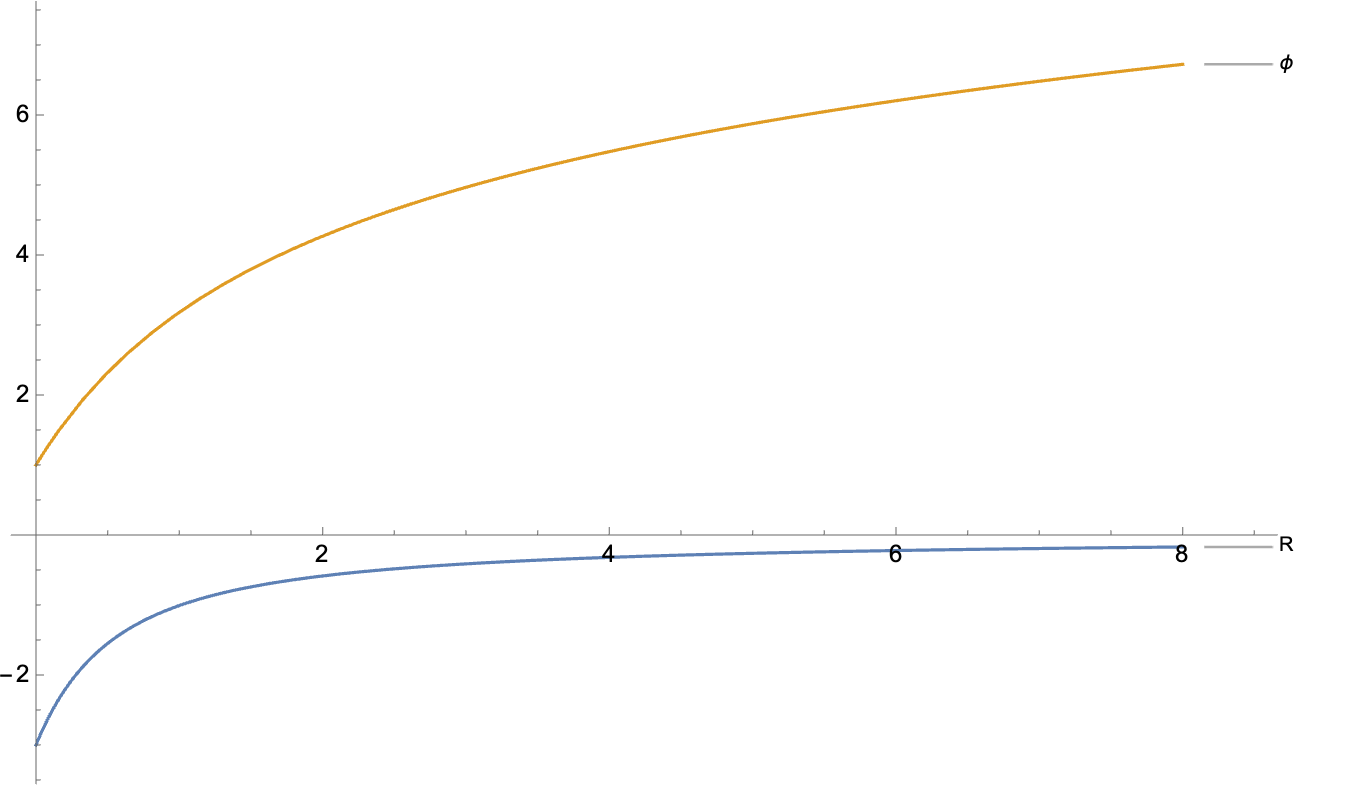}
    \caption{Plot of $\phi$ and $R$ with respect to the flow parameter $s$, when starting from the initial on-shell configuration $\left(\phi_0,R_0\right)$ and having taken $g_0=g_1=g_2=1$.}
    \label{fig:double}
\end{figure}

\begin{figure}[H]
    \centering
    \includegraphics[width=0.87\linewidth]{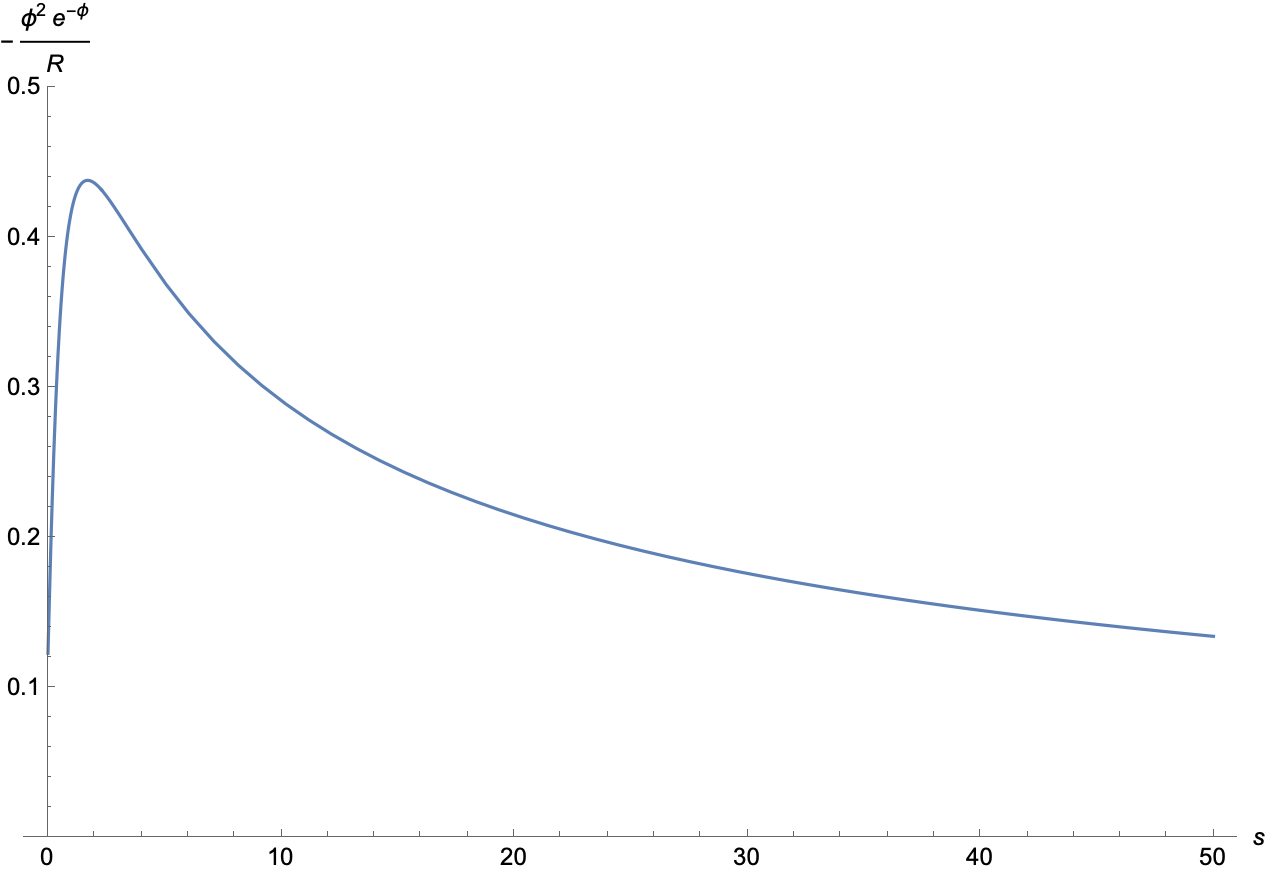}
    \caption{Plot of the ratio between $\phi^2e^{-\phi}$ and $-R$ with respect to $s$, when starting from the initial on-shell configuration $\left(\phi_0,R_0\right)$ and having taken $g_0=g_1=g_2=1$.}
    \label{fig:ratius}
\end{figure}

By imposing to work in four space-time dimensions, the flow equations reduce to
\begin{equation}\label{acteq4DG}
\begin{split}
        \frac{\diff R}{\diff s}&=\frac{R}{2}\left[R+G_4\left(\phi\right)\right]\ ,\quad
        \frac{\diff\phi}{\diff s}=-\left[R+G_4\left(\phi\right)\right]\ ,
\end{split}
\end{equation}
where the function $G_4\left(\phi\right)$ is equal to:
\begin{equation}
    G_4\left(\phi\right)=2e^{-\phi}\left[g_1-g_0+\left(g_2-g_1\right)\phi-\frac{g_{2}}{2}\phi^2\right]\ .
\end{equation}
As far as the initial condition is concerned, we have:
\begin{equation}
    \phi_0=\frac{g_1}{g_2}\ ,\quad R_0=-2\left(g_0+\frac{g^2_1}{2g_2}\right)\ .
\end{equation}
In order to solve the flow equations explicitly, we must make some assumptions on the potential parameters $g_0$, $g_1$ and $g_2$. As we want to focus on the simplest possible example, without considering a trivial one, we choose:
\begin{equation}
    g_0=g_1=g_2=1\ .
\end{equation}
Therefore, the initial value of the scalar is positive, while that of the curvature, which is associated with an Anti-de Sitter space-time geometry, is negative. In particular, we have:
\begin{equation}
        \phi_0=1\ ,\quad R_0=-3\ .
\end{equation}
The function $G_4\left(\phi\right)$, instead, reduces to:
\begin{equation}
    G_4\left(\phi\right)=-\phi^2e^{-\phi}\ .
\end{equation}
The action-induced flow equations, under the above assumptions, can be solved by employing numerical methods. 
The subsequent evolution is represented in figure \ref{fig:double}. 
Even if writing down an analytic solution might be too complicated, we can still greatly understand the general properties of the flow by observing at the right-hand side of the equations \eqref{acteq4DG}. In particular, as long as $\phi$ is positive and $R$ is negative, we have that both their derivatives in the flow parameter $s$ must themselves be strictly positive. Furthermore, since flat space-time corresponds to metric flow fixed point, starting from an Anti-de Sitter solution forces us never to reach de Sitter ones along the flow. 
Therefore, starting from an initial condition in which $\phi$ is positive and $R_0$ ensures not only that they both grow indefinitely with the flow parameter $s$, but that their signs are conserved along the action-induced geometric evolution. Numerically, we can observe that, as we proceed along the flow, the absolute value of the scalar curvature dominates over that of the other contribution, containing a negative exponential of the scalar field. The flow behaviour of their ratio is shown in figure \ref{fig:ratius}. 
Hence, for large flow times we have:
\begin{equation}\label{acteq4DGapprox}
\begin{split}
        \frac{\diff R}{\diff s}&\sim\frac{R^2}{2}\ ,\quad
        \frac{\diff\phi}{\diff s}\sim-R\ .
\end{split}
\end{equation}
The first equation can be easily solved by:
\begin{equation}
    R\left(s\right)\sim R_1\left[1-\frac{R_1}{2}\left(s-s_1\right)\right]^{-1}\ .
\end{equation}
By plugging such a solution into the flow equation for the scalar and solving it, we get:
\begin{equation}
    \phi\left(s\right)\sim \phi_1+2\log{\left[1-\frac{R_1}{2}\left(s-s_1\right)\right]}\ .
\end{equation}
In the above, $\left(\phi_1,R_1\right)$ are computed at a large enough value $s_1$ of the flow parameter. We can once more compute the distance behaviour for large flow times as was done in \eqref{distonf2}, by removing the volume term after a proper regularisation and obtaining
\begin{equation}\label{distonf25}
    \Delta\left(s,s_0\right)\sim\log{\frac{\mathcal{K}\left(s_0\right)}{\mathcal{K}\left(s\right)}}\ ,
\end{equation}
in which the flow-dependent quantity $\mathcal{K}\left(s\right)$ is equal to:
\begin{equation}
\begin{split}
        \mathcal{K}\left(s\right)&=e^{-\phi\left(s\right)}\left[R\left(s\right) -F_4\left(\phi\right)\right]\\
        &=e^{-\phi\left(s\right)}\left[R\left(s\right) -e^{-\phi\left(s\right)}\left(1+\phi\left(s\right)+\frac{\phi^2\left(s\right)}{2}\right)\right]\ .
\end{split}
\end{equation}
By plugging the approximate solutions into the above expression, we get that:
\begin{equation}
    \lim_{s\to\infty} \mathcal{K}\left(s\right)=0\ .
\end{equation}
Therefore, the corresponding behaviour of the distance is given by:
\begin{equation}
    \lim_{s\to\infty}\Delta\left(s,s_0\right)=\infty\ .
\end{equation}
Thus, the swampland distance conjecture would suggest it to be accompanied by an infinite tower of asymptotically massless supplementary states.

\subsubsection{Three-dimensional example}
\begin{figure}[H]
    \centering
    \includegraphics[width=0.87\linewidth]{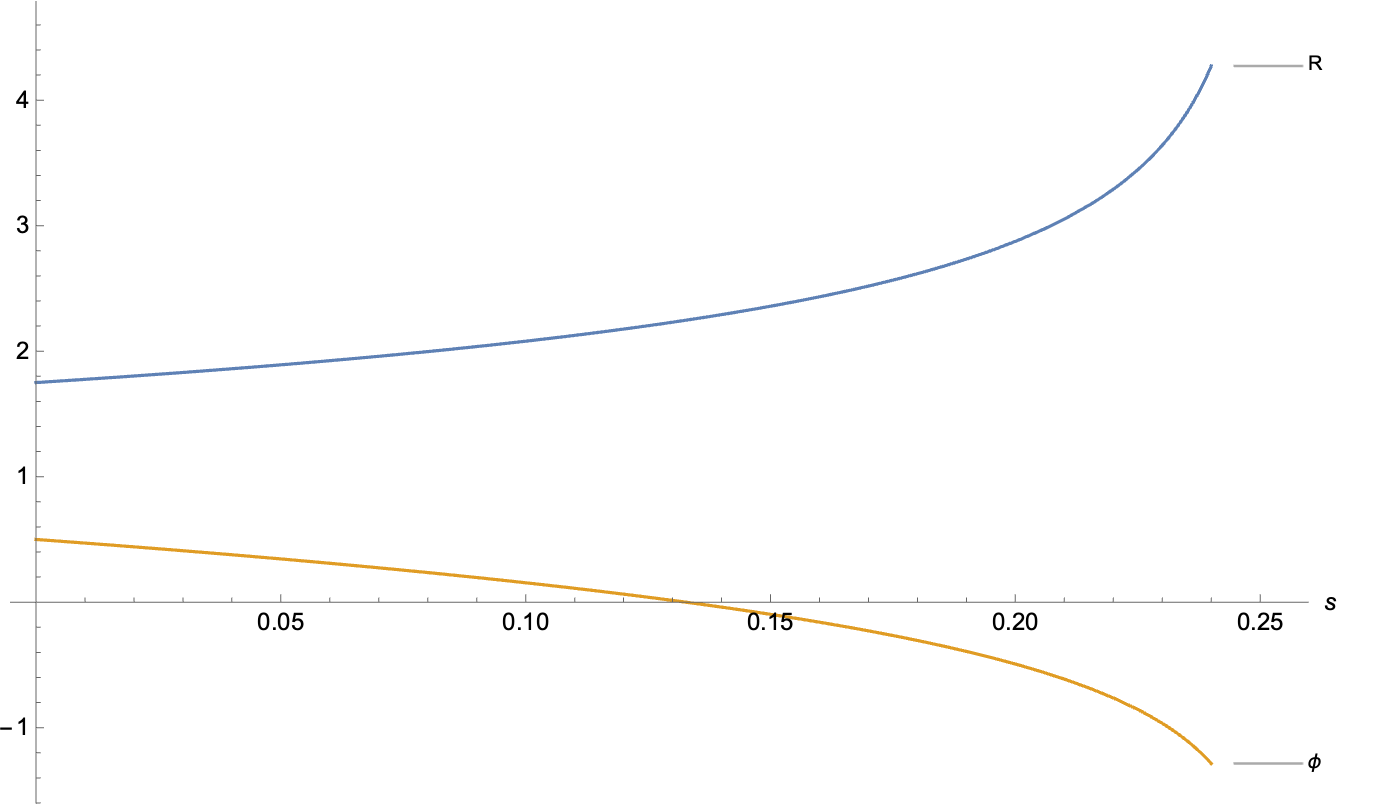}
    \caption{Plot of $\phi$ and $R$ with respect to the flow parameter $s$, when starting from the initial on-shell configuration $\left(\phi_0,R_0\right)$.}
    \label{fig:double3}
\end{figure}

By imposing to work in three space-time dimensions, the flow equations reduce to
\begin{equation}\label{acteq3DG}
\begin{split}
        \frac{\diff R}{\diff s}&=\frac{R}{2}\left[R+G_3\left(\phi\right)\right]\ ,\quad
        \frac{\diff\phi}{\diff s}=-\left[R+G_3\left(\phi\right)\right]\ ,
\end{split}
\end{equation}
where the function $G_3\left(\phi\right)$ is equal to:
\begin{equation}
\begin{split}
        G_3\left(\phi\right)&=-3g_0e^{-2\phi}\ .
\end{split}
\end{equation}
As far as the initial condition is concerned, we have:
\begin{equation}
    \phi_0=\frac{g_1}{g_2}\ ,\quad R_0=-2\left(g_0+\frac{g^2_1}{2g_2}\right)\ .
\end{equation}
In order to solve the flow equations explicitly, we must make some assumptions on the potential parameters $g_0$, $g_1$ and $g_2$. As we want to focus on the simplest possible example, without considering a trivial one, we choose:
\begin{equation}
    g_2=-g_0=1\ ,\quad g_1=\frac{1}{2}\ .
\end{equation}
Therefore, the initial value of the scalar is positive, while that of the curvature, which is associated with a de Sitter space-time geometry, is negative. In particular, we have:
\begin{equation}
        \phi_0=\frac{1}{2}\ ,\quad R_0=\frac{7}{4}\ .
\end{equation}
The function $G_3\left(\phi\right)$, instead, reduces to:
\begin{equation}
    G_3\left(\phi\right)=3e^{-2\phi}\ .
\end{equation}
The action-induced flow equations, under the above assumptions, can be solved by employing numerical methods. The first section of the subsequent evolution is represented in figure \ref{fig:double3}. From the numerical analysis, it can be shown that the flow encounters a singularity at a finite value $s_{\text{s}}$ of the flow parameter, in which the curvature blows to plus infinity, while the scalar blows to minus infinity.
As far as the distance along the flow is concerned, we again have
\begin{equation}\label{distonf255}
    \Delta\left(s,s_0\right)\sim\log{\frac{\mathcal{K}\left(s_0\right)}{\mathcal{K}\left(s\right)}}\ ,
\end{equation}
in which the flow-dependent quantity $\mathcal{K}\left(s\right)$ is equal to:
\begin{equation}
\begin{split}
        \mathcal{K}\left(s\right)&=e^{-\phi\left(s\right)}\left[R\left(s\right) -F_3\left(\phi\right)\right]=e^{-\phi\left(s\right)}\left[R\left(s\right) +e^{-2\phi}\right]\ .
\end{split}
\end{equation}
Therefore, it can be easily observed that $\mathcal{K}\left(s\right)$ blows up at the point $s\to s_{\text{s}}$ at which $R\to\infty$ and $\phi\to-\infty$. Hence, such singularity sits at infinite distance
\begin{equation}
    \lim_{s\to s_{\text{s}}}\Delta\left(s,s_0\right)=\infty
\end{equation}
and is hence expected to be accompanied by an infinite tower of asymptotically massless states. As both the absolute value of $R$ and that of $\phi$ grow, the absolute value of the derivative of $R$ with respect to the flow parameter, which contains an extra factor of $R$, grows quicker than that of the absolute value of the derivative of $\phi$. Hence, moving towards the infinite distance singularity, the flow equations can be approximated as
\begin{equation}\label{acteq3DGapprox}
\begin{split}
        \frac{\diff R}{\diff s}&\sim\frac{R^2}{2}\ ,\quad
        \frac{\diff\phi}{\diff s}\sim-R\ .
\end{split}
\end{equation}
and, as was done in the previous example, easily solved by
\begin{equation}
    R\left(t\right)\sim R_1\left(1-\frac{R_1}{2}t\right)^{-1}\ ,\quad \phi\left(t\right)\sim \phi_1+2\log{\left(1-\frac{R_1}{2}t\right)}\ ,
\end{equation}
where the values $\left(\phi_1,R_1\right)$ have been computed numerically at $s_1=0.252006$, with
\begin{equation}
    \phi_1\sim-6.69502\ ,\quad R_1\sim 63.8877\ ,
\end{equation}
and the new flow parameter $t$ has been defined as $ t\equiv s-s_1$, with domain $t\in\left(0,t_{\text{s}}\right)$. For what concerns the distance close to the singularity, we have the approximation
\begin{equation}\label{distimport}
    \Delta\left(s_1+t,s_0\right)=\Delta_0+3\log{\left(1-\frac{R_1}{2}t\right)}
\end{equation}
in which $\Delta_0$ is nothing more than a finite onset, in which all the constants have been absorbed for the sake of simplicity. 
\subsubsection{Keeping the flow on-shell}
In order to address the issue of preserving the on-shell conditions for the metric along the flow in a general setting, we would have to compute the tensors $B_{\mu\nu}$ and $E_{\mu\nu}$ whose properties were discussed in \eqref{act-inon}. In this case, nonetheless, the flow has already been solved numerically. Even more interestingly, it has been provided with an analytic approximation in the large $s$ regime. We can hence directly study the flow behaviour of the energy-momentum tensor $\Bar{T}_{\mu\nu}$ from those of the scalar field and the geometry, impose Einstein field equations not to be broken by the flow and read off the form of the additional energy-momentum contribution $\Hat{T}_{\mu\nu}$. As far as $\Bar{T}_{\mu\nu}$ is concerned, we obtain:
\begin{equation}
    \kappa_4^2\Bar{T}_{\mu\nu}\left(s\right)=-\frac{1}{2}V\left[\phi\left(s\right)\right]g_{\mu\nu}\left(s\right)\ .
\end{equation}
Focusing on its trace, which will be enough for the present analysis, we are left with:
\begin{equation}
    \kappa_4^2\Bar{T}\left(s\right)=-2V\left[\phi\left(s\right)\right]\ .
\end{equation}
On the other hand, we have that the scalar curvature $R$ flows as previously described. It is obvious that, for the geometry to be on-shell along the flow, the trace of the additional energy-momentum contribution must satisfy
\begin{equation}
    R\left(s\right)=-\kappa^2_4\left[\Bar{T}\left(s\right)+\Hat{T}\left(s\right)\right]\ .
\end{equation}
for every value of the flow parameter $s$. Therefore, moving to the flow parameter $t$ for which the near-singularity approximations fr $R$ and $\phi$ take simple forms, we have:
\begin{equation}
\begin{split}
        \kappa^2_4\Hat{T}\left(t\right)&=2V\left[\phi\left(t\right)\right]-R\left(t\right)=2+\phi^2\left(t\right)-\phi\left(t\right)-R\left(t\right)\\
        &\sim 4\log^2{\left(1-\frac{R_1}{2}t\right)}-R_1\left(1-\frac{R_1}{2}t\right)^{-1}\sim-R_1\left(1-\frac{R_1}{2}t\right)^{-1}\ .
\end{split}
\end{equation}
Therefore, for large enough values of $t$, we have that the following approximation holds:
\begin{equation}\label{flowaddem}
    \kappa^2_4\Hat{T}\left(t\right)\sim -R\left(t\right)\ .
\end{equation}
The additional energy-momentum tensor term, included for the flow to keep the geometry on-shell, displays a blow-up for large enough values of the flow parameter, close to the singularity. Let's now assume that the theory we are dealing with is defined below a certain energy cut-off $\Lambda_{\text{EFT}}$, above which its effective description breaks down. If the supplementary energy-momentum contribution is really to be interpreted as coming from an infinite tower of new states, getting exponentially lighter along the flow, then those must lie above $\Lambda_{\text{EFT}}$ at the beginning of the flow. This is consistent with $\Hat{T}=0$. Moreover, the $s$ dependence in \eqref{flowaddem} should be accounted for by the progressive appearance of states in the effective descriptions, as their masses decrease and cross the energy cut-off. Labelling such fields as $\phi_n$, the swampland distance conjecture suggests that their masses should be controlled by
\begin{equation}\label{expdrop}
    M_n\left(s\right)\sim \Bar{M}_ne^{-\beta\cdot\Delta\left(s\right)}\ ,
\end{equation}
where $\Delta\left(s\right)\equiv\Delta\left(s,0\right)$ is the usual distance and $\beta$ is an order one positive constant. Nonetheless, it must not be neglected that the behaviour described above should be characteristic of superstring low energy effective field theories, as widely discussed in \ref{sec:distanceconj}. There are, indeed, strong arguments \cite{Basile:2022sda} hinting at the fact that the typical exponential drop \eqref{expdrop} should solely be encountered in such scenarios, while towers emerging in apparently consistent theories not coming from superstrings should display power law decays in the distance. There is hence no reason to believe that a generic solution, as the one constructed in the current section, should be characterised by a tower of asymptotically massless states precisely fulfilling \eqref{expdrop}. It would, instead, be the case if and only if our original construction, with a quadratic potential \eqref{2pot}, belonged to the string theory landscape. At the same time, imposing the geometry to always be on-shell already provided us with an explicit formula for the flow behaviour of the additional energy-momentum tensor term, precisely associated to the tower of states. This, subsequently, poses a strong constraint on the flow behaviour of their masses. An incompatibility between such a constraint and an exponential drop of the form outlined in \eqref{expdrop} could, hence, be interpreted as a signal of our original $3$-dimensional theory not belonging to the landscape in the first place. We will therefore consider the results produced by an ansatz \eqref{expdrop} for the mass drop with the distance, assessing whether it correctly delivers the energy-momentum contribution described by \eqref{flowaddem}. In order to do so, we will assume the fields in the tower to be space-time constant, equal to each other, free and not to change along the flow, so that the only relevant flowing quantities will be their masses.
\paragraph{Exponential decay ansatz}
First of all, as previously anticipated, we focus on an exponential ansatz as the one presented in \eqref{expdrop}. For large flow times, the approximate formula \eqref{distimport} can be employed and such a scaling translates to:
\begin{equation}
    M_n\left(t\right)\sim \Bar{M}_n\left(1-\frac{R_1}{2}t\right)^{3\beta}\ .
\end{equation}
Hence, the state with initial mass $M_n\left(0\right)=\Bar{M}_n$, in which all constants have been absorbed, enters the low energy theory, thus contributing to $\Hat{T}$, when we have:
\begin{equation}
    M_n\left(s\right)\leq \Lambda_{\text{EFT}}\ .
\end{equation}
This is approximately achieved at the flow time $t_n$, with:
\begin{equation}
   t_n=\frac{2}{R_1}\left[1-\left(\frac{\Lambda_{\text{EFT}}}{\Bar{M}_n}\right)^{-3\beta}\right]\ .
\end{equation}
Assuming, for the sake of the argument, the masses to be distributed according to a relation

\begin{equation}
    \Bar{M}_n\sim\Lambda_{\text{EFT}}\frac{n}{\gamma} 
\end{equation}
inspired by Kaluza-Klein compactification \ref{circlecompactification} and with $\gamma\in\left(0,1\right]$, in which the energy scale was singled out and $\gamma$ was introduced to be a constant, we obtain:
\begin{equation}
   t_n=\frac{2}{R_1}\left[1-\left(\frac{n}{\gamma}\right)^{3\beta}\right]\ .
\end{equation}
We can therefore infer that the number of supplementary states which are expected to have entered the theory after a large flow time $t$, thus contributing to $\Hat{T}$, is given by:
\begin{equation}
     N\left(t\right)=\gamma\left(1-\frac{R_1}{2}t\right)^{-3\beta}\ .
\end{equation}
For the masses of such supplementary states to represent the main contribution to the extra energy-momentum tensor, we must have:
\begin{equation}
\begin{split}
        \kappa^2_4\Hat{T}\left(t\right)&\sim-\sum_{n=1}^{N\left(t\right)}M^2_n\left(t\right)\cdot\phi^2_n\left(t\right)=-\sum_{n=1}^{N\left(t\right)}\Bar{M}^2_n\left(1-\frac{R_1}{2}t\right)^{6\beta}\cdot\phi^2_n\left(t\right)\\
        &=-\left(\frac{\Lambda_{\text{EFT}}}{\gamma}\right)^2\left(1-\frac{R_1}{2}t\right)^{6\beta}\sum_{n=1}^{N\left(t\right)}n^2\cdot\phi^2_n\left(t\right)\ .
\end{split}
\end{equation}
The simple possible ansatz, now, is to take all fields to converge to the same long term value. This can be naturally achieved by requiring
\begin{equation}
    \phi_n\left(s\right)\sim \Hat{\phi}
\end{equation}
for large values of the flow parameter $s$. With such an assumption, the above expression simply turns to: 
\begin{equation}
\begin{split}
        \kappa^2_4\Hat{T}\left(t\right)&\sim-\left(\frac{\Lambda_{\text{EFT}}}{\gamma}\Hat{\phi}\right)^2\left(1-\frac{R_1}{2}t\right)^{6\beta}\sum_{n=1}^{N\left(t\right)}n^2\\
        &=-\frac{1}{6}\left(\frac{\Lambda_{\text{EFT}}}{\gamma}\Hat{\phi}\right)^2\left(1-\frac{R_1}{2}t\right)^{6\beta}\cdot\Bigl[2N^3\left(t\right)+3N^2\left(t\right)+N\left(t\right)\Bigr]\ .
\end{split}
\end{equation}
Neglecting sub-leading terms, we have:
\begin{equation}
\begin{split}
        \kappa^2_4\Hat{T}\left(t\right)&\sim-\frac{1}{3}\left(\frac{\Lambda_{\text{EFT}}}{\gamma}\Hat{\phi}\right)^2\left(1-\frac{R_1}{2}t\right)^{6\beta}N^3\left(t\right)\\
        &=-\frac{\gamma}{3}\left(\Lambda_{\text{EFT}}\cdot\Hat{\phi}\right)^2\left(1-\frac{R_1}{2}t\right)^{-3\beta}\ .
\end{split}
\end{equation}
Such a behaviour can be matched to \eqref{flowaddem}, by imposing:
\begin{equation}
\begin{split}
       \left(1-\frac{R_1}{2}t\right)^{-1}\propto\left(1-\frac{R_1}{2}t\right)^{-3\beta}\ .
\end{split}
\end{equation}
Namely, a tower of states displaying a mass exponential drop as the one suggested in \eqref{expdrop}, consistent with the swampland distance conjecture, would require the constant controlling the exponential to be equal to:
\begin{equation}
    \beta=\frac{1}{3}\ .
\end{equation}
Therefore, the theory for a dynamical geometry and a scalar field outlined in \ref{parabolicpot}, when selecting the initial point $\left(\phi_0,R_0\right)$ and imposing such values to evolve according to the action-induced geometric flow equations, approaches a singularity. Such point, albeit corresponding to a finite value of the flow parameter, actually sits an infinite distance in the generalised moduli space. By, moreover, imposing the equations of motion for the geometry to be kept on-shell along the flow, an extra energy-momentum tensor must be introduced. In the previous discussion we have shown that an infinite tower of states, progressively entering the theory and displaying an exponential mass drop with the moduli space distance, can precisely fulfil that role. At least, as long as the constant controlling the exponential is set to be equal to one third. We have hence found a behaviour consistent with the exponential drop predicted by the swampland distance conjecture.

%% file: Sections/Conclusions/conclusions.tex
After a brief introduction, chapter \ref{stringchap} straightforwardly presented the reader with a discussion of superstring theory. Albeit being far from complete, such analysis allowed to outline the broad conceptual framework in which the phenomenological inquiries developed within the swampland program are rooted. In particular, starting from the classical world-sheet action for a supersymmetric relativistic string, the corresponding quantum theory was derived by employing the standard old covariant methods. Remarkably, the condition of unitarity not to be broken at a quantum level forced us to impose the number of space-time dimension to be equal to ten. By scrutinising the spectrum of excited string states and decomposing those pertaining to its massless level into irreducible representations of the Poincaré group, it was therefore possible the identify the corresponding space-time fields. The equations of motion for the resulting theory were subsequently obtained by requiring the absence of conformal anomalies. From a technical perspective, this translated into forcing the world-sheet non-linear $\sigma$-model $\beta$-functions to vanish. The action for $10$-dimensional type IIA supergravity was thereafter stated and compactified, after having taken a specific and extremely simplified ansatz, on a circle, obtaining an effective $9$-dimensional theory for a dynamical metric, a radion field and an infinite tower of Kaluza-Klein states, associated to Fourier modes of the $10$-dimensional dilaton with respect to the compact dimension. Discussing such reduced description from the point of view of the world-sheet formulation, T-duality was introduced as a first counter-intuitive feature space-time displays in a superstring theory setting. Chapter \ref{swampchap} began by presenting the standard effective field theory approach to ultraviolet phenomenology and arguing why gravitational effects are expected to jeopardise it. Hence, the swampland program was introduced as a systematic attempt to formulate the supplementary constraints enforced on low energy effective field theories by the quantum gravitational dynamics captured by superstring theory. Particular attention was devoted to the swampland distance conjecture, as it could be connected to the previously presented compactification example. In chapter \ref{chap:geometricflows}, a swift overview of the Anti-de Sitter distance conjecture served as a suitable gateway towards extending the distance conjecture intuition to the space-time geometry. This objective was pursued by formally addressing the problem of defining a generalised moduli space for the metric tensors a given space-time manifold can be endowed with, that was itself provided with appropriate notion of geodesic distance. Geometric flows were thus established as natural mathematical tools with which specific moduli space paths could be singled out. The Ricci flow conjecture was therefore stated, shown to properly reproduce the large distance behaviour expected for Anti de-Sitter space-time and further generalised by means of Perelman's combined flow, which was obtained as the volume-preserving gradient flow of an entropy functional. The notion of distance induced by Perelman's entropy $\mathcal{F}$-functional on the moduli space of space-time geometries, which does not generally coincide with the one associated to geodesic paths, was moreover formulated and discussed in the context of the swampland program. While the above-mentioned chapters contained some original results, together with generalisations and formalisations of previously understood ones, it was not until part \ref{novel}, comprised of chapters \ref{chap:bubblesflow} and \ref{chap:onshell}, that the actual, novel outcomes of this thesis were properly derived and assessed. In the former, the behaviour of a scalar bubble solution embedded in various space-time backgrounds under Perelman's combined flow was analysed in detail. Afterwards, the more complicated example of a cosmological constant bubble, for which new and more apt geometric flow equations were derived from an altered entropy functional, was studied and revealed to evolve towards infinitely distant configurations. Chapter \ref{chap:onshell} opened, at last, with the derivation of a broad class of geometric flow equations from a generalised version of Perelman's $   \mathcal{F}$-entropy functional, in which both a Laplacian and a potential term for the scalar field were included. The latter, in particular, was expressed as a polynomial. Considering a space-time Lorentzian theory for a general relativistic space-time metric and a scalar field and rephrasing it in its string-frame version, an entropy functional of such kind was hence obtained. Therefore, a way of associating a set of geometric flow equations to given space-time theory without postulating any unphysical functional was outlined. This framework was thereupon applied to a circle-compactified model, recovering Perelman's combined flow in the case in which the full theory is taken to be five-dimensional, to the example of a sextic potential and to that of a quartic one. In such cases, solutions characterised by constant space-time curvature and scalar field were studied. Then, the issue of preserving the metric field equations along a geometric flow trajectory was faced and solved. In fact, drawing inspiration from the infinite towers expected to enter the low energy spectrum due to the swampland distance conjecture, the appearance of an additional energy-momentum contribution along the corresponding generalised moduli space path was postulated. The simple example of a scalar field in a parabolic potential was thus considered, construing space-time constant solutions to its associated equations of motion and making them evolve under the correct action-induced flow. Finally, it was shown how the the extra energy-momentum term required for Einstein's equations not to be violated by the flow, in a $3$-dimensional case, was the one associated to an infinite tower of space-time constant states, whose masses displayed the correct string-theoretic exponential drop. 

%% file: Sections/Conclusions/outlook.tex
As long as probing the microscopic behaviour of space-time will remain beyond our technological reach, any proposed theory of quantum gravity will have to be judged through its indirect phenomenological implications. Indeed, whether or not a framework provides us with clear-cut predictions regarding high-energy observables, such as the cross-sections of putative ultraviolet excitations, it might still be regarded as ambiguous or vain if no conceivable experiment can question its validity. This pragmatic perspective lies at the core of the scientific enterprise and represents its main methodological peculiarity. It is therefore not surprising that the discovery of a huge landscape of equally consistent superstring vacua was greeted with considerable concern, as it posed a serious threat to the theory's predictive power. If superstrings could give rise to all thinkable quantum field theories coupled to a dynamical space-time geometry, without discerning among them, there would have been no low energy experiment able to evaluate its usefulness. Fortunately, this appears not to be the case. As was broadly discussed in the previous chapters, the central tenet of the swampland program is precisely the fact that the features of low energy, four-dimensional effective theories coming from superstring theory are strongly constrained. While a general and top-down analysis of their shared properties is yet to be achieved, some of its aspects seem to be captured by various swampland conjectures. In this thesis, it was suggested that the action-induced flow equations should be employed when examining if a model belongs to the superstring theory landscape. Such a procedure allows to study a class of moduli space paths without the necessity of introducing any unphysical functional. In some sense, it tries to embody the principle, typically attributed to William of Ockham \cite{Ariew1976-ARIORA-2}, according to which \textit{entities must not be multiplied beyond necessity}, since it only makes use of those mathematical structures already offered by the low energy effective theory. Along the same line of thought, it was suggested that, at least as far as the space-time metric is considered, the equations of motion should be preserved by the evolution associated to the action-induced flow equations. It was hence shown how this can be realised by allowing for the gradual appearance of a new energy-momentum contribution, which could be traced back to an infinite tower of states displaying an exponential mass drop in a specific example. The most direct and natural prosecution of this thesis would be, first of all, to assess whether such interpretation can be generalised to a large class of low energy effective theories. It would moreover be important to search for reasons not to consider any alternative realisation of the extra energy-momentum tensor, together with a more solid \textit{first principles} argument supporting the usage of action-induced flow equations. No particular obstacle should hinder the inclusion of supplementary fields, like the vector bosons appearing in gauge theories. For what concerns fermionic matter, geometric flow equations might be instead derived after having expressed the metric in the \textit{tetrad} formalism. The consequences of preserving the on-shell conditions on the matter content of the theory are furthermore left to be investigated, as well as any physical implication, within the context of the on-shell flow, of the flow singularities analysed in \cite{Velazquez:2022eco}.